\input harvmac
\input epsf


\ifx\epsfbox\UnDeFiNeD\message{(NO epsf.tex, FIGURES WILL BE IGNORED)}
\def\figin#1{\vskip2in}
\else\message{(FIGURES WILL BE INCLUDED)}\def\figin#1{#1}\fi
\def\ifig#1#2#3{\xdef#1{Fig.~\the\figno}
\goodbreak\midinsert\figin{\centerline{#3}}%
\smallskip\centerline{\vbox{\baselineskip12pt
\advance\hsize by -1truein\noindent\footnotefont{\bf Fig.~\the\figno:} #2}}
\bigskip\endinsert\global\advance\figno by1}


\def\p{\partial}

\def\half{{1\over 2}}

\def\sh{\hat{\sigma}}

\def\rl{\rho_\Lambda}


\Title{}{\vbox{\centerline{Dark Energy}}}


\centerline{Miao Li$^{1,2}$, Xiao-Dong Li$^{2,3}$, Shuang Wang$^{2,3}$ and Yi Wang$^4$ }

\medskip
\centerline{\it $^1$ Institute of Theoretical Physics, Chinese
Academy of Sciences}
\centerline{\it  Beijing 100190, China}
\medskip
\centerline{\it $^2$
Kavli Institute for Theoretical Physics, Key Laboratory
of Frontiers in Theoretical Physics}
\centerline{\it Beijing 100190, China}
\medskip
\centerline{\it $^3$ Department of Modern
Physics, University of Science and Technology of China}
\centerline{\it Hefei 230026, China}
\medskip
\centerline{\it $^4$ Physics Department, McGill University}
\centerline{\it Montreal, H3A2T8, Canada}

\bigskip

\centerline{\tt mli@itp.ac.cn}
\centerline{\tt renzhe@mail.ustc.edu.cn}
\centerline{\tt swang@mail.ustc.edu.cn }
\centerline{\tt wangyi@hep.physics.mcgill.ca}
\bigskip


We review the problem of dark energy, including a survey of theoretical models and some aspects of numerical studies.

\Date{March 2011}

%



\lref\riess{
A. G. Riess, {\it et al.}, AJ. {\bf 116} (1998) 1009.
}
\lref\perl{
S. Perlmutter, {\it et al.}, ApJ {\bf 517} (1999) 565.
}
\lref\weinrev{
S. Weinberg, Rev. Mod. Phys. {\bf 61} (1989) 1.
}
\lref\DEReviewCarroll{
S. M. Carroll, W. H. Press, and E. L. Turner, Ann. Rev. Astron. Astrophys. {\bf 30} (1992) 499.
}
\lref\DEReviewPadmanabhanA{
T. Padmanabhan, Phys. Rept. {\bf 380} (2003) 235.
}
\lref\DEReviewPeebles{
P. J. E. Peebles and B. Ratra, Rev. Mod. Phys. {\bf 75} (2003) 559.
}
\lref\CopelandWR{
E. J. Copeland, M. Sami, and S. Tsujikawa, Int. J. Mod. Phys. D {\bf 15} (2006) 1753.
}
\lref\DEReviewUzan{
J. P. Uzan, Gen. Rel. Grav. {\bf 39} (2007) 307.
}
\lref\DEReviewLinder{
E. V. Linder, Rept. Prog. Phys. {\bf 71} (2008) 056901.
}
\lref\ReviewofTurner{
J. Frieman, M. Turner, and D. Huterer, Ann. Rev. Astron. Astrophys {\bf 46} (2008) 385.
}
\lref\DEReviewDurrer{
R. Durrer and R. Maartens, arXiv:0811.4132.
}
\lref\DEREviewTsujikawa{
S. Tsujikawa, arXiv:1004.1493.
}
\lref\DEbookYWang{
Y. Wang, {\it Dark Energy}, Wiley-VCH (2010).
}
\lref\DEReviews{
V. Sahni, A. Starobinsky, Int. J. Mod. Phys. D {\bf 9} (2000) 373;
S. M. Carroll, Living Rev. Rel. {\bf 4} (2001) 1;
V. Sahni, Lect. Notes. Phys. {\bf 653} (2004) 141;
J. A. S. Lima, BJP, {\bf 34} (2004) 1;
D. H. Weinberg, New. Astron. Rev. {\bf 49} (2005) 337;
N. Straumann, Mod. Phys. Lett. A {\bf 21} (2006) 1083;
M. S. Turner, D. Huterer, J. Phys. Soc. Jap. {\bf 76} (2007) 111015;
J. P. Uzan, arXiv:0912.5452;
A. Silvestri, M. Trodden, Rept. Prog. Phys. {\bf 72} (2009) 096901;
R. P. Caldwell, M. Kamionkowski, Ann. Rev. Nucl. Part. Sci. {\bf 59} (2009) 397;
P. Peter and J. P. Uzan, {\it Primordial Cosmology}, Oxford university Press, Oxford (2009);
D. Sapone, Int. J. Mod. Phys. A {\bf 25} (2010) 5253.
}
\lref\StraumannTV{
N. Straumann, arXiv:gr-qc/0208027.
}
\lref\EinsteinCE{
A. Einstein, Sitz. Preuss. Akad. Wiss. Phys-Math {\bf 142} (1917) 87.
}
\lref\EinsteinNoCC{
A. Einstein, Sitz. Preuss. Akad. Wiss. Phys-Math {\bf 235} (1931) 37.
}
\lref\PetrosianQH{
V. Petrosian, E. Salpeter, and P. Szekeres, ApJ. {\bf 147} (1967) 1222.
}
\lref\ZeldovichGD{
Y. B. Zeldovich, JETP Lett. {\bf 6} (1967) 316.
}
\lref\SahniZZA{
V. Sahni, A. Krasinski, and Y. B. Zeldovich, Sov. Phys. Usp. {\bf 11} (1968) 381.
}
\lref\anthrwein{
S. Weinberg, Phys. Rev. Lett. {\bf 59} (1987) 2607.
}
\lref\bp{
R. Bousso and J. Polchinski, JHEP {\bf 0006} (2000) 006.
}
\lref\suss{
L. Susskind, hep-th/0302219.
}
\lref\DETF{
A. Albrecht, {\it et al.}, astro-ph/0609591.
}




\lref\wb{
J. Wess and J. Bagger, {\it Supersymmetry and supergravity}, Princeton University Press, Princeton, New Jersey (1982).
}
\lref\ZuminoBG{
B. Zumino, Nucl. Phys. B {\bf 89} (1975) 535.
}
\lref\CremmerIV{
E. Cremmer, {\it et al.}, Phys. Lett. B {\bf 79} (1978) 231.
}
\lref\EllisSF{
J. R. Ellis, {\it et al.}, Phys. Lett. B {\bf 134} (1984) 429.
}


\lref\carter{
B. Carter, IAU Symp. {\bf 63} (1974) 291.
}
\lref\DickeNature{
R. H. Dicke, Nature {\bf 192} (1961) 440.
}
\lref\DickeGZ{
R. H. Dicke, Phys. Rev. {\bf 125} (1962) 2163.
}
\lref\GarrigaEE{
J. Garriga, A. Vilenkin, Prog. Theor. Phys. Suppl. {\bf 163} (2006) 245.
}
\lref\WeinbergFH{
S. Weinberg, In {\it Universe or multiverse}, ed. Bernard Carr, Cambridge University Press, Cambridge, UK(2007) pp. 29-42.
}
\lref\Vaucouleurs{
G. de Vaucouleurs, ApJ. {\bf 268} (1983) 468 ; Nature (London) {\bf 299} (1982) 303.
}
\lref\PeeblesCC{
P. J. E. Peebles, ApJ. {\bf 284} (1984) 439.
}
\lref\TurnerNF{
M. S. Turner, G. Steigman, and L. M. Krauss, Phys. Rev. Lett. {\bf 52} (1984) 2090.
}


\lref\DolgovGH{
A. D. Dolgov, In Cambridge 1982, Proceedings, The Very Early Universe, pp. 449-458.
}


\lref\vanderBijYM{
J. J. van der Bij, H. van Dam, and Y. J. Ng, Physica A {\bf 116} (1982) 307.
}
\lref\UnruhIN{
W. G. Unruh, Phys. Rev. D {\bf40} (1989) 1048.
}
\lref\EllisAs{
G. F. R. Ellis, {\it et al.}, arXiv:1008.1196.
}
\lref\latestMGreview{
T. Clifton, {\it et al.}, arXiv:1106.2476.
}

\lref\DeWittYK{
B. S. DeWitt, Phys. Rev. {\bf 160} (1967) 1113.
}
\lref\WheelerAA{
R. Penrose, {\it Battlelle Rencontres}, ed. C. DeWitt and J. A. Wheeler, Benjamin, New York, (1968).
}
\lref\hh{
J. B. Hartle and S. W. Hawking, Phys. Rev. D {\bf 28} (1983) 1960.
}
\lref\sh{
S. W. Hawking, Phys. Lett. B {\bf 134} (1984) 403.
}
\lref\tbanks{
T. Banks, Nucl. Phys. B {\bf 249} (1985) 332.
}
\lref\scoleman{
S. R. Coleman, Nucl. Phys. B {\bf 310} (1988) 643.
}



\lref\ewittenSC{
E. Witten, Mod. Phys. Lett. A {\bf 10} (1995) 2153.
}
\lref\bbs{
K. Becker, M. Becker, and A. Strominger, Phys. Rev. D {\bf 51} (1995) 6603.
}


\lref\thooft{
G. 't Hooft and S. Nobbenhuis, Class. Quant. Grav. {\bf 23} (2006) 3819.
}


\lref\ks{
D. E. Kaplan and R. Sundrum, JHEP {\bf 0607} (2006) 042.
}


\lref\ErdemYD{
R. Erdem, Phys. Lett. B {\bf 621} (2005) 11.
}


\lref\wett{
C. Wetterich, Phys. Rev. Lett. {\bf 102} (2009) 141303; C. Wetterich, Phys. Rev. D {\bf 81} (2010) 103507.
}



\lref\ClineHU{
J. M. Cline, hep-th/0612129.
}

\lref\BrownDD{
J. D. Brown and C. Teitelboim, Phys. Lett. B {\bf 195} (1987) 177.
}
\lref\BrownKG{
J. D. Brown and C. Teitelboim, Nucl. Phys. B {\bf 297} (1988) 787.
}


\lref\gkp{
S. B. Giddings, S. Kachru, and J. Polchinski, Phys. Rev. D {\bf 66} (2002) 106006.
}
\lref\cvafa{
C. Vafa, Nucl. Phys. B {\bf 469} (1996) 403.
}
\lref\randall{
L. Randall and R. Sundrum, Phys. Rev. Lett. {\bf 83} (1999) 3370;
Phys. Rev. Lett. {\bf 83} (1999) 4690.
}
\lref\kklt{
S. Kachru, {\it et al.}, Phys. Rev. D {\bf 68} (2003) 046005.
}
\lref\DouglasUM{
M. R. Douglas, JHEP {\bf 0305} (2003) 046.
}
\lref\SusskindKW{
L. Susskind, In {\it Universe or multiverse}, ed. Bernard Carr, Cambridge University Press, Cambridge, UK(2007) pp. 247.
}


\lref\TegmarkDB{
M. Tegmark, Sci. Am. {\bf 288} (2003) 30.
}
\lref\HawkingUR{
S. W. Hawking and T. Hertog, Phys. Rev. D {\bf 73} (2006) 123527.
}
\lref\SteinhardtEternal{
P. J. Steinhardt, {
\it The Very Early Universe}, Proceedings of the Nuffield Workshop,
Campridge, ed. G. Gibbons, S. W. Hawking, S. T. C. Siklos, Cambridge University Press, (1982).
}
\lref\VilenkinXQ{
A. Vilenkin, Phys. Rev. D {\bf 27} (1983) 2848.
}
\lref\PageNT{
D. N. Page, Phys. Lett. B {\bf 669} (2008) 197.
}
\lref\ArkaniHamedYM{
N. Arkani-Hamed, {\it et al.}, JHEP {\bf 0803} (2008) 075.
}
\lref\HuangZT{
Q. G. Huang, M. Li, and Y. Wang, JCAP {\bf 0709} (2007) 013.
}
\lref\WangCS{
Y. Wang, arXiv:0805.4520.
}
\lref\ArkaniHamedDZ{
N. Arkani-Hamed, {\it et al.}, hep-th/0601001.
}
\lref\DysonPF{
L. Dyson, M. Kleban, and L. Susskind, JHEP {\bf 0210} (2002) 011.
}
\lref\GoheerVF{
N. Goheer, M. Kleban, and L. Susskind, JHEP {\bf 0307} (2003) 056.
}
\lref\GarrigaAV{
J. Garriga, {\it et al.}, JCAP {\bf 0601} (2006) 017.
}
\lref\DeSimoneBQ{
A. De Simone, {\it et al.}, Phys. Rev. D {\bf 78} (2008) 063520.
}
\lref\BoussoEV{
R. Bousso, Phys. Rev. Lett. {\bf 97} (2006) 191302.
}
\lref\BoussoGE{
R. Bousso, B. Freivogel and I. S. Yang, Phys. Rev. D {\bf 74} (2006) 103516.
}
\lref\GibbonsXK{
G. W. Gibbons, S. W. Hawking, and J. M. Stewart, Nucl. Phys. B {\bf281} (1987) 736.
}
\lref\GibbonsPA{
G. W. Gibbons and N. Turok, Phys. Rev. D {\bf 77} (2008) 063516.
}
\lref\LiRP{
M. Li and Y. Wang, JCAP {\bf 0706} (2007) 012.
}
\lref\LiUC{
M. Li and Y. Wang, JCAP {\bf 0708} (2007) 007.
}
\lref\PageBT{
D. N. Page, arXiv:0707.4169.
}
\lref\PageMX{
D. N. Page, Phys. Rev. D {\bf 78} (2008) 023514.
}
\lref\HartleZV{
J. B. Hartle and M. Srednicki, Phys. Rev. D {\bf 75} (2007) 123523.
}
\lref\LiDH{
M. Li and Y. Wang, arXiv:0708.4077.
}
\lref\BoussoKQ{
R. Bousso, {\it et al.}, Phys. Rev. D {\bf76} (2007) 043513.
}
\lref\Boltzmann{
L. Boltzmann, Nature {\bf 51} (1985) 413.
}
\lref\BoussoYN{
R. Bousso, {\it et al.}, Phys. Rev. D {\bf 83} (2011) 023525.
}



\lref\KachruHF{
S. Kachru, M. B. Schulz, and E. Silverstein, Phys. Rev. D {\bf 62} (2000) 045021.
}

\lref\ClineAK{
J. M. Cline, {\it et al.}, JHEP {\bf 0306}, 048 (2003).
}

\lref\VinetBK{
J. Vinet and J. M. Cline, Phys. Rev. {\bf D70}, 083514 (2004).
}

\lref\NeupaneAs{I. P. Neupane, Phys. Rev. D {\bf 83} (2011) 086004;
I. P. Neupane, Int. J. Mod. Phys. D {\bf 19} (2010) 2281.}


\lref\CsakiDM{
C. Csaki, J. Erlich, and C. Grojean, Nucl. Phys. B {\bf 604} (2001) 312.
}



\lref\BuchdahlZZ{
H. A. Buchdahl, MNRAS {\bf 150} (1970) 1.
}
\lref\Palatini{
A. Palatini, Rend. Circ. Mat. Palermo, {\bf 43} (1919) 203.
}
\lref\PalatiniEinstein{
A. Einstein, Sitzung-bet. Pruess. Akad. ICiss. {\bf 414} (1925) 37.
}
\lref\FerrarisVF{
M. Ferraris and M. Francaviglia, Gen. Rel. Grav. {\bf 14} (1982) 243.
}
\lref\SotiriouRP{
T. P. Sotiriou and V. Faraoni, Rev. Mod. Phys. {\bf 82} (2010) 451.
}
\lref\WhittPD{
B. Whitt, Phys. Lett. B {\bf 145} (1984) 176.
}
\lref\BarrowXH{
J. D. Barrow and S. Cotsakis, Phys. Lett. B {\bf 214} (1988) 515.
}
\lref\MukhanovME{
V. F. Mukhanov, H. A. Feldman, and R. H. Brandenberger, Phys. Rept. {\bf 215} (1992) 203.
}
\lref\YGMatwo{
B. Huang, S. Li, and Y. G. Ma, Phys. Rev. D {\bf 81} (2010) 064003.
}
\lref\CarrollWY{
S. M. Carroll, {\it et al.}, Phys. Rev. D {\bf 70} (2004) 043528.
}
\lref\ChibaIR{
T. Chiba, Phys. Lett. B {\bf 575} (2003) 1.
}
\lref\CarrollJN{
S. M. Carroll, {\it et al.}, New J. Phys. {\bf 8} (2006) 323.
}

\lref\BambaAS{
K. Bamba, C. Q. Geng, and S. Tsujikawa, Phys. Lett. B {\bf 688} (2010) 101;
K. Bamba and C. Q. Geng, arXiv:1005.5234.
}


\lref\MilgromCA{
M. Milgrom, ApJ. {\bf 270} (1983) 365.
}
\lref\BekensteinTV{
J. Bekenstein and M. Milgrom, ApJ. {\bf 286} (1984) 7.
}
\lref\BrunetonGF{
J. P. Bruneton, Phys. Rev. D {\bf 75} (2007) 085013.
}
\lref\BekensteinPC{
J. D. Bekenstein, In: Proceedings of the second Canadian Conference on General Relativity and Relativistic Astrophysics,
ed.Dyer, C., (Singapore: World Scientific, 1987).
}
\lref\BekinsteinAC{
J. D. Bekenstein, in Second Canadian Conference on General Relativity and Relativistic Astrophysics,
A. Coley, C. Dyer and T. Tupper, eds. (World Scientific, Singapore 1988).
}
\lref\BekensteinDFM{
J. D. Bekenstein, in Proceedings of the Sixth Marcel Grossman Meeting on General Relativity, H. Sato and T. Nakamura,
eds. (World Scientific, Singapore 1992).
}
\lref\BekensteinNE{
J. D. Bekenstein, Phys. Rev. D {\bf 70} (2004) 083509.
}


\lref\DGPDGP{
G. R. Dvali, G. Gabadadze, and M. Porrati, Phys. Lett. B {\bf 485} (2000) 208.
}
\lref\KiritsisAS{
E. Kiritsis, {\it et al.}, JHEP {\bf 0302} (2003) 035;
E. Kiritsis, JCAP {\bf 0510} (2005) 014;
K. Umezu, {\it et al.}, Phys. Rev. D {\bf 73} (2006) 063527.
}
\lref\NicolisIN{
A. Nicolis, R. Rattazzi, and E. Trincherini, Phys. Rev. D {\bf 79} (2009) 064036.
}


\lref\BransSX{
C. Brans and R. H. Dicke, Phys. Rev. {\bf 124} (1961) 925.
}
\lref\AmendolaQQ{
L. Amendola, Phys. Rev. D {\bf 60} (1999) 043501.
}
\lref\YGMaone{
L. E. Qiang, {\it et al.}, Phys. Rev. D {\bf 71} (2005) 061501(R);
L. E. Qiang, {\it et al.}, Phys. Lett. B {\bf 681} (2009) 210.
}
\lref\ZwiebachUQ{
B. Zwiebach, Phys. Lett. B {\bf156} (1985) 315.
}
\lref\NojiriVV{
S. Nojiri, S. D. Odintsov, and M. Sasaki, Phys. Rev. D {\bf 71} (2005) 123509.
}
\lref\LovelockYV{
D. Lovelock, J. Math. Phys. {\bf 12} (1971) 498.
}
\lref\DehghaniXF{
M. H. Dehghani and S. Assyyaee, Phys. Lett. B {\bf 676} (2009) 16.
}
\lref\HoravaUW{
P. Horava, Phys. Rev. D {\bf 79} (2009) 084008.
}
\lref\SaridakisBV{
E. N. Saridakis, Eur. Phys. J. C {\bf 67} (2010) 229.
}
\lref\BengocheaGZ{
G. R. Bengochea and R. Ferraro, Phys. Rev. D {\bf 79} (2009) 124019.
}
\lref\LinderPY{
E. V. Linder, Phys. Rev. D {\bf 81} (2010) 127301.
}
\lref\FRT{
T. Harko, {\it et al.}, arXiv:1104.2669.
}
\lref\MannheimBFA{
P. D. Mannheim, Prog. Part. Nucl. Phys. {\bf 56} (2006) 340.
}
\lref\MannheimXW{
P. D. Mannheim and J. G. O'Brien, arXiv:1011.3495.
}
\lref\MannheimHK{
P. D. Mannheim, arXiv:1005.5108.
}
\lref\SundrumJQ{
R. Sundrum, Phys. Rev. D {\bf 69} (2004) 044014.
}


\lref\FirouzjahiMX{
H. Firouzjahi, S. Sarangi, and S. H. H. Tye, JHEP {\bf 0409} (2004) 060.
}
\lref\SarangiCS{
S. Sarangi and S. H. H. Tye, hep-th/0505104.
}
\lref\HartleGI{
J. B. Hartle, S. W. Hawking, and T. Hertog, Phys. Rev. Lett. {\bf 100} (2008) 201301.
}
\lref\HawkingVF{
S. W. Hawking, arXiv:0710.2029.
}


\lref\MotlsSeesaw{
http://motls.blogspot.com/2005/12/cosmological-constant-seesaw.html
}
\lref\CarrollSeesaw{
http://blogs.discovermagazine.com/cosmicvariance/2005/12/05/duff-on-susskind/
}
\lref\McGuiganHS{
M. McGuigan, hep-th/0602112.
}
\lref\LindeGJ{
A. D. Linde, hep-th/0211048.
}
\lref\SSXueone{
S. S. Xue, Phys. Rev. D {\bf 82} (2010) 064039.
}
\lref\SSXuetwo{
S. S. Xue, Int. J. Mod. Phys. A {\bf 24} (2009) 3865.
}

\lref\HenryTyeTG{
S. H. H. Tye, hep-th/0611148.
}
\lref\TyeJA{
S. H. H. Tye, arXiv:0708.4374.
}
\lref\CopelandQF{
E. J. Copeland, A. Padilla, and P. M. Saffin, JHEP {\bf 0801} (2008) 066.
}
\lref\SaffinVI{
P. M. Saffin, A. Padilla, and E. J. Copeland, JHEP {\bf 0809} (2008) 055.
}
\lref\TyeRB{
S. H. H. Tye and D. Wohns, arXiv:0910.1088.
}
\lref\Anderson {
P. W. Anderson, Phys. Rev. {\bf 109} (1958) 1492.
}
\lref\PodolskyVG{
D. Podolsky and K. Enqvist, JCAP {\bf 0902} (2009) 007.
}
\lref\PodolskyDU{
D. I. Podolsky, J. Majumder, and N. Jokela, JCAP {\bf 0805} (2008) 024.
}
\lref\KieferPB{
C. Kiefer, F. Queisser, and A. A. Starobinsky, arXiv:1010.5331.
}


\lref\tHooftGX{
G. 't Hooft, gr-qc/9310026.
}
\lref\SusskindVU{
L. Susskind, J. Math. Phys. {\bf 36} (1995) 6377.
}
\lref\IsraelWQ{
W. Israel, Phys. Rev. {\bf 164} (1967) 1776.
}
\lref\IsraelZA{
W. Israel, Commun. Math. Phys. {\bf 8} (1968) 245.
}
\lref\HawkingTU{
S. W. Hawking, Phys. Rev. Lett. {\bf 26} (1971) 1344.
}
\lref\BekensteinTM{
J. D. Bekenstein, Lett. Nuovo Cim. {\bf 4} (1972) 737.
}
\lref\BekensteinUR{
J. D. Bekenstein, Phys. Rev. D {\bf 7} (1973) 2333.
}
\lref\BekensteinAX{
J. D. Bekenstein, Phys. Rev. D {\bf 9} (1974) 3292.
}
\lref\HawkingRV{
S. W. Hawking, Nature {\bf 248} (1974) 30.
}
\lref\CohenZX{
A. G. Cohen, D. B. Kaplan, and A. E. Nelson, Phys. Rev. Lett. {\bf 82} (1999) 4971.
}
\lref\LiRB{
M. Li, Phys. Lett. B {\bf 603} (2004) 1.
}


\lref\HsuRI{
S. D. H. Hsu, Phys. Lett. B {\bf 594} (2004) 13.
}
\lref\ChenQY{
B. Chen, M. Li, and Y. Wang, Nucl. Phys. B {\bf774} (2007) 256.
}
\lref\MyungPN{
Y. S. Myung, Phys. Lett. B {\bf 652} (2007) 223.
}
\lref\LiZQ{
M. Li, C. S. Lin, and Y. Wang, JCAP {\bf 0805} (2008) 023.
}
\lref\WangJX{
B. Wang, Y. G. Gong, and E. Abdalla, Phys. Lett. B {\bf 624} (2005) 141.
}


\lref\LiPM{
M. Li, R. X. Miao, and Y. Pang, Phys. Lett. B {\bf 689} (2010) 55.
}
\lref\LiZY{
M. Li, R. X. Miao, and Y. Pang, Opt. Express {\bf 18} (2010) 9026.
}
\lref\HoganRZ{
C. J. Hogan, astro-ph/0703775.
}
\lref\HoganHC{
C. J. Hogan, arXiv:0706.1999.
}
\lref\LeeZQ{
J. W. Lee, J. Lee, and H. C. Kim, JCAP {\bf 0708} (2007) 005.
}
\lref\VerlindeHP{
E. P. Verlinde, arXiv:1001.0785.
}
\lref\LiCJ{
M. Li and Y. Wang, Phys. Lett. B {\bf 687} (2010) 243.
}
\lref\LiQH{
M. Li, {\it et al.}, Commun. Theor. Phys. {\bf 51} (2009) 181.
}

\lref\ElizaldeAS{
E. Elizalde, {\it et al.}, Phys. Rev. D 71 (2005) 103504.
}
\lref\EHDE{
Y. G. Gong, Phys. Rev. D {\bf 70}, (2004) 064029.
H. Wei, Commun. Theor. Phys. {\bf 52}, 743 (2009).
K. Karami and J. Fehri, Phys. Lett. B {\bf 684}, 61 (2010).
F. Adabi, {\it et al.}, arXiv:1105.1008.
}

\lref\CaiUS{
R. G. Cai, Phys. Lett. B {\bf 657} (2007) 228.
}
\lref\WeiTY{
H. Wei and R. G. Cai, Phys. Lett. B {\bf660} (2008) 113.
}

\lref\KarolyhazyZZ{
F. Karolyhazy, Nuovo Cim. A {\bf 42} (1966) 390.
}

\lref\MaziashviliDK{
M. Maziashvili, Phys. Lett. B {\bf652} (2007) 165.
}


\lref\RDEPaper{
C. J. Gao, {\it et al.}, Phys. Rev. D {\bf 79} (2009) 043511.
}
\lref\NojiriASb{
S. Nojiri and S. D. Odintsov, Gen. Rel. Grav. {\bf 38} (2006) 1285.
}
\lref\RDEGO{
L. N. Granda and A. Oliveros, Phys. Lett. B {\bf 671} (2009) 199
}
\lref\RDEXLX{
L. X. Xu and Y. T. Wang, JCAP {\bf 06} (2010) 002;
Y. T. Wang and L. X. Xu, Phys. Rev. D {\bf 81} (2010) 083523.
}
\lref\RDECFR{
L. P. Chimento, M. Forte and M. G. Richarte, arXiv:1106.0781.
}


\lref\EllisBR{
G. F. R. Ellis, Gen. Rel. Grav. {\bf 41} (2009) 581.
}
\lref\EllisZZ{
G. F. R. Ellis and W. Stoeger, Class. Quant. Grav. {\bf 4} (1987) 1697.
}


\lref\RasanenKI{
S. Rasanen, arXiv:1102.0408.
}
\lref\RaychaudhuriYV{
A. Raychaudhuri, Phys. Rev. {\bf 98} (1955) 1123.
}
\lref\BuchertER{
T. Buchert, Gen. Rel. Grav. {\bf 32} (2000) 105.
}
\lref\IshibashiSJ{
A. Ishibashi and R. M. Wald, Class. Quant. Grav. {\bf 23} (2006) 235.
}
\lref\RasanenUW{
S. Rasanen, JCAP {\bf 1003} (2010) 018.
}
\lref\RasanenBE{
S. Rasanen, JCAP {\bf 0902} (2009) 011.
}
\lref\BuchertFZ{
T. Buchert and J. Ehlers, Astron. Astrophys. {\bf 320} (1997) 1.
}
\lref\ParanjapeAI{
A. Paranjape and T. P. Singh, JCAP {\bf 0803} (2008) 023.
}
\lref\ChuangYI{
C. H. Chuang, J. A. Gu, and W. Y. Hwang, Class. Quant. Grav. {\bf 25} (2008) 175001.
}
\lref\LTBL{
G. Lema\^\i tre, Annales Soc. Sci. Brux. Ser. ISci. Math. Astron. Phys. A {\bf 53} (1933) 51.
}
\lref\LTBT{
R. C. Tolman, Proc. Nat. Acad. Sci. {\bf 20} (1934) 169.
}
\lref\LTBB{
H. Bondi, Mon. Not. Roy. Astron. Soc. {\bf 107} (1947) 410.
}
\lref\MarraKMR{
V. Marra, {\it et al.}, Phys. Rev. D {\bf 76} (2007) 123004.
}
\lref\MarraKM{
V. Marra, E. W. Kolb and S. Matarrese, Phys. Rev. D {\bf 77} (2008) 023003
}
\lref\BiswasUB{
T. Biswas, R. Mansouri, A. Notari, JCAP {\bf 0712} (2007) 017.
}
\lref\BiswasGI{
T. Biswas, A. Notari, JCAP {\bf 0806} (2008) 021.
}


\lref\MukhanovAK{
V. F. Mukhanov, L. R. W. Abramo, and R. H. Brandenberger, Phys. Rev. Lett. {\bf 78} (1997) 1624.
}
\lref\AbramoHU{
L. R. W. Abramo, R. H. Brandenberger, and V. F. Mukhanov, Phys. Rev. D {\bf 56} (1997) 3248.
}
\lref\TsamisSX{
N. C. Tsamis and R. P. Woodard, Phys. Lett. B {\bf 301} (1993) 351.
}
\lref\TsamisQQ{
N. C. Tsamis and R. P. Woodard, Nucl. Phys. B {\bf 474} (1996) 235.
}
\lref\UnruhIC{
W. Unruh, astro-ph/9802323.
}
\lref\KodamaQW{
H. Kodama and T. Hamazaki, Phys. Rev. D {\bf 57} (1998) 7177.
}
\lref\GeshnizjaniWP{
G. Geshnizjani and R. Brandenberger, Phys. Rev. D {\bf 66} (2002) 123507.
}
\lref\GeshnizjaniCN{
G. Geshnizjani and R. Brandenberger, JCAP {\bf 0504} (2005) 006.
}
\lref\AbramoDB{
L. R. Abramo and R. P. Woodard, Phys. Rev. D {\bf 65} (2002) 043507.
}
\lref\AbramoDD{
L. R. Abramo and R. P. Woodard, Phys. Rev. D {\bf 65} (2002) 063516.
}
\lref\OnemliHR{
V. K. Onemli and R. P. Woodard, Class. Quant. Grav. {\bf 19} (2002) 4607.
}
\lref\OnemliMB{
V. K. Onemli and R. P. Woodard, Phys. Rev. D {\bf 70} (2004) 107301.
}
\lref\BrunierSB{
T. Brunier, V. K. Onemli, and R. P. Woodard, Class. Quant. Grav. {\bf 22} (2005) 59.
}
\lref\KahyaHC{
E. O. Kahya and V. K. Onemli, Phys. Rev. D {\bf 76} (2007) 043512.
}

\lref\OnemliAS{
E. O. Kahya, V. K. Onemli, and R. P. Woodard, Phys. Rev. D {\bf 81}, (2010) 023508.
}

\lref\DeserJK{
S. Deser and R. P. Woodard, Phys. Rev. Lett. {\bf 99} (2007) 111301.
}

\lref\DeffayetCA{
C. Deffayet and R. P. Woodard, JCAP {\bf 0908} (2009) 023.
}

\lref\TsamisPH{
N. C. Tsamis and R. P. Woodard, Phys. Rev. D {\bf 81} (2010) 103509.
}

\lref\SolaDH{
I. L. Shapiro and J. Sola, JHEP {\bf 0202} (2002) 006;
I. L. Shapiro, {\it et al.}, Phys. Lett. B {\bf 574} (2003) 149;
J. Sola and H. Stefancic, Phys. Lett. B {\bf 624} (2005) 147;
I. L. Shapiro and J. Sola, Phys. Lett. B {\bf 682} (2009) 105;
F. Bauer, J. Sola, and H. Stefancic, JCAP {\bf 1012} (2010) 030.
}

\lref\PolyakovMM{
A. M. Polyakov, Nucl. Phys. B {\bf 797} (2008) 199.
}
\lref\PolyakovNQ{
A. M. Polyakov, Nucl. Phys. B {\bf 834} (2010) 316.
}
\lref\KrotovMA{
D. Krotov and A. M. Polyakov, arXiv:1012.2107.
}
\lref\KumarUK{
N. Kumar and E. E. Flanagan, Phys. Rev. D {\bf 78} (2008) 063537.
}
\lref\WeinbergVY{
S. Weinberg, Phys. Rev. D {\bf72} (2005) 043514.
}



\lref\CaiZP{
Y. F. Cai, {\it et al.}, Phys. Rept. {\bf 493} (2010) 1.
}
\lref\WetterichFM{
C. Wetterich, Nucl. Phys. B {\bf 302} (1988) 668.
}
\lref\ZlatevTR{
I. Zlatev, L. M. Wang, and P. J. Steinhardt, Phys. Rev. Lett. {\bf 82} (1999) 896.
}
\lref\CaldwellEW{
R. R. Caldwell, Phys. Lett. B {\bf 545} (2002) 23.
}
\lref\ClineGS{
J. M. Cline, S. Jeon, and G. D. Moore, Phys. Rev. D {\bf 70} (2004) 043543.
}
\lref\FengAD{
B. Feng, X. L. Wang, and X. M. Zhang, Phys. Lett. B {\bf 607} (2005) 35.
}
\lref\VikmanDC{
A. Vikman, Phys. Rev. D {\bf 71} (2005) 023515.
}
\lref\HuKH{
W. Hu, Phys. Rev. D {\bf 71} (2005) 047301.
}
\lref\PandaAS{S. Panda, Y. Sumitomo, and S. P. Trivedi, arXiv:1011.5877.}

\lref\CaldwellAI{
R. R. Caldwell and M. Doran, Phys. Rev. D {\bf 72} (2005) 043527.
}
\lref\XiaKM{
J. Q. Xia, {\it et al.}, Int. J. Mod. Phys. D {\bf 17} (2008) 1229.
}
\lref\TurnerHE{
M. S. Turner, Phys. Rev. D {\bf 28} (1983) 1243.
}
\lref\DamourCB{
T. Damour and V. F. Mukhanov, Phys. Rev. Lett. {\bf 80} (1998) 3440.
}
\lref\SahniQE{
V. Sahni and L. M. Wang, Phys. Rev. D {\bf 62} (2000) 103517.
}
\lref\JohnsonSE{
M. C. Johnson and M. Kamionkowski, Phys. Rev. D {\bf78} (2008) 063010.
}
\lref\AmendolaER{
L. Amendola, Phys. Rev. D {\bf62} (2000) 043511.
}
\lref\ZimdahlAR{
W. Zimdahl and D. Pavon, Phys. Lett. B {\bf521} (2001) 133.
}

\lref\BoisseauPho{
B. Boisseau {\it et al.}, Phys. Rev. Lett. {\bf 85} (2000) 2236.
}

\lref\ENO{
E. Elizalde, S. Nojiri, and S. D. Odintsov, Phys. Rev. D {\bf 70} (2004) 043539.
}

\lref\GengASa{
K. Bamba {\it et al.}, Phys. Rev. D {\bf 79} (2009) 083014; K. Bamba and C. Q. Geng, Phys. Lett. B {\bf 679} (2009) 282;
K. Bamba and C. Q. Geng,  Prog. Theor. Phys. 122 (2009) 1267; K. Bamba et al., Mod. Phys. Lett. A {\bf 25} (2010) 900;
K. Bamba, C. Q. Geng, and C. C. Lee, arXiv:1007.0482.
}

\lref\Chime{
L. P. Chimento {\it et al.}, Phys. Rev. D {\bf 79} (2009) 043502.
}

\lref\StoAsa{ D. Stojkovic, G. D. Starkman, and R. Matsuo,
Phys. Rev. D {\bf 77} (2008) 063006.}
\lref\StoAsb{E. Greenwood, {\it et al.},
Phys. Rev. D {\bf 79} (2009) 103003.}
\lref\StoAsc{D. C. Dai, S. Dutta, and D. Stojkovic,
Phys. Rev. D {\bf 80} (2009) 063522.}

\lref\DeffayetAs{C. Deffayet, {\it et al.}, JCAP {\bf 1010} (2010) 026.}
\lref\DLSVAs{
E. A. Lim, I. Sawicki, and A. Vikman, JCAP {\bf 1005} (2010) 012.}

\lref\UzanAs{J. P. Uzan, Phys. Rev. D {\bf 59} (1999) 123510.}

\lref\UzanAsa{J. Martin, C. Schimd, and J. P. Uzan,
Phys. Rev. Lett. {\bf 96} (2006) 061303.}

\lref\OverduinAs{J. M. Overduin and F. I. Cooperstock, Phys. Rev. D {\bf 58} (1998) 043506.}

\lref\WangMengAs{P. Wang and X. H. Meng, Class. Quant. Grav. {\bf 22} (2005) 283.}


\lref\ChibaKA{
T. Chiba, T. Okabe, and M. Yamaguchi, Phys. Rev. D {\bf 62} (2000) 023511.
}
\lref\ArmendarizPiconDH{
C. Armendariz-Picon, V. F. Mukhanov, and P. J. Steinhardt, Phys. Rev. Lett. {\bf 85} (2000) 4438.
}
\lref\ArmendarizPiconAH{
C. Armendariz-Picon, V. F. Mukhanov, and P. J. Steinhardt, Phys. Rev. D {\bf 63} (2001) 103510.
}
\lref\ArmendarizPiconRJ{
C. Armendariz-Picon, T. Damour, and V. F. Mukhanov, Phys. Lett. B {\bf 458} (1999) 209.
}
\lref\GarrigaVW{
J. Garriga and V. F. Mukhanov, Phys. Lett. B {\bf 458} (1999) 219.
}
\lref\YamaguchiAS{
J. Martin and M. Yamaguchi, Phys. Rev. D {\bf 77} (2008) 123508.
}
\lref\AfshordiAD{
N. Afshordi, D. J. H. Chung, and G. Geshnizjani, Phys. Rev. D {\bf 75} (2007) 083513.
}
\lref\AfshordiYX{
N. Afshordi, {\it et al.}, Phys. Rev. D {\bf 75} (2007) 123509.
}
\lref\KobayashiCM{
T. Kobayashi, M. Yamaguchi, and J. Yokoyama, Phys. Rev. Lett. {\bf 105} (2010) 231302.
}
\lref\ChowFM{
N. Chow and J. Khoury, Phys. Rev. D {\bf80} (2009) 024037.
}
\lref\NesserisPC{
S. Nesseris, A. De Felice, and S. Tsujikawa, Phys. Rev. D {\bf 82} (2010) 124054.
}
\lref\DeffayetGZ{
C. Deffayet, {\it et al.}, arXiv:1103.3260.
}
\lref\ArkaniHamedUY{
N. Arkani-Hamed, {\it et al.}, JHEP {\bf 0405} (2004) 074.
}
\lref\PiazzaDF{
F. Piazza, S. Tsujikawa, JCAP {\bf 0407} (2004) 004.
}
\lref\CheungST{
C. Cheung, {\it et al.}, JHEP {\bf 0803} (2008) 014.
}

\lref\PiazzaAs{
F. Piazza, New. J. Phys. {\bf 11} (2009) 113050;
S. Nesseris, F. Piazza, and S. Tsujikawa, Phys. Lett. B {\bf 689} (2010) 122.
}

\lref\BMVAs{
E. Babichev, V. Mukhanov, and A. Vikman, JHEP {\bf 0802} (2008) 101.
}

\lref\DPSVAs{
O. Pujolas, I. Sawicki, and A. Vikman, arXiv:1103.5360.
}

\lref\JojiriAsx{
R. Saitou and S. Nojiri, arXiv:1104.0558.
}

\lref\GhoseDE{
F. R. Urban and A. R. Zhitnitsky, Phys. Lett. B {\bf 688} (2010) 9;
F. R. Urban and A. R. Zhitnitsky, Phys. Rev. D {\bf 80} (2009) 063001;
F. R. Urban and A. R. Zhitnitsky, JCAP {\bf 0909} (2009) 018;
F. R. Urban and A. R. Zhitnitsky, Nucl. Phys. B {\bf 835} (2010) 135.
}
\lref\GhoseDEstu{
N. Ohta, Phys. Lett. B {\bf 695} (2011) 41;
R. G. Cai, arXiv:1011.3212;
A. Sheykhi and M. S. Movahed, arXiv:1104.4713;
A. Rozas-Fernandez, arXiv:1106.0056.
}

\lref\RibasVR{
M. O. Ribas, F. P. Devecchi, and G. M. Kremer, Phys. Rev. D {\bf 72} (2005) 123502.
}
\lref\CaiGK{
Y. F. Cai and J. Wang, Class. Quant. Grav. {\bf 25} (2008) 165014.
}
\lref\YajnikMH{
U. A. Yajnik, arXiv:1102.2562.
}
\lref\ArmendarizPiconPM{
C. Armendariz-Picon, JCAP {\bf 0407} (2004) 007.
}
\lref\ZhaoBU{
W. Zhao and Y. Zhang, Class. Quant. Grav. {\bf 23} (2006) 3405.
}
\lref\KoivistoFB{
T. S. Koivisto and N. J. Nunes, Phys. Rev. D {\bf 80} (2009) 103509.
}
\lref\DasGuptaJY{
P. Das Gupta, arXiv:0905.1621.
}
\lref\TsybaAs{P. Tsyba, {\it et al.}, arXiv:1103.5918;
K. K. Yerzhanov, {\it et al.}, arXiv:1012.3031.}


\lref\KamenshchikCP{
A. Y. Kamenshchik, U. Moschella, and V. Pasquier, Phys. Lett. B {\bf511} (2001) 265.
}
\lref\BentoPS{
M. C. Bento, O. Bertolami, and A. A. Sen, Phys. Rev. D {\bf66} (2002) 043507.
}
\lref\HovaNA{
H. Hova and H. Yang, arXiv:1011.4788.
}
\lref\BrevikSJ{
I. H. Brevik, Phys. Rev. D {\bf 65} (2002) 127302.
}
\lref\BrevikBJ{
I. H. Brevik and O. Gorbunova, Gen. Rel. Grav. {\bf 37} (2005) 2039.
}
\lref\MengJY{
X. H. Meng, J. Ren, and M. G. Hu, Commun. Theor. Phys. {\bf 47} (2007) 379.
}
\lref\RenNW{
J. Ren and X. H. Meng, Phys. Lett. B {\bf 633} (2006) 1.
}

\lref\NojiriASa{
S. Nojiri and S. D. Odintsov, Phys. Rev. D {\bf 72} (2005) 023003.
}


\lref\DeDeoVI{
S. DeDeo, Phys. Rev. D {\bf 73} (2006) 043520.
}
\lref\BoehmerFD{
C. G. Boehmer and T. Harko, Found. Phys. {\bf 38} (2008) 216.
}

\lref\TakahashiAS{
R. Takahashi and M. Tanimoto, Phys. Lett. B {\bf 633} (2006) 675;
JHEP {\bf 0605} (2006) 021.
}

\lref\StichelAS{
P. C. Stichel and W. J. Zakrzewski, Phys. Rev. D {\bf 80} (2009) 083513;
P. C. Stichel and W. J. Zakrzewski, Eur. Phys. J. C {\bf 70} (2010) 713.
}

\lref\Das{
S. Das, {\it et al.}, JHEP. {\bf 0904} (2009) 115;
S. Das, {\it et al.}, arXiv:0906.1044.
}

\lref\CCDMLJO{
J. A. S. Lima, J. F. Jesus and F. A. Oliveira, JCAP {\bf 11} (2010) 027.
}
\lref\CCDMmore{
S. Basilakos and J. A. S. Lima, Phys. Rev. D. {\bf 82} (2010) 3504;
S. Basilakos M. Plionis and J. A. S. Lima, Phys. Rev. D {\bf 82} (2010) 083517,
J. F. Jesus, {\it et al.}, arXiv:1105.1027.
}

\lref\MaEY{
C. P. Ma and E. Bertschinger, ApJ. {\bf 455} (1995) 7.
}
\lref\BardeenKT{
J. M. Bardeen, Phys. Rev. D {\bf22} (1980) 1882.
}
\lref\ZhaoVJ{
G. B. Zhao, {\it et al.}, Phys. Rev. D {\bf 72} (2005) 123515.
}
\lref\XiaGE{
J. Q. Xia, {\it et al.}, Phys. Rev. D {\bf 73} (2006) 063521.
}
\lref\CDMPS{
W. Hu, Phys. Rev. D {\bf 65} (2001) 023003;
J. K. Erickson, {\it et al.}, Phys. Rev. Lett. {\bf 88} (2002) 121301.
}
\lref\DEPuse{
C. Gordon and W. Hu, Phys. Rev. D {\bf 70} (2004) 083003;
S. Hannestad, Phys. Rev. D {\bf 71} (2005) 103519;
S. Unnikrishnan, H. K. Jassal, and T. R. Seshadri, Phys. Rev. D {\bf 78} (2008) 123504;
J. C. B. Sanchez and L. Perivolaropoulos, Phys. Rev. D {\bf 81} (2010) 103505;
R. U. H. Ansari and S. Unnikrishnan, arXiv:1104.4609.
}

\lref\PolchinskiGY{
J. Polchinski, hep-th/0603249.
}

\lref\medianone{
J. Richard Gott III et al., ApJ. {\bf 549} (2001) 1.
}
\lref\mediantwo{
P. P. Avelino, C. J. A. P. Martins, and P. Pinto, ApJ. {\bf 575} (2002) 989.
}

\lref\medianthree{
A. Barreira and P. P. Avelino, arXiv:1107.3971.
}

\lref\MCMC{
A. Lewis and S. Bridle, Phys. Rev. D {\bf 66} (2002) 103511.
}

\lref\MCMCBasic{
R. L. Smith, Operations Research {\bf 32} (1984) 1296;
D. Gamerman, {\it Markov Chain Monte Carlo: Stochastic simulation for Bayesian inference}, Champman and Hall (1997).
}


\lref\SNIaone{
W. Hillebrandt and J. C. Niemeyer,  Ann. Rev. Astron. Astrophys. {\bf 38} (2000) 191.
}

\lref\HZearly{
B. Leibundgut, {\it et al.}, ApJ. {\bf 466} (1996) L21.
}

\lref\SCPearly{
S. Perlmutter, {\it et al.}, ApJ. {\bf 483} (1997) 565.
}

\lref\SNIatwo{
B. Leibundgut, Ann. Rev. Astron. Astrophys. {\bf 39} (2001) 67.
}

\lref\KirshnerPaper{
R. P. Kirshner, arXiv:0910.0257.
}

\lref\Tonry{
J. L. Tonry,, {\it et al.}, ApJ. {\bf 594} (2003) 1.
}
\lref\Barris{
B. Barris,, {\it et al.}, ApJ. {\bf 602} (2004) 571.
}
\lref\Goldfour{
A. G. Riess, {\it et al.}, ApJ. {\bf 607} (2004) 665.
}
\lref\Goldsix{
A. G. Riess, {\it et al.}, ApJ. {\bf 659} (2007) 98.
}
\lref\SNLS{
P. Astier, {\it et al.}, Astron. Astrophys. {\bf 447} (2006) 31;
S. Baumont, {\it et al.}, Astron. Astrophys. {\bf 491} (2008) 567.
}
\lref\SNLSThreeYear{
N. Regnault, {\it et al.}, arXiv:0908.3808;
J. Guy, {\it et al.}, arXiv:1010.4743.
}
\lref\ESSENCE{
G. Miknaitis, {\it et al.}, ApJ. {\bf 666} (2007) 674;
W. M. Wood-Vasey, {\it et al.}, ApJ. {\bf 666} (2007) 694.
}
\lref\NSF{
Y. Copin, {\it et al.}, New. Astronomy. Rev. {\bf 50} (2006) 436;
R. A. Scalzo, {\it et al.}, ApJ {\bf 713} (2009) 1073.
}
\lref\CSP{
G. Folatelli, {\it et al.}, AJ. {\bf 139} (2010) 120;
G. Folatelli, {\it et al.}, AJ. {\bf 139} (2010) 519.
}
\lref\LOSS{
J. Leaman, {\it et al.}, arXiv:1006.4611;
W. D. Li, {\it et al.}, arXiv:1006.4612;
W. D. Li, {\it et al.}, arXiv:1006.4613.
}
\lref\SDSSSN{
J. A. Holtzman,, {\it et al.}, AJ. {\bf 136} (2008) 2306;
R. Kessler, {\it et al.}, ApJS. {\bf 185} (2009) 32.
}
\lref\Union{
M. Kowalski, {\it et al.}, ApJ. {\bf 686} (2008) 749.
}
\lref\Constitution{
M. Hicken, {\it et al.}, ApJ. {\bf 700} (2009) 1097;
M. Hicken, {\it et al.}, ApJ. {\bf 700} (2009) 331.
}
\lref\UnionTwo{
R. Amanullah, {\it et al.}, ApJ. {\bf 716} (2010) 712.
}
\lref\UnionTwoPointOne{
N. Suzuki, {\it et al.}, arXiv:1105.3470.
}
\lref\SNtrouble{
P. Hoeflich, astro-ph/0409170;
T. Plewa, A. C. Calder, and D. Q. Lamb, ApJL {\bf 612} (2004) L37;
S. Jha, A. G. Riess, and R. P. Kirshner, ApJ. {\bf 659} (2007) 122.
}
\lref\Hamuy{
M. Hamuy, {\it et al.}, AJ. {\bf 112} (1996) 2408.
}
\lref\Phillips{
M. M. Phillips, ApJ. {\bf 413} (1993) L105.
}
\lref\PhillipsApp{
S. Perlmutter, {\it et al.}, ApJ. {\bf 483} (1997) 565;
M. Hamuy, {\it et al.}, AJ. {\bf 109} (1995) 1669;
A. G. Riess, W. H. Press, and R. P. Kirshner, ApJ. {\bf 473} (1996) 88.
}
\lref\Kim{
A. Kim, LBNL Report LBNL-56164 (2004).
}
\lref\SNNuisance{
S. Nesseris and L. Perivolaropoulos, Phys. Rev. D {\bf 72} (2005) 123519;
L. Perivolaropoulos, Phys. Rev. D {\bf 71} (2005) 063503;
S. Nesseris and L. Perivolaropoulos, JCAP {\bf 0702} (2007) 025.
}
\lref\UnionWeb{
http://supernova.lbl.gov/Union/
}
\lref\SNLSConley{
A. Conley, {\it et al.}, ApJS. {\bf 192} (2011) 1.
}
\lref\SNLSSullivan{
M. Sullivan, {\it et al.}, arXiv:1104.1444.
}
\lref\SNLSCode{
https://tspace.library.utoronto.ca/handle/1807/24512
}

\lref\WangMukherjee{
Y. Wang and P. Mukherjee, ApJ, {\bf 606} (2004) 654;
Y. Wang and M. Tegmark, Phys. Rev. D {\bf 71} (2005) 103513.
}
\lref\SNOther{
S. Perlmutter, {\it et al.}, Bull. Am. Astron. Soc. {\bf 29} (1997) 1351;
B. P. Schmidt, ApJ. {\bf 607} (1998) 46;
P. M. Garnavich, {\it et al.}, ApJ. {\bf 509} (1998) 74;
S. Perlmutter, M. S. Turner and M. J. White, Phys. Rev. Lett. {\bf 83} (1999) 670;
A. G. Riess, Publ. Astron. Soc. Pac. {\bf 112} (2000) 1284;
A. G. Riess, {\it et al.}, ApJ. {\bf 560} (2001) 49;
P. Nugent, A. Kim and S. Perlmutter, Publ. Astron. Soc. Pac. {\bf 114} (2002) 803;
R. A. Knop, {\it et al.}, ApJ. {\bf 598} (2003) 102;
S. Perlmutter and B. P. Schmidt, Lect. Notes Phys. {\bf 598} (2003) 195;
A. V. Filippenko, astro-ph/0410609;
C. Shapiro and M. S. Turner, ApJ. {\bf 649} (2006) 563;
B. Leibundgut, Gen. Rel. Grav. {\bf 40} (2008) 221.
}


\lref\PWCMB{
A. A. Penzias and R. W. Wilson, ApJ. {\bf 142} (1965) 419.
}
\lref\BigBang{
R. A. Alpher, H. Bethe and G. Gamow, Phys. Rev. D {\bf 73} (1948) 803.
}
\lref\CMBCOBE{
G. F. Smoot, {\it et al.}, ApJ. {\bf 396} (1992) L1;
C. L. Bennett, {\it et al.}, ApJ. {\bf 396} (1992) L7.
}
\lref\WMAP{
C. L. Bennett, {\it et al.}, ApJS. {\bf 148} (2003) 1;
D. N. Spergel, {\it et al.}, ApJS. {\bf 148} (2003) 175;
D. N. Spergel, {\it et al.}, ApJS. {\bf 170} (2007) 377;
L. Page, {\it et al.}, ApJS. {\bf 170} (2007) 335.
}
\lref\WMAPFive{
E. Komatsu, {\it et al.}, ApJS. {\bf 180} (2009) 330.
}
\lref\CMBACBAR{
C. L. Reichardt, {\it et al.}, ApJ. {\bf 694} (2009) 1200.
}
\lref\CMBQuaD{
M. L. Brown, {\it et al.}, ApJ. {\bf 705} (2009) 978.
}
\lref\BAOPeeblesandYu{
P. J. E. Peebles and J. T. Yu, ApJ. {\bf 162} (1970) 815.
}
\lref\BAOBond{
J. R. Bond and G. Efstathiou, ApJ. {\bf 285} (1984) L45.
}
\lref\BAOHoltzman{
J. A. Holtzman, ApJS. {\bf 71} (1989) 1.
}
\lref\CMBAP{
W. Hu and N. Sugiyama, ApJ. {\bf 444} (1995) 489;
U. Seljak and M. Zaldarriaga, ApJ. {\bf 469} (1996) 437;
M. Zaldarriaga, U. Seljak, and E. Bertschinger, ApJ. {\bf 494} (1998) 491;
M. Zaldarriaga and U. Seljak, ApJS. {\bf 129} (2000) 431;
A. Lewis, A. Challinor, and A. Lasenby, ApJ. {\bf 538} (2000) 473;
N. W. Halverson, {\it et al.}, ApJL. {\bf 568} (2002) 38;
B. S. Mason, {\it et al.}, ApJ. {\bf 591} (2003) 549;
A. T. Lee, ApJ. {\bf 561} (2001) L1;
C. B. Netterfield, {\it et al.}, ApJ. {\bf 571} (2002) 604;
T. J. Pearson, {\it et al.}, ApJ. {\bf 591} (2003) 556;
S. Dodelson, {\it Modern Cosmology}, Academic Press (2003).
}
\lref\BAODefzd{
D. J. Eisenstein and W. Hu, ApJ. {\bf 496} (1998) 605.
}
\lref\HuSuDefZast{
W. Hu and N. Sugiyama, ApJ. {\bf 471} (1996) 542.
}
\lref\CMBDefR{
J. R. Bond, G. Efstathiou, and M. Tegmark, MNRAS {\bf 291} (1997) L33.
}
\lref\CMBAPExp{
A. D. Miller, {\it et al.}, ApJL. {\bf 524} (1999) L1;
P. de Bernardlis, {\it et al.}, Nature (London) {\bf 404} (2000) 955;
S. Hanany, {\it et al.}, ApJL. {\bf 545} (2000) L5.
}
\lref\WMAPSeven{
E. Komatsu, {\it et al.}, arXiv:1001.4538.
}
\lref\PlanckER{
Planck Collaboration, arXiv:1101.2022; arXiv:1101.2023; arXiv:1101.2024.
}
\lref\CMBISW{
R. K. Sachs and A. M. Wolfe, ApJ. {\bf 147} (1967) 73.
}
\lref\CMBISWDE{
K. Coble, S. Dodelson, and J. A. Frieman, Phys. Rev. D {\bf 55} (1997) 1851;
R. Scranton, {\it et al.}, astro-ph/0307335;
P. S. Corasaniti, {\it et al.}, Phys. Rev. Lett. {\bf 90} (2003) 091303;
P. S. Corasaniti, T. Giannantonio, and A. Melchiorri, Phys. Rev. D {\bf 71} (2005) 123521;
W. Hu, {\it et al.}, Phys. Rev. D {\bf 57} (1998) 3290;
P. Fosalba, E. Gaztanaga, and F. Castander, ApJ. {\bf 597} (2003) L89;
A. Rassat, {\it et al.}, MNRAS {\bf 377} (2007) 1085.
}
\lref\CMBCaldwell{
R. R. Caldwell, R. Dave, and P. J. Steinhardt, Phys. Rev. Lett. {\bf 80} (1998) 1582.
}
\lref\CMBGiannantonio{
T. Giannantonio, {\it et al.}, Phys. Rev. D {\bf 77} (2008) 123520.
}
\lref\CMBShirley{
S. Ho, {\it et al.}, Phys. Rev. D {\bf 78} (2008) 043519.
}
\lref\CMBOtherOne{
G. Jungman, {\it et al.}, Phys. Rev. D {\bf 54} (1996) 1332;
A. H. Jaffe, {\it et al.}, Phys. Rev. Lett. {\bf 86} (2001) 3475;
E. Komatsu and D. N. Spergel, Phys. Rev. D {\bf 63} (2001) 063002;
R. Stompor, {\it et al.}, ApJ. {\bf 561} (2001) L7;
J. Kovac, {\it et al.}, Nature {\bf 420} (2002) 772;
A. Benoit, {\it et al.}, Astron. Astrophys. {\bf 399} (2003) L25;
A. Kogut, {\it et al.}, ApJS. {\bf 148} (2003) 161;
K. M. Gorski, {\it et al.}, ApJ. {\bf 622} (2005) 759;
U. Seljak, A. Slosar, and P. McDonald, JCAP {\bf 0610} (2006) 014;
J. Dunkley, {\it et al.}, ApJS. {\bf 180} (2009) 306;
H. Liu and T. P. Li, arXiv:0907.2731;
H. Liu and T. P. Li, arXiv:1003.1073.
}
\lref\CMBOtherTwo{
S. Dodelson and L. Knox, Phys. Rev. Lett. {\bf 84} (2000) 3523;
P. S. Corasaniti, {\it et al.}, Phys. Rev. D {\bf 70} (2004) 083006;
R. Bean and O. Dore, Phys. Rev. D {\bf 69} (2004) 083503;
N. Afshordi, Y. S. Loh and M. A. Strauss, Phys. Rev. D {\bf 69} (2004) 083524;
H. K. Jassal, J. S. Bagla, and T. Padmanabhan, MNRAS {\bf 405} (2010) 2639.
}
\lref\CMBOtherThree{
M. Kamionkowski and A. Kosowsky, Ann. Rev. Nucl. Part. Sci. {\bf 49} (1999) 77;
W. Hu and S. Dodelson, Ann. Rev. Astron. Astrophys. {\bf 40} (2002) 171;
X. H. Fan, C. L. Carilli, and B. G. Keating, Ann. Rev. Astron. Astrophys. {\bf 44} (2006) 415;
D. Samtleben, S. Staggs, and B. Winstein, Ann. Rev. Nucl. Part. Sci. {\bf 57} (2007) 245;
M. Tristram and K. Ganga, Rept. Prog. Phys. {\bf 70} (2007) 899;
N. Aghanim, S. Majumdar, and J. Silk, Rept. Prog. Phys. {\bf 71} (2008) 066902.
}

\lref\BAOSilk{
J. Silk, ApJ. {\bf 151} (1968) 459.
}
\lref\BAOSunyaevandZeldovich{
R. A. Sunyaev, Y. B. Zeldovich, and B. Ya., Astrophys $\&$ Space Science {\bf 7} (1970) 3;
P. J. Zhang and U. L. Pen, ApJ. {\bf 577} (2002) 555.
}
\lref\BAOEisenstein{
D. J. Eisenstein et. al. ApJ. {\bf 633} (2005) 560.
}
\lref\BAODodelson{
S. Dodelson, Modern Cosmology, Academic Press, San Francisco (2003).
}
\lref\BAOGMSSW{
D. M. Goldberg and M.A. Strauss, ApJ. {\bf 495} (1998) 29;
A. Meiksin, M. White, and J. A. Peacock, MNRAS {\bf 304} (1999) 851;
V. Springel, {\it et al.}, Nature {\bf 435} (2005) 485;
H. J. Seo and D. J. Eisenstein, ApJ. {\bf 633} (2005) 575;
M. White, Astroparticlephys. {\bf 24} (2005) 334;
D. J. Eisenstein, H. J. Seo, and M. White, ApJ. {\bf 664} (2007) 660.
}
\lref\BAOTwentyOne{
J. S. B. Wyithe and A. Loeb, ApJ. {\bf 588} (2003) L59;
R. Cen, ApJ. {\bf 591} (2003) L5;
Z. Haiman and G. P. Holder, ApJ. {\bf 595} (2003) 1;
M. Zaldarriaga, S. R. Furlanetto, and L. Hernquist, ApJ. {\bf 608} (2004) 622;
S. Furlanetto, S. P. Oh, and F. Briggs, Phys. Rept. {\bf 433} (2006) 181;
M. McQuinn, {\it et al.}, MNRAS {\bf 377} (2007) 1043;
M. McQuinn, {\it et al.}, ApJ. {\bf 653} (2006) 815;
J. R. Pritchard and A. Loeb, Phys. Rev. D {\bf 78} (2008) 103511;
X. C. Mao and X. Wu, ApJ. {\bf 673} (2008) L107;
X. L. Chen and J. Miralda-Escude, ApJ. {\bf 684} (2008) 18.
}
\lref\BAOPrinciple{
A. Cooray, {\it et al.}, ApJ. {\bf 557} (2001) L7;
H. J. Seo and D. J. Eisenstein, ApJ. {\bf 598} (2003) 720;
C. Blake and K. Glazebrook, ApJ. {\bf 594} (2003) 665;
W. Hu and Z. Haiman, Phys. Rev. D {\bf68} (2003) 3004;
T. Matsubara, ApJ. {\bf 615} (2004) 573;
B. A. Bassett and R. Hlozek, arXiv:0910.5224.
}
\lref\BAOError{
J. Guzik, G. Bernstein, and R. E. Smith, MNRAS {\bf 375} (2007) 1329;
H. J. Seo and D. J. Eisenstein, ApJ. {\bf 665} (2007) 14;
R. E. Smith, R. Scoccimarro and R. K. Sheth, Phys. Rev. D {\bf 77} (2008) 043525.
}
\lref\BAOPCEH{
W. J. Percival, {\it et al.}, MNRAS {\bf 327} (2001) 1297;
S. Cole, {\it et al.}, MNRAS {\bf 362} (2005) 505;
G. Huetsi, Astron. Astrophys. {\bf 449} (2006) 891.
}
\lref\BAOColless{
M. Colless, {\it et al.}, astro-ph/0306581.
}
\lref\BAOYorkDG{
D. G. York,, {\it et al.}, AJ. {\bf 120} (2000) 1579.
}
\lref\BAOTegmark{
M. Tegmark, {\it et al.}, Phys. Rev. D {\bf 74} (2006) 123507.
}
\lref\BAOSDSSDReight{
http://www.sdss3.org/dr8/
}
\lref\BAONewestData{
W. J. Percival, {\it et al.}, MNRAS {\bf 401} (2010) 2148.
}
\lref\BAOChuang{
C. H. Chuang and Y. Wang, arXiv:1102.2251.
}


\lref\WLTyson{
J. A. Tyson, R. A. Wenk, and F. Valdes, ApJL. {\bf 349} (1990) L1.
}
\lref\WLBraninerd{
T. G. Brainerd, R. D. Blandford, and I. Smail, ApJ. {\bf 466} (1996) 623.
}
\lref\WLBacon{
D. J. Bacon, A. R. Refregier, and R. S. Ellis, MNRAS {\bf 318} (2000) 625.
}
\lref\WLKaiser{
N. Kaiser, G. Wilson, and G. A. Luppino,astro-ph/0003338.
}
\lref\WLWaerbeke{
L. van Waerbeke, {\it et al.}, Astron. Astrophys. {\bf 358} (2000) 30.
}
\lref\WLWittman{
D. M. Wittman, {\it et al.}, Nature {\bf 405} (2000) 143.
}
\lref\WLHuterer{
D. Huterer, Phys. Rev. D. {\bf 65} (2002) 063001.
}
\lref\WLCFHTLS{
http://www.cfht.hawaii.edu/Science/CFHLS/
}
\lref\WLReview{
A. Heavens, Nucl. Phys. Proc. Suppl. {\bf 194} (2009) 76.
}
\lref\WLCorFun{
J. Benjamin, {\it et al.}, MNRAS {\bf 381} (2007) 702;
T. Hamana, S. Colombi, and Y. Mellier, astro-ph/0009459;
L. Fu, {\it et al.}, Astron. Astrophys. {\bf 479} (2008) 9;
H. Hoekstra, {\it et al.}, ApJ. {\bf 577} (2002) 595;
H. Hoekstra, {\it et al.}, ApJ. {\bf 647} (2006) 116.
}
\lref\WLEBMode{
R. G. Crittenden, {\it et al.}, ApJ. {\bf 568} (2002) 20.
}
\lref\WLBModeSmall{
P. Schneider, L. Van Waerbeke, and Y. Mellier, Astron. Astrophys. {\bf 389} (2002) 729.
}
\lref\WLErrors{
T. D. Kitching, {\it et al.}, MNRAS {\bf 376} (2007) 771;
A. Amara and A. Refregier, MNRAS {\bf 391} (2008) 228;
T. D. Kitching, {\it et al.}, arXiv:0812.1966.
}
\lref\WLerror{
D. Huterer, {\it et al.}, MNRAS {\bf 366} (2006) 101;
Z. Ma, W. Hu, and D. Huterer, ApJ. {\bf 636} (2006) 21;
C. Heymans, {\it et al.}, MNRAS. {\bf 371} (2006) 750.
}
\lref\WLZhan{
H. Zhan, JCAP {\bf 0608} (2006) 008;
H. Zhan, {\it et al.}, ApJ. {\bf 675} (2008) L1;
H. Zhan, L. Knox and J. A. Tyson, ApJ. {\bf 690} (2009) 923.
}
\lref\WLerrorFour{
A. R. Zentner, D. H. Rudd, and W. Hu, Phys. Rev. D {\bf 77} (2008) 043507.
}
\lref\WLReviews{
Y. Mellier, Ann. Rev. Astron. Astrophys. {\bf 37} (1999) 127;
M. Bartelmann and P. Schneider, Phys. Rept. {\bf 340} (2001) 291;
H. Hoekstra, H. Yee, and M. Gladders, New Astron. Rev. {\bf 46} (2002) 767;
T. Hamana, M. Takada, and N. Yoshidae, MNRAS {\bf 350} (2004) 893;
A. Lewis and A. Challinor, Phys. Rept. {\bf 429} (2006) 1;
H. Hoekstra and B. Jain, Ann. Rev. Nucl. Part. Sci. {\bf 58} (2008) 99;
D. Munshi, {\it et al.}, Phys. Rept. {\bf 462} (2008) 67;
D. Huterer, Gen. Rel. Grav. {\bf 42} (2010) 2177.
}


\lref\CLXMMNewton{
G. Lamer, {\it et al.}, arXiv:0805.3817.
}
\lref\CLBorgani{
S. Borgani, astro-ph/0605575.
}
\lref\CLWarrent{
M. S. Warren, {\it et al.}, ApJ. {\bf 646} (2006) 881.
}
\lref\CLHaiman{
Z. Haiman, J. J. Mohr, and G.P. Holder, ApJ. {\bf 553} (2001) 545.
}
\lref\CLWang{
L. M. Wang and P. J. Steinhardt, ApJ. {\bf 508} (1998) 483.
}
\lref\CLSWang{
S. Wang, {\it et al.}, Phys. Rev. D {\bf 70} (2004) 123008.
}
\lref\CLReview{
N. A. Bahcall, Ann. Rev. Astron. Astroph. {\bf 15} (1977) 505;
C. L. Sarazin, {\it X-ray emission from clusters of galaxies}, Cambridge Astrophysics Series, Cambridge: Cambridge University press (1988);
P. Rosati, S. Borgani, and C. Norman, Ann. Rev. Astron. Astroph. {\bf 40} (2002) 539;
A. Cooray and R. K. Sheth, Phys. Rep. {\bf 372} (2002) 1;
G. M. Voit, Rev. Mod. Phys. {\bf 77} (2005) 207;
S. Borgani and A. Kravtsov, arXiv:0906.4370;
A. J. Benson, Phys. Rep. {\bf 495} (2010) 33;
H. Bohringer and N. Werner, Ann. Rev. Astron. Astroph. {\bf 18} (2010) 127;
S. W. Allen, A. E. Evrard, and A. B. Mantz, arXiv:1103.4829.
}


\lref\GRBPaczynski{
B. Paczynski, ApJ. {\bf 308} (1986) L43;
B. Paczynski, PASP. {\bf 107} (1995) 1167.
}
\lref\GRBParadijs{
E. Costa, {\it et al.}, Nature {\bf 387} (1997) 783;
J. Paradijs, {\it et al.}, Nature {\bf 386} (1997) 686;
D. A. Frail, {\it et al.}, Nature {\bf 389} (1997) 261.
}
\lref\GRBSwift{
N. Gehrels, {\it et al.}, ApJ. {\bf 611} (2004) 1005;
P. Romano, arXiv:1010.2206.
}
\lref\GRBFermi{
W. B. Atwood, {\it et al.}, ApJ. {\bf 697} (2009) 1071.
}
\lref\GRBGCN{
http://gcn.gsfc.nasa.gov
}
\lref\GRBSchaefer{
B. E. Schaefer, ApJ. {\bf 660} (2007) 16.
}
\lref\GRBBromm{
V. Bromm and A. Loeb, ApJ. {\bf 575} (2002) 111;
J. R. Lin, S. N. Zhang, and T. P. Li, ApJ. {\bf 605} (2004) 819.
}
\lref\GRBCos{
T. P. Li and Y. Q. Ma, ApJ. {\bf 272} (1983) 317;
L. Amati, {\it et al.}, Astron. Astrophys. {\bf 390} (2002) 81;
G. Ghirlanda, G. Ghisellini, and D. Lazzati, ApJ. {\bf 616} (2004) 331;
B. Zhang, Chin. J. Astron. Astrophys. {\bf 7} (2007) 1;
B. Zhang, astro-ph/0611774;
P. Meszaros, Rept. Prog. Phys. {\bf 69} (2006) 2259;
S. E. Woosley and J. S. Bloom, Ann. Rev. Astron. Astrophys. {\bf 44} (2006) 507.
}
\lref\GRBGhirlandaB{
G. Ghirlanda, G. Ghisellini, and C. Firmani, New J. Phys. {\bf 8} (2006) 123.
}
\lref\GRBStat{
G. Ghirlanda, {\it et al.}, ApJ. {\bf 613} (2004) L13;
H. Li, {\it et al.}, ApJ. {\bf 680} (2008) 92;
H. Wei and S. N. Zhang, Eur. Phys. J. C {\bf 63} (2009) 139.
}
\lref\GRBYunWangZeroEight{
Y. Wang, Phys. Rev. D {\bf78} (2008) 123532.
}
\lref\GRBLiang{
E. W. Liang and B. Zhang, MNRAS {\bf 369} (2006) L37;
N. Liang, {\it et al.}, ApJ. {\bf 685} (2008) 354.
}
\lref\DEManyModelsHWei{
H. Wei, JCAP {\bf 1008} (2010) 020.
}
\lref\GRBAmatiB{
L. Amati, {\it et al.}, MNRAS {\bf 391} (2008) 577.
}


\lref\XRayGiacconi{
R. Giacconi, {\it et al.}, Phys. Rev. Letters {\bf 9} (1962) 439.
}
\lref\XRayUhuru{
http://heasarc.gsfc.nasa.gov/docs/uhuru/uhuru.html
}
\lref\XRayEinstein{
http://heasarc.gsfc.nasa.gov/docs/einstein/heao2.html
}
\lref\XRayChandra{
http://www.nasa.gov/mission pages/chandra/main/index.html
}
\lref\XRayXMMNewton{
http://xmm.esac.esa.int/
}
\lref\XRayFukugita{
M. Fukugita, C.J. Hogan, and P. J. E. Peebles, ApJ. {\bf 503} (1998) 518.
}
\lref\XRayWhite{
S. D. M. White, Nature {\bf 366} (1993) 429.
}
\lref\XRayEke{
V. R. Eke, J. F. Navarro, and C. S. Frenk, ApJ. {\bf 503} (1998) 569.
}
\lref\XRayOmegaM{
S. D. M. White and C. S. Frenk, ApJ. {\bf 379} (1991) 52;
A. C. Fabian, MNRAS {\bf 253} (1991) L29;
D. A. White and A. C. Fabian, MNRAS {\bf 273} (1995) 72;
L. P. David, C. Jones, and W. Forman, ApJ. {\bf 445} (1995) 578;
A. E. Evrard, MNRAS {\bf 292} (1997) 289;
A. J. R. Sanderson and T.J. Ponman, MNRAS. {\bf 345} (2003) 1241;
Y. T. Lin, J. J. Mohr, and S. S. Stanford, ApJ. {\bf 591} (2003) 794;
S. J. LaRoque, {\it et al.}, ApJ. {\bf 657} (2006) 917.
}
\lref\XRaySasaki{
S. Sasaki, Publ. Astron. Soc. Jpn. {\bf 48} (1996) L119.
}
\lref\XRayPen{
U. Pen, New Astronomy {\bf 2} (1997) 309.
}
\lref\XRayAllenZeroTwo{
S. W. Allen, R. W. Schmidt, and A. C. Fabian, MNRAS {\bf 334} (2002) L11.
}
\lref\XRayAllenZeroThree{
S. W. Allen, {\it et al.}, MNRAS {\bf 342} (2003) 287.
}
\lref\XRayEttori{
S. Ettori, P. Tozzi, and P. Rosati, Astron. Astrophys. {\bf 398} (2003) 879.
}
\lref\XRayAllenZeroSeven{
S. W. Allen, {\it et al.}, MNRAS {\bf 383} (2008) 879.
}
\lref\XRayRapetti{
D. Rapetti, S. W. Allen, and J. Weller, MNRAS {\bf 360} (2005) 555.
}
\lref\XRayRapettiTwo{
D. Rapetti and S. W. Allen, MNRAS {\bf 388} (2008) 1265.
}
\lref\XRayDetails{
S. Borgani, {\it et al.}, ApJ. {\bf 561} (2001) 13;
T. H. Reiprich and H. Bohringer, ApJ. {\bf 567} (2002) 716;
P. Schuecker, {\it et al.}, Astron. Astrophys. {\bf 402} (2003) 53;
A. Voevodkin and A. Vikhlinin, ApJ. {\bf 601} (2004) 610;
J. P. Henry, ApJ. {\bf 609} (2004) 603;
S. W. Allen, {\it et al.}, MNRAS {\bf 353} (2004) 457;
R. W. Schmidt and S. W. Allen, MNRAS {\bf379} (2007) 209;
A. Mantz, {\it et al.}, MNRAS {\bf 387} (2008) 1179.
}


\lref\HzHubble{
E. P. Hubble, PNAS, {\bf 15} (1929) 168.
}
\lref\HzHubbleHumason{
E. Hubble and M. L. Humason, ApJ. {\bf 74} (1931) 43.
}
\lref\HzHST{
http://hubblesite.org/
}
\lref\HzFreedman{
W. L. Freedman, {\it et al.}, ApJ. {\bf 533} (2001) 47.
}
\lref\HzSandage{
A. Sandage, {\it et al.}, ApJ. {\bf 653} (2006) 843.
}
\lref\HzRiess{
A. G. Riess, {\it et al.}, ApJ. {\bf 699} (2009) 539.
}
\lref\HzRiessnew{
A. G. Riess, {\it et al.}, ApJ. {\bf 730} (2011) 119.
}
\lref\HzSpitzer{
http://www.spitzer.caltech.edu/
}
\lref\HzGAIA{
http://www.esa.int/export/esaSC/120377 index 0 m.html
}
\lref\HzJWST{
http://www.jwst.nasa.gov/
}
\lref\HzReview{
W. L. Freedman and B. F. Madore, arXiv:1004.1856, (Annu. Rev. Astron. Astrophys. in press).
}
\lref\HzHu{
W. Hu, ASP Conf. Ser. {\bf 339} (2005) 215.
}
\lref\HzSimon{
J. Simon, L. Verde, and R. Jimenez, Phys. Rev. D {\bf 71} (2005) 123001.
}
\lref\HzStern{
D. Stern, {\it et al.}, JCAP {\bf 1002} (2010) 008.
}
\lref\HzGazta{
E. Gazta\~naga, A. Cabr\'e, and L. Hui, MNRAS {\bf 399} (2009) 1663.
}
\lref\TJZhangH{
H. Y. Wan, {\it et al.}, Phys. Lett. B {\bf 651} (2007) 352;
Y. Gong, {\it et al.}, arXiv:0810.3572;
Z. X. Zhai, H. Y. Wan and T. J. Zhang, Phys. Lett. B {\bf 689} (2010) 8;
H. R. Yu, {\it et al.}, Research in Astronomy and Astrophysics {\bf 11} (2011) 125.
}
\lref\TJZhangHz{
H. Lin, {\it et al.}, Mod. Phys. Lett. A {\bf 24} (2009) 1699;
C. Ma and T. J. Zhang, ApJ. {\bf 730} (2011) 74;
Z. X. Zhai, T. J. Zhang and W. B. Liu, JCAP {\bf 08} (2011) 019.
}
\lref\TJZhangreview{
T. J. Zhang, C. Ma and T. Lan, Advances in Astronomy {\bf 2010} (2010) 184284.
}

\lref\Chaboyer{
B. Chaboyer, Phys. Rept. {\bf 307} (1998) 23.
}
\lref\CosmicAgeA{
H. B. Richer, {\it et al.}, ApJ. {\bf 574} (2002) L151.
}
\lref\CosmicAgeB{
L. M. Krauss and B. Chaboyer, Science {\bf 299} (2003) 65.
}
\lref\CosmicAgeC{
B. M. S. Hansen, {\it et al.}, ApJ. {\bf 574} (2002) L155.
}
\lref\CosmicAgeD{
J. Dunlop, {\it et al.}, Nature {\bf381} (1996) 581;
H. Spinrad, {\it et al.}, ApJ. {\bf 484} (1997) 581.
}
\lref\AlcanizLima{
J. S. Alcaniz and J. A. S. Lima, ApJ. {\bf 521} (1999) L87.
}
\lref\Ma{
J. Ma, {\it et al.}, Astron. J. 137 (2009) 4884;
S. Wang, {\it et al.}, Astron. J. 139 (2010) 1438.
}
\lref\AgeSWone{
S. Wang, X. D. Li, and M. Li, Phys. Rev. D {\bf 82} (2010) 103006.
}
\lref\Hasinger{
G. Hasinger, N. Schartel, and S. Komossa, ApJ. 573 (2002) L77;
S. Komossa and G. Hasinger, {
\it Proceedings of the Workshop XEUS¡ªStudying the Evolution of the Universe}, ed. G. Hasinger, eds, (to be published).
}
\lref\HzMys{
A. Friaca, J. S. Alcaniz, and J. A. S. Lima, Mon.Not.Roy.Astron.Soc. {\bf 362} (2005) 1295;
D. Jain and A. Dev, Phys Lett B {\bf 633} (2006) 436;
N. Pires, Z. H. Zhu, and J. S. Alcaniz, Phys. Rev. D {\bf 73} (2006) 123530;
H. Wei and S. N. Zhang, Phys. Rev. D {\bf 76} (2007) 063003.
}
\lref\HzBinWang{
B. Wang, {\it et al.}, Phys. Lett. B {\bf 662} (2008) 1.
}
\lref\HzBWone{
B. Wang, {\it et al.}, Nucl. Phys. B {\bf 778} (2007) 69.
}
\lref\HzTammann{
G. A. Tammann, {\it et al.}, arXiv:astro-ph/0112489.
}
\lref\AgeSWtwo{
S. Wang and Y. Zhang, Phys. Lett. B {\bf 669} (2008) 201.
}
\lref\CuiZhang{
J. L. Cui and X. Zhang, Phys. Lett. B {\bf 690} (2010) 233.
}


\lref\GrowFacPeebles{
P. J. E. Peebles, {
\it The Large-Scale Structure of the Universe}, Princeton University Press, Princeton, New Jersey (1980).
}
\lref\GrowFacLuScSt{
A. Lue, R. Scoccimarro, and G. Starkman, Phys. Rev. D {\bf 69} (2004) 124015.
}
\lref\DGPStructureFormation{
K. Koyama and R. Maartens, JCAP {\bf 0610} (2006) 016.
}
\lref\GrowFacSol{
D. Polarski and R. Gannouji, Phys. Lett. B {\bf 660} (2008) 439;
Y. G. Gong, Phys. Rev. D {\bf 78} (2008) 123010;
W. Hu and I. Sawicki, Phys. Rev. D {\bf 76} (2007) 104043;
S. Daniel, {\it et al.}, Phys. Rev. D {\bf 77} (2008) 103513.
}
\lref\GrowFacFry{
J. N. Fry, Phys. Lett. B {\bf 158} (1985) 211.
}
\lref\GrowFacLightman{
A. P. Lightman and P. L. Schechter, ApJ. {\bf 74} (1990) 831.
}
\lref\GrowFacLahav{
D. Lahav, {\it et al.}, MNRAS {\bf 251} (1991) 128.
}
\lref\GrowFacFormular{
A. J. S. Hamilton, MNRAS. {\bf 322} (2001) 419.
}
\lref\GrowFacredshift{
E. V. Linder and R. N. Cahn, Astropart. Phys. {\bf 28} (2007) 481;
Y. G. Gong, M. Ishak, and A. Z. Wang, Phys. Rev. D {\bf 80} (2009) 023002;
M. J. Mortonson, W. Hu, and D. Huterer, Phys. Rev. D {\bf 79} (2009) 023004;
M. Ishak and J. Dossett, Phys. Rev. D {\bf 80} (2009) 043004;
J. Dossett, {\it et al.}, JCAP. {\bf 1004} (2010) 022.
}
\lref\GrowFacPoAm{
C. Di Porto and L. Amendola, Phys. Rev. D {\bf 77} (2008) 083508.
}
\lref\GrowFacNesseris{
S. Nesseris and L. Perivolaropoulos, Phys. Rev. D {\bf 77} (2008) 023504.
}
\lref\GrowFacGuzzo{
L. Guzzo, {\it et al.}, Nature {\bf 451} (2008) 541.
}
\lref\GrowFacRoss{
N. P. Ross, {\it et al.}, MNRAS. {\bf 381} (2007) 573.
}
\lref\GrowFacJDa{
J. Da \^Angela, {\it et al.}, MNRAS {\bf 383} (2008) 565.
}
\lref\GrowFacMcDonald{
P. McDonald, {\it et al.}, ApJ. {\bf 635} (2005) 761.
}
\lref\GrowFacVHS{
M. Viel, M. G. Haehnelt, and V. Springel, MNRAS. {\bf 354} (2004) 684.
}
\lref\GrowFacMViel{
M. Viel, {\it et al.}, MNRAS {\bf 365} (2006) 231.
}
\lref\GrowFacMasGalClu{
D. Papetti, {\it et al.}, arXiv:0911.1787.
}
\lref\GrowFacWHu{
M. J. Mortonson, W. Hu and D. Huterer, Phys. Rev. D {\bf 81} (2010) 063007.
}


\lref\SLSandage{
A. Sandage, ApJ. {\bf 139} (1962) 319.
}
\lref\SLLoeb{
A. Loeb, ApJ. {\bf 499} (1998) L111.
}
\lref\SLCorasaniti{
P. S. Corasaniti, D. Huterer, and A. Melchiorri, Phys. Rev. D {\bf 75} (2007) 062001.
}

\lref\GWSchutz{
B. Schutz, Nature (London) {\bf 323} (1986) 310.
}
\lref\GWHHD{
D. E. Holz and S. A. Hughes, ApJ {\bf 629} (2005) 15;
N. Dalal, {\it et al.}, Phys. Rev. D {\bf 74} (2006) 063006.
}
\lref\GWSG{
K. G. Arun, {\it et al.}, Phys. Rev. D {\bf 76} (2007) 104016;
C. L. MacLeod and C. J. Hogan, Phys. Rev. D {\bf 77} (2008) 043512;
K. G. Arun and C. M. Will, Class. Quant. Grav. {\bf 26} (2009) 155002;
C. Cutler and D. E. Holz, Phys. Rev. D {\bf 80} (2009) 104009;
B. S. Sathyaprakash, B. F. Schutz, and C. Van Den Broeck, Class. Quant. Grav. {\bf 27} (2010) 215006;
S. Nissanke, {\it et al.}, ApJ. {\bf 725} (2010) 496;
W. Zhao, {\it et al.}, Phys. Rev. D {\bf 83} (2011) 023005.
}
\lref\GWSteinhardt{
L. A. Boyle and P. J. Steinhardt, Phys. Rev. D {\bf 77} (2008) 063504.
}
\lref\GWGrishchuk{
L. P. Grishchuk, Class. Quant. Grav. {\bf 14} (1997) 1445;
L. P. Grishchuk, {\it et al.}, Phys. Usp. {\bf 44} (2001) 1;
L. P. Grishchuk, Lect. Notes. Phys. {\bf 562} (2001) 167;
L. P. Grishchuk, {\it et al.}, Class. Quant. Grav. {\bf 22} (2005) 245.
}
\lref\GWZhang{
Y. Zhang, {\it et al.}, Class. Quant. Grav. {\bf 22} (2005) 1383;
Y. Zhang, {\it et al.}, Class. Quant. Grav. {\bf 23} (2006) 3783;
W. Zhao and Y. Zhang, Phys. Rev. D {\bf 74} (2006) 043503;
H. X. Miao and Y. Zhang, Phys. Rev. D {\bf 75} (2007) 104009;
S. Wang, {\it et al.}, Phys. Rev. D {\bf 77} (2008) 104016;
S. Wang, Phys. Rev. D {\bf 81} (2010) 023002.
}

\lref\TESTGRAVConfGRExp{
C. M. Will, Living Rev. Rel. {\bf 9} (2005) 3.
}
\lref\TESTGRAVCosmoTestofGrav{
B. Jain and J. Khoury, Annals Phys. {\bf 325} (2010) 1479.
}
\lref\TESTGRAVUzan{
J. P. Uzan, Rev. Mod. Phys. {\bf 75} (2003) 403;
J. P. Uzan, Living Rev. Real. {\bf 4} (2011) 2.
}
\lref\TESTGRAVPPN{
K. Nordtvedt, Phys. Rev. {\bf 169} (1968) 1027.
}
\lref\TESTGRAVCassini{
S. S. Shapiro, {\it et al.}, Phys. Rev. Lett. {\bf 92} (2004) 121101.
}
\lref\TESTGRAVPeriShiftMercury{
K. Nordtvedt, Phys. Rev. {\bf 169} (1968) 1014.
}
\lref\TESTGRAVLightDeflection{
I. I. Shapiro, in N.
Ashby, D. F. Bartlett, and W. Wyss, eds., General Relativity and
Gravitation, {\it Proceedings of the 12th International Conference
on General Relativity and Gravitation}, University of Colorado at
Boulder, July 2-8, 1989, 313-330, (Cambridge University Press,
Cambridge, U.K., New York, U.S.A., 1990).
}
\lref\TESTGRAVImproveLLR{
J. G. Williams, S. G. Turyshev, and T. W. Murphy, Int. J. Mod. Phys. D {\bf 13} (2004) 567.
}
\lref\TESTGRAVBBNReview{
F. Iocco, {\it et al.}, Phys. Rep. {\bf 472} (2009) 1.
}
\lref\TESTGRAVCMBConstrainG{
S. Galli, {\it et al.}, Phys. Rev. D {\bf 80} (2009) 023508.
}


\lref\LHC{
http://lhc.web.cern.ch/lhc
}
\lref\FOMWang{
Y. Wang, Phys. Rev. D {\bf 77} (2008) 123525.
}


\lref\DEObjSDSSBOSS{
http://www.sdss3.org/cosmology.php
}
\lref\DEObjLAMOST{
http://www.lamost.org
}
\lref\DEObjPanSTARRS{
http://pan-starrs.ifa.hawaii.edu/public/;
N. Kaiser, {\it et al.}, Proc. SPIE, {\bf 4836} (2002) 154.
}
\lref\DEObjSPT{
http://pole.uchicago.edu/;
J. E. Ruhl, {\it et al.}, Proc. SPIE Int. Soc. Opt. Eng. {\bf 5498} (2004) 11.
}
\lref\DEObjHST{
http://hubblesite.org/
}
\lref\DEObjHSTNew{
K. S. Dawson, {\it et al.}, ApJ {\bf 138} (2009) 1271;
K. Barbary, {\it et al.}, arXiv:1010.5786.
}
\lref\DEObjPLANCK{
http://www.rssd.esa.int/index.php?project=PLANCK
}


\lref\DEObjHETDEX{
http://hetdex.org/
}
\lref\DEObjDES{
http://www.darkenergysurvey.org/;
T. Abbott, {\it et al.}, AIP Conf. Proc. {\bf 842} (2006) 989.
}
\lref\DEObjALPACA{
http://www.astro.ubc.ca/LMT/alpaca/index.html/
}
\lref\DEObjBigBOSS{
http://bigboss.lbl.gov/;
D. J. Schlegel, {\it et al.}, arXiv:0904.0468.
}


\lref\DEObjLSST{
http://www.lsst.org/lsst
}
\lref\AstroLatest{
http://sites.nationalacademies.org/bpa/BPA 049810
}
\lref\DEObjSKA{
http://www.skatelescope.org/
}
\lref\DEObjWFIRST{
http://wfirst.gsfc.nasa.gov/
}
\lref\DEObjWFIRSTscience{
http://wfirst.gsfc.nasa.gov/science/
}
\lref\DEObjEuclid{
http://sci.esa.int/science-e/www/area/index.cfm?fareaid=102
}
\lref\DEObjIXO{
http://ixo.gsfc.nasa.gov/
}



\lref\QuinStarobinsky{
A. A. Starobinsky, JETP Lett. {\bf 68} (1998) 757.
}
\lref\QuinNakamura{
T. Nakamura and T. Chiba, MNRAS {\bf 306} (1999) 696.
}
\lref\QuinSaini{
T. D. Saini, {\it et al.}, Phys. Rev. Lett. {\bf 85} (2000) 1162.
}
\lref\QuinChiba{
T. Chiba and T. Nakamura, Phys. Rev. D {\bf 62} (2000) 121301.
}
\lref\QuinEaryWorks{
C. P. Ma, {\it et al.}, ApJ. {\bf 521} (1999) L1;
Z. Haiman and J. J. Mohr, ApJ. {\bf 553} (2000) 545;
P. Brax, J. Martin, and A. Riazuelo, Phys. Rev. D {\bf 62} (2000) 103505;
B. Barger and D. Marfatia, Phys. Lett. B {\bf 498} (2001) 67;
P. S. Corasaniti and E. J. Copeland, Phys. Rev. D {\bf 65} (2002) 043004;
R. Bean and A. Melchiorri, Phys. Rev. D {\bf 65} (2002) 041302;
B. F. Gerke and G. Efstathiou, MNRAS {\bf 335} (2002) 33;
P. S. Corasaniti and E. J. Copeland, Phys. Rev. D {\bf 67} (2003) 063501;
E. D. Pietro and J. F. Claeskens, MNRAS {\bf 341} (2003) 1299.
}
\lref\DEReviewSahniB{
V. Sahni, Class. Quant. Grav. {\bf 19} (2002) 3435.
}
\lref\QuinSahni{
V. Sahni and A. Starobinsky, Int. J. Mod. Phys. D {\bf 15} (2006) 2105.
}
\lref\SNIaHamuy{
M. Hamuy, {\it et al.}, AJ. {\bf 112} (1996) 2391.
}
\lref\QuinPara{
Z. K. Guo, N. Ohta, and Y. Z. Zhang, Phys. Rev. D {\bf 72} (2005) 023504;
Z. K. Guo, N. Ohta, and Y. Z. Zhang, Mod. Phys. Lett. A {\bf 22} (2007) 883.
M. Sahlen, A. R. Liddle, and D. Parkinson, Phys. Rev. D {\bf 75} (2007) 023502.
}
\lref\DEBeanCaroll{
R. Bean, S. M. Carroll, and M. Trodden, astro-ph/0510059.
}
\lref\DEReviewPadmanabhanB{
T. Padmanabhan, Curr. Sci. {\bf 88} (2005) 1057.
}
\lref\Perivolaropoulos{
R. Lazkoz, S. Nesseris, and L. Perivolaropoulos, JCAP {\bf 0511} (2005) 010.
}
\lref\AASen{
G. Gupta, S. Panda and A. A. Sen, arXiv:1108.1322.
}
\lref\QuinRefs{
V. B. Johri, Phys. Rev. D {\bf 70} (2004) 041303;
S. Lee, G. C. Liu, and K. W. Ng, Phys. Rev. D {\bf 73} (2006) 083516;
X. Zhang, Phys. Rev. D {\bf 74} (2006) 103505;
R. Crittenden, E. Majerotto, and F. Piazza, Phys. Rev. Lett. {\bf 98} (2007) 251301;
X. Zhang, Phys. Lett. B {\bf 648} (2007) 1;
C. Schimd, {\it et al.}, Astron. Astrophys. {\bf 463} (2007) 405;
E. V. Linder, Gen. Rel. Grav. {\bf 40} (2008) 329;
I. P. Neupane and C. Scherer, JCAP {\bf 0805} (2008) 009;
B. J. Li, D. F. Mota, and J. D. Barrow, ApJ. {\bf 728} (2011) 109;
H. Wei, Nucl. Phys. B {\bf 845} (2011) 381;
O. Luongo and H. Quevedo, arXiv:1104.4758;
A. Sheykhi, arXiv:1106.5697.
}
\lref\NoScalar{
T. Padmanabhan, AIP Conf. Proc. {\bf 861} (2006) 179.
}
\lref\NonScalar{
V. V. Kiselev, Class. Quant. Grav. {\bf 21} (2004) 3323;
H. Wei and R. G. Cai, Phys. Rev. D {\bf 73}, (2006) 083002;
H. Wei and R. G. Cai, JCAP. {\bf 0709} (2007) 015;
J. B. Jimenez and A. L. Maroto, Phys. Rev. D {\bf 78} (2008) 063005;
T. S. Koivisto and D. F. Mota, JCAP. {\bf 0808} (2008) 021;
K. Bamba, S. Nojiri, and S. D. Odintsov, Phys. Rev. D {\bf 77} (2008) 123532;
J. B. Jimenez, R. Lazkoz, and A. L. Maroto, arXiv:0904.0433;
V. A. De Lorenci, Class. Quant. Grav. {\bf 27} (2010) 065007.
}
\lref\NonnScalar{
Y. Zhang, T. Y. Xia, and W. Zhao, Class. Quantum Grav. {\bf 24} (2007) 3309;
T. Y. Xia and Y. Zhang, Phys. Lett. B {\bf 656} (2007) 19;
S. Wang, Y. Zhang, and T. Y. Xia, JCAP. {\bf 10} (2008) 037.
}
\lref\Quintom{
B. Feng, {\it et al.}, Phys. Lett. B {\bf 634} (2006) 101;
S. Nesseris and L. Perivolaropoulos, JCAP {\bf 0701} (2007) 018;
H. Li, {\it et al.}, Phys. Lett. B {\bf 658} (2008) 95.
}
\lref\DEIntWetterich{
C. Wetterich, Astron. Astrophys. {\bf 301} (1995) 321.
}
\lref\CQNumWorks{
L. Amendola, {\it et al.}, ApJ. {\bf 583} (2003) L53;
L. P. Chimento, {\it et al.}, Phys. Rev. D {\bf 67} (2003) 083513;
L. Amendola and C. Quercellini, Phys. Rev. D {\bf 68} (2003) 023514;
L. Amendola, MNRAS {\bf 342} (2003) 221;
G. Olivares, F. Atrio-Barandela, and D. Pavon, Phys. Rev. D {\bf 71} (2005) 063523;
J. H. He and B. Wang, JCAP {\bf 0806} (2008) 010;
J. Q. Xia, Phys. Rev. D {\bf 80} (2009) 103514;
J. Valiviita, R. Maartens, and E. Majerotto, MNRAS {\bf 402} (2010) 2355;
M. J. Mortonson, W. Hu, and D. Huterer, Phys. Rev. D {\bf 83} (2011) 023015.
}
\lref\CQAmendolaD{
L. Amendola and D. Tocchini-Valentini, Phys. Rev. D {\bf 66} (2002) 043528.
}
\lref\CQMaccio{
A. V. Maccio, {\it et al.}, Phys. Rev. D {\bf 69} (2004) 123516.
}
\lref\CQGuo{
Z. K. Guo, N. Ohta, and S. Tsujikawa, Phys. Rev. D {\bf 76} (2007) 023508.
}
\lref\CQBean{
R. Bean, {\it et al.}, Phys. Rev. D {\bf 78} (2008) 123514.
}
\lref\MBaldi{
M. Baldi, {\it et al.}, MNRAS {\bf 413} (2010) 1684;
M. Baldi, arXiv:1109.5695.
}
\lref\CQOther{
R. G. Cai and A. Z. Wang, JCAP {\bf 0503} (2005) 002;
T. Koivisto, Phys. Rev. D {\bf 72} (2005) 043516;
M. Manera and D. F. Mota, MNRAS {\bf 371} (2006) 1373;
G. Olivares, F. A. Barandela, and D. Pavon, Phys. Rev. D {\bf 74} (2006) 043521;
J. H. He, B. Wang, and Y. P. Jing, JCAP {\bf 0907} (2009) 030;
J. H. He, B. Wang, and P. J. Zhang, Phys. Rev. D {\bf 80} (2009) 063530;
X. M. Chen, {\it et al.}, Phys. Lett. B {\bf 695} (2011) 30;
G. Caldera-Cabral, R. Maartens, and B. M. Schaefer, JCAP {\bf 0907} (2009) 027;
E. Abdalla, {\it et al.}, Phys. Lett. B {\bf 673} (2009) 107;
G. Caldera-Cabral, Roy Maartens and L. A. Urena-Lopez, Phys. Rev. D {\bf 79} (2009) 063518;
A. Coc, {\it et al.}, Phys. Rev. D {\bf 79} (2009) 103512;
M. Baldi, {\it et al.}, MNRAS {\bf 403} (2010) 1684;
J. H. He, B. Wang, and E. Abdalla, arXiv:1012.3914.
}
\lref\Urena{
L. A. Urena-Lopez and T. Matos, Phys. Rev. D {\bf 62} (2000) 081302;
L. A. Urena-Lopez, JCAP {\bf 0509} (2005) 013;
L. A. Urena-Lopez, arXiv:1108.4712;
G. Caldera-Cabral, Roy Maartens and L. A. Urena-Lopez, Phys. Rev. D {\bf 79} (2009) 063518.
}

\lref\GCGasOrdinaryPaperA{
N. Bilic, G. B. Tupper, and R. D. Viollier, Phys. Lett. B {\bf 535} (2002) 17.
}
\lref\NCGasNumWorks{
P. T. Silva and O. Bertolami, ApJ. {\bf 599} (2003) 829;
M. Makler, {\it et al.}, Phys. Lett. B. {\bf 555} (2003) 1;
M. C. Bento, O. Bertolami, and A. A. Sen, Phys. Rev. D {\bf 67} (2003) 063003;
M. C. Bento, O. Bertolami, and A. A. Sen, Phys. Lett. B {\bf 575} (2003) 172;
A. Dev, J. S. Alcaniz, and D. Jain, Phys. Rev. D {\bf 67} (2003) 023515;
L. Amendola, {\it et al.}, JCAP {\bf 0307} (2003) 005;
R. Colistete Jr, {\it et al.}, Int. J. Mod. Phys. D {\bf 13} (2004) 669;
O. Bertolami, {\it et al.}, MNRAS {\bf 353} (2004) 329;
T. Multamaki, M. Manera, and E. Gaztanaga, Phys. Rev. D {\bf 69} (2004) 023004;
A. Dev, D. Jain, and J. S. Alcaniz, Astron. Astrophys. {\bf 417} (2004) 847;
Y. G. Gong, JCAP {\bf 0503} (2005) 007;
R. J. Colistete, J. C. Fabris, and S. V. B. Goncalves, Int. J. Mod. Phys. D {\bf 14} (2005) 775;
J. S. Alcaniz and J. A. S. Lima, ApJ. {\bf 618} (2005) 16;
M. Biesiada, W. Godlowski, and M. Szydlowski, ApJ. {\bf 622} (2005) 28;
X. Zhang, F. Q. Wu, and J. F. Zhang, JCAP {\bf 0601} (2006) 003;
O. Bertolami and P. T. Silva, MNRAS {\bf 365} (2006) 1149;
T. Giannantonio and A. Melchiorri, Class. Quant. Grav. {\bf 23} (2006) 4125;
P. X. Wu and H. W. Yu, JCAP {\bf 0703} (2007) 015;
Z. Li, P. Wu, and H. Yu, JCAP {\bf 0909} (2009) 017.
}
\lref\NCGasSandvik{
H. Sandvik, {\it et al.}, Phys. Rev. D {\bf 69} (2004) 123524.
}
\lref\NCGasZhu{
Z. H. Zhu, Astron. Astrophys. {\bf 423} (2004) 421.
}
\lref\NCGasDaly{
R. A. Daly and S. G. Djorgovski, ApJ. {\bf 597} (2003) 9.
}
\lref\DEManyModelsTDavis{
T. M. Davis, {\it et al.}, ApJ. {\bf 666} (2007) 716.
}
\lref\NCGasMakerB{
M. Makler, S. Q. de Oliveira, and I. Waga, Phys. Rev. D {\bf 68} (2003) 123521.
}
\lref\NCGasUDM{
L. M. G. Beca, {\it et al.}, Phys. Rev. D {\bf 67} (2003) 101301;
M. C. Bento, O. Bertolami, and A. A. Sen, Phys. Rev. D {\bf 70} (2004) 083519;
V. Gorini, {\it et al.}, JCAP {\bf 0802} (2008) 016;
J. C. Fabris, {\it et al.}, Phys. Lett. B, {\bf 694} (2011) 289.
}
\lref\NCGasPark{
C. G. Park, {\it et al.}, Phys. Rev. D {\bf 81} (2010) 063532.
}


\lref\HDEHuangSNIa{
Q. G. Huang and Y. G. Gong, JCAP {\bf 0408} (2004) 006.
}
\lref\HDEZhangX{
X. Zhang and F. Q. Wu, Phys. Rev. D {\bf 72} (2005) 043524.
}
\lref\HDEZhangTJ{
Z. L. Yi and T. J. Zhang, Mod. Phys. Lett. A {\bf 22} (2007) 41.
}
\lref\HDEWorks{
Q. G. Huang and M. Li, JCAP {\bf 0408} (2004) 013;
J. Y. Shen, {\it et al.}, Phys. Lett. B {\bf 609} (2005) 200;
Z. Chang, F. Q. Wu, and X. Zhang, Phys. Lett. B {\bf 633} (2006) 14;
X. Zhang and F. Q. Wu, Phys. Rev. D {\bf 76} (2007) 023502;
M. R. Setare, J. F. Zhang, and X. Zhang, JCAP {\bf 0703} (2007) 007;
Y. T. Wang and L. X. Xu, Phys. Rev. D {\bf 81} (2010) 082523;
X. Zhang, Phys. Lett. B {\bf 683} (2010) 81;
S. del Campo, J. C. Fabris, and R. Herrera, arXiv:1103.3441.
}
\lref\HDECompareHDEModels{
M. Li, {\it et al.}, JCAP {\bf 0906} (2009) 036.
}
\lref\HDEMaGongChen{
Y. Z. Ma, Y. Gong, and X. L. Chen, Eur. Phys. J. C {\bf 60} (2009) 303.
}
\lref\HDEYZMa{
Y. Z. Ma, AIP Conf. Proc. {\bf 1166} (2009) 44.
}
\lref\HDEGong{
Y. G. Gong, B. Wang, and Y. Z. Zhang, Phys. Rev. D {\bf 72} (2005) 043510.
}
\lref\DEManyModelsMLi{
M. Li, X. D. Li, and X. Zhang, Sci. China Phys. Mech. Astron. {\bf 53} (2010) 1631.
}
\lref\HDEIntWangA{
B. Wang, C. Y. Lin, and E. Abdalla, Phys. Lett. B {\bf 637} (2006) 357.
}
\lref\HDEInt{
D. Pavon and W. Zimdahl, Phys. Lett. B {\bf 628} (2005) 206;
H. M. Sadjadi and M. Honardoost, Phys. Lett. B {\bf 647} (2007) 231;
M. R. Setare, Phys. Lett. B {\bf 654} (2007) 1;
W. Zimdahl and D. Pavon, Class. Quant. Grav. {\bf 24} (2007) 5461;
J. Zhang, X. Zhang, and H. Liu, Phys. Lett. B {\bf 659} (2008) 26.
}
\lref\HDEIntXLC{
Y. Z. Ma, Y. Gong and X. L. Chen, Eur. Phys. J. C {\bf 69} (2010) 509.
}
\lref\HDEIntFeng{
C. Feng, {\it et al.}, JCAP {\bf 0709} (2007) 005.
}
\lref\Barrave{
A. Barreira and P.P. Avelino, Phys. Rev. D 83 (2011) 103001
}
\lref\HDECurvInt{
M. Li, {\it et al.}, JCAP {\bf 0912} (2009) 014.
}
\lref\ADEPaperThree{
H. Wei and R. G. Cai, Eur. Phys. J. C {\bf 59} (2009) 99;
I. P. Neupane, Phys. Lett. B {\bf 673} (2009) 111;
I. P. Neupane, Phys. Rev. D {\bf 76} (2006) 123006.
}
\lref\RDEXu{
L. X. Xu, W. B. Li, and J. B. Lu, Mod. Phys. Lett. A {\bf 24} (2009) 1355.
}
\lref\RDEZhang{
X. Zhang, Phys. Rev. D {\bf 79} (2009) 103509.
}
\lref\RDEFeng{
C. J. Feng, Phys. Lett. B {\bf 670} (2008) 231.
}
\lref\ADEWeiCai{
H. Wei and R. G. Cai, Phys. Lett. B {\bf 663} (2008) 1.
}
\lref\ADEKyoung{
K. Y. Kim, H. W. Lee, and Y. S. Myung, Phys. Lett. B {\bf 660} (2008) 118.
}


\lref\DGPGenDGP{
G. Dvali and M. S. Turner, astro-ph/0301510.
}
\lref\DEManyModelsDRubin{
D. Rubin, {\it et al.}, ApJ. {\bf 695} (2009) 391.
}
\lref\DGPHorScalGrowGeo{
W. J. Fang, {\it et al.}, Phys. Rev. D {\bf 78} (2008) 103509.
}
\lref\DGPGuo{
Z. K. Guo, {\it et al.}, ApJ. {\bf 646} (2006) 1.
}
\lref\DGPFairbairn{
M. Fairbairn and A. Goodbar, Phys. Lett. B {\bf 642} (2006) 432.
}
\lref\DGPmaartens{
R. Maartens and E. Majerotto, Phys. Rev. D {\bf 74} (2006) 023004.
}
\lref\DGPAlam{
U. Alam and V. Sahni, Phys. Rev. D {\bf 73} (2006) 084024.
}
\lref\DGPJQXia{
J. Q. Xia, Phys. Rev. D {\bf 79} (2009) 103527.
}
\lref\DGPReview{
A. Lue, Phys. Rept. {\bf 423} (2006) 1.
}
\lref\DGPStru{
I. Sawicki and S. M. Carroll, astro-ph/0510364;
K. Yamamoto, {\it et al.}, Phys. Rev. D {\bf 74} (2006) 063525;
J. P. Uzan, Gen. Rel. Grav. {\bf 39} (2007) 307;
H. Wei, Phys. Lett B {\bf 664} (2008) 1;
L. Lombriser, {\it et al.}, Phys. Rev. D {\bf 80} (2009) 063536;
K. C. Chan and R. Scoccimarro, Phys. Rev. D {\bf 80} (2009) 104005.
}
\lref\DGPLSTests{
Y. S. Song, I. Sawicki, and W. Hu, Phys. Rev. D {\bf 75} (2007) 064003.
}
\lref\DGPCosPert{
A. Cardoso, {\it et al.}, Phys. Rev. D {\bf 77} (2008) 083512.
}
\lref\DGPPPF{
W. J. Fang, W. Hu, and A. Lewis, Phys. Rev. D {\bf 78} (2008) 087303.
}
\lref\DGPSelfConsis{
F. Schmidt, Phys. Rev. D {\bf 80} (2009) 043001.
}


\lref\fRReviewA{
S. Nojiri and S. D. Odintsov, hep-th/0601213.
}
\lref\fRReviewB{
A. D. Felice and S. Tsujikawa, Living. Rev. Rel. {\bf 13} (2010) 3.
}
\lref\fRReviewC{
S. Capozziello, M. D. Laurentis, and V. Faraoni, arXiv:0909.4672.
}
\lref\fRReviewD{
S. Nojiri and S. D. Odintsov, arXiv:1011.0544, (Phys. Rept. in press).
}
\lref\fRTheoryCapozzielloA{
S. Capozziello, Int. J. Mod. Phys. D {\bf 11} (2002) 483.
}
\lref\fRTheoryCapozzielloB{
S. Capozziello, {\it et al.}, Int. J. Mod. Phys. D {\bf 12} (2003) 1969.
}
\lref\fRTheoryNojiriA{
S. Nojiri and S. D. Odintsov, Phys. Rev. D {\bf 68} (2003) 123512.
}
\lref\fRMatInst{
A. D. Dolgov and M. Kawasaki, Phys. Lett. B {\bf 573} (2003) 1;
V. Faraoni, Phys. Rev. D {\bf 74} (2006) 104017.
}
\lref\fRSolarTest{
G. J. Olmo, Phys. Rev. Lett. {\bf 95} (2005) 261102;
G. J. Olmo, Phys. Rev. D {\bf 72} (2005) 083505;
I. Navarro and K. V Acoleyen, JCAP {\bf 0702} (2007) 022;
A. L. Erickcek, T. L. Smith, and M. Kamionkowski, Phys. Rev. D {\bf 74} (2006) 121501;
T. Chiba, T. L. Smith, and A. L. Erickcek, Phys. Rev. D {\bf 75} (2007) 124014.
}
\lref\fRCognola{
G. Cognola, {\it et al.}, JCAP {\bf 0502} (2005) 010.
}
\lref\fRDynamical{
V. Faraoni, Phys. Rev. D {\bf 70} (2004) 04437.
}
\lref\fRDynamicalHu{
W. Hu and I. Sawicki, Phys. Rev. D {\bf 76} (2007) 064004.
}
\lref\fRMullerA{
V. Muller, H. J. Schmidt, and A. A. Starobinsky, Phys. Lett. B {\bf 202} (1988) 198.
}
\lref\fRAmendolaC{
L. Amendola, {\it et al.}, Phys. Rev. D {\bf 75} (2007) 083504.
}
\lref\fRStarobinB{
A. A. Starobinsky, J. Exp. Theor. Phys. Lett. {\bf 86} (2007) 157.
}
\lref\fRAmendolaA{
L. Amendola, D. Polarski, and S. Tsujikawa, Phys. Rev. Lett. {\bf 98} (2007) 131302;
L. Amendola, D. Polarski and S. Tsujikawa, Int. J. Mod. Phys. D {\bf 16} (2007) 1555.
}
\lref\frNojiriOdintsov{
S. Nojiri and S. D. Odintsov, Phys. Rev. D {\bf 74} (2006) 086005.
}
\lref\fRNojiriOdintsov{
S. Nojiri and S. D. Odintsov, Phys. Lett. B {\bf 657} (2007) 238;
S. Nojiri and S. D. Odintsov, Phys. Rev. D {\bf 77} (2008) 026007;
G. Cognola et al., Phys. Rev. D {\bf 77} (2008) 046009.
}
\lref\fRCosTest{
P. J. Zhang, {\it et al.}, Phys. Rev. Lett. {\bf 99} (2007) 141302;
R. Bean, {\it et al.}, Phys. Rev. D {\bf 75} (2007) 064020;
S. Nojiri and S.D. Odintsov, J. Phys. Conf. Ser. {\bf 66} (2007) 012005;
T. Faulkner, {\it et al.}, Phys. Rev.  D {\bf 76} (2007) 063505;
L. M. Sokolowski, Class. Quantum. Grav. {\bf 24} (2007) 3713;
P. J. Zhang, Phys. Rev. D {\bf 76} (2007) 024007;
S. Tsujikawa, {\it et al.}, Phys. Rev. D {\bf 80} (2009) 084044;
A. De Felice, S. Mukohyama, and S. Tsujikawa, Phys. Rev. D {\bf 82} (2010) 023524.
}
\lref\fRSongA{
Y. S. Song, W. Hu, and I. Sawicki, Phys. Rev. D {\bf 75} (2007) 044004.
}
\lref\fRSongB{
Y. S. Song, H. Peiris, and W. Hu, Phys. Rev. D {\bf 76} (2007) 063517.
}
\lref\fRSchmidt{
F. Schmidt, {\it et al.}, Phys. Rev. D {\bf 79} (2009) 083518;
F. Schmidt, A. Vikhlinin, and W. Hu, Phys. Rev. D {\bf 80} (2009) 082505;
F. Schmidt, Phys. Rev. D {\bf 81} (2010) 103002.
}
\lref\fRCamera{
S. Camera, A. Diaferio and V. F. Cardone, JCAP. {\bf 07} (2011) 016.
}
\lref\fRLombriser{
L. Lombriser, {\it et al.}, arXiv:1003.3009.
}
\lref\fRWL{
S. Tsujikawa and T. Tatekawa, Phys. Lett B {\bf 665} (2008) 325;
F. Schmidt, Phys. Rev. D {\bf 78} (2008) 043002;
A. Borisov and B. Jain, Phys. Rev. D {\bf 79} (2009) 103506;
T. Narikawa and K. Yamamoto, Phys. Rev. D {\bf 81} (2010) 043528.
}
\lref\fRMasui{
K. W. Masui, {\it et al.}, Phys. Rev. D {\bf 81} (2010) 062001.
}
\lref\fRWebb{
J. K. Webb, {\it et al.}, Phys. Rev. Lett. {\bf 82} (1999) 884;
J. K. Webb, {\it et al.}, Phys. Rev. Lett. {\bf 87} (2001) 091301;
J. K. Webb, {\it et al.}, arXiv:1008.3907.
}
\lref\fRfsc{
K. A. Olive, M. Peloso and J. P. Uzan, Phys. Rev. D {\bf 83} (2011) 043509;
T. Chiba and M. Yamaguchi, JCAP {\bf 1103} (2011) 044;
K. Bamba, S. Nojiri and S. D. Odintsov, arXiv:1107.2538.
}
\lref\TESTGRAV{
J. P. Uzan and F. Bernardeau, Phys. Rev. D {\bf 64} (2001) 083004;
H. Oyaizu, M. Lima, and W. Hu, Phys. Rev. D {\bf 78} (2008) 123524;
K. Koyama, A. Taruya, and T. Hiramatsu, Phys. Rev. D {\bf 79} (2009) 123512;
P. J. Zhang, Phys. Rev. D {\bf 73} (2006) 123504;
K. N. Ananda, S. Carloni, and P. K. S. Dunsby, Calss. Quant. Grav. {\bf 26} (2009) 235018;
T. Giannantonio, {\it et al.}, JCAP {\bf 1004} (2010) 030;
S. Capozziello and M. Francaviglia, Gen. Rel. Grav. {\bf 40} (2008) 357;
B. Jain and P. J. Zhang, Phys. Rev. D {\bf 78} (2008) 063503;
M. C. Martino and R.K. Sheth, arXiv:0911.1829;
I. Tereno, E. Semboloni, and T. Schrabback, arXiv:1012.5854;
J. P. Uzan, Gen. Rel. Grav. {\bf 42} (2010) 2219;
S. Unnikrishnan, S. Thakur and T. R. Seshadri, arXiv:1106.6353.
}
\lref\fRCurvSinProAba{
M. C. B. Abdalla, S. Nojiri, and S. D. Odintsov, Class. Quant. Grav. {\bf 22} (2005) L35.
}
\lref\fRCurvSinProBris{
F. Briscese, {\it et al.}, Phys. Lett. B {\bf 646} (2007) 105.
}
\lref\fRCurvSinProB{
A. V. Frolov, Phys. Lett. {\bf 101} (2008) 061103;
T. Kobayashi and K. I. Madeda, Phys. Rev. D {\bf 78} (2008) 064019.
}
\lref\fRCurSinProCure{
S. Nojiri and S. D. Odintsov, Phys. Rev. D {\bf 78} (2008) 046006;
K. Bamba, S. Nojiri, and S. D. Odintsov, JCAP {\bf 0810} (2008) 045.
}
\lref\fRCurvPro{
A. Dev, {\it et al.}, Phys. Rev. D {\bf 78} (2008) 083515;
T. Kobayashi and K. I. Maeda, Phys. Rev. D {\bf 79} (2009) 024009;
S. Appleby, R. Battye, and A. Starobinsky, JCAP {\bf 1006} (2010) 005;
E. Babichev and D. Langlois, Phys. Rev. D {\bf 81} (2010) 124051.
}
\lref\fRAmarzguioui{
M. Amarzguioui, {\it et al.}, Astron. Astrophys. {\bf 454} (2006) 707.
}
\lref\fRTKoivisto{
T. Koivisto, Phys. Rev. D {\bf 73} (2006) 083517.
}
\lref\BAOSDSSTegmark{
M. Tegmark, {\it et al.}, ApJ. {\bf 606} (2004) 702.
}
\lref\fRPalatini{
E. E. Flanagan, Phys. Rev. Lett. {\bf 92} (2004) 071101;
E. E. Flanagan, Class. Quan. Grav. {\bf 21} (2004) 417;
E. E. Flanagan, Class. Quan. Grav. {\bf 21} (2004) 3817.
}
\lref\fRPalatinitwo{
X. H. Meng and P. Wang, Class. Quant. Grav. {\bf 20} (2003) 4949;
X. H. Meng and P. Wang, astro-ph/0308284;
X. H. Meng and P. Wang, Phys. Lett. B {\bf 584} (2004) 1;
X. H. Meng and P. Wang, Gen. Rel. Grav. {\bf 36} (2004) 1947;
X. H. Meng and P. Wang, Class. Quant. Grav. {\bf 21} (2004) 951.
}
\lref\fRPalatinithree{
E. Barausse, T. P. Sotiriou, and J. C. Miller, Class. Quan. Grav. {\bf 25} (2008) 105008;
E. Barausse, T. P. Sotiriou, and J. C. Miller, Class. Quan. Grav. {\bf 25} (2008) 062001;
G. J. Olmo, Phys. Rev. D {\bf 78} (2008) 104026;
G. J. Olmo and P. Singh, JCAP {\bf 0901} (2009) 030;
C. Barragan, G. J. Olmo, and H. Sanchis-Alepuz, Phys. Rev. D {\bf 80} (2009) 024016;
B. Li, D. F. Mota, and D. J. Shaw, Class. Quan. Grav. {\bf 26} (2009) 055018;
A. Borowiec, {\it et al.}, arXiv:1109.3420.
}


\lref\GBTerm{
K. S. Stelle, Gen. Rel. Grav. {\bf 9} (1978) 353;
N. H. Barth and S. M. Christensen, Phys. Rev. D {\bf 28} (1983) 1876;
A. De Felice, M. Hindmarsh, and M. Trodden, JCAP {\bf 0608} (2006) 005;
G. Calcagni, B. de Carlos, and A. De Felice, Nucl. Phys. B {\bf 752} (2006) 404.
}
\lref\GBKovistoA{
T. Koivisto and D. F. Mota, Phys. Lett. B {\bf 644} (2007) 104.
}
\lref\GBKovistoB{
T. Koivisto and D. F. Mota, JCAP {\bf 0701} (2007) 006.
}
\lref\GBLi{
B. Li, J. D. Barrow, and D.F. Mota, Phys. Rev. D {\bf 75} (2007) 023520.
}
\lref\GBFelice{
A. De Felice, D. F. Mota, and S. Tsujikawa, Phys. Rev. D {\bf 81} (2010) 023532.
}
\lref\BDOther{
A. Riazuelo and J. P. Uzan, Phys. Rev. D {\bf 62} (2000) 083506;
A. Riazuelo and J. P. Uzan, Phys. Rev. D {\bf 66} (2002) 023525;
A. Coc, {\it et al.}, Phys. Rev. D {\bf 73} (2006) 083525;
C. Schimd, J. P. Uzan, and A. Riazuelo, Phys. Rev. D {\bf 71} (2005) 083512.
}
\lref\BDWuFQ{
F. Q. Wu, {\it et al.}, Phys. Rev. D {\bf 82} (2010) 083002;
F. Q. Wu and X. L. Chen, Phys. Rev. D {\bf 82} (2010) 083003.
}
\lref\fTWu{
P. X. Wu and H. W. Yu, Phys. Lett. B {\bf 693} (2010) 415.
}
\lref\fTBengochea{
G. R. Bengochea, Phys. Lett. B {\bf 695} (2011) 405.
}
\lref\fTWei{
H. Wei, X. P. Ma, and H. Y. Qi, Phys. Lett. B {\bf 703} (2011) 74.
}
\lref\fTWeitwo{
H. Wei, H. Y. Qi, and X. P. Ma, arXiv:1108.0859.
}
\lref\fTBamba{
K. Bamba, {\it et al.}, JCAP {\bf 1101} (2011) 021.
}
\lref\fTChen{
S. H. Chen, {\it et al.}, Phys. Rev. D {\bf 83} (2011) 023508.
}
\lref\fTDent{
J. B. Dent, S. Dutta, and E. N. Saridakis, JCAP {\bf 1101} (2011) 009.
}
\lref\fTZheng{
R. Zheng and Q. G. Huang, JCAP {\bf 1103} (2011) 002.
}
\lref\fTSotiriou{
B. Li, T. P. Sotiriou, and J.D. Barrow, arXiv:1010.1041.
}


\lref\LTBModel{
A. Paranjape and T. P. Singh, Class. Quant. Grav. {\bf 23} (2006) 6955.
}
\lref\LTBTheory{
H. Alnes, M. Amarzguioui, and O. Gron, Phys. Rev. D {\bf 73} (2006) 083519.
}
\lref\LTBTom{
K. Tomita, ApJ. {\bf 529} (2000) 38;
K. Tomita, MNRAS {\bf 326} (2001) 287;
M. N. Celerier, Astron. Astrophys. {\bf 353} (2000) 63.
}
\lref\LTBReCon{
N. Mustapha, C. Hellaby, and G. F. R. Ellis, MNRAS. {\bf 292} (1997) 817;
M. N. C\'{e}l\'{e}r\.{i}\'{e}r, K. Bolejko, and A. Krasi\'{e}ski, arXiv:0906.0905;
E. W. Kolb and C.R. Lamb, arXiv:0911.3852;
A. E. Romano, JCAP {\bf 1001} (2010) 004;
A. E. Romano, JCAP {\bf 05} (2010) 020.
}
\lref\DEManyModelsSollerman{
J. Sollerman, {\it et al.}, ApJ. {\bf 703} (2009) 1374.
}
\lref\LTBBellidoB{
J. G. Bellido and T. Haugbolle, JCAP {\bf 4} (2008) 3.
}
\lref\LTBCenterA{
H. Alnes and M. Amarzguioui, Phys. Rev. D {\bf 74} (2006) 103520.
}
\lref\LTBCenterB{
C. Quercellini, M. Quartin, and L. Amendola, Phys. Rev. Lett. {\bf 102} (2009) 151302.
}
\lref\LTBBiswas{
T. Biswas, A. Notari, and W. Valkenburg, JCAP {\bf 1011} (2010) 030.
}
\lref\LTBMoss{
A. Moss, J. P. Zibin, and D. Scott, arXiv:1007.3725.
}
\lref\LTBBlomqvist{
M. Blomqvist and E. Mortsell, JCAP {\bf 1005} (2010) 6.
}
\lref\LTBkSZ{
J. G. Bellido and T. Haugboelle, JCAP {\bf 0809} (2008) 016.
}
\lref\LTBWiltshire{
D. L. Wiltshire, Phys. Rev. Lett. {\bf 99} (2007) 251101.
}
\lref\LTBZibin{
J. P. Zibin, A. Moss, and D. Scott, Phys. Rev. Lett. {\bf 101} (2008) 251303;
J. P. Zibin, Phys. Rev. D {\bf 78} (2008) 043504.
}
\lref\LTBNadathur{
S. Nadathur and S. Sarkar, Phys. Rev. D {\bf 83} (2011) 063506.
}
\lref\LTBTimeDrift{
J. P. Uzan, C. Clarkson, and G. F. R. Ellis, Phys. Rev. Lett. {\bf 100} (2008) 191303;
P. Dunsby, {\it et al.}, JCAP {\bf 1006} (2010) 017.
}
\lref\LTBYoo{
C. M. Yoo, K. I. Nakao, and M. Sasaki, JCAP {\bf 1007} (2010) 012.
}
\lref\LTBQuartin{
M. Quartin and L. Amendola, Phys. Rev. D {\bf 81} (2010) 043522.
}
\lref\LTBGoodman{
J. Goodman, Phys. Rev. D {\bf 52} (1995) 1821.
}
\lref\LTBMirror{
R. R. Caldwell and A. Stebbins, Phys. Rev. Lett. {\bf 100} (2008) 191302.
}
\lref\LTBDistantSN{
T. Clifton, P. G. Ferreira, and K. Land, Phys. Rev. Lett {\bf 101} (2008) 131302.
}
\lref\LTBClarksonA{
C. Clarkson, B. Bassett, and T. H. C. Lu, Phys. Rev. Lett. {\bf 101} (2008) 011301.
}
\lref\LTBSmallScaleCMB{
T. Clifton, P. G. Ferreira, and J. Zuntz, JCAP {\bf 0907} (2009) 029.
}
\lref\LTBNeutrino{
J. J. Jia and H. B. Zhang, JCAP {\bf 0810} (2008) 002.
}
\lref\CosmicAgeLTB{
M. X. Lan, {\it et al.}, Phys. Rev. D {\bf 82} (2010) 023516.
}
\lref\DEInhomUniverBuchert{
T. Buchert, Gen. Rel. Grav. {\bf 40} (2008) 467.
}
\lref\MarraPaakkonen{
V. Marra and M. Paakkonen, JCAP {\bf 12} (2010) 021.
}
\lref\LTBOther{
J. G. Bellido and T. Haugboelle JCAP {\bf 0804} (2008) 003;
S. February, {\it et al.}, MNRAS {\bf 405} (2010) 2231;
R. A. Vanderveld, E. E. Flanagan, and I. Wasserman, Phys. Rev. D {\bf 78} (2008) 083511;
S. Alexander, {\it et al.}, JCAP {\bf 0909} (2009) 025;
K. Bolejko and J.S.B. Wyithe, JCAP {\bf 0902} (2009) 020;
V. Marra and M. Paakkonen, arXiv:1105.6099;
C. Clarkson and M. Regis, arXiv:1007.3443.
}
\lref\MarraRev{
V. Marra and A. Notari, Class. Quantum Grav. {\bf 28} (2011) 164004.
}
\lref\BRRasanen{
S. Rasanen, JCAP {\bf 0402} (2004) 003.
}
\lref\BRKolbA{
E. W. Kolb, {\it et al.}, Phys. Rev. D {\bf 71} (2005) 023524.
}
\lref\BRKolbB{
E. W. Kolb, S. Matarrese, and A. Riotto, New J. Phys. {\bf 8} (2006) 322.
}
\lref\BRNLi{
N. Li and D. J. Schwarz, Phys. Rev. D {\bf 78} (2008) 083531.
}
\lref\BRSeikel{
M. Seikel and D. J. Schwarz, arXiv:0912.2308.
}
\lref\BRLarena{
J. Larena, {\it et al.}, Phys. Rev. D {\bf 79} (2009) 083011.
}
\lref\BROther{
T. Buchert and M. Carfora, Class. Quant. Grav. {\bf 25} (2008) 195001;
C. Clarkson, K. Ananda, and J. Larena, Phys. Rev. D {\bf 80} (2009) 083525;
E. W. Kolb, V. Marra, and S. Matarrese, Gen. Rel. Grav. {\bf 42} (2010) 1399;
O. Umeh, J. Larena, and C. Clarkson, arXiv:1011.3959;
R. A. Sussman, arXiv:1102.2663.
}


\lref\DEManyModelsNesseris{
S. Nesseris and L. Perivolaropoulos, Phys. Rev. D {\bf 70} (2004) 043531.
}
\lref\DEManyModelsSzydlowski{
M. Szydlowski, A. Kurek, and A. Krawiec, Phys. Lett. B {\bf 642} (2006) 171.
}
\lref\DEManyModelsAKurek{
A. Kurek and M. Szydlowski, ApJ. {\bf 675} (2008) 1.
}
\lref\DEManyModels{
Y. G. Gong and C. K. Duan, MNRAS {\bf 352} (2004) 847;
M. Li, X. D. Li, and S. Wang, arXiv:0910.0717;
S. Basilakos, M. Plionis, and J. A. S. Lima, Phys. Rev. D {\bf 82} (2010) 083517.
}
\lref\ModelCompBIC{
G. Schwarz, The Annals of Statistics {\bf 6} (1978) 461.
}
\lref\ModelCompAIC{
H. Akaike, IEEE Transactions on Automatic Control {\bf 19} (1974) 716.
}
\lref\ModelCompIC{
W. Godlowski and M. Szydlowski, Phys. Rev. Lett. B {\bf 623} (2005) 10;
M. Biesiada, JCAP {\bf 0702} (2007) 003;
J. Magueijo and R. D. Sorkin, MNRAS {\bf 377} (2007) L39.
}
\lref\ModelCompBICLiddle{
A. R. Liddle, MNRAS {\bf 351} (2004) L49.
}
\lref\ModelCompBE{
T. D. Saini, J. Weller, and S. L. Bridle, MNRAS {\bf 348} (2004) 603;
A. R. Liddle, {\it et al.}, Phys. Rev. D {\bf 74} (2006) 123506;
O. Elgaroy and T. Multamaki, JCAP {\bf 0609} (2006) 002;
P. Marshall, N. Rajgru, and A. Slosar, Phys. Rev. D {\bf 73} (2006) 067302.
}
\lref\ModelCompBELZ{
Y. Gong and X. L. Chen, Phys. Rev. D {\bf 76} (2007) 123007.
}
\lref\ModelCompNestA{
P. Mukherjee, D. R. Parkinson, and A. R. Liddle, ApJ. {\bf 638} (2006) L51.
}
\lref\LCDMPerivolaropoulos{
L. Perivolaropoulos and A. Shafieloo, Phys. Rev. D {\bf 79} (2009) 123502;
L. Perivolaropoulos, arXiv:0811.4684; arXiv:1002.3030; arXiv:1104.0539.
}

\lref\WellerAlbrecht{
J. Weller and A. Albrecht, Phys. Rev. D {\bf 65} (2002) 103512.
}
\lref\GongWang{
Y. G. Gong and A. Wang, Phys. Rev. D {\bf 75} (2007) 043520.
}
\lref\HT{
D. Huterer and M. S. Turner, Phys. Rev. D {\bf 64} (2001) 123527.
}
\lref\Wetterich{
C. Wetterich, Phys. Lett. B {\bf 594} (2004) 17.
}
\lref\JBP{
H. K. Jassal, J. S. Bagla, and T. Padmanabhan, MNRAS {\bf 356} (2005) L11.
}
\lref\Starobinsky{
A. Shafieloo, V. Sahni, and A. A. Starobinsky, Phys. Rev. D, {\bf 80} (2009) 101301.
}
\lref\Wellergroup{
S. Benitez-Herrera, {\it et al.}, arXiv:1109.0873.
}
\lref\DEModInde{
G. Efstathiou, MNRAS {\bf 310} (1999) 842;
P. Astier, Phys. Lett. B {\bf 500} (2001) 8;
J. Weller and A. Albrecht, Phys. Rev. Lett. {\bf 86} (2001) 1939;
M. Tegmark, Phys Rev D {\bf 66} (2002) 103507;
D. N. Spergel and G. D. Starkman, astro-ph/0204089;
S. Corasaniti and E. J. Copeland, Phys. Rev. D {\bf 67} (2003) 063521;
Y. G. Gong and A. Wang, Phys. Rev. D {\bf 73} (2006) 083506;
C. Clarkson and C. Zunckel, arXiv:1002.5004;
F. C. Solano and U. Nucamendi, arXiv:1109.1303.
}
\lref\TurnerWhite{
M. S. Turner and M. J. White, Phys. Rev. D {\bf 56} (1997) 4439.
}
\lref\WangGarnavich{
Y. Wang and P. Garnavich, ApJ. {\bf 552} (2001) 445.
}
\lref\WangTegmark{
Y. Wang and M. Tegmark, Phys. Rev. Lett {\bf 92} (2004) 241302.
}
\lref\WangFreese{
Y. Wang and K. Freese, Phys. Lett. B {\bf 632} (2006) 449.
}
\lref\WangMukherjeetwo{
Y. Wang and P. Mukherjee, Phys. Rev. D {\bf 76} (2007) 103533.
}
\lref\YWangtwo{
Y. Wang, Phys Rev D {\bf 80} (2009) 123525.
}
\lref\YWanglatest{
Y. Wang, C. H. Chuang and P. Mukherjee, arXiv:1109.3172.
}
\lref\Maor{
I. Maor, R. Brustein, and P. J. Steinhardt, Phys. Rev. Lett. {\bf 86} (2001) 6;
I. Maor, {\it et al.}, Phys. Rev. D {\bf 65} (2002) 123003.
}
\lref\Linderargue{
E. V. Linder, Phys. Rev. D {\bf 70} (2004) 061302.
}
\lref\WDQ{
F. Y. Wang, Z. G. Dai and S. Qi, Astron. Astrophys. {\bf 507} (2009) 53.
}
\lref\LWY{
Z. X. Li, P. X. Wu, and H. W. Yu, Phys. Lett. B {\bf 695} (2011) 1;
Z. X. Li, P. X. Wu, and H. W. Yu, arXiv:1011.2036.
}
\lref\CA{
J. C. Carvalho and J. S. Alcaniz, arXiv:1102.5319.
}
\lref\SCA{
B. Santos, J. C. Carvalho, and J. S. Alcaniz, arXiv:1009.2733.
}
\lref\statefinder{
V. Sahni, {\it et al.}, JETP Lett. {\bf 77} (2003) 201.
}
\lref\Visser{
M. Visser, Class. Quant. Grav. {\bf 21} (2004) 2603.
}
\lref\Rapetti{
D. Rapetti, {\it et al.},  MNRAS. {\bf 375} (2007) 1510.
}
\lref\DEModIndeZ{
C. Zunckel and C. Clarkson, Phys. Rev. Lett. {\bf 101} (2008) 181301.
}
\lref\Omone{
V. Sahni, A. Shafieloo, and A. A. Starobinsky, Phys. Rev. D {\bf 78} (2008) 103502.
}
\lref\PanAlam{
A. V. Pan and U. Alam, arXiv:1012.1591.
}


\lref\CPLLensSysSuyu{
S. H. Suyu, {\it et al.}, ApJ. {\bf 711} (2010) 201.
}
\lref\CPLBE{
M. Chevallier and D. Polarski, Int. J. Mod. Phys. D {\bf 10} (2001) 213.
}
\lref\CPL{
E. V. Linder, Phys. Rev. Lett. {\bf 90} (2003) 091301.
}
\lref\CPLtwo{
E. V. Linder, Phys. Rev. D {\bf 70} (2004) 023511.
}
\lref\CPLPerivolaropoulos{
R. Lazkoz, S. Nesseris, and L. Perivolaropoulos, JCAP {\bf 0807} (2008) 012;
S. Basilakos, S. Nesseris, and L. Perivolaropoulos, MNRAS {\bf 387} (2008) 1126.
}
\lref\CPLpar{
E. V. Linder and D. Huterer, Phys. Rev. D {\bf 72} (2005) 043509.
}
\lref\HutererTurner{
D. Huterer and M. S. Turner, Phys. Rev. D {\bf 60} (1999) 081301.
}
\lref\Efstathiou{
G. Efstathiou, MNRAS {\bf 342} (2000) 810.
}
\lref\DeParUSeljak{
U. Seljak, {\it et al.}, Phys. Rev. D {\bf 71} (2005) 103515.
}
\lref\Upadhye{
A. Upadhye, M. Ishak, and P.J. Steinhardt, Phys. Rev. D {\bf 72} (2005) 063501;
K. Ichikawa and T. Takahashi, JCAP {\bf 0702} (2007) 001;
K. Ichikawa and T. Takahashi, JCAP {\bf 0804} (2008) 027;
J. Z. Ma and X. Zhang, arXiv:1102.2671;
Y. G. Gong, B. Wang, and R. G. Cai, JCAP {\bf 1004} (2010) 019;
N. N. Pan, {\it et al.}, Class. Quantum. Grav. {\bf 27} (2010) 155015;
}
\lref\YGGongtwo{
Y. G. Gong, {\it et al.}, arXiv:1008.5010.
}
\lref\BABassett{
B. A. Bassett, {\it et al.}, MNRAS {\bf 336} (2002) 1217;
B. A. Bassett, P. S. Corasaniti and M. Kunz, ApJ. {\bf 617} (2004) L1;
B. A. Bassett, {\it et al.}, JCAP {\bf 0807} (2008) 007.
}
\lref\SilvaAlcaniz{
R. Silva, {\it et al.}, arXiv:1104.1628.
}
\lref\ansatzZhang{
H. Li and X. Zhang, Phys. Lett. B {\bf 703} (2011) 119.
}

\lref\HutererStarkman{
D. Huterer and G. Starkman, Phys. Rev. Lett. {\bf 90} (2003) 031301.
}
\lref\HutererCooray{
D. Huterer and A. Cooray, Phys. Rev. D {\bf 71} (2005) 023506.
}
\lref\Sullivan{
S. Sullivan, A. Cooray, and D. E. Holz, JCAP. {\bf 09} (2007) 004.
}
\lref\Qi{
S. Qi, F. Y. Wang, and T. Lu, Astron. Astrophys. {\bf 483} (2008) 49.
}
\lref\Ournewpaper{
X. D. Li, {\it et al.}, JCAP. {\bf 07} (2011) 011.
}
\lref\YGGongone{
Y. G. Gong, {\it et al.}, JCAP {\bf 01} (2010) 019.
}
\lref\WangMPLA{
Y. Wang, Mod. Phys. Lett. A {\bf 25} (2010) 3093.
}
\lref\FreeZione{
Q. G. Huang {\it et al.}, Phys. Rev. D {\bf 80} (2009) 083515.
}
\lref\FreeZitwo{
S. Wang, X. D. Li, and M. Li, Phys. Rev. D {\bf 83} (2011) 023010.
}
\lref\FreeZithree{
X. D. Li {\it et al.}, JCAP {\bf 07} (2011) 011.
}
\lref\wavelet{
A. Hojjati, L. Pogosian, and G. B Zhao, JCAP, {\bf 04} (2010) 1007.
}
\lref\ShAlSaStr{
A. Shafieloo, {\it et al.}, MNRAS {\bf 366} (2006) 1081.
}


\lref\ASSSBefore{
U. Alam, {\it et al.}, MNRAS. {\bf 344} (2003) 1057.
}
\lref\ASSS{
U. Alam, {\it et al.}, MNRAS. {\bf 354} (2004) 275.
}
\lref\ASS{
U. Alam, V. Sahni, and A. A. Starobinsky, JCAP {\bf 06} (2004) 008.
}
\lref\Jonsson{
J. Jonsson, {\it et al.}, JCAP {\bf 0409} (2004) 007.
}
\lref\ASSStwo{
U. Alam, {\it et al.}, arXiv:astro-ph/0406672.
}


\lref\GPPRL{
T. Holsclaw, {\it et al.}, Phys. Rev. Lett. {\bf 105} (2010) 241302.
}
\lref\GPPRDOne{
T. Holsclaw, {\it et al.}, Phys. Rev. D {\bf 82} (2010) 103502.
}
\lref\GPBaRa{
S. Banerjee, B. P. Carlin, and A. E. Gelfand,
{\it Hierarchical Modeling and Analysis for Spatial Data}, New York: Chapman and Hall (2004);
C. E. Rasmussen and K. I. Williams,
{\it Gaussian Processes for Machine Learning}, Boston: MIT Press (2006).
}
\lref\GPPRDTwo{
S. Habib, {\it et al.}, Phys. Rev. D {\bf 76} (2007) 083503.
}



\listtoc

\writetoc

\newsec{Introduction}

Since its discovery in 1998 \refs{\riess,\perl},
dark energy has become one of the central problems in theoretical physics and cosmology
\refs{\weinrev,\DEReviewCarroll,\DEReviewPadmanabhanA,\DEReviewPeebles,\CopelandWR,\DEReviewUzan,
\DEReviewLinder,\ReviewofTurner,\DEReviewDurrer,\DEREviewTsujikawa,\DEbookYWang,\DEReviews}.
Thousands of papers have been written on this subject, while it is still as dunting as ever to understand the nature of dark energy.

We start with a brief history of the dark energy problem in time order
\refs{\weinrev,\StraumannTV}.

\noindent {\it 1917:} Einstein added a cosmological constant term
in his field equations, for the following reasons: Firstly, for isolated
mass not to impose a structure on space at infinity in closed
universe; secondly to obtain a static universe \EinsteinCE.

\noindent {\it 1920s:} Pauli realized that for a radiation field
the vacuum energy is too large to gravitate. As we shall review, the
vacuum energy density of a radiation field is proportional to the
cutoff to the fourth power. Pauli (unpublished) showed that using the
classical electron radius as an ultraviolet cutoff, the curvature of
the universe will be so large that the universe ``could not even reach
to the moon''.

\noindent {\it 1931:} Einstein removed the cosmological constant
\EinsteinNoCC\ because of the discovery of the cosmic expansion.
Although in 1923 on a postcard to Weyl he already wrote: ``If there
is no quasi-static world, then away with the cosmological term!'',
it seems that he does not believe in ``no quasi-static world'' until
\EinsteinNoCC.

\noindent {\it 1960s:} To explain why there are so many quasars
centering around red-shift around $z=1.95$, some people suggested to
use the Lamaitre model (with $\Lambda >0$, $k=1$), and around
$a=1/(1+z)=1/2.95$, the universe is like the Einstein's static
universe \PetrosianQH.

\noindent {\it 1967:} Zeldovich reintroduced the cosmological
constant problem by taking the vacuum fluctuations into account. This
introduced the old cosmological constant problem, Zeldovich used the
word ``fine-tuning'' \refs{\ZeldovichGD, \SahniZZA}.

\noindent {\it 1987:} Weinberg ``predicted" a non-vanishing and
small cosmological constant \anthrwein, and two years later
published by now the famous review article \weinrev.

\noindent {\it 1998:} Based on the analysis of 16 distant and 34 nearby supernovae,
Riess {\it et al.} first discovered the acceleration of expanding universe \riess.
Soon after, utilizing 18 nearby supernovae from the Calan-Tololo sample and 42 high-redshift supernovae,
Perlmutter {\it et al.} confirmed the discovery of cosmic acceleration \perl.

\noindent {\it 2000s:} String theorists reintroduced the anthropic principle when discovered the string landscape \refs{\bp,\suss}.

\noindent {\it 2011:} Because of the great discovery of cosmic acceleration,
Adam Riess, Brian Schmidt, and Saul Perlmutter win the Nobel prize in physics 2011.

We will review a set of theoretical ideas and phenomenological
models of dark energy, we will also review some numerical works done
so far. Thus the body of this review consists of two parts, the
first part reviews some of major theoretical ideas and models of
dark energy, and the second part is devoted to reviewing some
numerical works.

Summary of review on theoretical work: We will review briefly the history of the
problem of the cosmological constant. After this, we follow Weinberg's classical
review to divide the old approaches reviewed by Weinberg into five categories. Then,
we add three more categories in order to include most of the more recent ideas and
models.

Summary of review on numerical work:
First, we will review the mainstream cosmological observations probing dark energy,
including the type Ia supernovae (SNIa), the cosmic microwave background (CMB), the baryon acoustic oscillations (BAO),
the weak lensing (WL), the galaxy clusters (CL), the gamma-ray burst (GRB), the X-ray observations,
the Hubble parameter measurements, the cosmic age test, the growth factor (GF), and so on.
We will introduce the basic principles of these observations,
describe how these observations are included into the $\chi^2$ statistics,
and introduce the related research works of the observational groups.

Next, we will provide a brief overview of the present
and future dark energy projects. According to the Dark Energy
Task Force (DETF) report \DETF, the dark energy projects can be
classified into four stages: completed projects are Stage I;
on-going projects, either taking data or soon to be taking data, are
Stage II; near-future, intermediate-scale projects are Stage III;
larger-scale, longer-term future projects are Stage IV. Some most
representative dark energy projects of Stage II, Stage III, and
Stage IV will be introduced in this work.

Lastly, we will review the current numerical studies on dark energy.
Faced with the plethora of DE scenarios, the cosmologists currently
have two options: (1) they can test every single theoretical model
against observations and make a comparison of different dark energy
models; (2) they can reconstruct the evolution of dark energy based
on the observational data and explore the nature of dark energy in a
model independent manner. The related research works of these two
routes will be introduced in section 15 and section 16, respectively.

The conventions used in this review are as follows: we use the metric
convention ($-$,+,+,+), and use natural units $c=\hbar=1$. The
Newton's constant and the reduced Planck mass are kept explicit and they are
related by $M_p^2 = 1/(8\pi G)$.

\newsec{The theoretical challenge}

Einstein was the first to introduce the famous cosmological constant term
in his equations. In the Einstein equations
\eqn\ein{R_{\mu\nu}-\half g_{\mu\nu}R=8\pi GT_{\mu\nu},}
 one adds a term to the
stress tensor on the R.H.S. of the above equations
\eqn\tens{T_{\mu\nu}\rightarrow T_{\mu\nu}-{1\over 8\pi G}\Lambda g_{\mu\nu},}
where $\Lambda$ is a constant, the cosmological constant. The Einstein equations
can be rewritten as
\eqn\eins{R_{\mu\nu}-\half g_{\mu\nu}R+\Lambda g_{\mu\nu}=8\pi GT_{\mu\nu}.}

Before trying to understand the nature of $\Lambda$, we note that it
has a couple of ready interpretations. First, if we take the
additional term in \tens\ as coming from some ideal fluid, whose
stress tensor is given by \eqn\TCemtFluid{ T_{\mu\nu}=(\rho_\Lambda
+p_\Lambda )u_\mu u_\nu +p_\Lambda g_{\mu\nu},} where $\rho_\Lambda$
is the energy density and $p_\Lambda$ the pressure, then the
cosmological constant can be interpreted as a fluid with
\eqn\TCemtprho{ p_\Lambda=-\rho_\Lambda,\quad \rho_\Lambda={1\over
8\pi G}\Lambda.} This is certainly an unusual fluid, since if
$\Lambda > 0$, the pressure is negative, and the strong energy
condition is violated since $(T_{\mu\nu}-\half Tg_{\mu\nu})u^\mu
u^\nu=-\Lambda<0$. Of course the strong energy condition is not
something sacred. The null energy condition is marginally satisfied.

Second, since the cosmological constant term is proportional to $g_{\mu\nu}$
thus is Lorentz invariant, it can be interpreted as the vacuum energy. And indeed
Lorentz invariance forces upon us the condition $p=-\rho$. In principle, before
we understand the origin of the vacuum energy, the energy density can be positive,
negative and zero.

In the Friedmann-Robertson-Walker cosmology, the two Friedmann
equations read \eqn\frie{\eqalign{3M_p^2H^2&=\rho_m+\rho_\Lambda
-{3M_p^2k\over a^2},\cr 6M_p^2{\ddot{a}\over a}&=2\rho_\Lambda-\rho_m
,}} where $M_p^2=1/(8\pi G)$ and $M_p$ is usually called the reduced
Planck mass, $H =\dot{a}/a$ is the Hubble ``constant", $\rho_\Lambda$
is the vacuum energy density $\rho_\Lambda=M_p^2\Lambda$, and $\rho_m$
is the matter energy density (with $p_m=0$).  Einstein introduced a
positive cosmological constant motivated by a static universe.  To
have a static universe $H=0$ and without the cosmological constant, we
deduce from the first equation above $\rho_m=3M_p^2/a^2$ for a
positive spatial curvature $k=1$. But this does not work, since the
second Friedmann equation we have $\ddot{a}/a<0$ for $\rho_m$. Namely
the universe will collapse due the attractive force of matter. Thus,
to have both $H=0$ and $\ddot{a}/a=0$ we need to have
$2\rho_\Lambda=\rho_m$ and $\rho_\Lambda=M_p^2/a^2$. This is why the
cosmological constant has to be positive and fine-tuned to balance
matter. The spatial curvature also has to be positive.

However, due to the cosmic expansion, Einstein himself later abandoned
the cosmological constant. Nevertheless, as the cosmological constant
is the simplest extension of the original Einstein equations, the
theoretical possibility is left open.

Up to now, we have only discussed the classical story. Without
quantum mechanics, a very small cosomological constant poses no
problem, one simply regards it as a parameter in theory. However, in
quantum mechanics, we know that vacuum fluctuations make contribution
to the energy of a vacuum, thus the vacuum energy density receives two
contributions, one may be called the bare vacuum energy, the classical
one without quantum contribution, the other comes from the zero-point
fluctuations of all quantum fields (in quantum field theory), and the
zero-point fluctuations whatever dynamic degrees of freedom there are.

For a quantum field with a given mode of frequency $\omega$, the
zero-point energy is $\pm \half \omega$, with the plus sign for a
bosonic field and the minus sign for a ferminic field.  The total
zero-point energy is then $\half\sum_i(\pm)\omega_i$. In the continuum
limit, we have for a free field
\eqn\zeropint{\half\sum_i\omega_i=\half\int {d^3xd^3k\over
(2\pi)^3}(k^2+m^2)^{\half}= V\int{k^2dk\over 4\pi^2}(k^2+m^2)^\half,}
this integral is divergent, thus we need to introduce a physical
cutoff. A cutoff is reasonable, for example, if the zero-point
energy density is infinite, then our universe is infinitely curved and
the space size is infinitely small. Thus by examining the Friedmann
equations, we know that this cutoff must be the Planck energy at
largest.

Let the cutoff be $\lambda\gg m$, the energy density of a bosonic
field is then \eqn\enden{\rho_b= \int_0^\lambda {k^2dk\over 4\pi^2}
(k^2+m^2)^\half\approx \int_0^\lambda {k^3dk\over
4\pi^2}={\lambda^4\over 16\pi^2}.}  Take, for example, $\lambda=M_p$,
then
$$\rho= {1\over 2^{10}\pi^4G^2}\approx 2\times 10^{71} {\rm GeV}^4.$$
But in reality, $\rl<\rho_c\approx (3\times 10^{-12}{\rm GeV})^4\approx
10^{-46}{\rm GeV}^4$. If there is no supersymmetry, $|\rl|$ should be greater
than $\rho_b$ from a single bosonic field, namely $\rho_b<\rho_c$, but
what we have is \eqn\edpro{{\rho_b\over\rho_c}\approx 10^{117}.}  This
is the famous cosmological constant problem. Because if the
cosmological constant is much larger than $\rho_c$, the universe will
never look like what we observe today.

With exact supersymmetry, the bosonic contribution to cosmological
constant $\rho_b$ is canceled by its fermionic counterpart. However,
we know that our world looks not supersymmetric. Supersymmetry, if exists,
has to be broken above or around 100GeV scale. Even if we take the
$\Lambda_{SM}=$100GeV cutoff, below which physics is believed to
be well-described by the particle physics standard model, still
\eqn\qcdcut{\rho_b\approx {\Lambda_{SM}^4\over 16\pi^2}=10^{6}{\rm
GeV}^4, \quad {\rho_b\over \rho_c}\approx 10^{52}.}

One could use fine-tuning to solve the problem in some sense, by
introducing a bare cosmological constant and letting it cancel with the
quantum contribution using renormalization. However, one has to make
two independent numbers cancel by the accuracy of one part in
$10^{117}$ or at least $10^{52}$. This is extremely unlikely to
happen. Thus it remains a problem why the cosmological constant is not
large. This is known as the old cosmological constant problem \ZeldovichGD.

The cosmological constant problem remained the above statement until crucial
experiments came in.  Riess {\it et al.} and Perlmutter {\it et al.} in 1998
discovered the accelerating expansion of our universe \refs{\riess,\perl}. The
simplest explanation of this phenomenon is the return of a positive
cosmological constant. Let $\rho_c=3M_p^2H^2$ be the critical energy
density, data available today tell us that $\rho_\Lambda=0.73\rho_c$
and $\rho_m=0.27\rho_c$. Thus not only there is a small positive
cosmological constant, but also its energy density is the same order of matter
energy density. These discoveries lead to a new version of the
cosmological constant problem, or the problem of dark energy.

The modern version of the cosmological problem divides itself into two
parts:

\noindent (a) Why $\rl\approx 0$, namely why it is so small?

\noindent (b) Why $\rl\sim \rho_m$, this is called the coincidence
problem.

Our writing of the theoretical part of this review is motivated by
Weinberg's review article \weinrev, we will first review the
theoretical efforts before the discovery of acceleration \refs{\riess, \perl},
following Weinberg's classification:

\noindent 1. Supersymmetry and superstring.

\noindent 2. The anthropic principle.

\noindent 3. The tuning mechanisms.

\noindent 4. Modifying general relativity.

\noindent 5. Ideas of quantum gravity.

Twenty years have passed since Weinberg's classification, all the new
ideas more or less still belong to the above five categories. For
example, category 1 and category 2 combine to become the string
landscape + the anthropic principle, and there are new ideas in the
category of symmetries.  The tuning mechanisms now include ideas
associated with brane-world models. The category of modifying gravity
now includes $f(R)$ models, $f(T)$ models, MOND and TeVes models and
the Dvali-Gabadadze-Porrati (DGP) model. There are also some progress in the category of quantum
gravity.

Further, we can add three more categories: 6. The holographic
principle; 7. Back-reaction of gravity; 8. Phenomenological models.
Thus, we will also write about:

\noindent 1. Symmetry.

\noindent 2. The anthropic principle.

\noindent 3. The tuning mechanisms.

\noindent 4. Modifying gravity.

\noindent 5. Quantum gravity

\noindent 6. The holographic principle

\noindent 7. Back-reaction of gravity.

\noindent 8. Phenomenological models.

\newsec{Weinberg's classification}

\seclab\secWeinberg

In this section, we will briefly review Weinberg's classification of
ideas about the cosmological constant problem as given in \weinrev.

\subsec{Supersymmetry and superstring}

In any supersymmetric theory in 4 dimensions, the supersymmetry algebra contains at least
${\cal N}=1$ generators $Q_\alpha$ and their conjugate $Q_\alpha^{\dagger}$ \wb\ such that
\eqn\superal{\{Q_\alpha, Q_\beta^{\dagger}\}=(\sigma_\mu)_{\alpha\beta}P^\mu,}
where $\sigma_i$ are the Pauli matrices and $\sigma_0=1$. If supersymmetry is unbroken, then
\eqn\unbs{Q_\alpha |\Omega\rangle =Q_\alpha^{\dagger}|\Omega\rangle =0,}
the supersymmetry algebra leads to
\eqn\zeroe{H|\Omega\rangle =P^0|\Omega\rangle = 0,}
namely, the unbroken vacuum has exactly zero energy \ZuminoBG.

In a supersymmetric quantum field theory, there are a number of chiral super-fields for which
the potential $V$ is determined by the super-potential $W$. $W$ is a function of complex scalar
fields $\phi^i$ in the chiral multiplets,
\eqn\spoten{V=\sum_i|\p_i W|^2,}
where $\p_i W=\p W/\p \phi^i$. For a vacuum with unbroken supersymmetry $V=0$ thus
$\p_iW=0$.

Of course any supersymmetry must be broken in our world, thus in general
$\sum_\alpha \{Q_\alpha, Q_\alpha ^{\dagger}\}=2H>0$. If there is translational symmetry in
space-time, we must have $H\sim V\rl$ thus $\rl>0$. So a positive cosmological constant
is a consequence of broken supersymmetry in quantum field theory, as
long as this is the whole contribution to the effective
cosmological constant.
This conclusion agrees with observations since 1998, but what is the exact value of $\rl$?
For a typical quantum field theory, one expects $\rl\sim M_{SUSY}^4\gg\rho_c$. Thus we
would say that supersymmetry does not solve the cosmological constant problem.

It was commonly thought that it is better to explain $\rl=0$ first, then take the next step
to explain why $\rl\sim\rho_c$.

The potential in a super-gravity theory is determined by both the super-potential and the
K\"{a}hler potential $K(\phi^i,\bar{\phi}^i)$ \CremmerIV
\eqn\sugrap{V=e^K\left[G^{i\bar{j}}D_iW\overline{D_j\phi}-3|W|^2\right],}
where we already set $8\pi G=1$, and
\eqn\sugrad{D_iW=\p_iW+\p_iK W,\quad G_{i\bar{j}}=\p_{\phi^i}\p_{\bar{\phi}^j}W,
\quad G^{i\bar{j}}G_{\bar{j}k}=\delta^i_k.}

In a class of the so-called no-scale super-gravity models \EllisSF,
one can fine-tune parameters to break supersymmetry, meanwhile keep the
vacuum energy vanishing. There are three classes of complex scalar
fields, $C^a, S^n$ and $T$. Let \eqn\supgrakw{K=-3\ln
(T+\bar{T}-h(C,\bar{C}))+\tilde{K}(S,\bar{S}), \quad W=W_1(C)+W_2(S),}
then
\eqn\sugrav{V=e^{\tilde{K}}\left[(T+\bar{T}-h)^{-3}N^{a\bar{b}}\p_aW\overline{\p_bW}
+G^{m\bar{n}}D_mW\overline{D_nW}\right],} where $N=\p_a\p_{\bar{b}}h$.
Since both $N$ and $G$ are positive definite, so $V\ge 0$. To have
$V=0$ there ought to be \eqn\sugrac{\p_aW=D_mW=0.}  The point where
$V=0$ is a minimum, so there is no instability problem.  For
\eqn\sugramc{D_aW=\p_aW+\p_aKW=\p_aKW\sim \p_ahW,} if $W\ne 0$, then
$D_aW\ne 0$ and by definition supersymmetry is broken.

In such a model, $T$ is not fixed, this is why a model like this is called no-scale
super-gravity model. The problems for this model include:

\noindent $\bullet$ The coefficient of the first term must be -3, this is fine-tuning.

\noindent $\bullet$ The form of $W$ is fine-tuned.

\noindent $\bullet$ Quantum corrections usually spoil those fined-tuned coefficients.

The no-scale super-gravity models continue to attract attention today.

\subsec{Anthropic principle}

The terminology of  the anthropic principle is due to Brandon Carter, who articulated the anthropic principle
in reaction to the Copernican Principle, which states that humans do not occupy a privileged position
in the Universe. Carter said: ``Although our situation is not necessarily central, it is inevitably privileged
to some extent." \carter.

As Weinberg formulated in his review \weinrev, there are three different kinds of anthropic principle:
Very strong version, very weak version and weak version.

The very strong version states that everything in our universe has something to do with humankind, this is
of course absurd.

The very weak version takes the very existence of our humankind as a
piece of experimental data. For instance, in order
not to kill a person with the products of radio-decay, the life-time of a proton must be at least $10^{16}$
years.

Now the weak version. This is the version that prevails in certain circle
of people. In this version, it is postulated that there are many
regions in the universe. In these regions physical laws are in
different forms. It just happens that in the region we are dwelling
all physical laws, physical constants and cosmological parameters are
such that clusters of galaxies, galaxies and our solar system can
form, and humankind can appear. It appears that all the conditions are
fine-tuned but they are not, because all different sorts of regions
exist in the universe, such a universe is called multiverse. We found
ourselves in our part of the multiverse simply because this is all we
can observe.

In such a weak version, we are not supposed to expect every physical law and every physical constant
be designed for the existence of humankind. For instance, we do not know whether the fact that the life-time of proton is
more than $10^{33}$ years has anything to do with us, or whether it is connected with other conditions which
are necessary for the existence of us.

Dicke \refs{\DickeGZ,\DickeNature} may be one of the first persons making use of this weak version
of the anthropic principle, when he was considering Dirac's large number
problem. Let $T$ be the age of the universe. The death date of
the sun is greater than $T$, and $T$ is greater than the formation
time of the second and the third generations of stars. Thus $T$ is
about $10^{10}$ years. This is a nice explanation of a large
number. However, other time scales such as $1/(m_\pi^3l_p^2)\sim
10^{10}$ years are completely incomprehensible.

Weinberg's anthropic consideration of the cosmological constant \anthrwein\  is considered to be a genuine prediction of
the anthropic principle by some people. The simplified version of his argument is the following.
Let $z$ be the red-shift when galaxies form, the matter density is $\rho_m(z)=(1+z)^3\rho_m^0\sim 100\rho_m^0$.
The formation of galaxies crudely requires $\rl\leq 100\rho_m^0$.

It is assumed that the primordial density fluctuation is about
$\delta\rho/\rho\sim 10^{-5}$ in Weinberg's calculation. But in a
full anthropic calculation, even this number is not to be presumed
\GarrigaEE\ (for a complete discussion on ``scanning parameters'', see
\WeinbergFH). If we let $\delta\rho/\rho$ be a free parameter, we
usually have $\rl\le X\rho_m^0$ and $X\gg 100$!, thus the currently
observed value of dark energy can not be considered a consequence of
the anthropic argument.

Weinberg also pointed that the age of the universe is a problem. He argues that if $\rl/\rho_m^0=9$ then
$T=1.1H_0^{-1}$. Actually, other people, including de Vauconleurs \Vaucouleurs, Peebles \PeeblesCC,
Turner, Steigman and Krauss \TurnerNF\ also considered this issue.

The anthropic argument is also applied to a possible negative cosmological constant. Let $\rl <0$, then
the Friedmann equation $3M_p^2H^2=\rho_m+\rl$ implies that when $\rho_m=|\rl|$, the universe starts to
contract, thus $|\rl|\le \rho_m^0$.

\subsec{Tuning mechanism}

The idea is to use a scalar to self-tune the stress tensor, which is a
function of this scalar \DolgovGH.
Assuming
\eqn\fixedp{\nabla^\mu\nabla_\mu\phi\sim T^\mu{}_\mu\sim R,}
where $R$ is the scalar curvature. If the trace of the stress tensor vanishes when $\phi$ rolls
to a certain value $\phi_0$, $\phi$ will stay at this value and the effective cosmological constant
becomes vanishing. However, it can be proven that
\eqn\dil{T_{\mu\nu}=e^{4\phi}g_{\mu\nu}{\cal L}_0(\hbox{other fields}),}
so this tuning mechanism can not be realized (otherwise this will force ${\cal L}_0=0$).

Another possibility is $\phi_0=-\infty$, but in this case the effective Newtonian constant vanishes.

\subsec{Modifying gravity}

This is a rather popular theme now, but at the time when Weinberg wrote his review, there was only
one proposal mentioned. This is the uni-modular metric theory
\refs{\vanderBijYM,\UnruhIN} (see \refs{\EllisAs,\latestMGreview} for recent progress).

The idea is very simple. Removing the trace part from the Einstein equations
\eqn\eineq{R_{\mu\nu}-\half g_{\mu\nu}R=8\pi GT_{\mu\nu},}
we will have
\eqn\tlein{R_{\mu\nu}-{1\over 4}g_{\mu\nu}R=8\pi G(T_{\mu\nu}-{1\over 4}g_{\mu\nu}T),}
where $T=T^\mu_\mu$. Assume $T_{\mu\nu}$ be conserved, namely $\nabla^\mu T_{\mu\nu}=0$, we
deduce from \tlein\ that
\eqn\gredt{\p_\mu R=-8\pi G\p_\mu T,}
thus
\eqn\tracet{8\pi GT=-R+4\Lambda,}
where $\Lambda$ is an integral constant, when we use \tlein\ instead of
\eineq\ as a starting point. Substituting this back to the traceless Einstein
equations, we obtain
\eqn\einlam{R_{\mu\nu}-\half g_{\mu\nu}R=8\pi GT_{\mu\nu}-\Lambda g_{\mu\nu},}
so $\Lambda$ is indeed the cosmological constant. That $\Lambda$ is an integral constant
is due to the fact that there are one fewer equations in the traceless Einstein equations. In fact
Einstein himself considered this theory for a while.

The traceless equations can be considered as a consequence of the requirement that the determinant
of the metric $g_{\mu\nu}$ is 1, so we have $g^{\mu\nu}\delta g_{\mu\nu}=0$ or $\delta\sqrt{g}=0$,
we can introduce a Lagrangian multiplier into the action to enforce this.

As an integral constant, now $\Lambda$ is  considered as a free parameter in the theory to be
determined by initial conditions, or to set a framework for utilizing the anthropic principle.

\subsec{Quantum cosmology}

An accurate definition of quantum cosmology does not exist since there is no theory of quantum gravity to provide an
appropriate framework.

The so-called quantum cosmology as advocated by Hawking can at most be considered as
a qualitative picture of more accurate underlying theory. One starts with the Hamiltonian
constraint
\eqn\halc{^{(3)}R-2\Lambda +N^{-2}(E_{ij}E^{ij}-E^2)-2N^{-2}T_{00}=0,}
where $N$ is the lapse function as in $g_{00}=-N^2$, and $E^{ij}=\delta/\delta h_{ij},$ $^{(3)}R$
is the scalar curvature of the three spatial geometry in the ADM splitting. The
wave equation governing the wave-function of the universe is called
the Wheeler-De Witt equation \refs{\DeWittYK, \WheelerAA}
\eqn\wavef{\left(-G^{ij,lk}{\delta^2\over\delta h^{ij}\delta h^{lk}}+^{(3)}R-2\Lambda
-2T_{00}\right)\Psi=0,}
where $\Psi$ is the wave function of the universe, it is a functional of the three geometry
$h_{ij}$ and other fields on the three dimensional spatial slice, however, it contains no time,
and time has no place in the constraints. Actually, the Hamiltonian constraint is a consequence
of the requirement of time-reparametrization invariance. As a consequence, we can not impose
the usual normalization condition on $\Psi$:
\eqn\normc{\int \left|\Psi (h_{ij},\phi)\right|^2[dhd\phi]=1.}

Weinberg proposed to use one of the dynamical variables buried in $h_{ij}$ or $\phi$ to
replace the role of time.

There are infinitely many solutions to the Wheeler-De Witt equation,
if we assume this equation be well-defined. Hartle and Hawking
proposed in \hh\ to select one out many by using the path integral
\eqn\hhw{\Psi(h_{ij},\phi)=\int [dgd\phi]_Me^{-S_E},} where we assume
that the three dimensional space $\Sigma$ is the boundary of the four
dimensional space $M$ and $g$ is a Euclidean metric on $M$, $S_E$ is
the Hilbert-Einstein action on this Euclidean four space obtained by
an prescription of Wick-rotation: \eqn\euca{S_E={1\over 16\pi
G}\int\sqrt{g}(R+2\Lambda )+S_E(\phi),} where $S_E(\phi)$ is the
Euclidean action of the matter fields $\phi$. Since the path integral
is determined by the value of $h_{ij}$ and $\phi$ on $\Sigma$, there
are no additional boundary conditions. This is called by Hartle and
Hawking the no-boundary proposal of the wave function of the universe.

To have a probabilistic interpretation, consider a physical quantity $A$, a function of $h_{ij}$
and $\phi$ on $\Sigma$. We define the probability of $A$ assuming value $A_0$ be
\eqn\pbba{P(A_0)=\int [dgd\phi]\delta (A(h,\phi)-A_0)e^{-S_E}.}
However, $\Lambda$ is a fixed parameter in action $S_E$. In order to compute the probability distribution
of $\Lambda$, we need to make it variable. Hawking introduced a four form field strength in \sh\ to
make the effective $\Lambda$ a dynamic variable (see also \tbanks)
\eqn\fours{F_{\mu\nu\lambda\rho}=4\p_{[\mu}A_{\nu\lambda\rho]},}
with the action
\eqn\foura{S(A)=-{1\over 2\times 4!}\int\sqrt{-g}F_{\mu\nu\lambda\rho}F^{\mu\nu\lambda\rho}.}
The equation of motion for $F$ is
\eqn\foure{\p_\mu F^{\mu\nu\lambda\rho}=0.}
Since $F$ is totally asymmetric, let $F^{\mu\nu\lambda\rho}=F\epsilon^{\mu\nu\lambda
\rho}/\sqrt{-g}$, the equation of motion leads to $\p_\mu F=0$, namely $F=$constant, and
\eqn\fourac{S(A)=\half \int\sqrt{-g}F^2,}
compared with the action
\eqn\lamac{S(g,A)={1\over 16\pi G}\int \sqrt{-g}(R-2\Lambda),}
we may conclude that $\Lambda\rightarrow \Lambda -4\pi GF^2$. This is incorrect, since we need
to consider the energy of this solution $E=\int d^3x \half F^2$, and this leads to
\eqn\effl{\Lambda\rightarrow \Lambda +4\pi GF^2.}

In the Euclidean action
\eqn\euclaa{S_E={1\over 16\pi G}\int (R+2\Lambda)\sqrt{g}+{1\over 2\times 4!}\int
\sqrt{g}F_{\mu\nu\lambda\rho}F^{\mu\nu\lambda\rho},}
let $F^{\mu\nu\lambda\rho}=F\epsilon^{\mu\nu\lambda\rho}/\sqrt{g}$, we have
\eqn\eculaffl{S_E={1\over 16\pi G}\int (R+2(\Lambda +4\pi GF^2))\sqrt{g},}
this agrees with the consideration of energy.

Now, according to \pbba, we compute the probability distribution of $\Lambda$:
\eqn\prol{P(\Lambda_0)=\int [dgdA]\delta (\Lambda-\Lambda_0)e^{-S_E},}
where $\Lambda=4\pi GF^2$ (for simplicity we let the bare cosmological constant be zero).
The most contribution to the above path-integral comes from a classical solution to
equation $3M_p^2H^2=\half F^2$. We find
\eqn\classa{S_E=-{3\pi\over G\Lambda (F)},}
thus
\eqn\probl{P(\Lambda_0)\sim e^{{3\pi\over G\Lambda_0}}.}
Hawking concluded that $\Lambda_0\rightarrow +0$ is the most probable value. But we now know
that this is a wrong prediction!

Coleman later pointed out that Hawking's consideration is not the complete story. He suggests that
one needs to take wormholes and baby universes into account \scoleman. Each type of baby universe
is characterized by its physics properties, and we use $i$ to label them. Let $a_i^{\dagger}$ be the
creation operator for the baby universe of type $i$. Local creation of wormholes has the effect
of modifying the action:
\eqn\babya{S\rightarrow \tilde{S}=S+\sum_i (a_i+a_i^{\dagger})\int {\cal O}_i(x),}
this effective action works for the parent universe, and ${\cal O}_i$ is an local operator.

Consider a state without any wormhole, so $a_i|B\rangle =0$ and
\eqn\worml{\eqalign{&|B\rangle=\int\prod_id\alpha_if(\alpha_i)|\alpha\rangle,\cr
&(a_i+a_i^{\dagger})|\alpha\rangle =\alpha_i|\alpha\rangle,}} and
\eqn\fofal{f(\alpha_i)=\prod_i\pi^{-1/4}e^{-{\alpha_i^2\over 2}}.}
When acting on $|\alpha\rangle$, the action becomes
\eqn\alac{S\rightarrow S+\sum_i\alpha_i\int{\cal O}_i,}
thus $\alpha_i$ becomes a coupling constant in the parent universe.  The
cosmological constant can be regarded as a coupling constant, its
corresponding operator is the lowest dimensional operator ${\cal
O}_1=\sqrt{-g}$. In this framework, coupling constants including the
cosmological constant become dynamical automatically. No additional
mechanisms such as four form fluxes are needed. When observation is
made on spacetime geometry, the observers will find the wave function
of the universe to be in an eigenstate of $|\alpha\rangle$.
Coleman suggested that any apparently disconnect universes could be
actually connected by wormholes. In this sense all the other apparently
disconnected universes, which Hartle and Hawking ignore, should be
summed over. This argument will not affect real constants in nature
but does affect the effective ``constants'' which come from the baby
universe creation operators \babya.  Using techniques well developed in quantum field
theory (the summation of all vacuum to vacuum Feynman diagrams is the
exponential of the summation of connected diagrams), one obtains
\eqn\propal{P(\alpha)=\exp\left(\int [dgdA]e^{-S_E}\right),}
and it leads to
\eqn\expexp{P(\Lambda)=\exp\left(\exp({3\pi\over G\Lambda})\right),}
again this predicts $\Lambda=0$, a wrong prediction.

We have briefly reviewed Weinberg's classification, now we turn to
more recent ideas and models about dark energy.

\newsec{Symmetry}

\subsec {Supersymmetry in 2+1 dimensions}

Witten pointed out in 1995 that supersymmetry in 2+1 dimensions may
help to solve the zero cosmological constant problem \ewittenSC.  If
such a theory exhibits the same phenomenon as the type IIA string
theory in 10 dimensions, an additional dimension may emerge and
becomes the third spatial dimension. The 2+1 dimensional supersymmetry
is smaller than the 3+1 dimensional supersymmetry. As a result the
unbroken 2+1 dimensional supersymmetry (in 3+1 spacetime with an
emergent spatial dimension) does not contradict with observations and
meanwhile still forces the cosmological constant be zero.

In 2+1 dimensions, one can argue that there is no boson-fermion degeneracy. The existence of a particle
of mass $m$ creates a deficit angle in 2 spatial dimensions, $\theta\sim mG_3$, where $G_3$
is the 3 dimensional Newtonian constant. The appearance of the deficit angle makes the definition of charge
impossible, thus the energy degeneracy becomes impossible too.

Becker, Becker and Strominger used an Abelian Higgs model to realize Witten's idea \bbs, but unfortunately
there is no emergent third dimension in their model. Their model is a supergravity theory with field
content $(\phi, A_\mu, N, \chi, \lambda)+(g_{\mu\nu}, \psi_\mu)$. BBS showed that the spectrum of vortices
in this theory does not exhibit boson-fermion degeneracy.

\subsec {'t Hooft-Nobbenhuis symmetry}

't Hooft and later 't Hooft and Nobbenhuis proposed to consider symmetry transformation \thooft
\eqn\tns{x^\mu\rightarrow iy^\mu ,\quad p^\mu_x\rightarrow -ip^\mu_y.}
As the simplest example, Consider a scalar field
\eqn\sac{S=\int d^4x \left(-\half (\p\phi)^2-V(\phi)\right),}
with the stress tensor
\eqn\sstress{T_{\mu\nu}=\p_\mu\phi\p_\nu\phi +g_{\mu\nu}{\cal L}(\phi).}
Under transformation \tns, or $x^\mu=iy^\mu$, $\p_\mu^y=i\p_\mu$,
\eqn\lagt{\eqalign{&{\cal L}_y=-{\cal L}=-\half (\p_\mu^y\phi)^2+V,\cr
&T_{\mu\nu}^y=-T_{\mu\nu}=\p_\mu^y\phi\p_\nu^y\phi+g_{\mu\nu}{\cal L}.}}
We have in particular $T^y_{00}=-T_{00}$, however since $d^{D-1}x=i^{D-1}d^{D-1}y$, the Hamiltonian
is not simply reversed in sign
\eqn\hct{H^y=\int T_{00}^yd^{D-1}y=-(-i)^{D-1}H.}

Let
\eqn\canq{\eqalign{&\phi(x,t)=\int d^{D-1}p\left(a(p)e^{ipx}+a^{\dagger}(p)e^{-ipx}\right),\cr
&\pi(x,t)=\int d^{D-1}p\left(-ia(p)e^{ipx}+ia^{\dagger}(p)e^{-ipx}\right),\cr
&p^0=(p^2+m^2)^\half.}}
We have
\eqn\canqy{\eqalign{&\phi(iy,it)=\int d^{D-1}q\left(a_y(q)e^{iqx}+a^{\dagger}_y(p)e^{-iqx}\right),\cr
&\pi_y=i\pi(iy,it)=\int d^{D-1}q\left(-ia_y(q)e^{iqx}+ia_y^{\dagger}(q)e^{-ipx}\right),}}
and
\eqn\cho{q^0=(q^2-m^2)^\half ,\quad a_y(q)=(-i)^{D-1}a(p),\quad a_y^{\dagger}(q)=(-i)^{D-1}a^{\dagger}(p),}
if $a^{\dagger}(p)$ is the Hermitian conjugate of $a(p)$, $a_y^{\dagger}$ is longer the Hermitian conjugate of
$a_y$.

Note that if we demand $T_{\mu\nu}\rightarrow -T_{\mu\nu}$ be a symmetry, then
$T_{00}|\Omega\rangle =0$. Upon introducing gravity, let $g^y_{\mu\nu}=g_{\mu\nu}(x=iy)$
thus $ds^2_x=-ds^2_y$, $R_{\mu\nu}\rightarrow -R_{\mu\nu}$, or
$R^y_{\mu\nu}=-R^x_{\mu\nu}(iy)$. Start with the Einstein equations with a cosmological
constant
\eqn\ceins{R_{\mu\nu}-\half g_{\mu\nu}R+\Lambda g_{\mu\nu}=8\pi GT_{\mu\nu},}
we obtain
\eqn\yein{R_{\mu\nu}^y-\half g^y_{\mu\nu}R^y-\Lambda g_{\mu\nu}^y=8\pi GT^y_{\mu\nu},}
where we used $T_{\mu\nu}\rightarrow -T_{\mu\nu}$. Demanding $|\Omega\rangle$ be invariant,
we deduce $\Lambda=0$, This transformation maps a de Sitter space to an anti-de Sitter space.

't Hooft and Nobbenhuis pointed out that a scalar field and an Abelian gauge field can realize
this symmetry transformation but

\noindent $\bullet$ The delta function $\delta^3(y)$ need be treated carefully in the second quantization
scheme.

\noindent $\bullet$ $m^2\rightarrow -m^2$, leading to tachyon.

\noindent $\bullet$ This symmetry can not be realized in a non-abelian gauge theory.

\noindent $\bullet$ The boundary conditions need be treated carefully. For instance, the boundary conditions
at $x=\infty$ makes $H^y<0$ in quantum mechanics, and makes it equal to $-iH^x$ in quantum field theory.

This concludes our discussion on the 't Hooft-Nobbenhuis symmetry.

\subsec {Kaplan-Sundrum symmetry}

This symmetry is quite similar to the 't Hooft-Nobbenhuis symmetry.
Kaplan and Sundrum \ks\ propose that to each matter field $\psi$ there is
a ghost companion $\tilde{\psi}$, the Lagrangian is
\eqn\ksl{{\cal L}=\sqrt{-g}\left({M_p^2\over 2}R-\Lambda\right)
+{\cal L}_{\rm matter}(\psi,
D_\mu)-{\cal L}_{\rm matter}(\tilde{\psi},D_\mu),}
where the form of ${\cal L}_{\rm matter}(\psi)$ and the form of ${\cal L}_{\rm matter}
(\tilde{\psi})$ are identical. We see that the name ghost is appropriate since
the kinetic term of $\tilde{\psi}$ has a wrong sign.

Ignoring gravity for a while, there is a symmetry between $\psi, \tilde{\psi}$:
\eqn\kss{P: \psi\rightarrow \tilde{\psi},\quad \tilde{\psi}\rightarrow \psi.}
Under this symmetry, $H\rightarrow -H$, since $H=H(\psi)-H(\tilde{\psi})$.
Namely $PH=-HP$ or $\{P, H\}=0$. For a vacuum $|0\rangle$, $P|0\rangle =
|0\rangle$, then
\eqn\zeroeks{\langle 0|\{P,H\}|0\rangle=2\langle 0|H|0\rangle=0,}
thus if $|0\rangle$ is an eigen-state of $H$, $H|0\rangle =0$.

When gravity is introduced, this symmetry is broken. Let $P: g_{\mu\nu}
\rightarrow g_{\mu\nu}$, the Hamiltonian
\eqn\hdha{{\cal H}=-G^{ij,lk}{\delta^2\over \delta h_{ij}\delta h_{lk}}
-2\Lambda -2T_{00}+^{(3)}R}
does not anti-commute with $P$. If the wave function $\Psi$ has $P\Psi=\Psi$
then
\eqn\invps{\eqalign{{\cal H}P\Psi&=(-G^{ij,lk}{\delta^2\over\delta h_{ij}\delta h_{lk}}
+^{(3)}R-2\Lambda -2T_{00})P\Psi\cr
&=(-G^{ij,lk}{\delta^2\over\delta h_{ij}\delta h_{lk}}
+^{(3)}R-2\Lambda +2T_{00})\Psi,}}
we deduce $T_{00}\Psi=0$.

As an example, consider a scalar $\phi$, then the ghost companion is $\tilde{\phi}$
\eqn\sgs{{\cal L}=\half (\p_\mu\phi)^2-\half m^2\phi^2-\lambda\phi^4+
\half(\p_\mu\tilde{\phi})^2-\half m^2\tilde{\phi}^2-\lambda\tilde{\phi}^4.}

Fixing $g_{\mu\nu}$, the path integral in a quantum theory is
\eqn\pathig{\int [d\phi d\tilde{\phi}]e^{iS(\phi)-iS(\tilde{\phi})}.}
The propagator of $\phi$ is  forward with positive energy, thus $i\epsilon$
prescription is used, while the propagator of $\tilde{\phi}$ is backward for
positive energy, $-i\epsilon$ prescription is used. We then infer
$S_{eff}(\phi,\tilde{\phi})=S_{eff}(\phi)-S_{eff}(\tilde{\phi}).$ In particular,
if there is a term $\Lambda\int\sqrt{-g}$ in $S_{eff}(\phi)$, this term
is canceled by a term in $S_{eff}(\tilde{\phi})$.

Now consider the effect of quantum gravity. The interesting aspect of
this symmetry is
that the effect of gravity will introduce a small $\Lambda$. Since there is no
ghost companion of $g_{\mu\nu}$, the quantum fluctuation of the metric introduces
a term $\Lambda\int\sqrt{-g}$. Let the cut-off be $\mu$, we must have
\eqn\eeflam{\Lambda\sim \mu^4.}
Since $\Lambda\sim (2\times 10^{-3}ev)^4$, $\mu\le 2\times 10^{-3}ev$, or
$\mu^{-1}\ge 30$ microns. We know that Newtonian gravity is tested above this
scale, so it is possible that quantum gravity may break Newtonian gravity below
this scale.

We see that the local qauntum effects of $\psi$ and $\tilde{\psi}$ cancel exactly,
while non-local effects may not cancel, so it is possible to have a contribution
to $\Lambda$: $\Lambda\rightarrow \Lambda +\mu^6/M_p^2$, this is much smaller
than $\mu^4$.

If $\psi$ and $\tilde{\psi}$ are coupled, the vacuum is unstable. For example, for
$g^2\phi^2\tilde{\phi}^2$, there is a process $|0\rangle \rightarrow
\phi+\phi+\tilde{\phi}+\tilde{\phi}$, the amplitude is divergent:
\eqn\divam{P_{0\rightarrow \phi^2\tilde{\phi}^2}=g^2\int\prod d^4p_id^4k_i\prod
\delta (p_i^2-m^2)\delta (k_i^2-m^2\delta^4(\sum (p_i+k_i))=\infty.}
Let $s=(p_1+p_2)^2$, the integral in the above amplitude is divergent, we need
to introduce a cut-off for $s$, $s_{\rm max}$, and a cut-off for $p_1^0\le \epsilon$,
then
\eqn\cutf{P_{0\rightarrow \phi^2\tilde{\phi}^2}\sim g^2\epsilon^2s_{\rm max}.}

Even without direct coupling between a field and a ghost field, their coupling to
the metric also triggers instability. For example, we estimate
\eqn\gast{P_{0\rightarrow \gamma\gamma\tilde{\gamma}\tilde{\gamma}}\sim
{1\over 4\pi}({1\over 8\pi})^2{\mu^8\over M_p^4}\sim
2\times 10^{-92}({\mu\over 2\times 10^{-3}ev})^8(cm^3\times 10Gyr)^{-1}.}
This term is negligible.

It is interesting to notice that the force between a field and its ghost
companion is repulsive. Next, we consider the effect of potential of scalar
fields. Suppose there are two local minima in $V(\psi)$, then there are two local
maxima in $-V(\tilde{\psi})$, if $\psi$ runs to the global minimum, $\tilde{\psi}$
runs to the global maximum, the values cancel exactly.

\subsec {Symmetry of reversing sign of the metric}

Recai Erdem considered this kind of symmetry \ErdemYD.
Consider action in D spatial dimensions
\eqn\ddein{S(g)={1\over 16\pi G}\int\sqrt{(-1)^Dg}R,}
this is a generalization of the Einstein-Hilbert action. We demand this action
be invariant under the reflection of the metric $g_{AB}\rightarrow -g_{AB}$.
Since $R\rightarrow -R$, we have $\sqrt{(-1)^Dg}R\rightarrow -\sqrt{(-1)^{D+D+1}g}
R$. Namely
\eqn\signc{(-1)^{D+\half+1}=(-1)^{D/2},}
or $(-1)^{(D+1)/2}=-1$, we deduce $D+1=2(2n+1)$. When $n=0$, the dimensionality
of space-time is 2, but $D+1=4$ is not a solution.

Since $(-1)^{(D+1)/2}=-1$, we know that $\sqrt{-g}$ changes sign, and the cosmological
constant term is not invariant, thus forbidden by this symmetry.

This symmetry is preserved in the action of a scalar field
\eqn\scaer{-\int d^{D+1}x\sqrt{-g}\half g^{AB}\p_A\phi\p_B\phi.}
The action of a fermionic field is not invariant, unless we demand $\gamma^A
\rightarrow -\gamma^A$. The worst thing is that the stress tensor
$T_{AB}=\p_A\phi\p_B\phi-\half g_{AB}{\cal L}$ has a invariant part so the
vacuum expectation of $T_{00}$ is not vanishing.

\subsec {Scaling invariance in $D>4$}

There may be many such approaches, a typical one was proposed by Wetterich \wett.
He postulates a dilatation field $\xi$, when we rescale the metric $g_{AB}
\rightarrow \lambda^{-2}g_{AB}$, $\chi\rightarrow \lambda^{(D-3)/2}\xi$, the action
\eqn\wetta{S(\xi)=\int \sqrt{-g}(-\half \xi^2R+{\alpha\over 2}\p^\mu\xi\p_\mu\xi)}
is invariant. If there is a potential $V(\xi)$, we require $\lambda^{-D}V(\lambda^{(D-2)/2}
\xi)=V(\xi)$. If $V(\xi)=\xi^\nu$, then $\nu=2D/(D-2)$. When $D=4, 6$, $\nu$
is an integer. If $D>6$, $V=0$.

Let $\Omega$ be the internal volume, $\Omega=\int d^{D-4}x\sqrt{h}$, let $\chi
=\Omega^\half\xi$, the effective four dimensional action is
\eqn\fwett{\Gamma=\half\sqrt{-g}\chi^2 R+{\alpha\over 2}\int\sqrt{-g}(\p\chi)^2.}
There is no term $\lambda\chi^4$, and $\chi=\chi_0$ is a solution. However,
there is a problem of stability.

\newsec{Anthropic principle}

If the anthropic principle is the reason for $\Lambda\sim 0$, one of the necessary
conditions is that, $\Lambda$ is either a continuous variable, or if
it is discrete, the interval $\Delta\Lambda$ must be sufficiently small.

Bousso and Polchinski are probably the first to point out that the second
possibility is realized in string theory \bp (see also \ClineHU\ for a review). This is the beginning of studies on the
string landscape.

\subsec{Bousso-Polchinski scenario}

There are many totally anti-symmetric 3-form fields in string theory, for instance,
there is a membrane coupled 3-form $A_{\mu\nu\rho}$ in M-theory. In the type IIA
string theory, the field $C^{(3)}_{\mu\nu\rho}$ is coupled to D2-branes. When IIA
(B) theory is compactified on a Calabi-Yau manifold $CY_3$, other $C$ fields
induces a number of 3-form fields. Let $\Sigma_{p-2}$ is a $p-2$ circle in $CY_3$,
we have
\eqn\indth{\int_{\Sigma_{p-2}}C^{p+1}\rightarrow C_{\mu\nu\rho},}
different $p$ and different $\Sigma_{p-2}$ gives rise to a different 3-form field
in 4 dimensions.

For instance, in IIA string theory, $C^{(5)}$ on $\Sigma_2$, $C^{(7)}$ on $\Sigma_4$;
in IIB string theory, $C^{(6)}$ on $\Sigma_3$, but there is no $\Sigma_5$ and
$\Sigma_1$, so $C^{(4)}$ and $C^{(8)}$ do not introduce 3-form fields.

Let us focus on a single 3-form field $C_{\mu\nu\rho}$, its four form strength is
$F_4=4dC$ with an action
\eqn\foura{S=\int\sqrt{-g}\left({1\over 2\kappa^2}(R-2\Lambda_0)-{Z\over 2\times
4!}F_4^2\right)+S_{brane},}
where $Z$ is a normalization constant. The equation of motion
\eqn\eomf{\p_\mu(\sqrt{-g}F^{\mu\nu\rho\sigma})=0,}
has a solution
\eqn\soluf{F^{\mu\nu\rho\sigma}=C\epsilon^{\mu\nu\rho\sigma}/\sqrt{-g},}
this leads to
\eqn\lamr{\rl=\rho_{\Lambda_0}+{Z\over 2}C^2.}

It may be not obvious that the constant $C$ is quantized. Indeed if space-time
is really four dimensional, $C$ is a continuous parameter. If space-time is
higher dimensional, $C$ is quantized. For example, in the 11 dimensional M
theory, we have
\eqn\mfour{S=2\pi M_{11}^9\int\sqrt{-g}(R-{1\over 2\times 4!}F_4^2).}
A M5-brane is coupled to $A_6$, the dual of $A_3$. The coupling is
\eqn\mfc{2\pi M_{11}^6\int A_6,}
this coupling leads to
\eqn\dqc{2\pi M_{11}^6\int_{M_7}F_7=2\pi n,}
this is the standard Dirac quantization condition. Now, $F_7$ is dual
to $F_4$
\eqn\sfd{F_7^{\mu_1\dots\mu_7}={1\over 4!}\epsilon^{\mu_1\dots\mu_7
\nu_1\dots\nu_4}F_{\nu_1\dots\nu_4}.}
Let $\mu_1\dots\mu_7\in M_7$, then $F_7=\epsilon^{\mu_1\dots\mu_7}F_0$,
and $2\pi M_{11}^6V_7F_0=2\pi n$, thus $F_0$ is quantized
$F_0=n/(M_{11}^6V_7)$. The reduced action is
\eqn\reda{S=2\pi M_{11}^9V_7\int \sqrt{-g}(R-{1\over 2\times 4!}F_4^2),}
thus
\eqn\kkaz{{1\over 2\kappa^2}=2\pi M_{11}^9V_7=Z, \quad
F_0={n\over M_{11}^6V_7}={2\pi nM_{11}^3\over Z}.}

For a M2-bane, the charge is $e=2\pi M_{11}^3$, namely the coupling is
$2\pi M_{11}^3\int A_3=e\int A_3$. We have $F_0=ne/Z$. Now, for solution
$F^{\mu_1\dots\mu_4}=C\epsilon^{\mu_1\dots\mu_4}/\sqrt{-g}$, we have
\eqn\fouq{F_0={1\over 4!}\epsilon_{\mu_1\dots\mu_4}F^{\mu_1\dots\mu_4}=C
={ne\over Z},}
we see that $C$ is quantized. Our conclusion is that, if the action is
\eqn\fouract{S(A)=-{Z\over 2\times 4!}\int F_4^2,}
and the charge $e\int A_3$, then $C=ne/Z$.

In formula
\eqn\enq{\rl=\rho_{\Lambda_0}+{Z\over 2}C^2=\rho_{\Lambda_0}+{n^2e^2\over
2Z},}
the dimensions of the constants are $[e]=M^3$, $[Z]=M^2$. If $e^2\sim
M_p^6$, $Z\sim M_p^6L^7$, then $\Delta \Lambda\ge M_p^{-3}L^{-7}$.
Let $M_p^{-3}L^{-7}\sim (2\times 10^{-3}{\rm eV})^4$, then $L^{-7}\sim 10^7
{\rm GeV}$, or $L^{-1}\sim 1{\rm GeV}$. This is possible, but there is a problem:
suppose the bare cosmological constant $\Lambda_0\sim -M_p^4$ thus
$n^2\sim M_p^4Z/e^2$ so $\Delta\Lambda\propto ne^2/Z\sim M_p^2e/\sqrt{Z}$.
Use $e\sim M_p^3$, $\sqrt{Z}\sim M_p^{9/2}L^{7/2}$ we infer
$\Delta\Lambda\sim \sqrt{M_p}L^{-7/2}\sim (2\times 10^{-12}{\rm GeV})^4$,
$L^{-7}\sim 10^{-64}{\rm GeV}^7$, or $L^{-1}\sim 10^{-9} {\rm GeV}=1$eV, this is too
small, or $L$ is too large.

To solve the above problem, Bousso and Polchinski propose to consider
multiple 3-form fields $C^a_{\mu\nu\lambda}$, $a=1,\dots , J$. Let $Z_i=1$
and $e\rightarrow q_i$, then
\eqn\mull{\Lambda=\Lambda_0+\half\sum_{i=1}^J n_i^2q_i^2.}
Thus, the additional term is a distance squared in the n-dimensional Euclidean
space with fundamental lattice spacing $q_i$. The volume of a fundamental cell
is
\eqn\fundc{\Delta V=\prod_{i=1}^Jq_i.}
Now, consider a shell in between $r$ and $r+\Delta r$, the volume of this shell
is
\eqn\shellv{\Omega_Jr^{J-1}\Delta r.}
If this volume is greater than $\Delta V$, then there is at least one lattice
point falling into this shell.

Let $\Lambda_0=-r^2/2$, and
\eqn\phyl{\Lambda=\half\left(\sum_{i=1}^Jn_i^2q_i^2-r^2\right)=\Delta\Lambda.}
Let $\{ n_i\}$ fall in between $r$ and $r+\Delta r$, there must be
$r\Delta r=\Delta\Lambda$, $\Delta r=r^{-1}\Delta\Lambda$,then
\eqn\manf{\Omega_Jr^{J-1}\Delta r=\Omega_J r^{J-2}\Delta\Lambda=
\Omega_J(2|\Lambda_0|)^{(J-2)/2}\Delta\Lambda.}
This volume must be no smaller than the volume of a fundamental cell, so
\eqn\fless{\prod_{i=1}^Jq_i\le \Omega_J(2|\Lambda_0|)^{(J-2)/2}\Delta\Lambda.}
The physical $\Lambda$ is equal to $\Delta\Lambda$, the smallest allowed
value is
\eqn\smlam{\Lambda={\prod_{i=1}^Jq_i\over \Omega_J(2|\Lambda_0|)^{J/2-1}},}
wee see that if $q_i<\sqrt{|\Lambda_0|}$, it is easy to have a very small
$\Lambda$. $\Omega_J$, the volume of the unit sphere in J-dimensional
Euclidean space, is $2\pi^{J/2}/\Gamma(J/2)$. Let $q_i=100^{-1}\sqrt{|\Lambda_0|}$,
we find that when $J\sim 100$, we have roughly
\eqn\smlamb{{\Lambda\over |\Lambda_0|}\sim 10^{-120}.}
This is how the Bousso-Polchinski scenario solves the cosmological constant
problem.

Just how to generate the right quantum numbers $n_i$? There is the Brown-Teitelboim
mechanism \refs{\BrownDD, \BrownKG} to use. The BT mechanism is similar to electron-positron creation in
a electric field in 1+1 dimensions.

In 1+1 dimensions , let there be a pair of charges $q$ and $-q$, $\p_xE=
-q\delta(x)+q\delta(x+L)$, the solution is
\eqn\SNfk{
E(x)=
\cases{
E & $x<0$;\cr
E-q & $0<x<L$;\cr
E & $X>L$.
}}
So before the pair-creation, the electric field strength is $E$ everywhere, and becomes
$E-q$ in between the two charges after creation. It decreases until $E$ reaches the minimum
$E_{\rm min}=E-[E/q]q\le q$. The rate of pair-creation is
\eqn\pairc{P\propto \exp(-{c\over q^2E^2}).}

Similarly, a spherical membrane is created in 4 dimensions in a background of $F_4$ and
$F_4\rightarrow F_4-C=F_4-q$, and the cosmological constant is shifted as
\eqn\lamshi{\Lambda\rightarrow \Lambda+\half (F_4\pm q)^2-\half F_4^2.}
But this process is very slow and in the end we need the help of the anthropic principle.
The difference between membrane creation and pair-creation is that the membrane creation
can either decrease $F_4$ or increase it, however the probability of decreasing $\Lambda$
is greater than that of increasing $\Lambda$.

In the original BT model, there is an empty universe problem. This is
because in the original BT, there is only one kind of membrane, thus
$q$ is required to be very small in order to have a very small $\Lambda$. But before the membrane
creation $\Lambda-\Delta\Lambda$ is also very small, thus the decay rate of this previous universe
is extremely small such that inflation in this phase makes the universe almost empty.

This problem is eliminated in the BP model, since before the last transition, the previous $\Lambda$
is large and the inflation period is short. The major problem remains in the BP model is the
moduli stablization.

\subsec{KKLT scenario}

In the Bousso-Polchinski scenario, charges $q_i$ as well as normalization constants $Z_i$ are all
moduli dependent.  To solve this problem, Giddings, Kachru and Polchinski proposed to introduce
D7-branes as well as fluxes of $H^{(3)}$ and $F^{(3)}$ in compactification \gkp.

A D7-brane is specified by a complex function $\tau$ over $CY_3$ \cvafa, $H^{(3)}$ and $F^{(3)}$ form
a complex 3-form $G^{(3)}=F^{(3)}-\tau H^{(3)}$. Let
\eqn\fform{\tilde{F}^{(5)}=F^{(5)}-\half C^{(2)}\wedge H^{(3)}+\half B\wedge F^{(3)}=
(1+*)[d\alpha\wedge dx^0\wedge\dots \wedge dx^3],}
then
\eqn\mofx{d\Lambda +{i\over {\rm Im} \tau}d\tau\wedge Re\Lambda =0,\quad \Lambda=e^{4A}_{*(6)}G^{(3)}
-i\alpha G^{(6)},}
where $_{*(6)}$ is the dual operation in the 6 manifold $CY_3$, $A$ is the warp factor in the warped
metric
\eqn\warpm{ds^2=e^{2A(y)}\eta_{\mu\nu}dx^\mu dx^\nu +e^{-2A(y)}ds_{CY}^2,}
and $y$ are coordinates on $CY_3$. This helps to fix many of the moduli except for all the Kahler moduli,
for example, the complex scalar
$\rho$ associated with the scale of the Calabi-Yau $CY_3$. This is the scalar appearing in a no-scale
supergravity theory.

The warp factor in the GKP model offers us a means to solve the hierarchy problem, as in the Randall-Sundrum
scenario \randall.

Consider a complex structure moduli, corresponding to a conifold ($z$ is the size of $S^3$ in $S^2\times S^3$),
let
\eqn\fluxess{{1\over 2\pi\alpha'}\int_{A=S^3} F^{(3)}=2\pi M, \quad {1\over 2\pi\alpha'}\int_B H^{(3)}=-2\pi K,}
then $z$ is stablized to
\eqn\zsab{z\sim e^{-{2\pi K\over M g_s}}, \quad e^{A_{\rm min}}=e^{-{2\pi K\over 3M g_s}}.}
KKLT first considered a single Kahler moduli \kklt, and pointed out that there are two effects to help to
fix the radial Kahler moduli.

\noindent (a) If there is a four dimensional complex sub-manifold in $CY_3$, then wrapping the
Euclidean D3-branes on this sub-manifold forms D-instantons. The contribution to the superpotential has
a form
\eqn\dinss{W\sim \exp (2\pi i\rho),}

\noindent (b) Fluxes induce D7-branes, and the field theory on D7-branes is $N=1$ super Yang-Mills
theory, and we have
\eqn\dscoup{{4\pi\over g_{YM}^2}={\rm Im}\rho.}
There is a superpotential
\eqn\dssp{W=A\exp({2\pi i\rho\over N_c}).}
Let
\eqn\insspp{W=W_0+Ae^{2\pi ia\rho},\quad K=-2\ln (-i(\rho-\bar{\rho})),}
then $DW=0$ leads to
\eqn\sppzero{W_0=-Ae^{-a\sigma_c}(1+{2\over 3}a\sigma_c),}
where $\sigma_c={\rm Im} \rho$, and we get
\eqn\poeti{V=-a^2A^2e^{-2a\sigma_c}/(6\sigma_c)<0,}
thus the cosmological constant is negative and the universe is an anti-de Sitter space $AdS_4$, supersymmetry
is unbroken.

If fluxes are not balanced, we need to introduce anti-branes, namely $\overline{D3}$ branes. The existence
of anti-branes breaks supersymmetry, and all moduli will be fixed. Let $a_0=\exp(2A_0)$ be the warp factor,
then  $\overline{D3}$ contribute to the energy density  by a factor
\eqn\delpoet{\Delta V={2a_0^4T_3\over g_s^4}{1\over ({\rm Im}\rho)^3},}
where $T_3$ is the D3-brane tension. Thus, $V\rightarrow V+\Delta V$. Although $\Delta V$ is small, but
$V+\Delta V>0$, the anti-de Sitter space is modified to become a de Sitter space.

Next, KKLT argue that although the de Sitter space is meta-stable, the life time is much longer than the age
of our universe, but smaller than the Poincare recurrence time $t\sim \exp(S_0)\sim \exp(10^{120})$.
All the stable and meta-stable anti-de Sitter spaces and de Sitter spaces form the string landscape.

There have been a lot of efforts invested to study statistics of the string landscape. There are in principle
infinitely many meta-stable vacua of the Bousso-Polchinski type, but many of them are not reliable. In the
KKLT models, are there upper limits for the flux numbers of $F^{(3)}$ and $H^{(3)}$? Some estimates
that there are at least \DouglasUM
\eqn\stland{10^{500}}
meta-stable vacua.

Susskind \SusskindKW\ invented terminology ``string landscape'', and argued that our universe may be multiverse consisting numerous
regions in which physics varies. The implication of the multiverse is
to be discussed in the following subsection.

\subsec{Populating the landscape and anthropic interpretations}
\subseclab\subsecAnthropic

The vast string landscape itself does not lead to an anthropic
interpretation for cosmological constant. Instead, the landscape must
be populated \foot{Except that, if one (much more aggressively) assumes every self-consistent
mathematical structure is automatically ``populated'' by a higher
level of the multiverse. See \TegmarkDB. }. In other words, one needs a mechanism to produce
different universes in a (at best connected patch of) multiverse.

There are different approaches to populate the landscape. For example,
Hawking and Hertog \HawkingUR\ pointed out that the wave function of
universe is one populating method. Different observers live in their
different histories, and they are summed over in the no boundary path
integral. In other words, observers in different universes live in
different decohered branches of a single wave function. On the other
hand, here we shall mainly discuss another better studied scenario:
eternal inflation.

When the quantum fluctuation of of the inflaton $\delta\varphi ={H\over2\pi}$ is larger that the rolling of
$\varphi$ in a Hubble time, namely $\delta\varphi ={H\over2\pi}>\Delta\varphi=\dot{\varphi}H^{-1}$, eternal
inflation occurs \refs{\SteinhardtEternal, \VilenkinXQ}. Whether
eternal inflation really happens is a matter of controversy
\refs{\PageNT, \ArkaniHamedYM, \HuangZT, \WangCS}. For instance,
the weak gravity conjecture \ArkaniHamedDZ\ may prohibit it to occur \HuangZT.

On the other hand, de Sitter space itself may not be eternal
\refs{\DysonPF,\GoheerVF}, it has a finite life time, for instance its
life time can not be longer than the Poincare recurrence time if we
view this space-time has finite dimension of the Hilbert space. It is
argued in \refs{\DysonPF,\GoheerVF} that a universe with conditions
all the same as our universe except the CMB temperature is higher is
more likely, the probability of its occurrence is $\sim e^{S_i-S}$
where $S=$ entropy of the pure de Sitter space, $S_i=$ entropy of a
particular universe. It is also pointed out that a de Sitter space is
a resonant state in the multiverse.

Anyway, eternal inflation is still a possibility which is
semi-classically well defined. Before a more complete quantum theory
of gravity clarifies all the subtleties, we have to take eternal
inflation seriously. If eternal inflation indeed happens, our universe
is in a small part of the eternal inflating universe. The situation is
like our earth is a small part of our observable universe. This
provides a playground for the anthropic principle.

The validity of the anthropic principle is very controversial, there
are a number of problems need to be addressed, including:

\noindent $\bullet$ The measure of the multiverse.

It is intuitive to imagine that a ``typical'' vacuum in the landscape
is kind of vacuum that is ``realized'' in the multiverse most
frequently. However, it is rather difficult to realize this idea. A
number of
measures have been proposed in the literature, including the volume
based measure \GarrigaAV\ (see \DeSimoneBQ\ with scale-factor cutoff),
the local measure \refs{\BoussoEV, \BoussoGE}, the Liouville measure
\refs{\GibbonsXK, \GibbonsPA, \LiRP}, the stochastic measure \LiUC,
and so on. However, there is so far no principle to determine which
measure to use. On the other hand, the measure problem is also closely
related with other problems, as discussed below.

\noindent $\bullet$ How to correctly apply Bayesian statistics?

The measure problem addresses the {\it a priori} probability from the
theoretical point of view. After that the anthropic probability must
be assigned to have anthropic predictions. The anthropic probability
is proportional to the product of the {\it a priori} probability and a
conditional probability for the theory to be observed, according to
the Bayesian statistics:
\eqn\ATBayes{P({\rm theory~}x|{\rm
selection}) = {P({\rm selection}|{\rm theory~}x) P({\rm theory~}x)
\over \sum_y P({\rm selection}|{\rm theory~}y) P({\rm theory~}x)}.}
The ``selection'' here can be understood as an anthropic
effect. Unfortunately, there is no principle stating how to properly
define the probability from anthropic selection effect. For example,
Page \refs{\PageNT, \PageBT, \PageMX} assumes typical observers to
make observations, Hartle and Srednicki \HartleZV\ argues it is
exactly us human observers instead of a typical observer that should
be used here. On the other hand, some of us \LiDH\ argue a compromise
treatment seems more natural, that the anthropic probability should be
assigned by the probability of existence, instead of the number of
observers. Alternatively, Bousso, Harnik, Kribs and Perez \BoussoKQ\ also
proposed a more operable entropic measure to mimic the anthropic measure.

\noindent $\bullet$ The problem of Boltzmann brains.

In the concluding part of \Boltzmann, Boltzmann described the idea of
his assistant Schuetz: we may come from thermal fluctuations in an
extremely large universe, whereas the whole universe is in a thermal
equilibrium state. Following this idea, one eventually arrives at a
paradox: the so called Boltzmann problem.

Now we are not going to talk about whether our galaxies, stars and
planets originate from thermal fluctuations or not. Instead, we refer
ourselves as ``human observers'', who live in a low entropy
environment and use entropy increasing to maintain their
lives. However, if the universe is large enough and in a thermal state
(indeed in an asymptotically de Sitter universe with Gibbons-Hawking
temperature), there will also be other observers originated completely
from thermal fluctuations, who are by pure chance be in a low entropy
state themselves, and isolated from the thermal equilibrium
environment. These observers are called Boltzmann brains, or freak
observers.

In principle, we can not assert  we are not Boltzmann brains but by the
next moment we will be almost sure, by observing that ourselves are
not returning to the thermal equilibrium. If we were Boltzmann brains,
this probability is exponentially small. However, why are we not
Boltzmann brains?

As an example, if our universe starts from a big bang and end up
asymptotically de Sitter, in a finite comoving volume the number of
human observers is finite, because the available matter entropy
difference will eventually be used up. \foot{To be more precise, for a
human observer, one at least must be able to remember something. To
prepare the memory to remember anything, there has to be entropy
increasing. When entropy is maximized, there is no available entropy
difference to remember anything.} However, the number of
Boltzmann brains is infinite, as long as de Sitter space does not
decay. If we were typical observers, we should have inferred that we
were Boltzmann brains, not human observers. This contradicts with the
observation in the last paragraph, that we are not Boltzmann
brains. Similar paradox happens in an eternally inflating universe.

Even worse, Bousso, Freivogel, Leichenauer and Rosenhaus \BoussoYN\
argued that time will end for eternal inflation with the assumption of
typicality. This is too bad since no observer is even worse than
too many observers.

\noindent $\bullet$ Is there any definite prediction of the anthropic principle?

It is in debate whether anthropic principle could make
predictions. Some may say that Weinberg has already predicted the
cosmological constant from anthropic principle and others may argue
that what Weinberg actually did is that he cannot calculate
cosmological constant from first principle, and simply assumes a
probable value from existence of human.

More generally, in principle one can construct a measure for the
whole landscape, and predict where is most probable for human
beings. If this prediction agrees with current experiments, we may
take the view point that anthropic principle indeed make
predictions. But again, the answer depends on the definition of
prediction, because the prediction that anthropic principle does is
logically different from predictions of traditional science.

To understand anthropic principle, it is good to return to the much
better understood question why we live on the earth, with an
environment surprisingly suitable for human beings. However, even in
this much easier case, we can not yet make predictions or answer
whether we live on a most typical planet suitable for
intelligence. The way for anthropic principle to make predictions for
fundamental physics, is thus much longer.

\noindent $\bullet$ There is evidence that there are just too many meta-stable vacua (Witten: M2-branes
$\rightarrow SU(n)\times SU(n)/({\rm discrete\ symmetry}) \rightarrow$ $C_{m+n-1}^m\times C_{m+n-1}^n$ vacua.
Take $m=n\sim 10^{100}$, then the number of vacua of this type is $(C_{2n}^n)^2\sim 2^{4n}\sim
10^{10^{100}}$, a googolplex!)

Although there are many problems with the anthropic principle, we can
not yet rule it out.

\newsec{Tuning mechanisms}

\subsec{Brane versus bulk mechanism}

We will mainly explain the work of Kachru, Schulz and Silverstein \KachruHF. They work in brane world embedded in a 5 dimensional
space-time. They introduce a tuning scalar field, but there is always a singularity in the bulk. As Witten
correctly pointed out, such a singularity can not be accepted based on general physics principle (Otherwise
one can introduce just about anything in our world, such as a monopole in the usual Maxwell theory).

Let us start with the action in 5 dimensions
\eqn\tunact{S=\int d^5x \sqrt{-G}(R-{4\over 3}(\nabla\phi)^2-\Lambda e^{a\phi})+\int d^4x\sqrt{-g}(-f(\phi)),}
where $f(\phi)=Ve^{b\phi}$.
If for any $V$, one can always find a flat 4 dimensional space-time solution, then the tuning mechanism is successful.
This is because $V$ is quantum corrected (with proof of the form of $f(\phi)$ be invariant). If $f(\phi)$ is
not invariant, one need to show a flat solution still exists.

The equations of motion derived from \tunact\ are
\eqn\eomtun{\eqalign{&\sqrt{-G}(R_{MN}-\half G_{MN}R)-{4\over 3}\sqrt{-G}(\nabla_M\phi\nabla_N\phi-\half
G_{MN}(\nabla\phi)^2)+\cr
&\half [\Lambda e^{a\phi}\sqrt{-G}G_{MN}-\sqrt{-g}g_{\mu\nu}\delta^\mu_M\delta^\nu_N\delta
(x_5)]=0,\cr
&\sqrt{-G}({8\over 3}\triangle \phi-a\Lambda e^{a\phi})-bV\delta(x_5)e^{b\phi}\sqrt{-g}=0.}}
Let the metric be
\eqn\fivdm{ds^2=e^{2A(x_5)}\eta_{\mu\nu}dx^\mu dx^\nu +dx_5^2,}
then
\eqn\eomsa{\eqalign{&{8\over 3}\phi''+{32\over 3}A'\phi'-a\Lambda e^{a\phi}-bV\delta(x_5)
e^{b\phi}=0,\cr
&6(A')^2-{2\over 3}(\phi')^2+\half \Lambda e^{a\phi}=0,\cr
&3A''+{4\over 3}\phi'^2+\half e^{b\phi}V\delta(x_5)=0.}}

To solve the above equations, let $A'=\alpha \phi'$, consider the following cases separately.

\noindent $\bullet$ $\Lambda=0$.

When $x_5\ne 0$, from
\eqn\scaeq{6\alpha^2\phi'^2={2\over 3}\phi'^2,}
we deduce $\alpha=\pm 1/3$ and
\eqn\quas{\phi''\pm {4\over 3}\phi'^2=0.}
Thus
\eqn\sfivs{\phi=\pm {3\over 4}\ln |{4\over 3}x_5+c|+d.}
This solution has a singularity at $x_5=-3c/4$.

\noindent Solution 1, let
\eqn\sanfen{\alpha=
\cases{{1\over 3}
& $x_5<0$;\cr
-{1\over 3} & $x_5>0$,
}}
thus
\eqn\sscaseo{\phi(x_5)=
\cases{{3\over 4}\ln |{4\over 3}x_5+c_1|+d_1 &$x_5<0$;\cr
-{3\over 4}\ln|{4\over 3}x_5+c_2|+d_2 & $x_5>0$.}}
The continuity at $x_5$ requires
\eqn\contx{{3\over 4}\ln |c_1|+d_1=-{3\over 4}\ln |c_2|+d_2.}
From equations of motion we obtain
\eqn\paraeq{\eqalign{&{2\over c_2}=[-{3b\over 8}-\half]Ve^{bd_1}|c_1|^{{3\over 4}d},\cr
&{2\over c_1}=[-{3b\over 8}+\half]Ve^{bd_1}|c_1|^{{3\over 4}d}.}}
The solution for the parameters always exists no matter what value of $V$ is. For an arbitrary
$f(\phi)$, we have
\eqn\arbss{\eqalign{&{8\over 3}(\phi'_2(0)-\phi_1'(0)=f'(\phi(0)),\cr
&3(\alpha_2\phi_2'(0)-\alpha_1\phi_1'(0)=-\half f(\phi(0)).}}
Or
\eqn\altarb{\eqalign{&-{8\over 3}(c_1^{-1}+c_2^{-1})=f'({3\over 4}\ln|c_1|+d_1),\cr
&c_2^{-1}-c_1^{-1}=-\half f({3\over 4}\ln |c_1|+d_1),}}
solution to these equations always exists.

Since the effective Newtonian constant is
\eqn\efffnew{G\sim \int dx_5e^{2A(x_5)}\sim \int dx_5|{4\over 3}x_5+c_1|^\half.}
This contant diverges if there is no cut-off on $x_5$, this is so when $c_1<0$ for $x_5<0$,
thus we let $c_1>0$, then the lower limit of the integral of $x_5$ is $\bar{x}_5=-3c_1/4$.
This lower limit is a singularity, however.
Similarly, for $x_5>0$, $c_2<0$, there is also a singularity.

For an arbitrary function $f(\phi)$. we require
\eqn\arbfc{f(\phi(0))>0,\quad -{4\over 3}<{f'\over f}<{4\over 3}.}

\noindent $\bullet$ Fluctuations

Let
\eqn\metf{g_{\mu\nu}=e^{2A}\eta_{\mu\nu}+h_{\mu\nu},}
then
\eqn\flucmet{h_{\mu\nu}\propto |{4\over 3}x_5+c|^\half,}
there is also a singularity, although $h_{\mu\nu}\rightarrow 0$, but
\eqn\singx{x_5\rightarrow
\cases{-{3\over 4}c_1
& $\phi\rightarrow -\infty$ weak coupling;\cr
-{3\over 4}c_2 & $\phi\rightarrow \infty$ strong coupling.
}}

\noindent Solution 2, in this case $\alpha_1=\alpha_2$

\eqn\phifive{\phi(x_5)=
\cases{\pm{3\over 4}\ln |{4\over 3}x_5+c_1|+d_1
& $x_5<0$ ;\cr
\pm{3\over 4}\ln |{4\over 3}x_5+c_2|+d_2 & $x_5>0$,
}}
and $b=\mp {4\over 3}$. Or
\eqn\arbf{f'(\phi(0))=\mp {4\over 3}f(\phi(0)).}
One finds
\eqn\ress{c_1=-c_2=c,\quad d_1=d_2=d,\quad e^{\mp {4\over 3}d}={4\over V}{c\over |c|}.}

\noindent $\bullet$ $\Lambda\ne0$.

In this case $\alpha=-{8\over 9a}$, and
\eqn\bphi{\phi=-{2\over a}\ln [{a\mp\sqrt{B}\over 2}x_5+d],\quad B={\Lambda\over
{4\over 3}-12\alpha^2}.}
The junction condition leads to
\eqn\vibe{V=-12\alpha \sqrt{B},\quad b={4\over 9\alpha},}
where $V$ is a function of $a$ and $\Lambda$, and is fined tuned. This is similar to the Randall-
Sundrum scenario.

When $a=0$ ($\Lambda\ne0$), let $h=\phi'$, $g=A'$, we have
\eqn\ghnonl{\eqalign{&h'+4hg=0,\cr
&6g^2-{2\over 3}h^2+\half \Lambda=0,\cr
&3g'+{4\over 3}h^2=0.}}

The case $\Lambda=0$ may be justified in string theory provided there is supersymmetry in the 5D
bulk. It is expected that we leave $f(\phi)$ to be a general function due to quantum corrections.

The major problem of this new tuning mechanism is the existence of singularities. The difference of this
mechanism from that of Randall-Sundrum is that we need to fine tune relationship between the 4D cosmological
constant and the 5D cosmological constant in the latter mechanism, while here the scalar field $\phi$
massages the correcstions to $V$ to the bulk.

There are also a number of other brane world approaches to dark
energy. For example, \refs{\ClineAK, \VinetBK} considers codimension
two branes. Other approaches include \NeupaneAs.

\subsec{Black hole self-adjustment}

This mechanism is due to Csaki, Erlich and Grojean \CsakiDM.

Introducing a black hole in the 5D bulk with metric
\eqn\fivdbl{ds^2=-h(r)dt^2+{r^2\over l^2}d\Sigma_k^2+h^{-1}(r)dr^2,}
where
\eqn\rfunc{h(r)=k+{r^2\over l^2}-{\mu\over r^2},}
and $d\Sigma_k^2$ is the metric on the maximally symmetric space with spatial curvature $k$.
Now suppose there is a three-brane, we need to require a $Z_2$ symmetry (so that there is one
black hole to the left and another black hole to the right), for $r<r_0$, the metric is given
by the functions $h(r)$, $r^2$ and $h^{-1}(r)$, and for $r>r_0$:
\eqn\newf{\tilde{h}(r)=h({r_0^2\over r}),\quad \tilde{h}^{-1}(r)=h^{-1}({r_0^2\over r}){r_0^2\over r^2},
\quad \tilde{r}=({r_0^2\over r})^2,}
namely, the metric when $r>r_0$ is
\eqn\rlarz{ds^2=-h({r_0^2\over r})dt^2+({r_0^2\over r})^2d\Sigma_1^2+h^{-1}({r_0^2\over r}){r_0^2\over r^2}dr^2.}

If the brane location $r_0=R(t)$ is a function of time, the equation of motion
\eqn\beof{\dot{\rho}+3(\rho +p){\dot{R}\over R}=0}
has a static solution $\dot{R}=\ddot{R}=\dot{\rho}=0$. The junction conditions are
\eqn\jucc{6\sqrt{h(r_0)}=\kappa_5^2\rho r_0,\quad 18h'(r_0)=-\kappa_5^4(2+3w)\rho^2r_0,}
where $w=p/\rho$. We have
\eqn\musl{\mu=-{1\over 24}\kappa_5^4(1+w)\rho^2r_0^2.}
If $\mu<0$, there is a naked singularity in the black hole solution. So we require $\mu>0$ thus
$w<-1$ and the positive energy condition is violated. Also
\eqn\rhoden{\rho=-{72\over 1+3w}{1\over l\kappa_5^2},}
this is fine-tuning. In order to avoid these drawbacks, we need to introduce new parameters. Assume that this
is a 5D Abelian gauge field $A_M$ and $A_r$ is even under $Z_2$ and $A_\mu$ odd under $Z_2$. The black hole
solution with charge $Q$ is
\eqn\chafivd{\eqalign{&h(r)=k+{r^2\over l^2}-{\mu\over r^2}+{Q^2\over r^4},\cr
&Q^4<{4\over 27}\mu^3 l^2.}}
The junction conditions are
\eqn\charjuc{\eqalign{&36({r_0^2\over l^2}-{\mu\over r_0^2}+{Q^2\over r_0^4})=\kappa_5^2\rho^2r_0^2,\cr
&36({r_0^2\over l^2}-{\mu\over r_0^2}-{2Q^2\over r_0^4})=-\kappa_5^4(2+2w)\rho^2r_0^2.}}
The solution to these equations is
\eqn\muqs{\eqalign{&\mu=2(l^{-2}+{1\over 36}\kappa_5^2w\rho^2)r_0^4,\cr
&Q^2=2(l^{-2}+{1\over 72}\kappa_5^2(1+3w)\rho^2)r_0^6..}}
The condition $Q^2\ge 0$ leads to
\eqn\condrho{\rho\le \rho_0=\sqrt{{-72\over 1+3w}}{1\over l\kappa_5^2},\quad w<-{1\over 3}.}
The condition for the existence of a horizon is $w>0$ or $w<-1$, and $\rho>\rho_-$ where
\eqn\minusrh{\rho_-={6\over l\kappa_5^2}({1\over 8w^3}(1+6w-3w^2+\sqrt{(1+w)^3(1+9w)})^\half.}
When $w<-1$, $\rho_-<\rho<\rho_0$, we find
\eqn\soluact{S=\int_{r_H}^{r_0}dr\sqrt{-g}[{1\over 2\kappa_5^2} R-{1\over 4}F^2-\Lambda +{\cal L}_{\rm matter}
\delta(\sqrt{g_{rr}(r-r_0)})].}
Since
\eqn\lmatter{\eqalign{&{\cal L}_{\rm matter}=p=(-{h'\over\kappa_5^2\sqrt{h}}-{4\sqrt{h}\over \kappa_5^2r_0})|_{r=r_0},\cr
&R=-h''(r)-6{h'\over h}-6{h\over r^2}+(2h'+12{h\over r}
)\delta(r-r_0),}}
we have
\eqn\actsol{S=\kappa_5^{-2}r_H^2h(r_H)=0,}
this tells us that the 4D effective cosmological constant is zero
$\Lambda_{\rm eff}=0$.

\noindent $\bullet$ Other cases

Take $r_0=R(t)$ and
\eqn\roft{R(t)=
\cases{R_0e^{Ht}
& $k=0$ ;\cr
\hbox{Sinh} (Ht)/H & $k=-1$;\cr
\hbox{Cosh} (Ht)/H & $k=1$,
}}
when $\Lambda_4>0$ and
\eqn\roftn{R(t)=\cos (Ht)/H,}
when $\Lambda_4<0$. The junction conditions are
\eqn\movjuc{\eqalign{&{\dot{R}^2\over R^2}={1\over 36}\kappa_5^2\rho^2-({k\over R^2}+l^{-2}-{\mu
\over R^4}+{Q^2\over R^6}),\cr
&\dot{\rho}+3H(1+w)\rho=0.}}
For $w=-1$, $\dot{\rho}=0$.

\noindent $\ast$ de Sitter

We have $w=-1$ and
\eqn\desmu{\eqalign{&\mu=0, \quad Q^2=0,\cr
&{1\over 36}\kappa_5^2\rho^2-l^{-2}=H^2>0.}}

\noindent $\ast$ Anti-de Sitter

When $w=-1$,
\eqn\andes{\eqalign{&H^2=l^{-2}-{1\over 36}\kappa_5^2\rho^2>0,\cr
&\mu=Q^2=0.}}

When $w=-{1\over 3}$,
\eqn\andessanfen{\eqalign{&\mu=-{1\over 36}\kappa_5^4\rho^2,\quad Q=0,\cr
&H^2=l^{-2}.}}

When $w=0$,
\eqn\andeswe{\mu=0,\quad Q^2={1\over 36}\kappa_5^2l^6\rho_0^2.}

For all the above cases a fine-tuning is required.

Finally, we note that Lorentz symmetry is violated. Since the metric is
\eqn\metricloren{ds^2=-hdt^2+{r^2\over l^2}d\Sigma_k^2+h^{-1}dr^2,}
the speed of gravitational wave is not equal to $c$, the speed of light. On the brane, the speed
of light is $c={dx\over dt}={\sqrt{h}l\over r}$. When transverse to the brane, the speed of light is
${dr\over dt}=h$. Detailed calculation shows that the speed of gravitational wave depends on the
parameter $E/|\vec{p}$, where $E$ and $\vec{p}$ are conserved quantities.

\newsec{Modified Gravity}

Modified gravity is now a huge category. Here we shall choose to
introduce several classes of models most related to dark
energy. Namely $f(R)$, MOND, DGP type models
and briefly mention other directions.

\subsec{$f(R)$ models}

Well before dark energy and inflation have been proposed, there have
been already attempts to replace the Ricci scalar $R$
in the gravitational action by a general function $f(R)$
\BuchdahlZZ.

It is well known that for the standard Einstein-Hilbert
action, one can either treat metric itself as dynamical variables, or
treat both metric and connection as variables when doing
variation. The former is known as the metric formulation, and the
latter is known as the Palatini formulation \refs{\Palatini,
\PalatiniEinstein} \foot{The so called Palatini formulation of
general relativity is perhaps also discovered by Einstein
\PalatiniEinstein. See a historical review \FerrarisVF\ for the story
for readers who do not speak German or Italian.}.

To generalize to $f(R)$, there are thus two possibilities, the metric
generalization or the Palatini generalization. Turns out that these
two are not identical for $f(R)$. Here we shall mainly discuss the
metric formulation. The readers interested in the Palatini formulation
are referred to the review \SotiriouRP.

It is also helpful to note that these $f(R)$ models can be related to
standard gravity with a non-minimally coupled scalar field, using
conformal transformation \refs{\WhittPD, \BarrowXH}. A conformal
transformation is not a symmetry of general relativity nor its $f(R)$
generalization. However there are ways to match observables such that
calculation on one side can be used on the other side. For these
aspects, the readers are referred to the review \MukhanovME.

Now we shall review the $f(R)$ gravity models applied to dark energy
\foot{Here we just introduce the four-dimensional $f(R)$ models.
For five-dimensional $f(R)$ gravity model, see \YGMatwo.}.
To be specific, we here review two simple models:

\noindent $\bullet$ The CDTT model

The whole thing started with the paper of Carroll, Duvvuri, Trodden
and Turner \CarrollWY. The idea is to use other Lagrangian terms
to generate accelerated solutions. Although these terms are more complicated and more unnatural than the
Einstein's cosmological term, we can not logically rule out these possibilities and need to test them.
The simplest example is
\eqn\iverser{S={1\over 16\pi G}\int d^4x\sqrt{-g}(R-{\mu^4\over R})+S_{\rm matter},}
The equation of motion is
\eqn\modiein{(1+{\mu^4\over R^2})R_{\mu\nu}-\half (1-{\mu^4\over R^2})g_{\mu\nu}R+\mu^4[ g_{\mu\nu}
-\nabla_{(\mu}\nabla_{\nu)}]R^{-2}=8\pi GT_{\mu\nu}.}
Consider matter as an ideal fluid $T_{\mu\nu}=(\rho+p)u_\mu u_\nu +pg_{\mu\nu}$. Since
\eqn\scalcur{R=6\left[{\ddot{a}\over a}+\left({\dot{a}\over a}\right)^2\right]=6\left(\dot{H}+2H^2\right),}
we obtain the modified Friedmann equation
\eqn\mfried{3M_p^2H^2-{\mu^4M_p^2\over 12(\dot{H}+2H^2)^3}(2H\ddot{H}+15H^2\dot{H}+2\dot{H}^2
+6H^4)=\rho.}
The other Friedmann equation is not independent and can be obtained from the above equation and the
continuity of equation of $\rho$.

We make a field redefinition
\eqn\redf{\eqalign{&e^{\alpha\varphi}=1+{\mu^4\over R^2},\cr
&\tilde{g}_{\mu\nu}=e^{\sqrt{{2\over 3}}M_p^{-1}\varphi}g_{\mu\nu}=e^{\alpha\varphi}g_{\mu\nu},}}
and
\eqn\modstress{\tilde{T}_{\mu\nu}=(\tilde{\rho}+\tilde{p})\tilde{u}_\mu\tilde{u}_\nu+\tilde{p}
\tilde{g}_{\mu\nu},}
where
\eqn\modden{\tilde{\rho}=e^{-2\alpha\varphi}\rho, \quad \tilde{p}=e^{-2\alpha\varphi}p,\quad
\tilde{u}_\mu=e^{\half \alpha\varphi}u_\mu,}
then
\eqn\newdact{S(\tilde{g},\varphi)={1\over 16\pi G}\int d^4
x\sqrt{-\tilde{g}}\tilde{R}+\int d^4 x \sqrt{-\tilde{g}}\left[-\half
(\p \varphi)^2-V(\varphi)\right]+S_{\rm matter}(\tilde{g},\varphi),}
where
\eqn\varpot{V(\varphi)=\mu^2M_p^2e^{-2\alpha\varphi}\sqrt{e^{\alpha\varphi}-1}.}
Thus, the modified theory is equivalent to a tensor-scalar theory, although matter is not
canonically coupled to the new metric.
The new cosmological equations are
\eqn\cosmof{\eqalign{&3\tilde{H}^2=M_p^{-2}(\tilde{\rho}+\rho_\varphi ),\cr
&\ddot{\varphi}+3\tilde{H}\dot{\varphi}+V'-{1-3w\sqrt{6}}\tilde{\rho}=0.}}
The last term in the equation of motion of $\varphi$ arises from the coupling
between matter and $\varphi$. Thus, the theory is equivalent to the theory of quintessence with
coupling to matter.

The evolution of matter is
\eqn\tildrho{\tilde{\rho}=\tilde{\rho}_0\tilde{a}^{-3(1+w)}e^{{3w-1\over\sqrt{6}}M_p^{-1}\varphi}.}
There are three separated cases.

\noindent 1. Eternal de Sitter, $V'=0$, but this solution is unstable.

\noindent 2. Power-law acceleration.

When $e^{\alpha\varphi}\gg 1$, $V\sim \mu^2M_p^2e^{-{\sqrt{3\over 2}}M_p^{-1}\varphi}$, then
\eqn\tildea{\tilde{a}(\tilde{t})\propto \tilde{t}^{{4\over 3}}, \quad a(t)\propto t^2.}

The problem of this model is obvious, it is similar to the Brans-Dicke theory thus in general
violate the equivalence principle.

Chiba \ChibaIR\ considered more generally the model
\eqn\frgravity{S=\half M_p^{2}\int d^4 x\sqrt{-g}f(R)+S_{\rm matter}(g),}
which is equivalent to
\eqn\tensca{S=\half M_p^2\int d^4 x\sqrt{-g}\left[f(\varphi)+f'(\varphi)(R-\varphi)\right].}
The variation of the above action with respect to $\varphi$ leads to $\varphi =R$ if
$f'(\varphi)\ne 0$.

Redefine
\eqn\fred{\tilde{g}_{\mu\nu}=f'(\varphi)g_{\mu\nu},}
then
\eqn\tenscala{S=\half M_p^2\int d^4 x\sqrt{-\tilde{g}}\left[\tilde{R}-{3\over 2f'^2}(\p\varphi)^2-
{1\over f'^2}(\varphi f'-f)\right]+S_{\rm matter}.}
As for the Brans-Dicke theory, we require
\eqn\gare{{\gamma-1}<2.8\times 10^{-4},\quad \gamma={\omega+1\over \omega +2}.}
For the CDTT model, $\omega=0$, so the equivalence principle is violated. For the Starobinsky
model $f(R)=R+M^{-2}R^2$, we need $M\sim 10^{12}$GeV, but the constraint $|\gamma-1|
< 2.8\times 10^{-4}$ is derived for $m\le 10^{-27}$GeV.

\noindent $\bullet$ Modified source gravity

To avoid the above problem, a new model was proposed
\CarrollJN. Again, the idea of conformal transformation is used. Let $\psi=\half \ln f'(\varphi)$, and
\eqn\rescga{\tilde{g}_{\mu\nu}=e^{2\psi}g_{\mu\nu},}
then the action becomes
\eqn\newacttens{\int d^4x\sqrt{-\tilde{g}}\left[\half M_p^2\tilde{R}-3\tilde{g}^{\mu\nu}(\nabla\psi)^2-
V(\psi)\right]+\int d^4 x{\cal L}(e^{-2\psi}\tilde{g}_{\mu\nu},\chi_m),}
where $\chi_m$ denotes matter fields collectively, and
\eqn\pispot{V(\psi)={\varphi f'(\varphi)-f(\varphi)\over 2f(\varphi)}M_p^2.}
There is still a problem to be consistent with the solar system, to avoid this problem, some
introduced Palatini formulaition,
or remove the kinetic term of $\psi$. The model without the kinetic term of $\psi$ is
the modified source gravity theory. $\psi$ in this model becomes a Lagrangian multiplier
\eqn\lagact{S=\int d^4 x\sqrt{-\tilde{g}}\left[\half M_p^2\tilde{R}-V(\psi)\right]+S_m(e^{-2\psi}\tilde{g},\chi_m).}
When there is no matter, $\psi$ is a just a number, and $V(\psi)$ becomes a constant.

Rescaling back to $g_{\mu\nu}$
\eqn\psiact{S=\int d^4 x\sqrt{-g}\left[\half M_p^2e^{2\psi}R+3e^{2\psi}(\p\psi)^2-e^{4\psi}V(\psi)\right]
+S_m(g,\chi_m).}
In the above action, the coupling between $g_{\mu\nu}$ and matter is simpler, and there is a
superficial kinetic term for $\psi$. The Einstein equations
\eqn\einpsi{M_p^2e^{2\psi}G_{\mu\nu}=T_{\mu\nu}(\chi)+T_{\mu\nu}(\psi),}
where
\eqn\tenpsi{T_{\mu\nu}(\psi)=-2\p_\mu\psi\p_\nu\psi +2\nabla_\mu\nabla_\nu\psi -g_{\mu\nu}
[e^{4\psi}V(\psi)+(\p\psi)^2+2\nabla\triangle\psi].}
The equation of motion for $\psi$
\eqn\eompsi{\triangle\psi+(\p\psi)^2+{1\over 6M_p^2}e^{-2\psi}\p_\psi (e^{4\psi}V)-{1\over 6}
R=0.}
In the following, we use $U(\psi)=e^{4\psi}V(\psi)$. There appears kinetics for $\psi$, but, taking
the trace of the Einstein equations we have
\eqn\einstrace{{1\over 6}R={e^{-2\psi}\over 6M_p^2}(-T+4U)+(\p\psi)^2+\triangle\psi,}
this combined with the e.o.m for $\psi$ we find
\eqn\algepsi{U'(\psi)-4U(\psi)=-T,}
this is an algebraic equation. We consider several cases in the
following \foot{The thermal dynamical properties of $f(R)$
gravity is different from Einstein gravity, which is discussed in \BambaAS}.

\noindent 1. The solar system

In this case $T_{\mu\nu}(\chi_m)=0$ thus
\eqn\vacpsi{4U(\psi_0)-U'(\psi_0)=0,}
the contribution of $\psi_0$ to $T_{\mu\nu}$ vansishes, so
\eqn\consein{M_p^2G_{\mu\nu}=-e^{2\psi_0}U(\psi_0)g_{\mu\nu},}
this is the Einstein equations with a cosmological constant, provided
the cosmological constant is small enough,
the solar system is fine.

\noindent 2. Interaction with matter

Integrating out $\psi$ results in a complicated and un-renormalizable action, so it is better to view the
model as a low energy effective model.

\noindent 3. Cosmology

Let $T^\mu_\nu=\hbox{diag}(-\rho, p, p, p)$, the Friedmann equation
\eqn\psifried{3H^2+{3k\over a^2}=M_p^{-2}e^{-2\psi}[\rho+U(\psi)]-3\dot{\psi}^2-6H\dot{\psi}.}
Let $x=\ln a$, $\dot{\psi}=H\psi,_x$,
\eqn\refridm{H^2=(1+\psi,_x)^{-2}[{1\over 3M_p^2}e^{-2\psi}(\rho +U(\psi))-{k\over a^2}],}
we see that the effective Newtonian constant is $8\pi G_{\rm eff}=M_p^{-2}e^{-2\psi}(1+\psi,_x)^{-2}$.
We have the freedom to choose the form of $U(\psi)$, amounting to choosing $\psi(\rho)$. Let
\eqn\psifunr{e^{-4\psi}=\alpha {\rho_0\over\rho}+1,}
so when $\rho\rightarrow\infty$, $\psi\rightarrow 0$ and when $\rho\rightarrow 0$, $\psi\rightarrow -\infty$.
Since $U'-4U=-T=\rho$ thus
\eqn\uofrho{\eqalign{&U(\rho)=-{\alpha\rho_0\rho\over 4(\alpha\rho_0+\rho)}\ln ({a\rho_0\over\rho}),\cr
&U(\psi)=\alpha\rho_0e^{4\psi}[\psi-{1\over 4}\ln (1-e^{4\psi})],\quad \psi,_x=-{3\alpha\rho_0\over
4(\alpha\rho_0+\rho)}.}}
The Friedmann equation is
\eqn\rhozefri{H^2=({4\alpha\rho_0+4\rho\over \alpha\rho_0+4\rho})^2{\sqrt{\rho}\over 3M_p^2}
{\alpha\rho_0+\rho-{\alpha\rho_0\over 4}\ln{\alpha\rho_0\over\rho}\over\sqrt{\alpha\rho_0+\rho}}
-{k\over a^2}.}
Let $\rho_{DE}=3M_p^2H^2-\rho$ then
\eqn\eosin{w_{DE}=-1-{1\over 3}{d\ln\rho_{DE}\over d\ln a}.}
Using the above equations in fitting the data, one finds
\eqn\datafit{\alpha=0.98, \quad h=0.72.}

\subsec{MOND and TeVes theories}

The MOND theory (the modified Newtonian dynamics) was proposed by Milgrom as a substitute for dark matter \MilgromCA.
It proposes to modify Newtonian dynamics at very large scales in order to explain the rotational curves of
galaxies.

Since MOND is aimed to modify Newtonian dynamics, the most appropriate
starting point is an equation for the Newtonian potential. The typical acceleration involved in the rotational curves is of
order $a_0=10^{-8} {\rm cm}/{\rm sec}^2$. To make use of this fact, the basic
assumption of MOND is the eqaution
\eqn\monde{\mu(a/a_0)a=-\nabla\Phi,}
where $\Phi$ is the Newtonian potential, $\mu(a/a_0)$ is a function with the property $\mu\rightarrow 1$ when $a\gg a_0$.
If $\mu(x)=x$ for $x\ll 1$, then we have
\eqn\mondd{{a^2\over a_0}=-\nabla\Phi={GM\over r^2},}
since $a=v^2/r$, the above leads to $v=\rm const$.

Bekenstein and Milgrom \BekensteinTV\ found that the MOND theory
coupled to matter can be derived from
a AQUadratic Lagrangian (AQUAL),
\eqn\mondact{{\cal L}=-{a_0^2\over 8\pi G}f({(\nabla\Phi)^2\over a_0^2})-\rho\Phi,}
where $f$ is given by $f'(x)=\mu (\sqrt{x})$. The equation of motion is
\eqn\mondeom{\nabla\cdot (\mu({\nabla\Phi\over a_0})\nabla\Phi)=4\pi G\rho,}
with
\eqn\mondfofy{f(y)=
\cases{y
& ($y\gg 1$) \cr
{2\over 3}y^{{3\over 2}}
& ($y\ll 1$).}}

\noindent $\bullet$  Relativistic MOND

The MOND action AQUAL can be easily generalized into the relativistic
case \BekensteinTV. Again a scalar field $\Psi$ is introduced in this model . Matter is coupled to $e^{2\Psi}g_{\mu\nu}$,
the Lagrangian of $\Psi$ and the action of a particle in the background are
\eqn\rmond{\eqalign{&{\cal L}_\Psi=-{1\over 8\pi GL^2}\tilde{f}(L^2g^{\mu\nu}\p_\mu\Psi\p_nu
\Psi),\cr
&S_m=-m\int e^\Psi\sqrt{-g_{\mu\nu}\dot{x}^\mu\dot{x}^\nu}dt.}}
The relativistic MOND model is also sometimes called relativistic
AQUAL, or simply AQUAL.
In the low velocity limit, the action of a particle is
\eqn\actofp{e^\Psi ds\sim (1+\Phi-\Psi -{v^2\over 2})dt,}
thus
\eqn\rmondeom{a=-\nabla (\Phi+\Psi),}
combined with the action of $\Psi$, we obtain MOND.

There are two problems with relativistic MOND:

\noindent 1. When $\tilde{f}=f$, causality is violated. The discussion
of causality and related modification of MOND, can be found in
\BrunetonGF.

\noindent 2. Light is almost decoupled from $\Psi$, dark matter
deflects light more than relativistic MOND does. This unfortunately
conflicts with the gravitational lensing experiments.
Thus the relativistic MOND is ruled out.

\noindent $\bullet$ PCG (The phase coupled gravity)

In addition to $\Psi$, another field $A$ is introduced with Lagrangian
\refs{\BekensteinPC, \BekinsteinAC}
\eqn\pcgl{{\cal L}(\Psi, A)=-\half [A,_\mu A,^\mu +\eta^{-2}A^2\Psi,_\mu\Psi,^\mu)
+V(A^2)],}
the equation of motion is
\eqn\eoma{\eqalign{&\triangle A-\eta^{-2}A(\p\Psi)^2-AV'(A)=0, \cr
&\nabla_\mu (A^2\nabla^\mu \Psi)=\eta^2e^\Psi M\delta (x).}}
If $\eta$ is very small, this model reduces to relativistic MOND, but without violation of causality. Still, it
contradicts gravitational lensing experiments.

\noindent $\bullet$ Deformed metric

Since both relativistic MOND and PCG contradict gravitational lensing experiment, a deformed metric theory was
introduced \BekensteinDFM. Let
\eqn\defmet{\eqalign{&\tilde{g}_{\mu\nu}=e^{-2\Psi}(A(X)g_{\mu\nu}+B(X)\p_\mu\Psi\p_\nu\Psi),\cr
&X=-\half (\p\Psi)^2.}}
The second is introduced to deflect light more. To make sure of causality is npt violated, the sign of $B$
is chosen such that the deflection of light is not enough.

It is thus motivated to introduce $U_\mu$ to replace $\Psi,_\mu$ and
\eqn\defom{\tilde{g}_{\mu\nu}=e^{-2\Psi}g_{\mu\nu}-2U_\mu U_\nu \sinh (2\Psi),}
where $U_\mu$ is time-like.

\noindent $\bullet$ TeVeS

TeVeS is by far the most complicated theory among the MOND like models
\BekensteinNE. Besides the metric, there are
new fields $\varphi$, $\sigma$, $U_\mu$, with $U^2=-1$ and
\eqn\tevesm{\tilde{g}_{\mu\nu}=e^{-2\varphi}g_{\mu\nu}-2U_\mu U_\nu \sinh 2\varphi ,}
and actions
\eqn\tevesact{\eqalign{&S={1\over 16\pi G}\int d^4 x\sqrt{-g}R,\cr
&S(\varphi, \sigma)=-\half \int  d^4 x [\sigma^2h^{\mu\nu}\varphi,_\mu\varphi,_\nu +\half Gl^{-2}F(k\sigma)]
\sqrt{-g},\cr
&h^{\mu\nu}=g^{\mu\nu}-U^\mu U^\nu ,\cr
&S(U)=-{K\over 32\pi G}\int  d^4 x[U_{[\mu ,\nu]}U^{[\mu,\nu]}-{2\lambda\over K}(U^2+1)]\sqrt{-g}.}}
The dynamics of $g_{\mu\nu}$ remains unchanged, however matter is minimally coupled to $\tilde{g}_{\mu\nu}$,
for example
\eqn\partcact{-m\int  d^4 x \sqrt{-\tilde{g}_{\mu\nu}\dot{x}^\mu\dot{x}^\nu}dt.}
The equations of motion
\eqn\teveseom{\eqalign{&G_{\mu\nu}=8\pi G (\tilde{T}_{\mu\nu}+(1-e^{-4\varphi}U^\alpha
\tilde{T}_{\alpha(\mu}U_{\nu)}+\tau_{\mu\nu})+\Theta_{\mu\nu},\cr
&\tau_{|mu\nu}=\sigma^2(\varphi,_\mu\varphi,_\nu+\dots ),\cr
&\Theta_{\mu\nu}=K[g^{\alpha\beta}U_{[\mu,\alpha]}U_{[\nu,\beta]}+\dots ].}}
In TeVeS, $\sigma$ is a Lagrangian multiplier. This model can avoid problems of violating causality and
contradicting gravitational lensing experiments, and reduces to MOND in a certain limit. Taking a proper
function $F$, it can also generate the effect of dark energy.

The major problem of TeVeS is that if $\tilde{g}_{\mu\nu}$ is taken as the physical metric, then the tensor,
vector and scalar perturbations always have propagation speed greater than speed of light. Also, the appearance
of dark energy in this model is unnatural.

\subsec{DGP model}

This theory is actually quite simple. It postulates that there are two independent gravity theories, one
on a 3+1 dimensional brane and another in the 4+1 dimensional bulk
\DGPDGP \foot{The idea of brane bulk energy exchange as an origin of
dark energy is also considered in \KiritsisAS}. The action is
\eqn\dgpact{S={1\over 2\kappa_5^2}\int  d^4 x dy \sqrt{-G}R+{1\over
2\kappa_4^2}\int  d^4 x\sqrt{-g}R
+S_m,}
where the metric on the brane is $g_{\mu\nu}=G_{\mu\nu}|_{\rm brane}$. Dvali, Gabadadze and Porratri found out the
3+1 propagator. To do so, they first studied the 3+1 propagator of a scalar with action
\eqn\dgps{S=M^3\int d^4xdy\p_A\phi\p^A\phi +M_p^2\int d^4xdy\delta(y)\p_\mu\phi\p^\mu\phi.}
The Green's function satisfies
\eqn\dgpgreen{(M^3\p_A\p^A +M_p^2\delta(y)\p_\mu\p^\mu)G(x,y;0)=\delta^4(x)\delta(y),}
Take $G(x,y;0)$ as the retarded Green's function, the potential generated by the scalar is
\eqn\dgppot{V(t)=\int G_R(t,\vec{x},y;0)dt.}
Let
\eqn\dgpfour{G_R(x,y;0)=\int{d^4p\over (2\pi)^4}e^{ipx}\tilde{G}_R(p,y),}
$\tilde{G}$ satisfies
\eqn\dgpp{(M^3(p^2-\p_y^2)+M_p^2p^2\delta(y))\tilde{G}_R(p,y)=\delta (y).}
In the Euclidean space (after Wick rotation), the retarded function is
\eqn\eucret{\tilde{G}_R(p,y)={1\over M_p^2p^2+2M^3p}e^{-p|y|},}
we have
\eqn\gdpsp{V(r)=-{1\over 8\pi^2M_p^2}{1\over r}\{\sin {r_0\over r}{\rm
Ci} ({r_0\over r})
+\half \cos{r_0\over r}[\pi -2{\rm Si}({r_0\over r})]\},}
where
\eqn\doulog{{\rm Ci} (z)=\gamma +\ln z +\int^z_0{dt\over t}(\cos t-1),\quad
{\rm Si} (z)=\int^z_0{dt\over t}\sin t ,}
where $\gamma=0.577\dots$, the Euler constant and $r_0=M_p^2/(2M^3)$. When $r\ll r_0$
\eqn\dgpsamm{V(r)=-{1\over 8\pi^2M_p^2}{1\over r}[{\pi\over 2}+(-1+\gamma +\ln
{r_0\over r}){r\over r_0}]+O(r),}
it has the correct $1/r$ form in 3+1 dimensions. When $r\gg r_0$
\eqn\dgplarg{V(r)=-{1\over 8pi^2M_p^2}{1\over r}[{r_0\over r}+O(1/r^2)],}
it goes like $1/r^2$, the correct potential form in 4+1 dimensions.

Thus, $r_0$ is a crucial
scale, the world behaves as a 3+1 dimensional one when $r$ is much smaller than this scale
and as a 4+1 dimensional one when $r$ is much larger than this scale. Let $r_0\sim 10^{28}$ cm.
we have $M\sim 10^{12} (cm)^{-1}\sim 10$MeV.

Back to gravity, there is one problem, namely there is one more degree of freedom. Let
\eqn\dgpdec{G_{AB}=\eta_{AB}+h_{AB},}
and take the gauge
\eqn\dgphar{\p^Ah_{AB}=\half \p_Bh,}
the equation for the Green's function is
\eqn\dgpmgreen{(M^3\p_A\p^A+M_p^2\delta(y)\p_\mu\p^\mu)h_{\mu\nu}=M_p^2\delta(y)\p_\mu
\p_\nu h_{55}+[T_{\mu\nu}-\half \eta_{\mu\nu}T]\delta(y).}
The Fourier transform of $h$
\eqn\dgpmfour{\tilde{h}_{\mu\nu}(p,y=0)\tilde{T}^{\mu\nu}(p)={\tilde{T}^{\mu\nu}\tilde{T}_{\mu\nu}
-{1\over 3}\tilde{T}\tilde{T}\over M_p^2p^2+2M^3p}.}
One reads the tensor structure from the above formula
\eqn\dgpten{D^{\mu\nu;\alpha\beta}=\half\eta^{\mu\alpha}\eta^{\nu\beta}+\half\eta^{\mu\beta}
\eta^{\nu\alpha}-{1\over 3}\eta^{\mu\nu}\eta^{\alpha\beta}.}
Although the tensor structure is not what we want, when $r\ll r_0$, one still gets
the correct Newtonian potential and when $r\gg r_0$ one gets the Newtonian potential in
4+1 dimensions.

\noindent $\bullet$ Cosmology

To study cosmology, let us consider the following form of time-dependent metric
\eqn\dgpcos{ds^2=-n^2(t,y)dt^2+b^2(t,y)dy^2+a^2(t,y)d\Sigma^2,}
with the Einstein equations
\eqn\eindgp{G_{AB}=\kappa^2T_{AB},\quad
T_{AB}={\rm diag} (-\rho_B,p_B,\dots, p_B)+{\delta(y)\over b}{\rm diag} (-\rho, p,p,p,0),}
Solving the Einstein equations with the junction conditions
\eqn\dgpsolv{{\kappa_5^2\over 2\kappa_4^2}(H^2+{k\over a^2})-{\kappa^2\over 6}\rho =\epsilon
(H^2-{\kappa^2\over 2}\rho_B -{C\over a^4}+{k\over a^2})^\half,}
where $\epsilon=\pm1$. Take $\rho_B=C=0$, we find
\eqn\dgpfrie{H^2+{k\over a^2}=[{\epsilon\over 2r_0}+({1\over 3M_p^2}\rho+{1\over 4r_0^2})^\half
]^2.}
Consider two limits of the above equation.

\noindent 1. ${1\over 3M_p^2}\rho\gg {1\over 4r_0^2}$

This is the case in the very early universe when the energy density is large. we have
\eqn\dgpearl{H^2+{k\over a^2}={\rho\over 3M_p^2}+\dots,}
this is the usual Friedmann equation.

\noindent 2. ${1\over 3M_p^2}\rho\ll {1\over 4r_0^2}$

We have
\eqn\dgpsamm{H^2+{k\over a^2}={1\over 4r_0^2}(1+\epsilon)^2.}
Take $\epsilon =1$, then
\eqn\dgpcc{H^2+{k\over a^2}={1\over r_0^2}={\rl \over 3M_p^2},}
this is the Friedmann equation with a cosmological constant $\rl =3M_p^2/r_0^2$, or $\Lambda^{-1}=r_0^2$. Thus, the
universe is automatically accelerated in the late stage in the DGP model.

What is the physical meaning of $\epsilon$? Consider $n$, $a$ and $b$ as functions of $y$ near
$y=0$
\eqn\dgpnear{\eqalign{&n(t,y)=1+\epsilon |y|\ddot{a}(\dot{a}^2+k)^{-\half},\cr
&a(t,y)=a+\epsilon |y|(\dot{a}^2+k)^{\half},\cr
&b(t,y)=1.}}

Let $\Omega_{r_0}=1/(4r_0^2H_0^2)$, $\Omega_k=-k/(H_0^2a_0^2)$, we have
\eqn\dgpmg{\Omega_k+(\sqrt{\Omega_{r_0}}+\sqrt{\Omega_{r_0}+\Omega_m})^2=1,}
this is different from the usual Friedmann equation
\eqn\friedeq{\Omega_k+\Omega_m+\Omega_X=1.}
$\Omega_{r_0}$ can be regarded as the fraction of dark energy.

Later, Nicolis, Rattazzi and Trincherini proposed a Galileon model of gravity
as a modification of DGP, aiming at curing the ghost instabilities of
the DGP self-accelerating solution \NicolisIN. A simplified version of
\NicolisIN, as a scalar field model is reviewed in the section of
phenomenological models.

\subsec{Other modified gravity theories}

There are perhaps too many modified gravity theories to review
here. In this section we briefly mention some other modified gravity
models.

\noindent $\bullet$ Brans-Dicke and scalar tensor theories

In the early 1960s, there have already been modified gravity theory known as
Brans-Dicke gravity \BransSX. The gravitational part of the action is
\eqn\MGBransDicke{S = \int d^4 x \sqrt{-g} \left[
{1\over 2}\varphi R -{\omega\over 2\varphi}\partial_\mu\varphi \partial^\mu\varphi
\right],}
where $\omega$ is a constant. The theory is later generalized to
$f(\varphi, R)$, known as the scalar tensor theory \AmendolaQQ.
In addition, this theory has also been extended to the case of 5-dimensional \YGMaone.

\noindent $\bullet$ Gauss-Bonnet gravity and Lovelock gravity

Gauss-Bonnet gravity \ZwiebachUQ\ is proposed in the mid 1980s, aiming
to derive a low energy effective gravitational action from string
theory. In $D$ dimensions, the action of Gauss-Bonnet gravity has an
Einstein-Hilbert part, plus a correction \eqn\MGGaussBonnet{\int d^D x
\sqrt{-g}~ {\cal G}
, \qquad {\cal G}\equiv
R^{\mu\nu\rho\sigma}R_{\mu\nu\rho\sigma}-4R^{\mu\nu}R_{\mu\nu}+R^2. }
This is the only ghost free combination at the $R^2$ order. However,
in four dimensions the Gauss-Bonnet term is a total derivative. In
order to have cosmological implications, in \NojiriVV, Nojiri,
Odintsov and Sasaki coupled the Gauss-Bonnet term $\cal G$ to a scalar field,
and obtained a model of dark energy.

Well before Gauss-Bonnet gravity becomes known in string
theory, its more general form is already there in the early
1970s. This is known as the Lovelock gravity \LovelockYV. The
Lagrangian of $k$-th order Lovelock gravity is the summation
\eqn\MDLovelock{{\cal L}=\sum_{m=0}^{k}c_m{\cal L}_m,}
where ${\cal L}_m$ is called the Euler density:
\eqn\MDLovelockE{{\cal L}_m = 2^{-m}
\delta^{a_1b_1\cdots a_mb_m}_{c_1d_1\cdots c_md_m}
R^{c_1d_1}_{a_1b_1}\cdots R^{c_md_m}_{a_mb_m}.}
The term $m=0$ corresponds to a cosmological constant; $m=1$
corresponds to the Einstein-Hilbert action and $m=2$ corresponds to the
Gauss-Bonnet term.

Unfortunately, in four dimensions the $m\geq 3$ terms are simply
zero. Thus Lovelock gravity cannot be directly applied to dark
energy. Nevertheless, 3rd order Lovelock dark energy models have been
proposed from a dimensional reduction \DehghaniXF.

\noindent $\bullet$ Horava-Lifshitz gravity

One of the biggest problems in quantum gravity is that the Einstein
gravity is nonrenormalizable. Horava \HoravaUW\ suggested that (at
least power counting) renormalizability can be obtained by giving up
Lorentz invariance at high energies. The idea is to introduce higher
order spatial derivatives while keep the time derivative at second
order in the equation of motion. For cosmological applications,
Saridakis proposed a model of Horava-Lifshitz dark energy \SaridakisBV.

\noindent $\bullet$ $f(T)$ gravity

Alternatively, torsion might also be useful in modified
gravity. Bengochea, Ferraro and Linder \refs{\BengocheaGZ, \LinderPY}
considered the cosmology of a so called $f(T)$ gravity theory. The
gravitational action is \eqn\MGfT{S={M_p^2\over 2}\int d^4 x ~eT,}
where $e\equiv \det e^\mu{}_a$, and $e^\mu{}_a$ is the tetrad field
satisfying $g_{\mu\nu}e^\mu{}_ae^\nu{}_b=\eta_{ab}$. The scalar $T$ is
defined as \eqn\MGfTT{T\equiv S_\rho{}^{\mu\nu} T^{\rho}{}_{\mu\nu},
\qquad T^{\rho}{}_{\mu\nu}\equiv -e^\rho{}_a(\partial_\mu e^a{_\nu}-
\partial_\nu e^a{_\mu}), }
\eqn\MGfTS{S_{\rho}{}^{\mu\nu}\equiv {1\over 2}(K^{\mu\nu}{}_\rho
+\delta^\mu{}_\rho T^{\sigma\nu}{}_\sigma -\delta^\nu{}_\rho
T^{\sigma\mu}{}_\sigma),\qquad K^{\mu\nu}{}_\rho \equiv -{1\over 2}
(T^{\mu\nu}{}_\rho-T^{\nu\mu}{}_{\rho}-T_\rho{}^{\mu\nu}).}
It is shown that the $f(T)$ theory is also a candidate of dark
energy \refs{\BengocheaGZ, \LinderPY}
\foot{In a latest work \FRT, a $f(R,T)$ gravity model,
where the gravitational Lagrangian is given by an arbitrary function of R and of T, is proposed.}.

\noindent $\bullet$ Conformal gravity

Mannheim \MannheimBFA\ suggests that the gravitational action could be
modified using the conformal Weyl tensor as
\eqn\MGMannheim{-\alpha_g\int d^4 x \sqrt{-g}
C_{\mu\nu\rho\lambda}C^{\mu\nu\rho\lambda}. } It is shown that this
model is on the one hand fits well the galactic rotation curves
\MannheimXW\ on galactic scales, on the other hand could behave as a
component of dark energy on cosmological scales \MannheimHK.

\noindent $\bullet$ Fat graviton

Sundrum \SundrumJQ\ proposed a scenario of fat graviton, in which the
graviton has a size instead of localized. Unlike the
other approaches where an action is proposed at the first place, the
fundamental principle of a fat graviton scenario is yet
unknown. Instead, the work \SundrumJQ\ aims to address how can a
graviton be fat, without violating known physics. Especially, with the
help of non-locality, the fat graviton is able to ``know'' whether a
loop diagram is a vacuum loop or a loop attached to external legs. In
this way, the equivalence principle is preserved. However, it is not
clear what detailed rules a fat graviton obeys. Also, the size of
graviton is required as an input parameter. To solve the cosmological
constant problem, the graviton size $l>20$ microns seems to be a large
number in the particle physics point of view.

There are also a number of scalar field models which can be viewed as
modified gravity models. This is because the scalar part of gravity is
coupled to matter. We shall review this class of models later, as part
of the
phenomenological models.

\newsec{Quantum cosmology}

As reviewed in Section \secWeinberg, the approach of wave function of
the universe either results in a probability $P(\Lambda) =
\exp({3\pi\over G\Lambda})$, or its exponential. The predicted
cosmological constant is not only too small, but also conflict with
the thermal history of the universe. Several ways to solve the problem
have been proposed.

For example, Firouzjahi, Sarangi and Tye \refs{\FirouzjahiMX,
\SarangiCS} brings the decoherence effect to the quantum creation of
universe, and found the preferred classical universe is born at string
scale, which is good for inflationary cosmology. As another approach,
Hartle, Hawking and Hertog \refs{\HartleGI, \HawkingVF} argues that from
the Bayesian point of view, the quantum creation probability should
multiply the spatial volume factor to get the observed
probability. Again, inflation after the born of the universe is
preferred.

These approaches on the one hand bring the Hartle-Hawking wave
function back to be consistent with observations, but on the other
hand make the whole subject more relevant to inflation, and less
relevant to dark energy.

Nevertheless, dark energy is very probably an effect from quantum gravity.
Indeed a lot of attempts on dark energy are related to quantum theories
\foot{For example, it is argued that cosmological constant is an intrinsic scale of quantum gravity \SSXueone,
and can be given by the energy-density of quantum black holes \SSXuetwo.}.
Here in this section by quantum cosmology we shall restrict our attention to approaches closely related to the wave
function of the universe. Other quantum approaches are reviewed in other related sections.

\subsec{Cosmological constant seesaw}

The seesaw mechanism provides a connection between particles between
large mass and small mass, which is extensively used in neutrino
physics. Non-technically speaking, the (type I) seesaw mechanism
relies on the fact that a matrix
\eqn\QCseesawMass{M=\pmatrix{0&x\cr x&y}}
has eigenvalue
$$\lambda_{\pm}={y\pm\sqrt{y^2+4x^2} \over 2}.$$
When $y\gg x$, the smaller eigenvalue is
\eqn\QCseesaw{\lambda_-\simeq -{x^2\over y},}
which is suppressed by the larger matrix element $y$.  Note that
$\lambda_+\lambda_- = -x^2$. Thus when adjusting $y$, one eigenvalue
goes up, another goes down, like a seesaw. In neutrino physics, the
analog of matrix $M$ is the neutrino mass matrix.

The seesaw mechanism is aimed to explain why the neutrino mass is
non-zero and small.  This saturation appears very similar to the
cosmological constant problem. Based on this observation, Motl and
Carroll proposed on their blogs \refs{\MotlsSeesaw, \CarrollSeesaw} the idea
that the smallness of the cosmological constant may be explained by
the seesaw mechanism as well. The observation is that the present
cosmological constant, the Planck scale and the lowest possible SUSY
scale has a hierarchy \eqn\QChierarchy{M_{\Lambda} \sim M_{\rm SUSY}^2
/ M_{p}, } where $M_\Lambda\equiv (\rho_\Lambda)^{1/4}$ is the energy
scale of the dark energy. This equation is similar to \QCseesaw.

This idea is realized using the wave function of the universe by
McGuigan \McGuiganHS. It is noticed that the cosmological constant
term acts exactly like a mass term in the Wheeler DeWitt
equation. Consider a model with a set of massless free scalar fields
$\varphi^I$ minimally coupled to gravity, the homogeneous and
isotropic background action can be written as \eqn\QCaction{S={1\over
2}\int dt a^3 N\left[ -9M_p^2 \left(\dot a \over
Na\right)^2+{\dot\varphi^I\dot\varphi^I\over N^2} - \lambda + M_p^2
{k\over a^2} \right],} where $N$ is the lapse function. In \QCaction\
several coefficients are different from the literature. This can be
understood as a different convention used in \McGuiganHS. Write down
the conjugate momentum and quantize the equation of motion, one gets
the Wheeler-DeWitt equation \eqn\QCWDW{\left( -{1\over
M_p^2}{\partial^2\over\partial V^2} +{1\over V^2}{\partial^2\over
\varphi^{I2}}+M_p^2{k\over V^{2/3}} -\lambda\right)\Phi = 0,} where
$V\equiv a^3$ is the volume of the universe, which can be understood
as a time variable. The scalar fields $\varphi^I$, on the other hand,
acts as spatial coordinates.  Note that the term proportional to
$\lambda$ acts effectively as a mass term in this equation.

Up to now, everything comes as conventional. Now consider two
universes with different cosmological constant and also with
``coupling'': \eqn\QCWDWa{ \left( -{1\over
M_p^2}{\partial^2\over\partial V^2} +{1\over V^2}{\partial^2\over
\varphi^{I2}}+M_p^2{k\over V^{2/3}} -\lambda_1 \right)\Phi_1
+\sqrt{\lambda_1\lambda_2}\Phi_2 = 0, } \eqn\QCWDWa{ \left( -{1\over
M_p^2}{\partial^2\over\partial V^2} +{1\over V^2}{\partial^2\over
\varphi^{I2}}+M_p^2{k\over V^{2/3}} -\lambda_1-\lambda_2 \right)\Phi_2
+\sqrt{\lambda_1\lambda_2}\Phi_1 = 0, } where the effective mass
matrix now is \eqn\QCmass{M^2=\pmatrix{0&\sqrt{\lambda_1}\cr
\sqrt{\lambda_1}&\sqrt{\lambda_2}}^2.} This form is similar to
\QCseesawMass. Now one can apply the seesaw
mechanism, such that when \eqn\QCparam{\lambda_1 \sim (10 {\rm
TeV})^4, \qquad \lambda_2 \sim M_p^4,} the smaller eigenvalue becomes
\eqn\QCsmallEigen{\lambda_- \sim {(10{\rm TeV}^8)\over M_p^4},} the
correct order of magnitude for the current cosmological
constant. However, it remains curious how the coupling constant is
derived from first principle. In \LindeGJ, Linde proposed that some
interactions between universes could follow from averaging effects. It
remains interesting to see whether these interactions are of the type
which we review in this subsection. Moreover, it remains not clear in
which form will interacting universes classically be. It is clear that
particles should stay in their mass eigenstates for energy and momentum
measurement but the analog is not clear for quantum universes.

\subsec{Wave function through the landscape}

In quantum mechanics, there is a well-established mechanism of resonant
tunneling, which is not only tested by experiments but also widely applied
in the industry of electronics. Tye \HenryTyeTG\ applied the resonant
tunneling mechanism to the string landscape, which provides a possible
solution for the cosmological constant problem.

As a pure quantum effect, the probability for tunneling is typically
exponentially suppressed. An intuitive understanding is that a
bump in the potential, even higher than the energy of the quantum
state, can not completely block the wave function. An exponentially
small part of the wave function is leaked into the classically
forbidden regime, which connects to another classically allowed
regime.

\ifig\FigResonant{A resonant tunneling in quantum mechanics. The
analog is applied to cosmology for the cosmological constant problem.}
{\epsfysize=1.5in \epsfbox{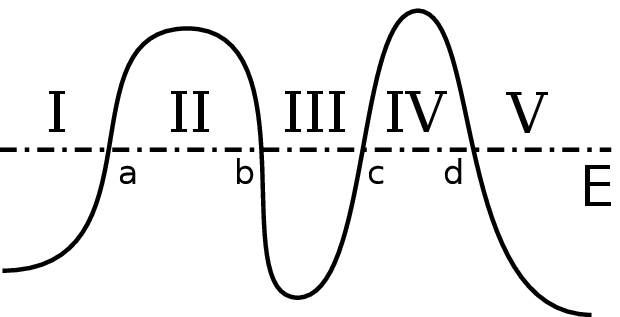}}

However, things get changed when there are multiple bumps in the
potential. As illustrated in Figure \FigResonant, When the incoming
wave coming from region I has an energy equal to a bound state in
region III, the probability for tunneling through the whole potential
to region V is no longer exponentially suppressed by both of the
barriers. Instead, the tunneling rate will equal to the larger
exponential suppression factor (of II or IV) divided by the smaller
exponential suppression factor. Especially, when these two factors are
of the same order, the tunneling rate becomes of order one.

The reason for the odd thing to happen is as follows: In region V, the
wave function only has outgoing component. In WKB approximation, the
mode is $\sim e^{i\int pdq}$. This outgoing component can be linked to
two modes in region IV, namely an exponentially decaying mode (here
decay means becomes smaller from III to IV) $\sim e^{-\int pdq}$ and
an exponentially growing mode $\sim e^{\int pdq}$. In conventional
tunneling events without resonance, only the decaying mode is
considered because the other one contributes an exponentially small
part to the wave function in region III. Similarly, there are two
modes in region II, only the decaying mode need to be considered for
tunneling without resonance. However, when the incoming energy
satisfies a bound state condition in region III
\eqn\QCResonantCond{\int p dq = \left(n+{1\over 2}\right)\pi, \qquad
n=0,1,2,\cdots,} the decaying mode in region IV can only link to the
growing mode in region II, and the growing mode in IV can only link to
the decaying mode in II, by requiring the wave function and its
derivative are continuous at points $a$, $b$, $c$ and $d$. Thus there
must be one barrier in which the tunneling amplitude is suppressed and
another barrier in which the amplitude is enhanced exponentially. This is
why resonant tunneling can happen in quantum mechanics.

Further calculation shows that, for resonant tunneling to happen, the
incoming energy should be exponentially close to
\QCResonantCond. Otherwise the enhancement disappears. For a random
incoming energy, the expectation value of tunneling coefficient
$T({\rm I}\rightarrow {\rm V})$ can be expressed as

\eqn\QCResonantCoef{T({\rm I}\rightarrow {\rm V}) \sim
{T({\rm I}\rightarrow {\rm III})T({\rm III}\rightarrow
{\rm V})\over T({\rm I}\rightarrow {\rm III})+T({\rm III}\rightarrow
{\rm V})}. }
It is also straightforward to generalize the above two step resonant
tunneling to $n$ steps.

Tye \HenryTyeTG\ proposed that the universe may have experienced such
resonant tunnelings in the string landscape. As reviewed in previous
sections, the string landscape is extremely complicated and in some
sense random. Using \QCResonantCoef, one can show that the
tunneling probability is enhanced. Moreover, the string landscape is
many-dimensional. The total tunneling rate $\Gamma$ can be calculated by
\eqn\QCResonantRate{\Gamma \sim n^d \Gamma_0,}
where $\Gamma_0$ is the tunneling rate for a single tunneling event,
$d$ is the dimension of the landscape and $n$ is the number of steps
for resonant tunneling \TyeJA. This is in sharp contrast with the case
discarding resonance, where the tunneling rate is $\Gamma \sim
 (d\times\Gamma_0)^n$.

In this picture, the universe starts with a site in the landscape with
large cosmological constant and tunnels though a series of sites to
the state with the current cosmological constant, with reasonably
large probability. It is assumed that the universe can not tunnel to
any of the vacua with negative cosmological constant. The argument is
that if the whole universe tunnels to a vacuum with negative
cosmological constant, the entropy would decrease; while if a portion
of the universe tunnels to such a vacuum, a singularity will develop.

The resonant tunneling picture gives an elegant understanding for
the cosmological constant problem in the context of cosmic landscape,
as long as one considers the mini-superspace model, where a closed
universe as a whole could tunnel to other sites with resonance. In
this case the quantum gravity problem reduces to a quantum mechanical
problem. However, it is quite difficult to figure out what resonant
tunneling looks like in quantum field theory. For example, in
\CopelandQF, it is proved that if the tunneling process, in terms of
bubbles, satisfies a static boundary condition, there will be no such
resonant tunneling in scalar quantum field theory. Later, it is shown
\SaffinVI\ that for contracting spherical bubbles, resonant tunneling
is possible. Moreover, in \TyeRB, it is shown that the requirements
in \CopelandQF\ are too restrictive. A most probable escape path does
not necessarily satisfy the no-go theorem in \CopelandQF, and a phase
diagram is given to show that some of the potentials have resonant
tunneling in scalar quantum field theory. For the case of gravity, and
for the estimation of tunneling probabilities in a landscape, further
investigations are deserved.

In \TyeJA, other rapid tunneling events are also considered, for
example the Hawking-Moss tunneling without slow roll
approximation. One reaches the conclusion that rapid tunneling can be
common in the complicated string landscape thus the tunneling process
become analogous to an electron in a random potential. Using
renormalization group approach, Tye shows that after a number of
successful tunneling events, the universe will settle down with
localized wave function due to Anderson localization
\Anderson. Similar ideas on Anderson localization of universe in a
landscape is also discussed in \PodolskyVG\ and \PodolskyDU. Besides
Anderson localization, decoherence of the quantum wave function may
also eventually localize the universe to one site in the landscape
\KieferPB.

\newsec{Holographic principle}

The essence of reality have been one of the central problems in
philosophy and physics since thousands of years ago. Among all the
discussions and debates on this subject, one of the most profound
allegory is the Plato's Cave. In the recent decades, the inverse
problem of Plato's Cave, under the name holographic principle
\refs{\tHooftGX, \SusskindVU}, has become one of the building blocks for a
modern understanding of theoretical physics.

Holographic principle asserts that the world can be understood as a
hologram. In other words a theory of gravity is dual to a boundary
field theory without dynamical gravity in one less dimensions.

The story of holographic principle tracks back to the investigation of
black hole physics. Starting from \refs{\IsraelWQ, \IsraelZA}, it is realized
that classically, a stationary black hole in four dimensions is
characterized by its mass, angular momentum and charge. A paradox
arises here that matter that collapses into a classical black hole
appears to lose almost all its entropy. This is a contradiction to
the second law of thermodynamics. Moreover, Hawking proved that
\HawkingTU\ the event horizon of a black hole never decreases with
time. The paradox becomes sharper in this sense because it becomes
hopeless for the ``lost'' entropy to come out.

The paradox was conceptually resolved by Bekenstein \refs{\BekensteinTM,
\BekensteinUR, \BekensteinAX}, by the conjecture that a black hole has a
entropy proportional to its horizon area. The conjecture was soon
proved by Hawking \HawkingRV, with the form
\eqn\BHentropy{S_{\rm BH}={A\over 4G}.}

Later, 't Hooft and Susskind \refs{\tHooftGX, \SusskindVU}  realized that the black
hole entropy can be understood as a dimensional reduction, or
holographic principle. The entropy is viewed as degrees of freedom
measured in Planck units, which lives on the surface of the strongly
gravitating system.

The right hand side of equation \BHentropy\ is
not only the entropy of a black hole, but also an entropy bound for
any form of matter localized in a spherical region
\eqn\SphericalBound{S_{\rm matter}\leq{A\over 4G}.}
For example, consider a spherical symmetric shell of photons which
falls towards the center. Before the photons forms a black hole, the
left hand side of \SphericalBound\ is always smaller than the right
hand side. When the photons hit the Schwarzschild radius, the bound may
be saturated when the photons has maximal possible entropy, in which
case the wave length of the photon equals to the Schwarzschild radius.

However, as pointed out by Cohen, Kaplan and Nelson \CohenZX,
the thermal entropy of an effective field theory can never saturate the entropy bound \SphericalBound.
This is because thermal energy $E$ and
entropy $S$ of a field theory scales with the length scale of the
system (infrared cutoff) and the temperature as
\eqn\EnergyEntropyScales{M \sim L^3T^4, \qquad S \sim L^3T^3.}  At the
point just before the formation of a black hole, the field theory has
Schwarzschild mass, however the entropy takes the form
\eqn\EntropyLimitFieldTheory{S \sim (M_p L)^{3/2} \sim S_{\rm BH}^{3/4},}
which is always smaller than the black hole entropy.
Thus, for thermal field theory matter, Schwarzschild mass
behaves as a tighter bound than the entropy bound. In other words, let
the ultraviolet cutoff of the system be $\Lambda_{\rm UV}$, the
maximum energy density in the effective field theory $\rho \sim \Lambda_{\rm UV}^4$ must satisfy
\eqn\CKNboundDerive{L^3\Lambda_{\rm
    UV}^4 \sim E \leq L M_p^2.}
The equation \EntropyLimitFieldTheory\ could also be written as follows: the
maximal allowed energy density $\rho$ satisfies
\eqn\CKNbound{\rho = 3c^2M_p^2 L^{-2}, }
where $c$ is a number introduced in \LiRB.

\subsec{Holographic dark energy}

The reasoning of last subsection can be applied to the vacuum, which
leads to a holographic model of dark energy. The question is how to
choose the infrared cutoff. As pointed out by Hsu \HsuRI, the simplest choice
$L=1/H$ does not work because it has a wrong equation of state. Li
\LiRB\ pointed out that if one take $L=1/R_h$, where $R_h$ is the
future event horizon defined as
\eqn\futureEventHorizon{R_h = a \int_t^\infty{dt\over a} = a
  \int_a^\infty {da \over Ha^2},}
the energy density \CKNbound\ becomes
\eqn\HDEenergyDensity{\rho_{de} = 3c^2M_p^2R_h^2,}
which does behave as dark energy, with the Friedmann equation\
\eqn\HDEFriedmann{3M_p^2H^2=\rho_{de}+\rho_m+\ldots,}
where $\rho_m+\ldots$ denotes matter and other components in the
universe.

To see this, note the index of equation of state $w_{de}$ can be defined as
\eqn\HDEwLambda{\rho'_{de}+3(1+w_{de})\rho_{de} = 0,}
where prime denotes derivative with respect to $\ln a$.
Take derivative of \HDEenergyDensity\ and use \HDEwLambda\ to substitute $\rho'_{de}$, one can get
\eqn\HDEeqos{w_{de} = -{1\over 3}-{2\over 3HR_h} = -{1\over 3}-{2\sqrt{\Omega_{de}}\over 3c}. }
This equation is about the nature of holographic dark energy itself,
which is independent of what form of matter is present in the universe.
Here $\Omega_{de}$ is the relative energy density of holographic dark energy, defined as
\eqn\HDEOmega{\Omega_{de}\equiv {\rho_{de} \over 3M_p^2H^2}={c^2 \over R_h^2H^2}.}
$\Omega_{de}$ turns out to be a convenient variable for solving equations of motion for holographic dark energy.

Before going to the equation of motion, now one can already find out qualitative behavior of holographic dark energy from \HDEOmega.
When the holographic dark energy is sub-dominant($\Omega_{de} \ll 1$), $w_{de} \simeq -1/3$ thus $\Omega_{de} \sim a^{-2}$.
When the holographic dark energy is dominant ($\Omega_{de}\simeq 1$), $w_{de} \simeq -1/3-2c/3$ thus the universe experiences accelerating expansion as long as $c>0$.

Now consider the universe dominated by holographic dark energy and pressureless matter.
Take derivative of \HDEOmega\ with respect to $\ln a$, and make use of the Friedmann equation \HDEFriedmann,
one get an equation for $\Omega_{de}$ as
\eqn\HDEOmegaEq{{\Omega'_{de}\over\Omega_{de}}
  =(1-\Omega_{de})\left(1+{2\sqrt{\Omega_{de}}\over c}\right).}
The equation can be solved as
\eqn\HDEsolution{\ln(a/a_0) = \ln\Omega_{de}
  +{c\ln(1+\sqrt{\Omega_{de} })\over 2-c}
  -{c\ln(1-\sqrt{\Omega_{de} })\over 2+c}
  -{8\ln(c+2\sqrt{\Omega_{de} })\over 4-c^2},}
where $a_0$ is a constant. The phenomenological implication of this solution will be discussed in later sections.

It is also noticed in \LiRB\ that during inflation holographic dark
energy is diluted. To have the correct fraction of dark energy at
present time, one requires about 60 e-folds of inflation. In other
words, holographic dark energy provides an explanation of the
coincidence problem, as long as inflation only last for about 60
e-folds, before which dark energy also dominates. The detailed
implication for inflation from holographic dark energy is considered
in \ChenQY.

As an energy component with negative pressure, one might question the
stability of holographic dark energy against fluctuations
\MyungPN. However, the dynamics of holographic dark energy is actually
not governed by that of a perfect fluid. Instead it is governed by the
dynamics of the future event horizon. The fluctuation of the future
event horizon can be written as
\eqn\HDEfluct{\delta\rho_{de} = -2 \rho_{de} {\delta R_h \over R_h}.}
This fluctuation can be
analyzed using cosmic perturbation theory. Spherical symmetric
perturbations are calculated in \LiZQ, and it is shown that the
perturbation approaches a constant at late times thus the background
is stable against the fluctuations.

Phenomenologically it is interesting to investigate interactions between
holographic dark energy and matter components \WangJX. The interaction
can be added to the continuity equation as
\eqn\HDEwLambdaInt{\rho'_{de}+3(1+w_{de})\rho_{de} = 3b\rho_i,}
where $\rho_i$ can be set to dark energy density $\rho_{de}$,
matter density $\rho_m$, critical energy density $\rho_c$, or their
combinations.

Finally, one should note that holographic dark energy solves both the
old and new cosmological problem in a consistent way. This is unlike a
number of other models, in which one have to assume the solution of
the old cosmological constant problem, i.e. assume the vacuum energy
to be zero, and propose a small dark energy on top of that.

\subsec{Complementary motivations}

Besides motivation from \CohenZX, there are also a number of other
theoretical motivations leading to the form of holographic dark
energy, among which some are motivated by holography and others from
other principles of physics. We shall briefly review some of the
motivations in this subsection.

\noindent $\bullet$ Casimir energy in de Sitter space

The Casimir energy of electromagnetic field in static de Sitter space
is calculated in \refs{\LiPM, \LiZY}. The Casimir energy can be written as
\eqn\HDECasimir{E_{\rm Casimir} = {1\over 2}\sum_\omega |\omega|,}
where the absolute value of $\omega$ is the energy with respect of the
time of the static patch. $E_{\rm Casimir}$ can be calculated using
heat kernel method with $\zeta$ function regularization. The result is
\eqn\HDECasimirRes{E_{\rm Casimir} = {3\over 8\pi}
  \left(\ln\mu^2-\gamma-{\Gamma'(-1/2)\over\Gamma(-1/2)}\right)
  \left({L\over l_p^2}-{1\over L}\ln\left({2L\over l_p^2}\right)\right) +{\cal
    O}(1/L),} where $L$ is the de Sitter radius, $\gamma$ is the Euler
constant and $\Gamma'(-1/2)\simeq -3.48$. Here a cutoff at stretched
horizon is imposed, which has a distance $l_p$ away from the classical
horizon. Note that the dominate term scales as $E_{\rm Casimir}\sim
L/l_p^2$. Thus the energy density scales as $\rho_{\rm Casimir}\sim
M_p^2L^{-2}$, which is the form of holographic dark energy.

\noindent $\bullet$ Quantum uncertainty of transverse position

At Planck scale, gravity becomes strongly coupled and the classical
spacetime picture breaks down. It is suggested by Hogan \refs{\HoganRZ,
\HoganHC} that the Planck scale quantum gravitational effect could be
modeled by a random noise. A particle moving a distance $l_p$ will
have a kick of the same order $l_p$ in the transverse direction. As
the kick in the transverse direction is random, the summation of $n$
kicks results in a random walk of distance $\sqrt{n} l_p$. Thus when a
particle moves distance $L$, there is a uncertainty in the transverse
direction of order
\eqn\HDEHogan{\Delta X = \sqrt{Ll_p}.}
This relation looks quite like the energy bound \CKNbound. Indeed one
can show that these two bounds are related. When one take $L$ to be
the scale of the universe, $\Delta X$ as an infrared cutoff gives the
energy density for holographic dark energy.

\noindent $\bullet$ Entanglement entropy from quantum information theory

The vacuum entanglement energy is considered in the cosmological
context by \LeeZQ. The entanglement entropy of the quantum field
theory vacuum with a horizon can be generically written as
\eqn\HDEentangEntropy{S_{\rm Ent} = {\beta R_h^2 \over l^2},} where
$\beta$ is an order one constant and $l$ is the ultraviolet cutoff
from quantum gravity. The entanglement energy is conjectured to
satisfy \eqn\HDEentangEnergy{dE_{\rm Ent} = T_{\rm Ent}dS_{\rm Ent},}
where $T_{\rm Ent}=1/(2\pi R_h)$ is the Gibbons-Hawking
temperature. Integrate equation \HDEentangEnergy, one gets
\eqn\HDEentangEnergyCalc{ E_{\rm Ent} = {\beta N_{\rm dof}R_h \over
    \pi l^2}, } where $N_{\rm dof}$ is the number of light fields
present in the vacuum. Thus the energy density is
\eqn\HDEentanEntropyy{\rho_{de} = 3c^2 M_p^2 R_h^{-2},
\qquad c = {\sqrt{\beta N_{\rm dof}}\over 2\pi l M_p}.}
Here $c$ is in principle calculable in the quantum information
theory.

\noindent $\bullet$ Dark energy from entropic force

Verlinde conjectured \VerlindeHP\ that gravity may be an entropic force,
instead of a fundamental force of nature. \LiCJ\ investigated the
implication of the conjecture for dark energy. It is suggested that
the entropy change of the future event horizon should be considered
together with the entropy change of the test holographic
screen. Consider a test particle with physical radial coordinate $R$,
which is the distance between the particle and the ``center'' of the
universe where the observer is located. The energy of the future event
horizon, using Verlinde's proposal, can be estimated as
\eqn\HDEentropicE{E_h \sim N_h T_h \sim R_h/G,} where $N_h\sim
R_h^2/G$ is number of degrees of freedom on the horizon, and $T_h\sim
1/R_h$ is the Gibbons-Hawking temperature. Following Verlinde's
argument (instead of Newtonian mechanics), the energy of the horizon
induces a force to a test particle of order $F_h \sim GE_hm/R^2$,
which can be integrated to obtain a potential
\eqn\HDEentropicV{V_h\sim -{R_h m \over R} = -c^2m/2,} where after the
integration one can take the limit $R\rightarrow R_h$, and $c$ is a
constant reflecting the order one arbitrarily. Using standard argument
leading to Newtonian cosmology, this potential term for a test
particle will show up in the Friedmann equation as a component of dark
energy $\rho_{de} = 3c^2M_p^2R_h^{-2}$. Again it is the form of
holographic dark energy.

\noindent $\bullet$ Holographic gas as dark energy

The nature of a general strongly correlated gravitational system is not well
understood. In \LiQH\ it is suggested that the quasi-particle
excitations of such a system may be described by holographic gas, with
modified degeneracy
\eqn\HDEhologasW{\omega = \omega_0 k^a V^{b}M_p^{3b-a},}
where $\omega_0$ is a dimensionnless constant. Inspired by holography,
when taking $T\propto V^{-1/3}$ and $S\propto V^{2/3}$, one needs
$b=(a+2)/3$ and the energy density $\rho$ can be written as
\eqn\HDEhologasRho{\rho = {a+3 \over a+4}{ST \over V},}
where $S$, $T$, $V$ are the entropy, temperature and volume of the
system. Applying to cosmology, $S=8\pi^2R^2M_p^2$ and $T=1/(2\pi R)$,
one obtains
\eqn\HDEhologasRhoDE{\rho = 3{a+3 \over a+4}M_p^2 R^{-2}.}
This has the same form as holographic dark energy with
\eqn\HDEhologasC{c^2 = {a+3 \over a+4}.}

There are some other alternative motivations for
holographic dark energy. For example, the relation between holographic
dark energy and vacuum decay is discussed in \ElizaldeAS.
In addition, there are also many extended versions of holographic
dark energy, see \EHDE\ and references therein for more details.

\subsec{Agegraphic dark energy}

In this subsection we review another dark energy model motivated form
holographic physics, named agegraphic dark energy by Cai \CaiUS\ and later Wei
and Cai \WeiTY. As
discussed in \refs{\CaiUS,\WeiTY}, there is a subtlety in the original version
of agegrephic dark energy model \CaiUS, where cosmic time is used as
the age cutoff. Thus here we mainly review the so called new
agegraphic dark energy model \WeiTY.

The agegraphic dark energy model is based on the Karolyhazy \KarolyhazyZZ
uncertainty principle \eqn\ADEKarolyhazy{\delta t = \beta
  t_p^{2/3}t^{1/3},} where $\beta$ is an order one constant and $t_p$
is Planck time. It is noticed in \CaiUS\ and \MaziashviliDK\ that this
relation has close relation with holographic principle and black hole
entropy bound. Based on the Karolyhazy relation, Maziashvili derived
an energy density of the vacuum energy
\eqn\ADEMaziashvili{\rho_{de} = {3n^2 m_p^2\over t^2},}
where $n$ is a numerical factor as introduced in \CaiUS.
Wei and Cai proposed that, when the time in the above formula takes the form of conformal time
\eqn\ADEWeiCaiTime{\eta = \int {dt \over a} = \int {da \over a^2 H},}
the energy density \ADEMaziashvili\ can be well behaved as a dark energy component.

The equation of motion for $\Omega_{de}$ takes the form
\eqn\ADEeom{{d\Omega_{de} \over da} =
  {\Omega_{de}\over a}(1-\Omega_{de})\left(1-{2\over n}{\sqrt{\Omega_{de}}\over
    a}\right).}
Note that the scale factor $a$ appears explicitly.

The index of equation of state for agegraphic dark energy takes the form
\eqn\ADEeos{w_{de} = -1+{2\over 3n}{\sqrt{\Omega_{de}}\over a},}
thus there can be accelerating solutions.
Especially at late times when $a\rightarrow \infty$, $w_{de}\rightarrow -1$.

Unlike the case of holographic dark energy, here $\Omega_{de}$, $n$ and $a$ should satisfy an additional consistency relation.
For example, in the matter dominated universe the conformal time $\eta\propto a^{1/2}$, thus $\rho_{de} \propto a^{-1}$.
Comparing with \ADEeos, one obtains $\Omega_{de} = n^2a^2/4$.
Similarly, in radiation dominated universe the corresponding relation is $\Omega_{de} = n^2 a^2$.

With the consistency relation, agegraphic dark energy is a model with
a single parameter. This is different from holographic dark energy
where there are two parameters $c$ and $\Omega_m$.

\subsec{Ricci dark energy}

Another natural choice for cosmological infrared cutoff is the
intrinsic curvature of the universe. Based on this, Gao, Chen and Shen
\RDEPaper\ proposed a model of Ricci dark energy \foot{In \NojiriASb, Nojiri
and Odintsov proposed a generalized framework of holographic dark
energy, which contains Ricci dark energy as a special case.}. In the Ricci dark energy model,
the energy density is
\eqn\RDErhoXeq{\rho_{de} = {3\alpha \over 8\pi}
  \left(\dot H+2H^2+{k\over a^2}\right) = -{\alpha\over 16\pi}R,}
where $R$ is the Ricci scalar.

With the proposed form of energy density, the energy density can be solved from the Friedmann equation as
\eqn\RDErhoX{\rho_{de} = {\alpha\over 2-\alpha}\Omega_{m0}e^{-3\ln a}
  +f_0 e^{-(4-{\alpha\over 2})\ln a},}
where $f_0$ is a integration constant. The pressure can be solved from
energy conservation,
\eqn\RDEp{p_{de} = -\left({2\over 3\alpha}-{1\over 3}\right)f_0
  e^{-\left(4-{2\over\alpha}\right)\ln a}.}
The model parameter $\alpha$ and $f_0$ are to be determined by
data fittings.
For some extended versions of Ricci dark energy, see \refs{\RDEGO,\RDEXLX,\RDECFR}.

\newsec{Back-reaction}

A universe without nonlinearity is simple and simply dull. Physicists
like to start their calculation from the linear case because it is
mathematically easy and often solvable. However, nobody likes to
live in such a universe where two waves always propagate through each
other without any impact.

In modern cosmology, one of the most important assumption is the
cosmological principle, which states that the universe is homogeneous and
isotropic on (observablely) large scales. In the standard setup, the
cosmological principle is encoded into the FRW metric and all the
subsequent conclusions, especially the Friedmann equation, are under
this assumption
\foot{The Friedmann equation, coupled to the
matter equation of motion, becomes already non-linear at the
background level. While the non-linearity we are discussing here is
on the perturbation level. By definition, linear perturbation will
not back-react the background. Back-reaction from perturbations is
possible to show up only when non-linear fluctuations are
considered.}.

The cosmological principle is indeed supported strongly by
cosmological experiments on CMB and large scale structure (LSS). However, at
late times and on small scales (sub-Hubble scales), the universe is
not homogeneous at all. There are all kinds of structures in the
universe. On the other hand, on scales larger than the observable
universe (super-Hubble scales), it is not known whether the universe
remains homogeneous and isotropic or not. Due to the nonlinearity of
gravity, these sub-Hubble or super-Hubble scales may back-react on to
the scale of the observable universe.

As early as in 1931, Einstein \EinsteinNoCC\ already mentioned that the
matter distribution is in reality inhomogeneous and the approximate
treatment (cosmological principle) may be illusionary, when he was
trying to explain why the Hubble's value of the Hubble parameter is
about ten times too large \StraumannTV. Half a century later, Ellis
re-examined the effect of clumpiness on the average, under the name
fitting problem \refs{\EllisBR,\EllisZZ}. This starts the modern story
of back-reaction as an effective component of dark energy.

\subsec{Sub-Hubble inhomogeneities}

In this subsection, we consider back-reaction from sub-Hubble scale
fluctuations. Typically, the aim of the sub-Hubble scale back-reaction
theory is that, assuming there is no cosmological constant (i.e. the
old cosmological constant problem is solved), and provide cosmological
acceleration from back-reaction from sub-Hubble scale inhomogeneities.
Most details of this section can be found in \RasanenKI. We assume
in this subsection that the universe is dominated by pressureless
dust.

To consider small scale fluctuations, it is useful to derive local
versions of the Friedmann equation and the continuity equation. To do
this, it is helpful to decompose the derivative of the velocity field
into components, where each component has clear meaning in the sense
of fluid mechanics:
\eqn\BRudiv{u_{\mu;\nu} = {1\over 3}\theta h_{\mu\nu}
+ \sigma_{\mu\nu}+\omega_{\mu\nu},} where $u^\mu$ is the four velocity
of the fluid, and $h_{\mu\nu} = g_{\mu\nu}-u_\mu u_\nu$ is the spatial
projection of the metric.

On the right hand side of \BRudiv, the scalar part $\theta$ is called
volume-expansion scalar, defined as $\theta \equiv u^\mu_{~;\mu}$,
which measures the local expansion of the fluid. In the familiar
homogeneous and isotropic FRW universe, $\theta$ is reduced to $3H$.
The symmetric part $\sigma_{\mu\nu}$ is defined by $\sigma_{\mu\nu} =
u_{(\mu;\nu)}-\theta h_{\mu\nu}/3$ and describes the shear of the
fluid. The anti-symmetric part $\omega_{\mu\nu}$ is defined by
$\omega_{\mu\nu} \equiv u_{[\mu;\nu]}$ and describes the
vorticity.

Take one more covariant derivative on equation \BRudiv, using the
commutative relation for covariant derivative to relate the
derivatives to curvature, and using the Einstein equations to relate
the curvature to stress tensor, one can obtain
\eqn\BRRaychaudhuri{\dot\theta + {1\over 3}\theta^2 = -4\pi
G\rho-2\sigma^2+2\omega^2,} This equation is known as the Raychaudhuri
equation \RaychaudhuriYV, which is widely used in general relativity. Another useful
equation from the combination of the Einstein equations and the
decomposition equation \BRudiv\ is \eqn\BRLocalFriedmann{{1\over
3}\theta^2 = 8\pi G\rho-{1\over 2}R_3+\sigma^2-\omega^2,} which is the
local version of the Friedmann equation, where $R_3$ is the spatial
curvature on the slice orthogonal to $u^\mu$, $\sigma^2\equiv
\sigma^{\mu\nu}\sigma_{\mu\nu}/2$, and $\omega^2 \equiv
\omega^{\mu\nu}\omega_{\mu\nu}/2$. Finally the continuous equation for
matter component takes the form \eqn\BRcontinuous{\dot\rho+\theta\rho
= 0.}  One can also derive equations for the time evolution of
$\sigma_{\mu\nu}$ and $\omega_{\mu\nu}$ from the Einstein equations,
but we will not need them here.

Now we are about to average these local analog of the Friedmann
equations. The spatial average operation is defined as
\eqn\BRaverage{\langle f \rangle (t) \equiv
    {\int d^3 x \sqrt{g_3(t,x)}f(t,x)
    \over \int d^3 x \sqrt{g_3(t,x)}},}
where $g_3$ is the determinant of the three dimensional metric.

Using this definition, the local equations \BRRaychaudhuri,
\BRLocalFriedmann\ and \BRcontinuous\ can be written as
\eqn\BRavgeRay{3{\ddot {a} \over {a}} = -4\pi
G\langle\rho\rangle+{\cal Q},} \eqn\BRavgeFried{3 H^2 = 8\pi
G\langle\rho\rangle-{1\over 2}\langle R_3\rangle-{1\over 2}{\cal Q},}
\eqn\BRavgeCont{\partial_t\langle\rho\rangle+3H\langle\rho\rangle=0.}
The equations \BRavgeRay, \BRavgeFried\
and \BRavgeCont\ are called the Buchert equations \BuchertER,
where the averaged scale factor $a$ is defined as \eqn\BRaDef{a(t)
\equiv\left( {\int d^3 x \sqrt{g_3(t,x)} \over \int d^3 x
\sqrt{g_3(t_0,x)}}\right)^{1/3}, \qquad H \equiv {\dot a \over
a}.}Note that $a$ and $H$ used in this subsection denote averaged
variables, not to confuse with those in other sections. $\cal Q$ is defined as \eqn\BRQDef{{\cal Q}
\equiv {2\over
3}\left(\langle\theta^2\rangle-\langle\theta\rangle^2\right)
-2\langle\sigma^2\rangle.}

In the Buchert equations \BRavgeRay\ and \BRavgeFried, the
back-reaction variable $\cal Q$ is the novel term compared from the
familiar FRW equations. $\cal Q$ can be thought of emergent in the
sense of coarse graining. When ${\cal Q}>0$, $\cal Q$ will behave as a
effective component in the universe, which drives the late time
acceleration.

So far we have not said anything about how to choose spatial slices to
do the average. In fact choosing spatial slices properly is extremely
important for the back-reaction calculation to make correct
predictions, both theoretically and phenomenologically.

Theoretically, in general relativity there is
no preferred choice for the spatial slices. In linear cosmic
perturbation theory, there is an elegant gauge invariant way to do
calculation. However, here the back-reaction one considers is beyond
linear order and a gauge invariant formalism is not available on
sub-Hubble scales.

Phenomenologically, cosmological experiments are
typically carried out by measuring redshift and distance. When the
universe is perturbed one has to make sure whether the calculated
scale factor is indeed the one which is used to calculate these
quantities.

For example, as pointed out by Ishibashi and Wald \IshibashiSJ, if one
averages two disconnected decelerating universes, one could get the
conclusion that the ``whole'' universe is accelerating. This absurd
conclusion shows that great care is needed to select the averaging
hypersurface. For this purpose, one is led to consider the light
propagation in a perturbed space.

In \RasanenUW, Rasanen showed that the redshift in a dusty universe
can be calculated as \eqn\BRrasanenZ{1+z = \exp
\left\{\int_\eta^{\eta_0}d\eta \left({1\over 3}\theta
+\sigma_{\mu\nu}e^{\mu}e^{\nu}\right) \right\},}
where the integral is along the null geodesic, and $e^\mu$ is along
the spatial direction of the geodesic. One can argue that when there
is no preferred directions in the sky, the integration of
$\sigma_{\mu\nu}e^{\mu}e^{\nu}$ should be suppressed by averaging
effect, as long as one choose the spatial slice with statistical
homogeneity and isotropy.

Similarly, it is shown in \RasanenBE\ that for angular diameter
distance and the luminosity distance, similar average effects occur
for statistical homogeneous and isotropic slicing. These results show
evidence that the statistical homogeneous and isotropic slicing is the
one that should be used when the calculation of back-reaction need to
be compared with observations.

Having determined the choice of slicing, one can focus on the
calculation of the back-reaction variable $\cal Q$. Unfortunately the
calculation is very difficult. The Newtonian calculation might be
oversimplified, as in Newtonian gravity the two terms in $\cal Q$
cancels up to a surface term \BuchertFZ. For calculations in general
relativity, only toy models are doable currently. Some calculations
report there is strong cancellation between the two terms in $\cal Q$,
although they do not exactly cancel \ParanjapeAI. Others show that it
is not the case \ChuangYI. Thus whether or not small scale back-reaction
could become important deserves further investigation.

Besides the approach of averaging, there is also a
Lema\^\i tre-Tolman-Bondi \refs{\LTBL,\LTBT,\LTBB} approach of sub-Hubble
back-reaction. The idea is that there are voids in the large scale
structure. We may live in a void in which the observables such as the
luminosity distance need to be calculated with more care. For
examples, in \refs{\MarraKMR,\MarraKM,\BiswasUB,\BiswasGI}, it is shown that the LTB type
back-reaction could mimick dark energy.

\subsec{Super-Hubble inhomogeneities}

Compared with the sub-Hubble theories, the theories of super-Hubble
back-reaction has a wider variety of goals. Some of the works aim to
provide a screening mechanism for the cosmological constant, and
completely solve the cosmological constant problem. While some other
works, like sub-Hubble theories, aims to give acceleration assuming
the cosmological constant is zero. The difference is that, screening
mechanism gives a negative energy density contribution from
backreaction, while the latter gives a positive energy density from
backreaction.

For the screening mechanism, Mukhanov, Abramo and Brandenberger
\refs{\MukhanovAK,\AbramoHU} have set up a gauge invariant formalism for
perturbations up to second order. The quantities they use are gauge
invariant after spatial integration (averaging). For scalar field
matter $\varphi = \varphi_0 + \delta\varphi$, the second order
perturbated energy momentum tensor is calculated. Interestingly, the
second order energy momentum tensor, in the slow roll approximation,
has an equation of state $p=-\rho>0$, in other words, it is a
cancellation or screening of cosmological constant. There are also
similar results from gravitational loop calculations by earlier
studies by Tsamis and Woodard \refs{\TsamisSX,\TsamisQQ}.

However, Unruh \UnruhIC\  (see also \KodamaQW) pointed out that gauge
invariance of spatial averaged variables does not guarantee that the
calculated effect is accessible or observable by a local observer.
This objection is supported by Geshnizjani and Brandenberger
\GeshnizjaniWP\  from an explicit calculation of a scalar field coupled
to gravity. It is shown that without isocurvature fluctuations, the
expansion $\theta$, which is locally accessible, do not receive any
backreactions from super-Hubble perturbations. The expansion $\theta$
including back-reactions, as a function of
$\varphi=\varphi_0+\delta\varphi$, has the same function dependence as
the background expansion $\theta_0$, as a function of
$\varphi_0$:
\eqn\BRnoAdiabic{\theta = \sqrt{3GV(\phi)}.}
Thus for super-Hubble perturbations, a local observer
will observe $\varphi$ and can not find a difference in dynamics
compared with the background.

However, with isocurvature perturbations, the situation changes. In
\GeshnizjaniCN, Geshnizjani and Brandenberger considered back
reaction in two scalar field model and found that in this case the
back-reaction does not vanish. The reason is that there are different
choices for local clocks. For different choice of clocks (proper time
clock and energy density for a scalar field, as considered in
\GeshnizjaniCN), the results of back-reaction are different (actually
the correction changes sign). Thus it is crucial to identify which is
the observable related to observations of dark energy.

Abramo and
Woodard \refs{\AbramoDB,\AbramoDD} proposed a local operator (before spatial
average) and showed that the back-reaction does not vanish.
Further calculation \refs{\OnemliHR,\OnemliMB,\BrunierSB,\KahyaHC,\OnemliAS} showed from
one-loop and two-loop calculation that
the probe scalar field obtains an equation of state $p<-\rho$, which
may drive a period of super-acceleration. Eventually the
super-acceleration will be turned off by the non-zero renormalized
mass. Based on these consideration, a scenario of non-local cosmology
is proposed to model the above behavior \refs{\DeserJK, \DeffayetCA,
\TsamisPH} (see also \SolaDH\ and references therein).

The back-reaction from scalar field to de Sitter space as a screening
mechanism is also considered in
\refs{\PolyakovMM,\PolyakovNQ,\KrotovMA}. The authors argue that the
de Sitter space analytically continued from a sphere should not be
used to describe realistic cosmology because it corresponds to de
Sitter space artificially kept at fixed Gibbons-Hawking temperature,
which corresponds to de Sitter space sourounded by reflecting
walls. Instead, They consider a period of de Sitter expansion
\eqn\polyakovSetup{ds^2 = -dt^2 + a(t)^2 d{\bf x}^2, \qquad a(t) =
e^{T\tanh(t/T)}, } where $T\gg 1/H$ is a time cutoff such that
interactions from de Sitter background are removed in the asymptotic
past and future. Correspondingly, the vacuum choice is that the field
approaches local Minkowski vacuum at $t\rightarrow -\infty$, which is
different from the Bunch-Davies vacuum.

The calculations in \KrotovMA\ show result as follows. For free scalar
fields, as a result of de Sitter symmetry, the correction to the one
point function is a trivial redefinition of the cosmological
constant. For interacting fields in the FRW patch of the de Sitter
space, two point function of the scalar correlator produces non-trivial
infrared divergence but the one point function is not affected. In
global de Sitter space, where the accelerating expansion of the
universe is initialized by a period of contraction, there are
non-trivial effects for the one point function for scalar
fields. However, for the case of dark energy, our universe is clearly
not contracting right before dark energy domination. Thus it remains
an open question whether this approach could provide a dynamical
explanation for the dark energy problem. Nevertheless, the
interaction which breaks the de Sitter symmetry is an interesting
candidate for dynamical dark energy.

Some other screening mechanisms are also reviewed in the previous
section on tuning mechanisms, where the issue of back-reaction is
not as relevant as discussed here.

As mentioned above, there is another class of super-Hubble
back-reaction models, which produce instead of screen a cosmological
constant. For example, Kolb, Matarrese, Notari and Riotto considered
the effect of super-Hubble perturbations from inflation. These
perturbations, viewed from a local observer, may look like a source of
acceleration. There are also counter arguments on these class of
mechanisms, see for example \KumarUK.

One should also note that there is a whole literature for loop
calculation during inflation, which is another accelerating epoch of
the universe. The techniques developed in those loop calculations
are also helpful in understanding the dynamics of dark energy. But
we shall not introduce these works here. Interested readers are
referred to \WeinbergVY\ and related works.

\newsec{Phenomenological models}
\seclab\secPhenoModels

In this section we introduce phenomenological models for dark energy,
focusing on modification for known matter components. \foot{Sometimes it is
difficult to classify whether a model belongs to modifying matter or
modifying gravity because they are coupled. Indeed some models reviewed
in this section can be thought as modification of gravity in the
infrared. But we include them here anyway because in these models
gravity is not modified from the action.}

\subsec{Quintessence, phantom and quintom}
The most well-studied parts of phenomenological models are models
with rolling fields. This is because such models are direct
generalizations of the cosmological constant: When the
fields are not rolling, their potential energy behaves as a
cosmological constant. Here we review these models in a logic
order and in a brief way. The readers can find more details in the reviews
\refs{\CopelandWR,\CaiZP}, and of course as well as the original papers.

\noindent $\bullet$ Quintessence

A quintessence field \refs{\WetterichFM,\ZlatevTR} is a scalar field
with standard kinetic term, minimally coupled to gravity. The scalar
field part action takes the form \eqn\PMquintessenceAction{S = \int
d^4 x \sqrt{-g} \left[-{1\over
2}g^{\mu\nu}\partial_\mu\varphi\partial_\nu\varphi
-V(\varphi)\right],} where the metric convention is ($-$, $+$, $+$,
$+$) such that the scalar field has standard kinetic term. Take
variation of $g^{\mu\nu}$, one obtains the stress tensor.
\eqn\PMquintessenceStress{T_{\mu\nu}
= \partial_\mu\varphi\partial_\nu\varphi -
g_{\mu\nu}\left[{1\over 2}\partial^\lambda\varphi\partial_\lambda\varphi+V(\varphi)\right].}
The energy density and pressure can be read off from the energy
momentum tensor as \eqn\PMquintessenceRhoP{\rho = {1\over
2}\dot\varphi^2+V(\varphi), \qquad p = {1\over
2}\dot\varphi^2-V(\varphi).}

As usual, the Friedmann equations are
\eqn\PMFriedmann{3M_p^2 H^2 = \rho, \qquad
-2M_p^2\dot H =\rho+p,}
where, as a reminder, $M_p^2 = 1/(8\pi G)$. In terms of $\varphi$, the
above equation reads
\eqn\PMFriedmannVarphi{3M_p^2 H^2 = {1\over 2}\dot\varphi^2+V, \qquad
-2M_p^2\dot H =\dot\varphi^2,}

Another useful equation comes out of a combination of these equations:
\eqn\PMacceleration{
-6M_p^2\left({\ddot a \over a}\right) = \rho + 3p = 2(\dot\varphi^2-V).}
From local energy conservation, the continuous equations is
\eqn\PMcontinuous{\dot\rho+3H(\rho+p)=0,\qquad
\ddot\varphi+3H\dot\varphi+V'(\varphi)=0,}
where the latter is a rewritten of the former in terms of the field $\varphi$.

The equation of state takes the form
\eqn\PMeqos{w={p\over\rho}={{1\over 2}\dot\varphi^2-V(\varphi) \over
{1\over 2}\dot\varphi^2+V(\varphi)}.}  Note that the kinetic term has
positive pressure and the potential term has negative pressure. When
the field rolls slowly, the potential dominates thus $w$ approaches
$-1$ from above.

What kind of potentials can give rise to acceleration? A simple
answer, is flat potentials. To make the answer more precise, one can
have a look at the critical case: cosmic expansion without
acceleration or deceleration. For this purpose, consider the power-law
expansion
\eqn\PMpowerLaw{a\propto t^p,}
where we have kept $p$ general, keeping in mind that $p=1$ corresponds
to zero acceleration and $p>1$ corresponds to acceleration. The
potential driving this kind of acceleration can be solved from
\PMFriedmann\ as
\eqn\PMcriticalPotential{V=V_0 \exp \left(
-\sqrt{2\over p}{\varphi\over M_p}\right)
.}
Thus potentials flatter than
\eqn\PMcriticalPotential{V=V_0 \exp \left(
-{\sqrt{2}\varphi\over M_p}\right)
}
have acceleration solution. Moreover, as the kinetic term has much
larger pressure than the potential term, the potential domination
epoch is an attractor solution as long as the potential is flat.

Quintessence may be realized using axions \PandaAS,
dilatons \UzanAs,
in QCD \StoAsa, in Higgs potential \StoAsb, or in unparticle theories \StoAsc.

\noindent $\bullet$ Phantom

A menace from phantom \CaldwellEW\ can be expressed in terms of the action
\foot{Alternatively, phantom dark energy can be also obtained in the scalar-tensor models \BoisseauPho.}
\eqn\PMquintessenceAction{S = \int d^4 x \sqrt{-g} \left[{1\over
2}g^{\mu\nu}\partial_\mu\varphi\partial_\nu\varphi -V(\varphi)\right],}
The action has a ``wrong'' sign kinetic term: ${\cal L}_{\rm kin}\propto -\dot\varphi^2$.
Here phantom is also often referred to as ghost in the literature.

When expressed in terms of $\rho$ and p, the
equations of motion for phantom are identical as written in the
quintessence case, while now $\rho$ and p are expressed in terms of
$\varphi$ as
\eqn\PMphantomRhoP{\rho = -{1\over
2}\dot\varphi^2+V(\varphi), \qquad p = -{1\over
2}\dot\varphi^2-V(\varphi).}
The equation of state is
\eqn\PMeqos{w={p\over\rho}={{1\over 2}\dot\varphi^2+V(\varphi) \over
{1\over 2}\dot\varphi^2-V(\varphi)}.}
Now there are two possibilities for the equation of state: $w>1$ for
the kinetic dominated regime and $w<-1$ for the potential dominated
regime. The latter behaves as a component dark energy with
super-acceleration. In other words, the universe will have an
acceleration faster than exponential. The energy density keeps growing
until it reaches infinity in finite proper time. When the energy
density reaches infinity, the expansion rate of the universe diverges
and every thing is tore off. This is called the ``big rip''
singularity, a physical singularity where all known physical laws
break down. To see this, consider for simplicity the constant $w$
case. In this case the scale factor can be written as a simple
power of time as
\eqn\PMrip{a=a_0 (t-t_0)^{2\over 3(1+w)}.}
When $w<-1$, the power of $t-t_0$ becomes negative and one concludes
$a\rightarrow \infty$ when $t\rightarrow t_0$, a finite proper time
for a comoving observer. The Hubble parameter $H\propto 1/(t-t_0)$
also blows up in the future. The reason is that when $w<-1$, $t_0$ is
in the future instead of in the past. Similarly the curvature blows up
thus the big rip $t=t_0$ is a physical singularity. The big rip is a
disaster not only for civilizations but also for physical laws, which
need to be avoided or resolved.

Moreover, there is a quantum instability in the phantom models
\foot{However, the statement $w<-1$ itself does not necessarily mean
an instability. For example, holographic dark energy has $w<-1$
solutions without an instability. Also in some modified gravity models
such a well-behaved $w<-1$ solution may be obtained \UzanAsa.}. Once
the phantom quanta interacts with other fields, even through gravity,
there will be an instability of the vacuum because energy is no longer
bounded from below. The ghost busters Cline, Jeon and Moore \ClineGS\
pointed out that for the phantom to be consistent with CMB
observations, the Lorentz symmetry must be broken for phantom at an
energy scale lower than $3$MeV. Otherwise the effect of vacuum decay
into phantom quanta and photons via gravitation could have been
observed on the CMB.

\noindent $\bullet$ Quintom

The quintessence field always have $w>-1$ while the phantom
field (as dark energy) always have $w<-1$. Is it possible for a field
theory model to cross the phantom divide $w=-1$? The answer is
positive, which is known as the quintom dark energy, a combination of
the quintessence and the phantom \FengAD.

Before reviewing what can be done, it is helpful to first have a look
at what is under no-go theorems. As is noticed from the beginning
\FengAD, it is not possible to have a single scalar field to cross the
phantom divide \foot{For counter examples on classical instabilities,
see \refs{\DeffayetAs, \DLSVAs}.}. To see this, consider a model with time varying kinetic
term $\sim f(t)\dot\varphi^2$. If one wants to have the field cross
the phantom divide when $f$ cross 0, at $f=0$ the field will have
vanishing kinetic term thus divergent sound speed. More general proofs
can be found in \refs{\VikmanDC,\HuKH,\CaldwellAI,\XiaKM}.

Thus one is forced to consider models with at least two fields, with
the matter action \eqn\PMquintessenceAction{S = \int d^4 x \sqrt{-g}
\left[-{1\over 2}\partial^\mu\varphi\partial_\mu\varphi
+{1\over 2}\partial^\mu\sigma\partial_\mu\sigma -V(\varphi,\sigma) \right],}

The two-component dark energy has an equation of state
\eqn\PMphantomW{w={{1\over 2}\dot\varphi^2-{1\over 2}\dot\sigma^2-V \over
{1\over 2}\dot\varphi^2-{1\over 2}\dot\sigma^2+V}.}

The model is extensively studied in \refs{\ENO,\CaiZP}, and the references
therein. Depending on the potential, the quintom equation of state may
across $-1$ from above, or across $-1$ from below \foot{Alternatively, in
$f(R)$ gravity models, $w$ may also cross $-1$ \GengASa.}.
For an extended version of quintom scenario, see \Chime.

\noindent $\bullet$ Fast oscillating fields

Here we consider a scalar
field with standard kinetic term and potential $V\propto \varphi^n$,
focusing on the case that the field is oscillating around its minima
and the oscillation rate is much more quickly than Hubble
parameter. Turner calculated the averaged equation of state of this
case \TurnerHE, with \eqn\PMOsc{\langle w\rangle={n-2\over n+2}.}
When $n\ll 1$, the fast oscillating field can drive cosmic
acceleration \refs{\DamourCB,\SahniQE}. To see this, one can apply the virial
theorem. Define $G\equiv \varphi\dot\varphi$. We have \eqn\PMVir{\dot
G = 2T-(3H\dot\varphi+V')\varphi\simeq 2T -nV,} where $T\equiv
\dot\varphi^2/2$ and we have neglected Hubble expansion in the last
expression because the field is oscillating fast. Note that $\dot G$
is a total derivative thus the time average vanishes when we take the
averaging time to be a multiple of the oscillation period. Thus
\eqn\PMavg{\langle T \rangle = {n\over 2}\langle V\rangle. }
Translating to the averaged equation of state, $w$ takes the form of
\PMOsc.

Unfortunately, the fast oscillating field has an instability, as discussed
in \JohnsonSE. The inhomogeneity due to the instability may get this
mechanism observationally disfavored.

\noindent $\bullet$ Dark energy interactions

It is possible that dark energy is dark but not lonely. Especially
dark energy may decay into dark matter or vice versa.  In \refs{\AmendolaER,
\ZimdahlAR, \WangJX}, the following class of interaction is
considered:
\eqn\PMintM{\dot\rho_m + 3H\rho_m=\delta,}
\eqn\PMintDE{\dot\rho_{de} + 3H(1+w_{de})\rho_{de}=-\delta,}
where $\delta$ could take the form $\delta = -b\rho_{de}$, $\delta =
-b\rho_m$, or $\delta = -b(\rho_m+\rho_{de})$, etc. The effective
equations of state for dark energy and matter are what one actually
measures and different from their actual equations of state:
\eqn\PMintEOS{w^{\rm eff}_m = -{\delta\over 3H}, \qquad
w^{\rm eff}_{de} = w_{de}+{\delta\over 3H}.}
In some sense, the interaction could provide an explanation for the
coincidence problem because if dark energy decays to dark matter in
the future, the fraction of dark energy may remain at the value one
observes now. However, why dark energy starts to dominate today but
not much earlier, as
the other half of the coincidence problem, remains unsolved.

Alternatively, the phenomenological time variation of the cosmological
constant can also be implemented by a $\Lambda(t)$CDM approach, as
described in \refs{\OverduinAs,\WangMengAs}.

\subsec{K-essense, custuton, braiding and ghost condensation}

Here we briefly review dark energy models from fields with modified
kinetic terms.

\noindent $\bullet$ K-essense

Chiba, Okabe and Yamaguchi \ChibaKA\ (see also
Armendariz-Picon, Mukhanov and Steinhardt \refs{\ArmendarizPiconDH,
\ArmendarizPiconAH}) proposed a more general framework on field
theoretic dark energy, named k-essense. The idea is to generalize the
kinetic term, as long as the perturbations still have second order
derivative and Lorentz symmetry is preserved \refs{\ArmendarizPiconRJ,
\GarrigaVW}. The corresponding action in this case is
\eqn\PMKessense{S=\int d^4 x \sqrt{-g} ~p(X, \varphi), \qquad X\equiv
-g^{\mu\nu}\partial_\mu\varphi\partial_\nu\varphi.}  Here $X$ is the
conventional kinetic term. Note that $p$ appears in the action is
exactly the pressure, and the energy density takes the form
\eqn\PMKrho{\rho = 2Xp_X-p,}
where $p_X\equiv \partial p /\partial X$. In terms of $\rho$ and $p$,
again one have the the Friedmann equations \PMFriedmann.

It is shown \GarrigaVW\ that the perturbations obey second order
differential equation of motion as usual (except a special case considered later). To be ghost free, the theory
should have
\eqn\PMKghostFree{p_X>0.} To be perturbatively stable, it is also
required that the sound speed is real. The condition is
\eqn\PMKcs{c_s^2 = {p_X \over
p_X + 2X p_{XX}}>0.}
Further, one might also require the sound speed $c_s\leq 1$. This is
satisfied when
\eqn\PMKcsLeqOne{p_{XX}\geq 0.}
There is a debate on whether $c_s\leq 1$ is necessary or not \BMVAs. Thus the
condition that k-essense is well behaved is \PMKghostFree\ plus either
\PMKcs\ or \PMKcsLeqOne.

To make predictions, more concrete forms of $p(X, \varphi)$ are
needed. The simplest case is perhaps power law k-essense, where
\eqn\PMKpowerLaw{p(X,\varphi)={4(1-3w)\over
9(1+w)^2\varphi^2}\left(-X+X^2\right).} It is shown that the parameter
$w$ is indeed $w=p/\rho$, with the late time behavior $a\propto
t^{2\over 3(1+w)}$. Another example is DBI-essense, considered by
Martin and Yamaguchi \YamaguchiAS. Also, k-essence may provide a
unified framework of inflation and dark energy \JojiriAsx.

\noindent $\bullet$ Cuscuton

The Cuscuton dark energy is introduced
by Afshordi, Chung and Geshnizjani \AfshordiAD, as a singular limit of
k-essense. In the cuscuton model, the matter action takes the from
\eqn\PMCuscuton{S = \int d^4 x \sqrt{-g} \left[ \mu^2
\sqrt{|g^{\mu\nu}\partial_\mu\varphi\partial_\nu\varphi|}
-V(\varphi)\right].}  The cuscuton model has an infinite propagating
speed for linear perturbations. However, the phase space volume of
linear perturbations is vanishing, thus no information is propagating
faster than speed of light.

Cuscuton is inspired by the plant Cuscuta, because cuscuton is
parasitic in the sense that the cuscuton itself does not have its own
dynamics. The equation of motion takes the form
\eqn\PMCuscuton{(3\mu^2 H) ~{\rm sgn}(\dot\varphi)+V'(\varphi)=0.}
The evolution of cuscuton follows from other energy components, which
can be derived from  combining \PMCuscuton\ and the Friedmann equation.
When adjusting $V(\varphi)$, cosmic acceleration can be obtained. For
example, when $V$ is an exponential potential, the expansion history
of cuscuton behaves like DGP (but perturbation theory behaves
different). Cuscuton may be a minimal dynamical dark energy
model because it has no dynamics, while remains the dynamical feature
for dark energy. The cosmic evolution in the cuscuton model is investigated
in \AfshordiYX.

\noindent $\bullet$ Kinetic gravity braiding

K-essense is not the most general form for a scalar field with second
order equation of motion. As shown in \refs{\DeffayetAs,\DPSVAs} (see also
\refs{\NicolisIN, \KobayashiCM} in terms of a simplified yet
generalized Galileon model), a more general form can be written
as \eqn\PMGalileon{S=\int d^4 x \left[ K(\varphi,
X)-G(\varphi,X)\partial^2\varphi \right]. } The cosmological
implications are studied in \refs{\ChowFM, \NesserisPC}. It is shown
that the field could behave as dark energy. This class of
actions can be further generalized to \DeffayetGZ, where cosmological
implications are so far not studied.

\noindent $\bullet$ Ghost condensation

As another generalization of k-essense, a model of ghost condensation
is aimed to cure the quantum instability of phantom dark energy, and
give an equation of state $p=-\rho$ to drive the late time
acceleration of the universe. The idea is that the instabilities come
form the perturbation level, where the gradient energy plays an
important role. On the other hand, the spatial gradient energy does
not show up in the homogeneous and isotropic background lever. Thus
one can propose a ghost like background action, and let the spatial
gradient terms cure the instability problem.

To realize this idea, Arkani-Hamed, Cheng, Luty and Mukohyama
\ArkaniHamedUY\ (see also \PiazzaDF) proposed
an effective field theory of rolling ghost, preserving
cosmological symmetries. The action takes the form
\eqn\PMGhostCondensation{S=\int d^4x \sqrt{-g} \left[
p(X)+(\nabla^2\varphi)^2+\cdots\right].}  The cosmological solution of
the model is either $\dot\varphi=0$ or $p_X=0$. In interesting models
analogous to tachyon condensation with a spontaneous symmetry breaking,
the $\dot\varphi=0$ solution is unstable and the universe is driven to
the $p_X=0$ solution dynamically. This results in $X={\rm constant}$, with
$\rho=-p$ (recall \PMKrho\ with $p_X=0$). Thus the condensation of
ghost behaves like a component of dark energy. The intuitive
understanding of a constant equation of state is that, there is a
shift symmetry of the system, thus rolling of $\varphi$ with a
constant speed results in a constant $w$. The model of ghost
condensation can be thought of as an IR modification of gravity (for
other IR modifications, see \PiazzaAs).

Inspired by ghost condensation, an inflation model named effective
field inflation \CheungST\ is proposed, with the most general action
preserving cosmological symmetries.
In addition, A ghost dark energy has also been proposed \GhoseDE \ and has been widely studied \GhoseDEstu.

\subsec{Higher spin fields}

In cosmology, going to higher spin mostly means spin higher than
zero. There are attempts to generalize the scalar field dark energy
model to spinors \refs{\RibasVR, \CaiGK, \YajnikMH, \TsybaAs}, vectors
\refs{\ArmendarizPiconPM, \ZhaoBU}, and $p$-form fields \refs{\KoivistoFB,
\DasGuptaJY}. Here we shall briefly mention the approaches of spinors.

In curved spacetime, the action of a spinor field can be written as
\eqn\PMspinorAction{S=\int d^4 x ~e
\left[{i\over 2}\left(\bar\psi \Gamma^\mu
D_\mu\psi-(D_\mu\bar\psi)\Gamma^\mu\psi\right)-V(\bar\psi\psi,\bar\psi\gamma^5\psi)\right],}
where $e\equiv \det e^\mu{}_a$, and $e^\mu{}_a$ is the tetrad field
satisfying $g_{\mu\nu}e^\mu{}_ae^\nu{}_b=\eta_{ab}$. The $\Gamma$ matrices
are defined as $\Gamma^\mu = e^\mu{}_a \gamma^a$. The covariant
derivative is defined as $D_\mu\psi=(\partial_\mu+{1\over
2}\omega_{\mu ab}\Sigma^{ab})$, with $\omega_{\mu
ab}=e^\nu{}_aD_{\mu}e_{\nu b}$, and $\Sigma^{ab}={1\over
4}[\gamma^a,\gamma^b]$. If the vacuum expectation
value transform as a scalar instead of a pseudo-scalar, the
$\bar\psi\gamma^5\psi$ term can be dropped.

When the field is homogeneous, the energy density and
pressure can be written as \eqn\PMspinorEos{\rho = V, \qquad p =
V'\bar\psi\psi-V.}  It can be shown from the equation of motion that
$\bar\psi \psi\propto a^{-3}$. In other words the spinor field rolls
very fast. Thus to have acceleration (and not coming simply from a
cosmological constant), one need a very flat potential. For example, a
potential $V\propto (\bar\psi\psi)^n$ with $n<1/2$ will have
acceleration. Moreover, as a spinor has multiple degrees of freedom
built in, the equation of state $w$ may cross $-1$ without
divergence of sound speed \CaiGK.

For vector fields, one have to take care of constraints and large
scale anisotropy. Also one has to break the conformal
invariance otherwise the energy density decays too quickly. We shall
not discuss these issues in details here.

\subsec{Chaplygin gas and viscous fluid}

The matter components in cosmology are usually and conveniently
written in terms of fluids. Most dark energy models have fluid
description. The fluid models we refer to here correspond to the
models that originate from direct modification of fluid properties.

\noindent $\bullet$ Chaplygin gas

The best studied fluid model for dark energy is the Chaplygin gas
model \refs{\AbramoDB,\KamenshchikCP}. The original Chaplygin gas model has the
equation of state
\eqn\PMChaplyginOriginal{p=-{A\over\rho}.}
This kind of fluid is first used in fluid mechanics which describes
the air flow near the wing of a aircraft.

In the presence of matter, the energy density and pressure of the
Chaplygin gas can be written as
\eqn\PMChaylyginRP{\rho \simeq \sqrt{A+B a^{-6}},
\qquad p=-{A\over \sqrt{A+B a^{-6}}}.}
where $B$ is an integration constant.

Although the Chaplygin gas itself is not motivated from field theory
models, it is interesting that the model can be mimicked by a scalar
field with a simple potential, in the matter or dark energy dominated
universe:
\eqn\PMChaplyginPotential{V(\varphi)={\sqrt{A}\over 2}\left(
\cosh 3\varphi + {1\over \cosh 3\varphi}\right).}

The Chaplygin gas model is later generalized to
\eqn\PMChaplyginGeneralized{p=-{A\over\rho^\alpha}} in \BentoPS. Other
kinds of generalization are also possible, for example the form
\HovaNA \eqn\PMChaplyginHY{p=-\rho\left[1-{\rm
sinc}(\rho_0/\rho)\right],} where ${\rm sinc}(x)\equiv \sin(x)/x$, and
$\rho_0$ is a parameter with dimension [mass]$^4$. The possible
theoretical embedding and phenomenology are also considered there.

\noindent $\bullet$ Viscous fluid

The local energy conservation equation \PMcontinuous\ is by far
satisfied by all the phenomenological models discussed above.
However, it is not always the case because energy can leak to other
forms such as the thermal motion from bulk viscosity. Brevik and
Gorbunova noticed that the modification due to bulk viscosity leads to
a model of viscous cosmology \refs{\BrevikSJ, \BrevikBJ}. The energy momentum
tensor of the fluid takes the form \eqn\PMviscous{T_{\mu\nu}=\rho
u_\mu u_\nu +(p-3H\zeta)h_{\mu\nu}, \qquad h_{\mu\nu}\equiv
g_{\mu\nu}+u_\mu u_\nu.}  The local energy conservation equation now
reads \eqn\PMviscous{\dot\rho+3H(\rho+p)=9H^2\zeta~,} where $\zeta$ is
the bulk viscosity. The equation for $\ddot a$ becomes
\eqn\PMviscosA{{\ddot a\over a} = 12\pi G H\zeta - {1+3w\over 2}H^2.}
Thus not only $w<-1/3$ could drive acceleration, a viscosity $\zeta>0$
will also be able to drive acceleration. Meng, Ren and Hu
\refs{\MengJY,\RenNW} pointed out that the viscous fluid may give a unified
description of dark energy and dark matter, under the name ``dark
fluid''. A more general form of such an inhomogeneous equation of
state is proposed in \NojiriASa.

\subsec{Particle physics models}

As is well known, non-relativistic particles behave as $w=0$ matter
and relativistic particles behave as $w=1/3$ radiation. However, when
one considers particles with exotic properties, an equation of state
other than $w=0$ or $w=1/3$ may arise.

For example, DeDeo \DeDeoVI\ considered a Lorentz violating fermion
field
\eqn\PMparticleDeDeo{S=\int d^4x ~ e \left[\bar\psi (i\Gamma^\mu D_\mu-m)\psi
-(\bar\psi\gamma_5\Gamma^\mu\psi)b_\mu\right],}
where $b_\mu=(b,0,0,0)$ is a time like vector, and other notations
have been clarified under equation \PMspinorAction. It is argued
that the Lorentz violation can be achieved by coupling the spinor to
a condensate of ghost.

The equation of
state of the fermion particle can be calculated as
\eqn\PMDeDeoEOS{w_\pm(k)={k(k\mp b)\over 3[m^2+(k\mp b)^2]},}
where $\pm$ denotes the two helicities of the spinor, and $k$ is the
momentum of the particle.
For the positive helicity particle, $w$ becomes negative and a gas of
such particles becomes a candidate for dark energy.

As another approach, Bohmer and Harko \BoehmerFD\ reported that when a
minimum length is considered, particle excitations could also arise as
a component of dark energy.

Mass variation could also affect the particles' equation of state. For
example, Takahashi and Tanimoto \TakahashiAS \ showed that neutrinos with time
varying mass could behave as dark energy.

Alternatively, one could modify the particles' action from symmetry
considerations. For example, Stichel and Zakrzewski \StichelAS\
derived an action as a dynamical realization of the zero-mass Galilean
algebra with anisotropic dilational symmetry. It is shown that when
the dynamical exponent is $z=5/3$, the particles described by such an
action (named darkons) behave as dark energy.

By constructing a generalized dynamics for particles,
Das {\it et al.} \Das \ presented a new framework to generate an effective negative pressure and to give rise to a source for dark energy.

Another particle dark energy model is holographic gas \LiQH, which
modifies the dispersion relation of the particle such that the
entropy of the gas is holographic. The model is already
briefly reviewed in the section of Holographic principle.

In addition, Lima, Jesus and Oliveira \CCDMLJO \ suggested that
the creation of cold dark matter can yield a negative pressure and is capable to accelerate the Universe.
This model can mimics the $\Lambda$CDM model, thereby can provide a good fit to current cosmological data.
For more studies about this model, see \CCDMmore.

\subsec{Dark energy perturbations}

A simple cosmological constant will not have perturbations simply
because it is a constant. However, when the cosmological constant is
replaced by other fields, fluids or particles, dark energy will
typically has perturbations, either originating from the fluctuation itself or
originating from the clustering of other components via gravitational
interaction.

Here we follow the analysis in \MukhanovME\ (see also \MaEY), where the analysis is performed in
the Newtonian gauge, or equivalently, in terms of the gauge invariant
Barden potentials \BardeenKT. The perturbed metric can be written as
\eqn\PMpertds{ds^2 = a(\eta)^2 \left[ (1+2\Phi)d\eta^2 -
(1-2\Psi)\delta_{ij}dx^idx^j \right],} where after gauge fixing ($B=0$
and $E=0$), $\Phi$ and $\Psi$ are the two remaining scalar
perturbation variables. Also, perturbe the matter and dark energy
components as
\eqn\PMpertMatter{\rho\rightarrow\rho+\delta\rho \equiv \sum_i
(\rho_i+\delta\rho_i), \qquad p\rightarrow p+\delta p \equiv \sum_i
(p_i+\delta p_i),
}
\eqn\PMpertVS{\Theta \equiv ik^i\delta T^0{}_i/(\rho+p),
\qquad \Sigma \equiv {\hat k}_i{\hat k}_j\left(\delta T^i{}_j-{1\over 3}\delta^i{}_j
\delta T^k{}_{k}\right)/(\rho+p).}

In Fourier space, the perturbed Einstein
equations at linear order can be written as the following set of
equations: \eqn\PMpertEinsteinA{-k^2\Psi-3{\cal H}(\Psi'+{\cal H}\Phi)
= 4\pi G a^2 \delta\rho,} \eqn\PMpertEinsteinB{k^2(\Psi'+{\cal H}\Phi)
= 4\pi G a^2 (\rho + p)\Theta,} \eqn\PMpertEinsteinC{\Psi'' + {\cal
H}(2\Psi'+\Phi')+(2{\cal H}'+{\cal H}^2)\Phi+{k^2\over 3}(\Psi-\Phi) =
4\pi G a^2 \delta p,} \eqn\PMpertEinsteinD{k^2(\Psi - \Phi) = 12\pi G
a^2 (\rho+p)\Sigma,}
where prime denotes derivative with respect to the conformal time
$\eta$, and ${\cal H}\equiv a'/a$. The above Einstein equations are
consistent with a total energy conservation equation. For multiple
fluids, these equations should be solved together with the equations
of motion for these fluids to get a closed set of equations.

For shear-less fluids (such as perfect fluid and scalar fields), the
perturbation $\Sigma$ vanishes and $\Psi = \Phi$. Then the Newtonian
potential $\Phi$ can be solved as functions of the matter energy
densities and velocities from \PMpertEinsteinA\ and
\PMpertEinsteinB\ as \eqn\PMpertPoisson{k^2\Phi = -4\pi G a^2 \left[
\delta\rho + {3{\cal H}(\rho+p)\Theta \over k^2} \right].}

The solution of the above perturbation equations depend on detailed
models. When $w$ crosses the phantom
divide, some equations becomes singular and special care are needed
for proper treatments \refs{\ZhaoVJ,\XiaGE}.
In addition, dark energy perturbations also affect the cold dark matter power spectrum at large scale \CDMPS.
This can be used to distinguish dark energy models \DEPuse.

\newsec{The theoretical challenge revisited}

Before getting to the part of experiments and fitting, it is helpful
to pause and revisit the theoretical challenge. We are having too many
dark energy models. On the other hand, are we close to a solution to
the dark energy problem?

Let us review the questions discussed by Polchinski
\PolchinskiGY. Firstly, Polchinski reminds us that vacuum will
gravitate, at least in a local theory of gravity.

\ifig\FigElectron{An electron self energy diagram, in the presence of
an atom (left panel) and in the vacuum (right panel). In a local field
theory, graviton $g$ can not tell the difference because it only feels
the electron. It is not aware of the fields attached to the electron. }
{\epsfysize=2.5in \epsfbox{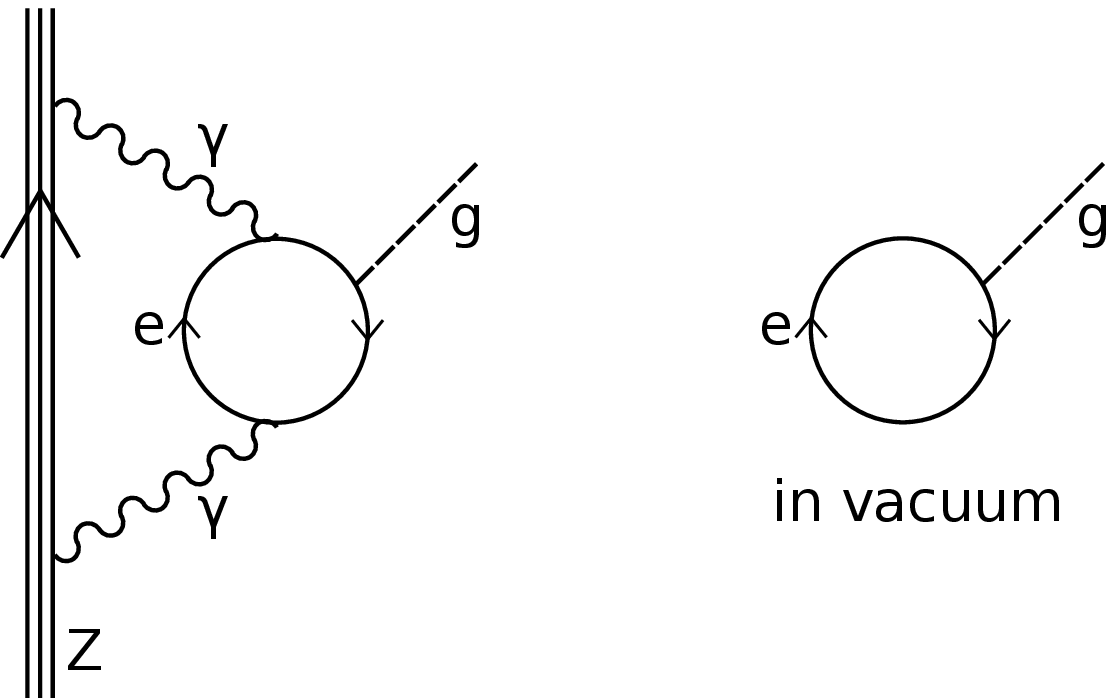}}

The reason is that the gravitational coupling of vacuum energy is
already locally measured. This comes from the accurate measurement of
equivalence principle. Aluminum and platinum have the same
gravitational mass to inertial mass ratio, up to an experimental error
of $10^{-12}$. On the other hand, the electronic loop (see the left
panel of \FigElectron)  contributions to their mass are about $10^{-3}$
and $3\times 10^{-3}$ to the rest energy of aluminum and platinum
respectively. Thus up to an accuracy of $10^{-9}$, gravity can feel
the energy from the electronic vacuum loop diagram.

On the other hand, the left panel and the right panel of
\FigElectron\ do not look locally different. As long as we have a local field
theory, gravity can not couple to one but not the other. Thus
one can not simply ignore the gravitational coupling to vacuum energy
in a serious consideration of dark energy models.

Secondly, the universe is not born empty. Instead it has a thermal
history. At least we expect the abundance of helium and other light
elements are explained by the big bang nucleosynthesis (BBN). Then the
early universe should at least as hot as MeV in temperature. How to
tune the current value of dark energy at that early time? Or if a
tracking mechanism operates, how does the mechanism know the energy
density will eventually fall to a small value instead of a large one?

Moreover, if we expect the universe originates from a higher energy
state, how does a solution of dark energy know we will eventually
settle down in the present vacuum, not the $SU(2)\times U(1)$
invariant one or any other one?

Thirdly, for modifications of gravity or matter on small scales, how
does this modification affect large scales while passing the solar
system tests; On the other hand, for modifications of gravity or
matter on larger scales, how can such a mechanism avoid the universe
explode or collapse at small scales? For example, for a UV cutoff
$\sim 100$GeV, the curvature of the universe will be only one meter!

After these considerations, Polchinski turned to anthropic
principle. However, as we have discussed in subsection
\subsecAnthropic, the anthropic principle itself has as many problems
as any of the other models, if not a lot more. Polchinski also
said, ``The cosmological constant is nonzero, therefore we can
calculate nothing.'' Anthropic principle not only has trouble itself,
but brings trouble to the whole field of theoretical physics.

With a great number of principles, ideas, mechanisms and models at
hand, we are still not able to find a model which can answer all these
questions, which at the same time is theoretically clearly
derived. Thus we are either still very far away from a solution to the
dark energy problem, or the solution hides behind one of the known
mechanisms above, but a lot more details need to be understood. Very
probably, the status of theoretical study needs to be substantially
driven by future experiments, as we shall review below.

\newsec{Cosmic probes of dark energy}
\seclab\secProbDE

The most common approach to probe dark energy is through its effect on the expansion history of the universe.
This effect can be detected via the luminosity distance $d_{L}(z)$ and the angular diameter distance $D_{A}(z)$.
In addition, the growth of large-scale structure can also provide useful constraints on dark energy.

Theoretical models and observational data can be related through the $\chi^2$ statistic
\foot{Another alternative is the median statistic. For more details, see \refs{\medianone,\mediantwo,\medianthree}}.
For a physical quantity $\xi$ with experimentally measured value $\xi_{\rm obs}$,
standard deviation $\sigma_{\xi}$, and theoretically predicted value $\xi_{\rm th}$,
$\chi^2$ is given by
\eqn\che{\chi_{\xi}^2={(\xi_{\rm obs}-\xi_{\rm th})^2\over\sigma_{\xi}^2}.}
Different cosmological observation gives different $\chi_{\xi}^2$,
and the total $\chi^2$ is the sum of all $\chi_{\xi}^2$s, i.e.
\eqn\chetotal{\chi^2=\sum_{\xi}\chi_{\xi}^2.}
By minimizing the total $\chi^{2}$, the best-fit model parameters can be determined.
Moreover, by calculating $\Delta \chi^2\equiv \chi^2-\chi_{\rm min}^2$,
one can also plot the 1$\sigma$, 2$\sigma$, and 3$\sigma$ confidence level (CL) ranges of a specific model.
Statistically, for models with different $n_p$ (denoting the number of free model parameters),
the 1$\sigma$, 2$\sigma$ and 3$\sigma$ CL correspond to different $\Delta \chi^2$.
In Table I, we list the relationship between $n_p$ and $\Delta \chi^2$.
In addition, the $\chi^2$ statistics are often performed by using the Markov Chain Monte-Carlo technology \refs{\MCMC,\MCMCBasic}.

\

\centerline{{\bf Table I : Relationship between $n_p$ and $\Delta \chi^2$}}

$$\vbox{\halign{\bf#\hfil&
\quad\hfil {$\displaystyle{#}$}\hfil & \quad\hfil {$\displaystyle{#}$}\hfil & \quad\hfil {$\displaystyle{#}$}\hfil &
\quad\hfil {$\displaystyle{#}$}\hfil & \quad\hfil {$\displaystyle{#}$}\hfil & \quad\hfil {$\displaystyle{#}$}\hfil  &
\quad\hfil {$\displaystyle{#}$}\hfil  & \quad\hfil {$\displaystyle{#}$}\hfil  & \quad\hfil {$\displaystyle{#}$}\hfil  &
\quad\hfil {$\displaystyle{#}$}\hfil  & \quad\hfil {$\displaystyle{#}$}\hfil  & \quad\hfil {$\displaystyle{#}$}\hfil  \cr
\noalign{\hrule\medskip}
& n_p & 1 & 2 & 3 & 4 & 5 & 6 & 7 & 8 & 9 & 10\ \ \ \cr
\noalign{\medskip\hrule\medskip}
& \Delta \chi^2(1\sigma) & 1.00 & 2.30 & 3.53 & 4.72 & 5.89 & 7.04 & 8.18 & 9.30 & 10.42 & 11.54\ \ \ \cr
\noalign{\medskip\hrule\medskip}
& \Delta \chi^2(2\sigma) & 4.00 & 6.18 & 8.02 & 9.72 & 11.31 & 12.85 & 14.34 & 15.79 & 17.21 & 18.61\ \ \ \cr
\noalign{\medskip\hrule\medskip}
& \Delta \chi^2(3\sigma) & 9.00 & 11.83 & 14.16 & 16.25 & 18.21 & 20.06 & 21.85 & 23.57 & 25.26 & 26.90\ \ \ \cr
\noalign{\medskip\hrule}}}$$

In this section, we will review some mainstream cosmological observations,
introduce the basic principles of these observations,
and describe how they are introduced into the $\chi^2$ statistics.

\subsec{Type Ia supernovae}

\ifig\FigSNA{Discovery data: Hubble diagram
of SNIa measured by the Supernova Cosmology Project and the High-z
Supernova Team. Bottom panel shows residuals in distance modulus
relative to an open universe with $\Omega_m=0.3$ and
$\Omega_{\Lambda} = 0$. From \ReviewofTurner, based on
\refs{\riess,\perl}.}
{\epsfysize=4.5in \epsfbox{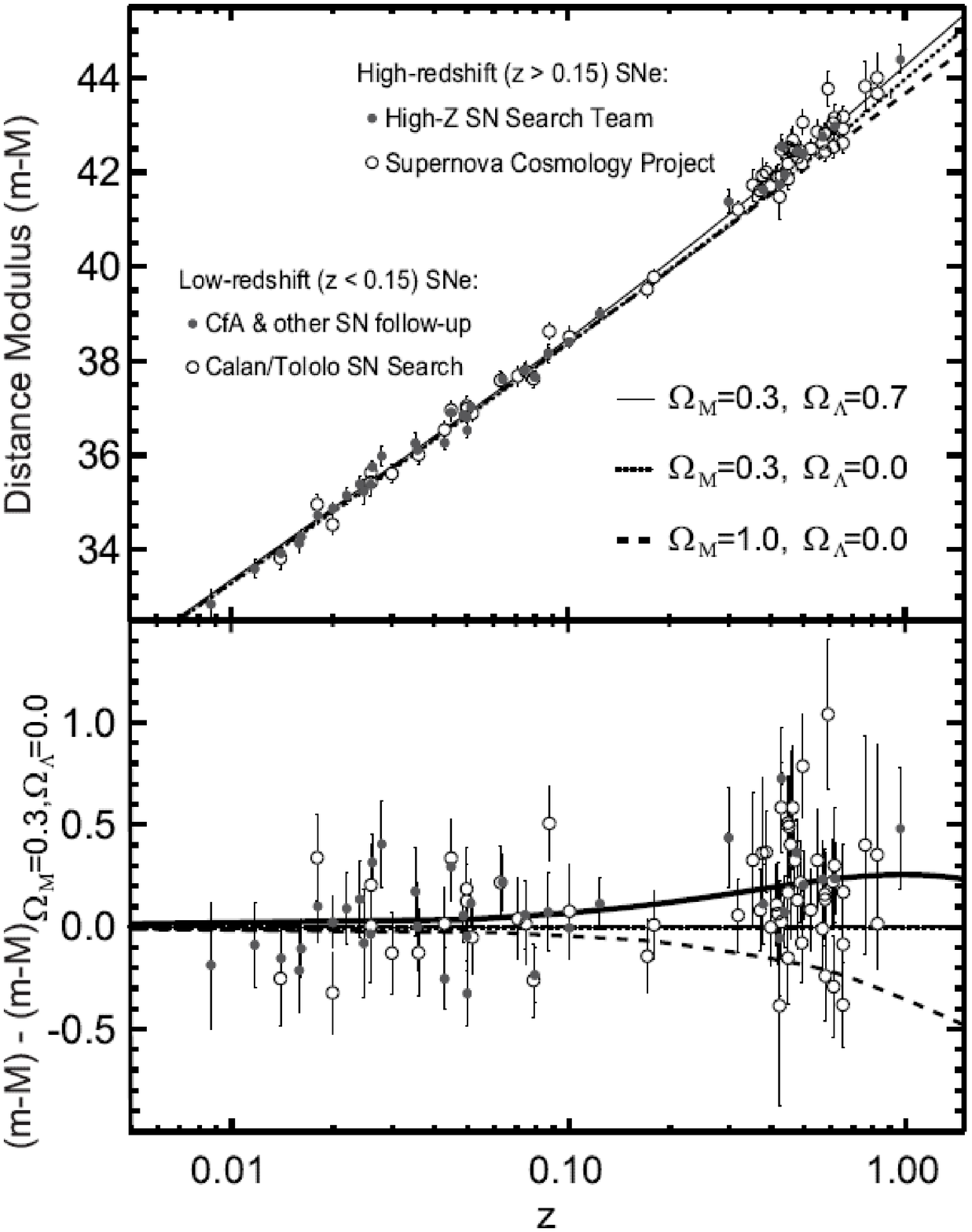}}

Type Ia supernovae (SNIa) is a sub-category of cataclysmic variable
stars that results from the violent explosion of a white dwarf star.
A white dwarf star can accrete mass from its companion star; as it
approaches the Chandrasekhar mass, the thermonuclear explosion will
occur \SNIaone. Therefore, SNIa can be used as standard candles to
measure the luminosity distance $d_L(z)$ \refs{\HZearly,\SCPearly,\SNIatwo}, and thus provides
a useful tool to measure the expansion history of the universe. In
1998, using 16 distant and 34 nearby SNIa from the Hubble space telescope (HST)
observations, Riess {\it et al.} \riess\ first discovered the acceleration of expanding universe.
Soon after, based on the analysis of 18 nearby supernova (SN) from the Calan-Tololo sample and 42 high-redshift SN,
Perlmutter {\it et al.} confirmed the discovery of cosmic acceleration \perl. The discovery of the
universe's accelerating expansion (see \FigSNA) was another big surprise since Edwin Hubble discovered the cosmic expansion in 1929.
Because of this great discovery, Adam Riess, Brian Schmidt, and Saul Perlmutter win the Nobel prize in physics 2011.
For a more detailed history about the discovery of cosmic acceleration, see \KirshnerPaper.

In recent years, the surveys of SNIa has drawn more and more
attention \refs{\ReviewofTurner,\Tonry,\Barris}. Many research groups focused
on this field, such as the Higher-Z Team \refs{\Goldfour,\Goldsix}, the
Supernova Legacy Survey (SNLS) \refs{\SNLS,\SNLSThreeYear}, the ESSENCE
(denoting ``Equation of State: SupErNovae trace Cosmic Expansion'')
\ESSENCE, the Nearby Supernova Factory (NSF) \NSF, the Carnegie
Supernova Project (CSP) \CSP, the Lick Observatory Supernova Search
(LOSS) \LOSS, and the Sloan Digital Sky Survey (SDSS) SN
Survey \SDSSSN, etc. In 2008, the Supernova Cosmology Project (SCP)
provided a framework to analyze these SNIa datasets in a homogeneous
manner. From 414 SNIa samples, they selected 307 high-quality SNIa
composing the ``Union'' dataset \Union. In 2009, the Center for
Astrophysics (CfA) SN project combined 90 low-redshift SNIa samples
with the Union dataset and obtained the ``Constitution'' sample
\Constitution. In 2010, the SCP updated their SNIa sample, and
released the ``Union2'' dataset \UnionTwo, which consisting of 557 SNIa.
Moreover, in a latest work \UnionTwoPointOne, the Union2.1 SNIa dataset,
which consisting of 580 SNIa, was released.

\ifig\FigSNB{{\it Top panel}: B-band light curves
of the Calan-Tololo SNIa sample before any duration-magnitude
correction. {\it Bottom panel}: Same light curves after applying the
``stretch'' duration-magnitude correction of Ref. \Kim.}
{\epsfysize=3.9in \epsfbox{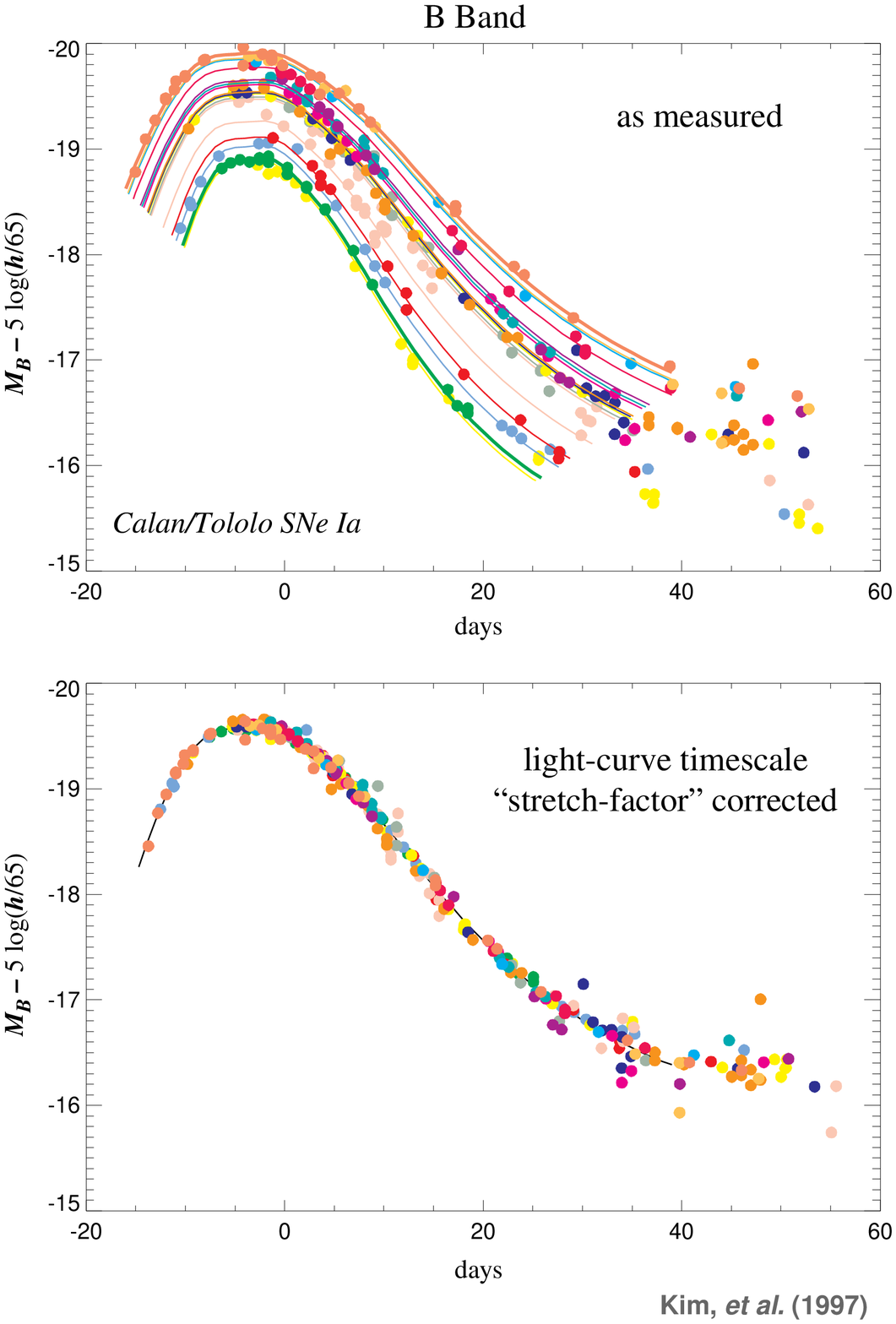}}

To constrain dark energy by using the SNIa data, the absolute magnitude of SNIa must be determined first.
Since the detailed mechanism of SNIa explosions remains uncertain \SNtrouble,
SNIa are not intrinsically standard candles, with a 1$\sigma$ spread of order 0.3 mag in peak $B$-band luminosity \Hamuy.
Fortunately, in 1992 Phillips found that SNIa has a clear correlation between their intrinsic brightness at maximum light and the duration of their light curve \Phillips.
This so-called ``Phillips correlation'' was then used to turn SNIa into standard candles \PhillipsApp.
In 2004, utilizing ``stretch'' duration-magnitude correction, Kim {\it et al.} \Kim\  reduced the dispersion on SNIa peak magnitude into only 0.10-0.15 mag.
\FigSNB\ shows a comparison of SNIa light curves before and after applying the ``stretch'' duration-magnitude correction.

After determining the absolute magnitude of the SNIa, one can obtain the observational distance modulus
\eqn\muobs{\mu_{\rm obs}=m-M,}
where $m$ is the apparent magnitude, and $M$ is the absolute magnitude.
On the other hand, the theoretical distance modulus can be calculated as
\eqn\eqmu{\mu_{\rm th}(z_i)\equiv 5 \log_{10} {d_L(z_i)} +25,}
and the luminosity distance $d_L(z)$ is
\eqn\dldl{d_L(z)={1+z\over H_0\sqrt{|\Omega_{k}|}}f_k\Big(H_0 \sqrt{|\Omega_{k}|}\int_0^z{dz'\over H(z')} \Big),}
where
\eqn\SNfk{f_k(x)= \cases{ \sin(x) & if $\Omega_k<0\ (k=1)$;\cr
x & if $\Omega_k=0\ (k=0)$;\cr
\sinh(x) & if $\Omega_k>0\ (k=-1)$.}}
The $\chi^2$ for the SNIa data is
\eqn\eqSN{\chi^2_{SN}(\theta)=\sum\limits_{i=1}^n{[\mu_{\rm obs}(z_i)-\mu_{\rm th}(z_i;\theta)]^2\over \sigma_i^2},}
where $\theta$ denotes the model parameters,
$\mu_{\rm obs}(z_i)$ and $\sigma_i$ are the observed value and the corresponding 1$\sigma$ error of distance modulus for each SNIa, respectively.
It should be mentioned that due to the uncertainty in the absolute magnitude of a SNIa,
the degeneracy between the Hubble constant and the absolute magnitude implies that one cannot quote constraints on either one.
Thus, when dealing with SNIa data,
people often analytically marginalize the nuisance parameter $H_0$ \SNNuisance.

It should be stressed that the Eq. \eqSN\ only includes the statistical errors of SNIa, while the systematic errors of SNIa are ignored.
The systematic errors come from various factors including the errors in the photometry, the calibration, the identification of SNIa,
the selection bias, the intrinsic variation of physical properties of SNIa, the host-galaxy extinction, the gravitational lensing, and so on.
Currently, the systematic errors in the SNIa data have been comparable with the statistical errors, thus they should be considered seriously.
To include the effect of systematic errors in the analysis, a prescription for using the Union2 compilation has been provided in \UnionWeb.
The key of this prescription is a $557 \times 557$ systematics covariance matrix, $C_{SN}$,
which captures the systematic errors from SNIa (This covariance matrix with systematics can be downloaded from \UnionWeb).
Utilizing $C_{SN}$, one can calculate the following quantities
\eqn\SNa{A(\theta)=(\mu_i^{\rm obs}-\mu_i^{\rm th}(\mu_0=0,\theta))(C_{SN}^{-1})_{ij}(\mu_j^{\rm obs}-\mu_j^{\rm th}(\mu_0=0,\theta)),}
\eqn\SNb{B(\theta)=\sum\limits_{i=1}^{557}{(C_{SN}^{-1})_{ij}(\mu_j^{\rm obs}-\mu_j^{\rm th}(\mu_0=0,\theta))},}
\eqn\SNc{C=\sum\limits_{i,j=1}^{557}{(C_{SN}^{-1})_{ij}},}
where $\mu_i^{\rm obs}=\mu_{\rm obs}(z_i)$, $\mu_i^{\rm th}=\mu_{\rm th}(z_i)$.
The $\chi^2$ for the SNIa data is given by
\eqn\SNMargChiSq{\chi^2_{SN}(\theta)=A(\theta)-{B(\theta)^2\over C}.}

In a latest work \SNLSConley, by using the SNLS3 sample of 472 SN,
Conley {\it et al.} gave a most careful and detailed study about the systematic uncertainties of SNIa.
After taking into account the systematic uncertainties of SNIa, the $\chi^2$ function can be written as
\eqn\SNLSchi{\chi^2_{SN}=\Delta \overrightarrow{m}^T \cdot {\bf C}^{-1} \cdot \Delta \overrightarrow{m},}
where {\bf C} is a $472 \times 472$ covariance matrix capturing the statistic and systematic uncertainties of the SNIa sample,
$\Delta {\overrightarrow {m}} = {\overrightarrow {m}}_{\rm B} - {\overrightarrow {m}}_{\rm mod}$ is a vector of model residuals,
$m_{\rm B}$ is the rest-frame peak $B$ band magnitude of an SN,
and $m_{\rm mod}$ is the predicted magnitude of the SN given by the theoretical model
and two other quantities (stretch and color) describing the light-curve of the particular SN.
The model magnitude $m_{\rm mod}$ is given by
\eqn\SNLSmb{m_{\rm mod} = 5\log_{10}D_L(z_{\rm hel},z_{\rm cmb})-\alpha (s-1)+\beta \cal{C} + \cal{M},}
where $D_L$ is the Hubble-constant free luminosity distance,
$z_{\rm cmb}$ and $z_{\rm hel}$ are the CMB frame and heliocentric redshifts of the SN,
$s$ is the stretch measure for the SN, and $\cal{C}$ is the color measure for the SN.
$\alpha$ and $\beta$ are nuisance parameters which characterize the stretch-luminosity and color-luminosity relationships, respectively.
$\cal{M}$ is another nuisance parameter representing some combination of the absolute magnitude of a fiducial SNIa and the Hubble constant.
The total covariance matrix {\bf C} in Eq. \SNLSchi \ captures both the statistical and systematic uncertainties of the SNIa data.
One can factor it as \SNLSConley,
\eqn\SNLSC{{\bf C} = {\bf D}_{\rm stat} + {\bf C}_{\rm stat} + {\bf C}_{\rm sys},}
where ${\bf D}_{stat}$ is the purely diagonal part of the statistical uncertainties,
${\bf C}_{\rm stat}$ is the off-diagonal part of the statistical uncertainties,
and ${\bf C}_{\rm sys}$ is the part capturing the systematic uncertainties.
It should be mentioned that, for different $\alpha$ and $\beta$, these covariance matrices are also different.
Therefore, one has to reconstruct the covariance matrix $\bf C$ for the corresponding values of $\alpha$ and $\beta$.
Here we do not describe these covariance matrices one by one.
One can refer to Refs \refs{\SNLSConley,\SNLSSullivan} and the public code \SNLSCode \
for more details about the explicit forms of these covariance matrices and the calculation of $\chi^2_{SN}$.

In addition, some interesting methods are also proposed to reduce
the systematic errors in the SNIa data. For instance, Wang and colleagues
\WangMukherjee\ developed a consistent frameworks for the
flux-averaging of SNIa to reduce the effect of weak lensing on SNIa data. Currently, the systematic
errors in the SNIa observations is the major factor that confining
their ability to precisely measure the properties of dark energy. To enhance
the precision of SNIa data, improvements on the photometric
technique, as well as better understandings of the dust absorption and the SN explosions, are needed. For more details on SNIa
observation and its cosmological applications, see \refs{\DEReviewLinder,\SNOther} and references therein.

\subsec{Cosmic microwave background}

\ifig\FigCMB{The WMAP 7-year temperature power spectrum, along with
the temperature power spectra from the ACBAR \CMBACBAR\ and QUaD
\CMBQuaD\ experiments. The solid line shows the best-fitting
$\Lambda$CDM model to the WMAP data, corresponding to
$\Omega_{\Lambda}$=0.738. From \WMAPSeven.} {\epsfysize=2.5in
\epsfbox{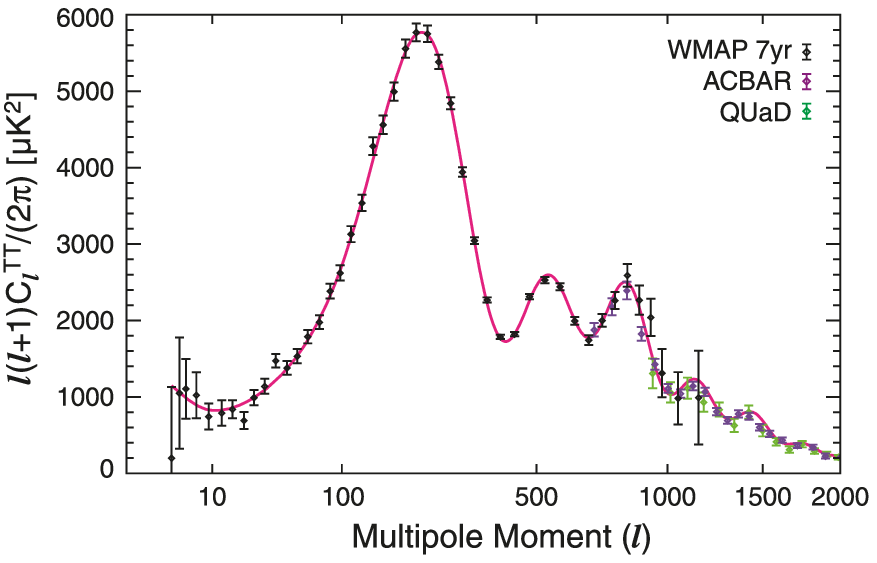}}

CMB is the legacy of the cosmic recombination epoch.
It contains abundant information of the early universe.
In 1964 CMB was firstly detected by Penzias and Wilson \PWCMB, who received the Nobel Prize in Physics 1978.
Their work provided strong evidence that supports the Big Bang theory of the universe \BigBang.
In 1989, the first generation of CMB satellite, the Cosmic Background Explorer (COBE), was launched.
It discovered the CMB anisotropy for the first time \CMBCOBE\ and opened the era of the precise cosmology.
Two of COBE's principal investigators, Smoot and Mather, received the Nobel Prize in Physics 2006.
In 1999, the TOCO, BOOMERang, and Maxima experiments \CMBAPExp\
firstly measured the acoustic oscillations in the CMB anisotropy angular power spectrum
\refs{\BAOPeeblesandYu,\BAOBond,\BAOHoltzman,\CMBAP,\BAODefzd,\HuSuDefZast,\CMBDefR}.
In 2001, the second generation of CMB satellite, the Wilkinson
Microwave Anisotropy Probe (WMAP) \refs{\WMAP,\WMAPFive}, was launched.
It precisely measured the CMB spectrum (for latest results of WMAP, see \FigCMB) and probed various cosmological parameters with a higher accuracy.
Recently, the Planck satellite, as the successor to WMAP, was launched in 2009.
The early results of Planck have been released recently \PlanckER.

The positions of the CMB acoustic peaks lie on the expand history from the decoupling epoch to the present epoch,
and contains the information of dark energy thereof.
Two distance ratios are often used to constrain dark energy.
The first distance ratio is the so-called ``acoustic scale'' $l_A$,
which represents the CMB multipole corresponds to the location of the acoustic peak.
It can be calculated as
\eqn\CMBDeflA{l_A=(1+z_{\ast}){\pi D_A(z_{\ast})\over r_s(z_{\ast})}.}
Here $z_{\ast}$ denotes the redshift of the photon decoupling epoch,
whose fitting formula is given by \HuSuDefZast
\eqn\CMBDefzast{z_{\ast}=1048 [1+0.00124(\Omega_b h^2)^{-0.738}] \left[1+g_1(\Omega_m h^2)^{g_2} \right],}
where
\eqn\CMBDefgOnegTwo{g_1={0.0783(\Omega_bh^2)^{-0.238}\over 1+39.5(\Omega_b h^2)^{0.763}}, \ \ \  g_2={0.560 \over 1+21.1(\Omega_bh^2)^{1.81}}.}
$D_A(z)$ is the proper angular diameter distance
\eqn\DefDA{D_A(z)={1\over H_0}{f_k\Big[H_0\sqrt{|\Omega_k|}\int^z_0{dz'\over H(z')}\Big]\over(1+z)\sqrt{|\Omega_k|}},}
and $r_s$ is the comoving sound horizon size
\eqn\Defrs{r_s(z)= \int_0^{t_{rec}}c_s(1+z)dt = {1\over \sqrt{3}}\int^{1/(1+z)}_0{da\over a^2H(a)\sqrt{1+(3\Omega_b/4\Omega_{\gamma})a}}.}
One can calculate $\chi^2_{{\rm
CMB}}=(l^{\rm obs}_{A}-l^{\rm th}_{A})^2/\sigma_{l_A}^2$ to include the CMB
data into the $\chi^2$ statistics. The second distance ratio is the
so-called ``shift parameter'' $R$ which takes the form \CMBDefR
\eqn\CMBDefR{R(z_{\ast})=\sqrt{\Omega_m}H_0(1+z_{\ast})D_A(z_{\ast}).}
One can also calculate
$\chi^2_{\rm CMB}=(R_{\rm obs}-R_{\rm th})^2/\sigma_{R}^2$ to reflect the contribution of the CMB data.

\bigskip

\centerline{{\bf Table II : Distance Priors from WMAP 7-year Fit}}
\medskip

$$\vbox{\halign{\bf#\hfil&\quad\hfil
{$\displaystyle{#}$}\hfil&\quad\quad\hfil {$\displaystyle{#}$}\hfil & \quad\hfil {$\displaystyle{#}$}\hfil\cr
\noalign{\hrule\medskip}
&  {\rm 7-year\ ML}^a & 7-{\rm year\ Mean}^b & {\rm Error,\ } \sigma \cr
\noalign{\medskip\hrule\medskip}
\noalign{\medskip}
$l_A(z_{\ast})$ & 302.09 & 302.69 & 0.76 \cr
\noalign{\medskip}
$R(z_{\ast})$ &  1.725 & 1.726 & 0.018 \cr
\noalign{\medskip}
$z_{\ast}$ &  1091.3 & 1091.36 & 0.91 \cr
\noalign{\medskip\hrule}}}$$
\centerline{\baselineskip12pt\advance\hsize by 01truein\noindent\footnotefont{ $\,^a$Maximum likelihood values; $\,^b$Mean of the likelihood.}}

Here we list the values of $l_A(z_{\ast})$, $R(z_{\ast})$ and
$z_{\ast}$ obtained from the WMAP 7-year observations (Table 9 of \WMAPSeven) in Table II.
One can use these data to calculate the
$\chi^2_{\rm CMB}$ defined above. Moreover, the $\chi^2_{\rm CMB}$ can be
calculated as
\eqn\CMBchisquare{\chi_{\rm CMB}^2=(x^{\rm obs}_i-x^{\rm th}_i)({\rm
Cov}^{-1}_{\rm CMB})_{ij}(x^{\rm obs}_j-x^{\rm th}_j)} where $x_i=(l_A, R, z_*)$
is a vector, and $({\rm Cov}^{-1}_{\rm CMB})$ is the inverse covariance
matrix. For the WMAP 7-year observations \WMAPSeven, the inverse
covariance matrix takes the following forms
$$
{\rm Cov}^{-1}_{\rm CMB}=\left(\matrix { 2.305 & 29.698 & -1.333 \cr  &
6825.270 & -113.180 \cr  &  & 3.414 \cr}\right).
$$

In addition, the presence of dark energy also affects the large
scale anisotropy of the CMB through the Integrated Sachs Wolfe (ISW)
effect \CMBISW, which is not captured by Eq. \CMBchisquare. The ISW
effect is caused by the variation of the gravitational potential
during the epoch of the cosmic acceleration. It provides independent
evidence for the existence of dark energy
\refs{\CMBISWDE,\CMBCaldwell}. The ISW effect can be detected
through the cross correlation between the CMB and LSS
\CMBGiannantonio. In \CMBShirley, Ho {\it et al.} reported a
$3.7\sigma$ detection of ISW by cross-correlating the SDSS LSS
observations with the WMAP CMB anistropies results. For more details
on CMB observation and its cosmological applications, see
\refs{\CMBOtherOne,\CMBOtherTwo,\CMBOtherThree} and references
therein.

\subsec{Baryon acoustic oscillations}

Baryon acoustic oscillation (BAO) refers to an overdensity or
clustering of baryonic matter at certain length scales due to
acoustic waves which propagated in the early universe
\refs{\BAOPeeblesandYu,\BAOSilk,\BAOSunyaevandZeldovich}. Similar to SNIa,
which provides a ``standard candle'' for astronomical observations,
BAO provides a ``standard ruler'' for length scale in cosmology to
explore the expansion history of the universe \BAOEisenstein. The
length of this standard ruler ($\sim$150 Mpc in today's universe
\BAODodelson) corresponds to the distance that a sound wave
propagating from a point source at the end of inflation would have
traveled before decoupling. BAO has a characteristic imprint on the
matter power spectrum \refs{\BAOBond,\BAOHoltzman}. So it can be measured at
low redshifts $z<1$ through the astronomical surveys of galaxy
clusters \BAOGMSSW. In addition, BAO scales can also be measured
through 21 cm emission from reionization, which provides abundant
information about the early universe at high redshifts $1.5\leq z
\leq 20$ \BAOTwentyOne. The apparent size of the BAO measured from
astronomical observations then leads to the measurements of the
Hubble parameter $H(z)$ and the angular diameter distance $D_A(z)$
\BAOPrinciple.

\ifig\FigBAO{The BAO power spectrum for the full luminous red galaxy
(LRG) and main galaxy samples, measured by the SDSS-II survey. The
solid curves correspond to the linear theory $\Lambda$CDM fits to
WMAP3. The dashed curves include the nonlinear corrections. Results
from LRG and main galaxy samples are consistent with each other, and
all provide a confirmation of the predicted large-scale $\Lambda$CDM
power spectrum. When combined with the WMAP3 result, the LRG data
yield to $\Omega_{\Lambda}=0.761^{+0.017}_{-0.018}$ and
$w=-0.941^{+0.087}_{-0.101}$. From \BAOTegmark.} {\epsfysize=3.4in
\epsfbox{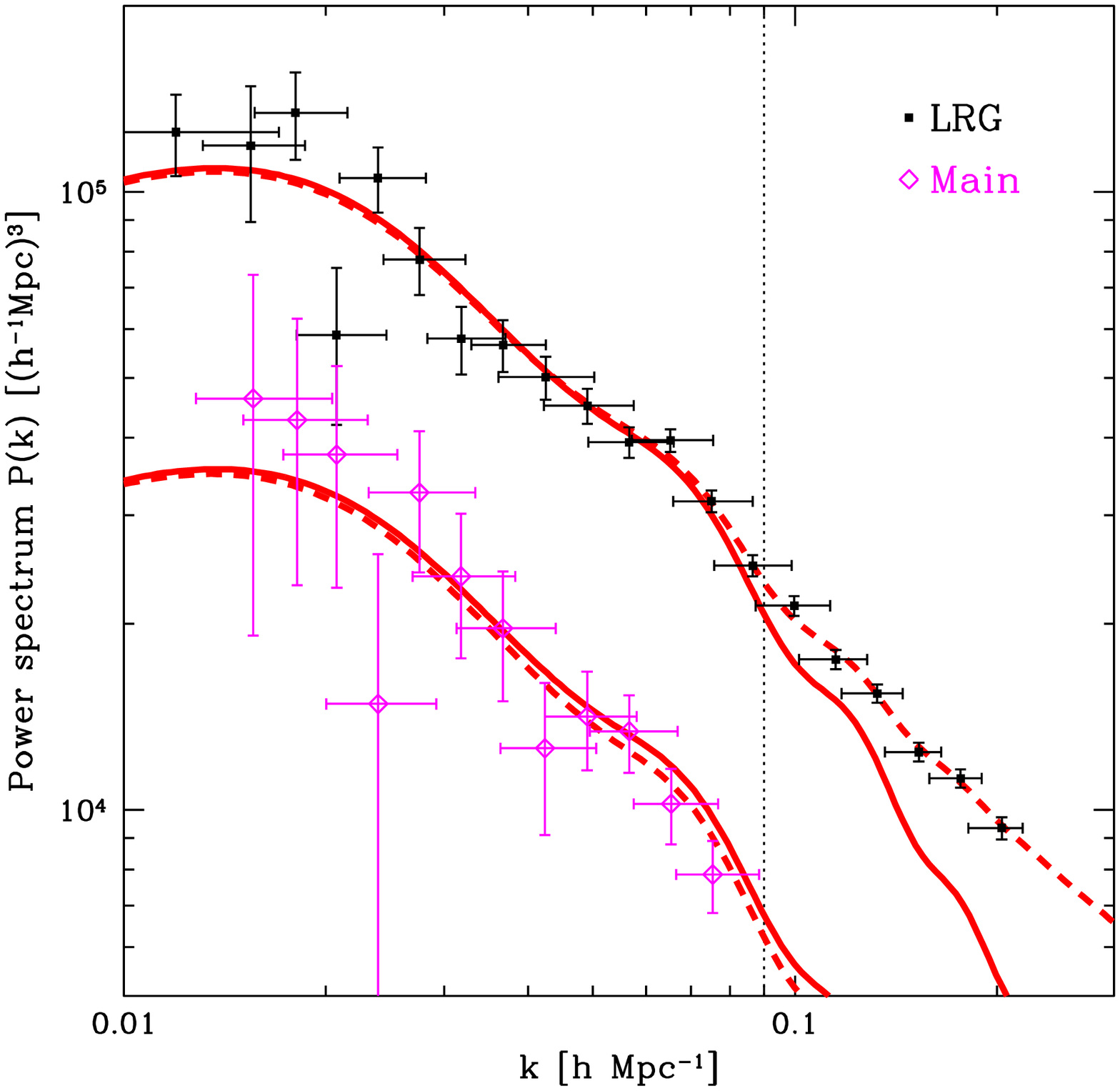}}

The BAO measurement does not require precision measurements of galaxy magnitudes,
nor did it require that galaxy images be resolved;
instead, only their three-dimensional positions need to be determined.
So the BAO observations are less affected by astronomical uncertainties than other probes of dark energy,
although it still suffers from some systematic uncertainties,
such the effects of non-linear gravitational evolution and redshift distortions of clustering \refs{\DETF,\BAOError}.
The acoustic signature of BAO has already been obtained in the galaxy power spectrum at low redshift \refs{\BAOPCEH,\BAOEisenstein}.
For examples, the Two-degree-Field Galaxy Redshift Survey (2dFGRS) \BAOColless,
which was operated by the Anglo-Australian Observatory (AAO) between 1997 and 2002, had made public their BAO data in 2003.
In addition, using a dedicated 2.5-m wide-angle optical telescope, the Sloan Digital Sky Survey (SDSS) \BAOYorkDG, was launched in 2000.
In 2006, the survey entered a new phase, the SDSS-II survey.
In 3 years, it completes the observations of a huge contiguous region of the northern skies.
The BAO power spectrum measured by the SDSS-II survey  \BAOTegmark\ is shown in \FigBAO.
In 2009. the SDSS-III survey begun. In January 2011, SDSS-III publicly release its eighth Data Release (DR8) \BAOSDSSDReight,
which is the latest BAO dataset to data.

From measurements of the galaxy clusters, the BAO scales in both
transverse and line-of-sight directions are obtained; they
correspond to the quantities $r(z)/r_s(z)$ and $r_s(z)/H(z)$,
respectively. In addition, three characteristic quantities of BAO,
including the $A$ parameter, $D_V(0.35)/D_V(0.2)$, and
$r_s(z_d)/D_V(z)$, are often used to constrain dark energy
parameters. In the following we will provide a rough introduction to
these three quantities.

The $A$ parameter is defined as \BAOEisenstein
\eqn\DefAPar
{A_{\rm th}={\sqrt{\Omega_{m}}\over {(H(z_{b})/H_0)}^{1\over3}}\left[{1\over
z_{b}\sqrt{|\Omega_{k}|}}f_k\Big(H_0\sqrt{|\Omega_{k}|}\int_0^{z_{b}}{dz'\over
H(z')} \Big)\right]^{2\over3},}
where $z_b=0.35$ is the redshift of the SDSS luminous red galaxies (LRG).
The SDSS BAO measurements \BAOEisenstein\ gave $A_{\rm obs} = 0.469 (n_s/0.98)^{-0.35}\pm0.017$,
where the WMAP 7-year results gave a best-fit value $n_s=0.963$.
One can calculate $\chi^2_{{\rm BAO}}=(A_{\rm obs}-A_{\rm th})^2/\sigma_{A}^2$ to reflect the impact of the BAO data.
The $A$ parameter is considered independent of dark energy models, and have been widely used in the literature.

Next, let us turn to the quantity $D_V(0.35)/D_V(0.2)$.
Since the current observations are not sufficient enough to measure the BAO scales in both transverse and line-of-sight directions independently,
alternatively people construct an effective distance ratio $D_V(z)$, which  is defined as \BAOEisenstein
\eqn\DefDV{D_V(z)=\Big[(1+z)^2D^2_A(z){z\over H(z)}\Big]^{1/3}.}
In 2005, Eisenstein {\it et al.} \BAOEisenstein\ provided a constraint $D_V(z)=1370\pm64$ Mpc at the redshift $z$=0.35.
In 2009, the SDSS DR7 sample \BAONewestData\ gives $D_V(0.35)/D_V(0.2)=1.736\pm0.065$.
One can use this quantity to reflect the impact of the BAO data.

Lastly, we introduce the quantity $r_s(z_d)/D_V(z)$.
From the SDSS and the 2dFGRS, one can extract a quantity $r_s(z_d)/D_V(z)$ at given $z$,
where $z_d$ denotes the redshift of the drag epoch, whose fitting formula is proposed by Eisensten and Hu \BAODefzd
\eqn\Defzd{z_d={1291(\Omega_mh^2)^{0.251}\over 1+0.659(\Omega_mh^2)^{0.828}}\left[1+b_1(\Omega_bh^2)^{b2}\right],}
where
\eqn\DefbOnebTwo{b_1=0.313(\Omega_mh^2)^{-0.419}\left[1+0.607(\Omega_mh^2)^{0.674}\right], \quad  b_2=0.238(\Omega_mh^2)^{0.223}.}
It is widely believed that $r_s(z_d)/D_V(z)$ contains more information of BAO than the previous two quantities.

In addition, one can also use the covariance matrix method to construct the $\chi^2$ function of the BAO data
\eqn\BAOChisqCovMatrix{\chi^2_{\rm BAO}=\Delta p_i[{\rm Cov}^{-1}_{\rm BAO}(p_i,p_j)]\Delta p_j,\ \ \ \Delta p_i=p^{\rm data}_i-p_i.}
For the latest BAO data from SDSS DR7 \BAONewestData, $p_1=d_{0.2}$ and $p_2=d_{0.35}$.
The covariance matrix is
$$
{\rm Cov}^{-1}_{\rm BAO}=\left(\matrix
{ 30124 & -17227  \cr  & 86977 \cr}\right),
$$
and the vales of $p_i$ are
\eqn\BAOValueofpi{p^{\rm data}_1=d^{\rm data}_{0.2}=0.1905,\ \ \ \ p^{\rm data}_2=d^{\rm data}_{0.35}=0.1097.}
There are also some other methods capturing the information of BAO from the SDSS data,
see \refs{\BAOEisenstein,\BAOChuang} and references therein for details.

\subsec{Weak lensing}

\ifig\FigWL{Cosmic shear field (white ticks) superimposed on the
projected mass distribution from a cosmological $N$-body simulation:
overdense regions are bright, underdense regions are dark. Note how
the shear field is correlated with the foreground mass distribution.
From \ReviewofTurner.} {\epsfysize=3in \epsfbox{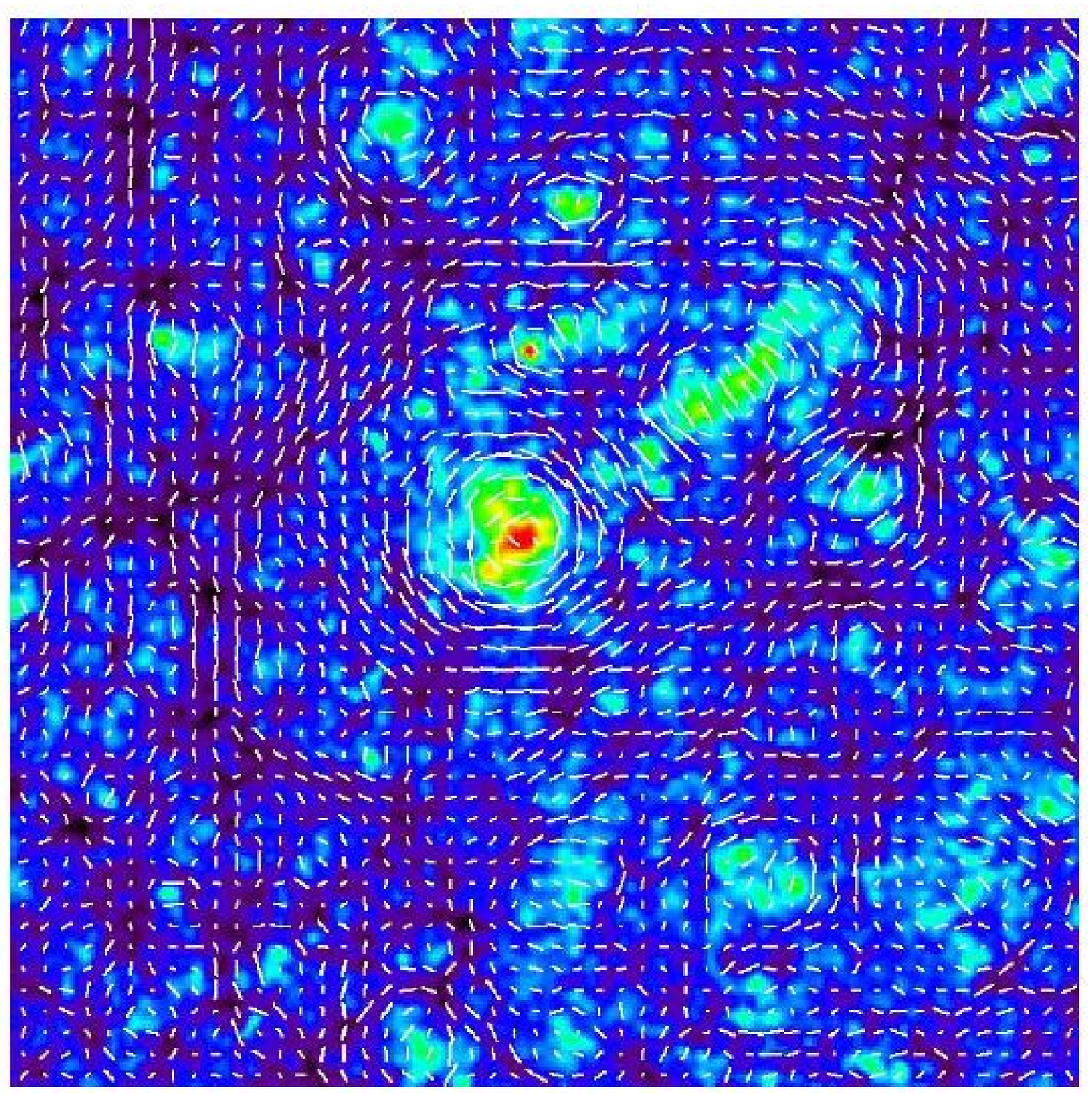}}

Weak lensing (WL) is the slight distortions of distant galaxies' images,
due to the gravitational bending of light by structures in the Universe (see \FigWL).
Utilizing the WL effect, the distribution of dark matter and its evolution with time can be measured,
thus providing a useful tool to probe dark energy through its influence on the growth of structure.

In the 1990s, WL around individual massive halos was measured \refs{\WLTyson,\WLBraninerd}.
Soon after, WL by LSS was detected by four research groups in 2000 \refs{\WLBacon,\WLKaiser,\WLWaerbeke,\WLWittman};
from then on, WL has grown into an increasingly accurate and powerful probe of dark matter and dark energy \WLHuterer.
The current survey project of WL is the Canada-France-Hawaii legacy survey (CFHTLS) \WLCFHTLS, covering $\sim$ 170 square degrees.

The effect of WL on the distant sources can represent on the distortions in the shapes, sizes and brightnesses \WLReview.
Current studies mainly focussed on the change in shapes (termed as ``cosmic shear''),
which can be more easily and precisely measured compared with the changes in the size and brightness.
The lensing effect on the shape of the galaxies is by general $\sim$0.01.
This effect is much smaller than the typical deviation in the galaxy shape, which is about $0.3-0.4$.
So a large number of galaxies are required to detect the cosmic shear signal with enough precision.

The cosmic shear analysis is a widely used method to relate the WL data with the dark energy.
The cosmic shear field is the weighted mass distribution integrated along the line of sight.
The Fourier transformed counterpart is the shear power spectrum $P_{\kappa}$.
Here $\kappa$ is the ``convergence'', which means a magnification of the source of $1+2\kappa$ at the linear order.
$P_{\kappa}$ can be inferred from the 3D distribution of matter, and it is related to the matter power spectrum $P(k,r)$ through
\eqn\WLPkappa{P_{\kappa}(l)={9H^4_0\Omega_m^2\over 4c^4}\int_0^{r_H}{dr\over a^2(r)}P\left(k={l\over f_k(r)};r\right)
\times\left[\int_r^{r_H}dr^\prime n(r^\prime){f_k(r^\prime-r)\over f_k(r^\prime)}\right].}
Here $r(z)=\int^z_0{dz\over H(z)}$ is the comoving distance, $n(z)$ is the mean redshift distribution of the source galaxies normalized to unity,
$r_H$ is the comoving horizon distance (i.e. the depth of the survey).

For statistical analysis of cosmic shear, it is common to use 2-point correlation functions \WLCorFun.
For example, a convenient way is to describe the shear field in terms of E/B-modes correlation functions
\WLEBMode
\eqn\WLxiEB{\xi_E(\theta)={\xi_+(\theta)+\xi'(\theta)\over 2},\ \xi_B(\theta)={\xi_+(\theta)-\xi'(\theta)\over2},}
where
\eqn\WLxiPlus{\xi_{+}(\theta)=\int^\infty_0{dl\over 2 \pi}lP_{\kappa}(l)J_0(l\theta),
\ \ \xi_{-}(\theta)=\int^\infty_0{dl\over 2 \pi}lP_{\kappa}(l)J_4(l\theta),}
and
\eqn\WLxiprime{\xi'(\theta)=\xi_{-}(\theta)+4\int^\infty_\theta d\theta'{\xi_-(\theta')\over\theta'}-12\theta^2\int^\infty_\theta d\theta'{\xi_-(\theta')\over\theta'^3}.}
Here $J_0$ and $J_4$ are the zeroth and forth order Bessel functions
of the first kind, respectively. The shear correlation functions can
then be compared with the measurements directly. The B-mode
correlation functions are expected to be very small \WLBModeSmall, so
it is common to use the E-mode correlation functions to construct
the weak lensing $\chi^2$ function
\eqn\WLLikelihood{\chi^2_{\zeta_E}=\left(\zeta_E(\theta_i)-m_i\right)({\rm
Cov}^{-1}_{\zeta_E})_{ij}\left(\zeta_E(\theta_j)-m_j\right).}

Systematic errors in weak lensing measurements arise from a number of sources,
including incorrect shear estimates, uncertainties in galaxy photometric redshift estimates,
intrinsic correlations of galaxy shapes, and theoretical uncertainties in the mass power spectrum on small scales   \refs{\WLErrors,\WLerror}.
Fortunately, future WL surveys have ability to internally constrain the impact of such effects  \refs{\DETF,\WLZhan,\WLerrorFour}.
For more details about the WL observation and its applications on the probe of dark energy,
see \refs{\CopelandWR,\WLReview,\WLReviews} and references therein.

\subsec{Galaxy clusters}

\ifig\FigCL{Massive galaxy cluster 2XMM J083026+524133 detected by
the X-ray telescope XMM-Newton (the green contours) and its the
optical image observed by the Large Binocular Telescope (LBT) in the
Arizona desert. From \CLXMMNewton.} {\epsfysize=3.0in
\epsfbox{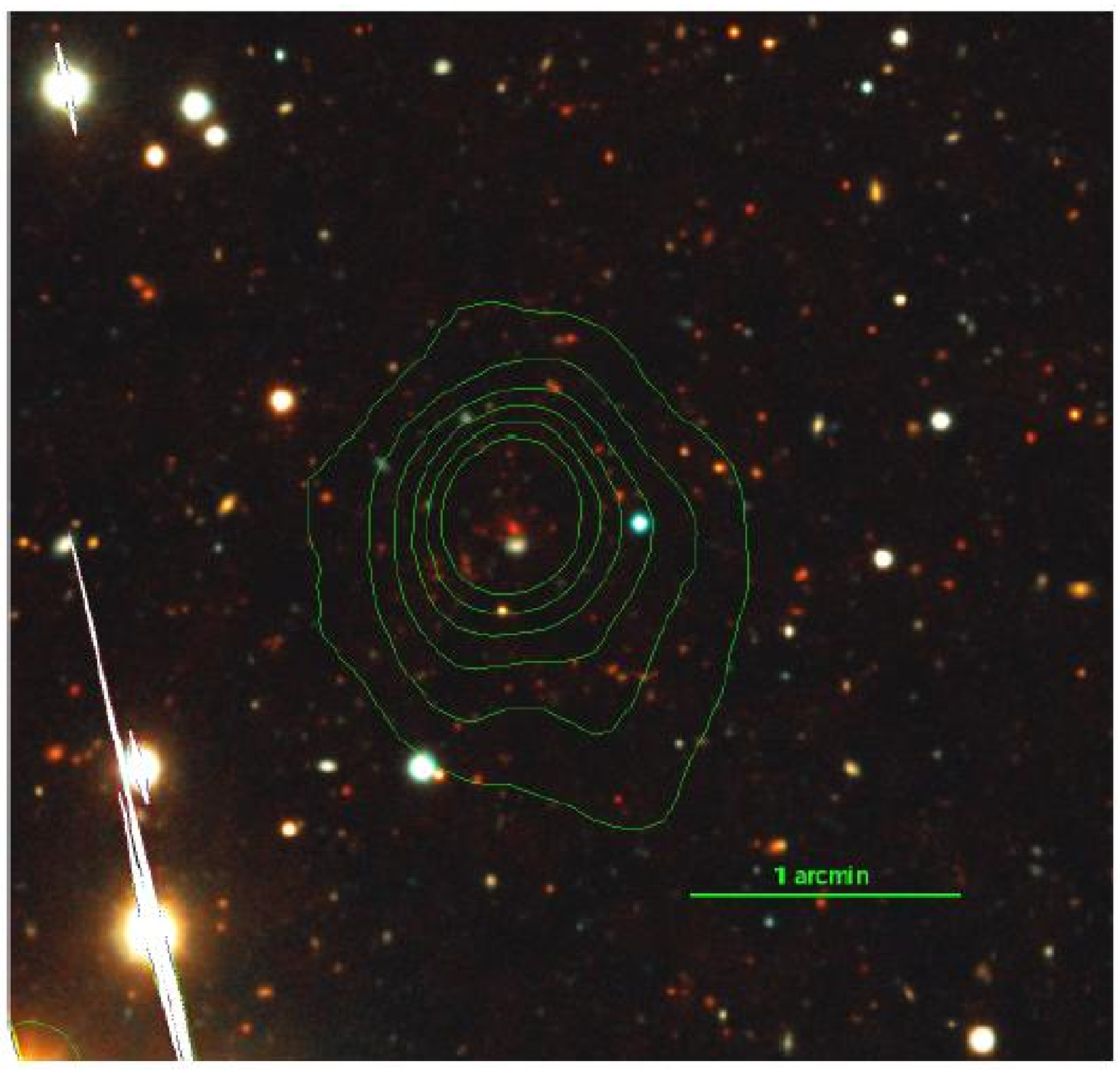}}

Galaxy clusters (CL) are the largest gravitationally bound objects in the universe.
They typically contain 50 to 1,000 galaxies and have a diameter from 2 to 10 Mpc.
CL can be detected through the following approaches \CLBorgani:
(1) optical or infrared imaging and spectroscopy (see \FigCL).
In comparison with optical surveys,
infrared searches are more useful for finding higher redshift clusters.
(2) X-ray imaging and spectroscopy (see \FigCL).
CL with active galactic nucleus (AGN) are the brightest X-ray emitting extragalactic objects,
so they are quite prominent in X-ray surveys.
(3) Sunyaev-Zel'dovich effect \BAOSunyaevandZeldovich.
The hot electrons in the intracluster medium scatter radiation from the cosmic microwave background through inverse Compton scattering.
This produces a ``shadow'' in the observed CMB at some radio frequencies.
(4) gravitational lensing.
CL bend the light from distance galaxies and distort the observed images.
The observed distortions can be used to detect the masses of clusters.

In principle, the number density of cluster-sized dark halos $n(z,M)$
as a function of redshift $z$ and halo mass $M$ can be accurately predicted from N-body simulations \CLWarrent.
Comparing these predictions to large area cluster surveys can provide precise constraints on the cosmic expansion history \refs{\CLHaiman,\CLWang,\CLSWang}.

In a survey that selects clusters according to some observable $O$ with redshift-dependent selection function $f(O,z)$,
the redshift distribution of CL is given by \ReviewofTurner
\eqn\CLMassFun{{d^2N(z)\over dzd\Omega}={r^2(z)\over H(z)}\int^{\infty}_0f(O,z)dO\int^{\infty}_0p(O|M,z){dn(z)\over dM}dM.}
Here $N$ is the number of CL, $\Omega$ is the solid angle, $r(z)$ is the comoving distance, $H(z)$ is the Hubble parameter,
$dn(z)/dM$ is the space density of dark halos in comoving coordinates,
and $p(O|M, z)$ is the mass-observable relation, the probability that a halo of mass $M$ at redshift $z$ is observed as a cluster with observable property $O$.
The utility of this probe depends on the ability to robustly associate cluster observables
such as cluster galaxy richness, X-ray luminosity, Sunyaev-Zel¡¯dovich effect flux decrement, or weak lensing shear with cluster mass \refs{\ReviewofTurner,\CLBorgani}.
As seen in this equation, the sensitivity of cluster counts to dark energy arises from two factors:
cosmic expansion history, $r^2(z)/ H(z)$ is the comoving volume element that contains information of the cosmic expansion history;
growth of structure, $dn(z) / dM$ depends on the evolution of density fluctuations,
and the cluster mass function is also determined by the primordial spectrum of density perturbations.
One can see \CLReview\ and references therein for more details about the surveys of CL and their applications on the probe of dark energy.

\subsec{Gamma-ray burst}

Gamma-ray bursts (GRB) are flashes of gamma rays associated with extremely energetic explosions in distant galaxies.
They are the most luminous electromagnetic events in the universe.

GRBs were first detected in 1967 by the U.S. Vela satellites.
They can be classified into ``long GRBs'' (their durations are longer than 2 seconds) and ``short GRBs'' (their durations are shorter than 2 seconds).
At the beginning, most astronomers believed that GRBs originate from inside the Milky Way galaxy,
only Paczynski insisted that GRBs originate from external galaxy \GRBPaczynski.
In 1997, an X-ray astronomy satellite BeppoSAX detected the ``afterglow'' of GRB 970228 \GRBParadijs,
and verified that GRBs indeed originate from external galaxy.
This discovery opened up a new era in the history of GRBs studies.

Many satellites had been launched to probe GRBs in the past decade.
The Swift mission \GRBSwift\ was launched in 2004.
This satellite is equipped with a very sensitive
gamma ray detector as well as on-board X-ray and optical telescopes,
which can be rapidly and automatically slewed to observe afterglow
emission following a burst. In 2008, the Fermi mission \GRBFermi\ was launched. Its main instruments include the Large Area Telescope
(LAT) and the Gamma-Ray Burst Monitor (GBM); the former is used to
perform an all-sky survey, and the latter is used to detect sudden
flares of gamma-rays. Meanwhile, on the ground, numerous optical
telescopes have been built or modified to incorporate robotic
control software that responds immediately to signals sent through
the Gamma-ray Burst Coordinates Network (GCN) \GRBGCN. This allows telescopes
to rapidly repoint towards a GRB, often within seconds of receiving
the signal and while the gamma-ray emission itself is still ongoing.

\ifig\FigGRB{The Hubble diagram for 69 GRBs, out to the redshift of
$z>6$. The curve is the luminosity distance in a flat $\Lambda$CDM
model with $\Omega_m$=0.27 and $w=-1$. It can be seen that the
observational data smoothly follows the curve. From \GRBSchaefer.}
{\epsfysize=3.5in \epsfbox{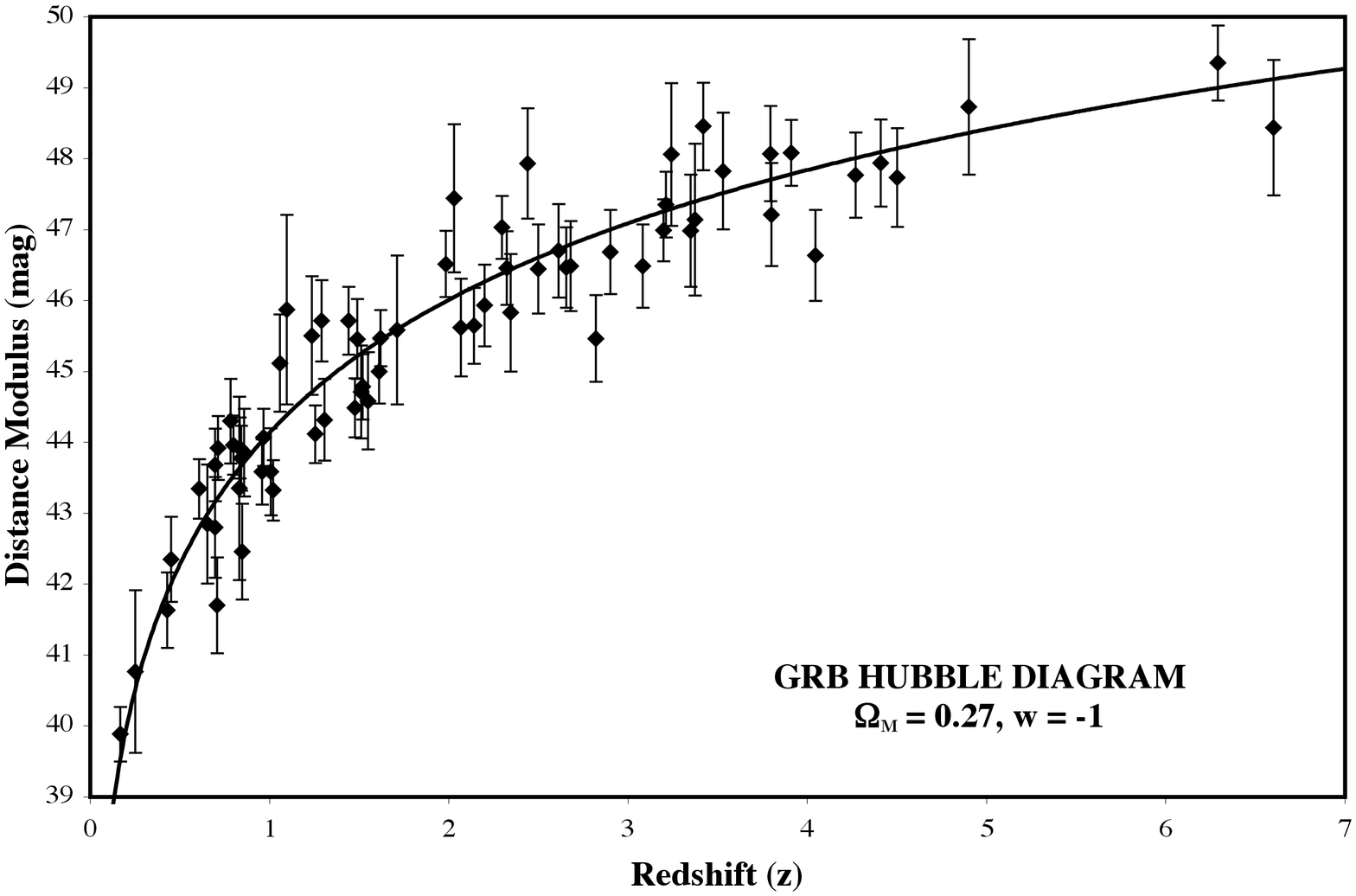}}

GRBs have been proposed to be a complementary probe to the SNIa in the studies of dark energy.
For example, in \GRBSchaefer, Schaefer presented 69 GRBs data points (see \FigGRB)
over a redshift range of $z=0.17$ to $z>6$ with half the bursts having a redshift larger than 1.7,
and showed that the GRB Hubble diagram is consistent with the existence of cosmological constant.
Compared to the SNIa, an advantage of GRB is that the high energy photons in the gamma-ray band are almost immune to dust extinction.
Moreover, the redshifts of observed GRBs are much higher than SNIa:
there have been many GRBs observed at $1 \leq z \leq 8$, whereas the maximum redshift of GRBs is expected to be 10 or even larger \GRBBromm.
Therefore, GRBs are considered to be a promising probe to
fill the redshift desert between the redshifts of SNIa and CMB. For
more details on the GRB cosmology, see \refs{\GRBCos,\GRBGhirlandaB} and references therein.

However, there still exist some troubles in the application of GRB data to the probe of dark energy.
A big problem is that since our knowledge on the mechanisms underlying the GRB emission is still limited,
treating them as standard candles is still suspicious.
Another well-known problem is the so called ``circularity" problem in the GRB calibration \refs{\GRBGhirlandaB,\GRBStat},
mainly due to the lack of low-redshift GRBs at z $<$ 0.1 which are cosmology-independent.
To alleviate this problem, various statistical methods have been proposed.
For examples, in \GRBYunWangZeroEight,
Wang summarized the GRB data by a set of model-independent distance measurements and provided a convenient method to use the GRB data in cosmological analysis.
In \GRBLiang, Liang {\it et al.} proposed a new method to estimate the distance modulus of GRBs by interpolating from the Hubble diagram of SNIa.
Using this method, in \DEManyModelsHWei, Wei obtained 59 calibrated high-redshift GRBs
from a total number of 109 long GRBs \refs{\GRBSchaefer,\GRBAmatiB} and the Union2 SNIa sample  \UnionTwo.
For more statistical methods to calibrate the GRB data, see e.g. \refs{\GRBGhirlandaB,\GRBStat,\GRBLiang}.
Due to the lack of a large amount of well observed GRBs,
the current GRB data are still not able to provide forceful constraint on dark energy.
Therefore, although the GRBs are considered to be a promoting probe to fill the redshift desert between SNIa and CMB,
there is still a long way to use them extensively and reliably to probe dark energy.

\subsec{X-ray observations}

X-ray is an important observational branch of astronomy which deals with the study of X-ray emission from celestial objects.
In 1962, utilizing a US Army V-2 rocket, Giacconi and colleagues firstly discovered cosmic X-ray source.
In addition, they also discovered the existence of astronomical X-ray background  \XRayGiacconi.
Due to his great contributions to this field, Giacconi received the Nobel Prize in Physics 2002.

Since Giacconi's great discovery, many X-ray satellites have been launched.
The first orbiting X-ray astronomy satellite, the Uhuru satellite \XRayUhuru, was launched in 1970.
Soon after, in 1978, the first fully imaging X-ray telescope, the Einstein Observatory \XRayEinstein, was launched.
In 1999, two important X-ray satellites, the Chandra observatory
\XRayChandra\ and the XMM-Newton observatory \XRayXMMNewton, were launched.
The Chandra observatory has high space resolution (less than 1 arc-second) and a wide wave band (0.1-1 keV),
while the XMM-Newton observatory has very high spectrum resolution.

\ifig\FigXRay{The 68.3\% and 95.4\% confidence constraints in the
$\Omega_m$-$\Omega_\Lambda$, using Chandra measurements of the X-ray
gas mass fraction $f_{\rm gas}$ data (red contours). A non-flat
$\Lambda$CDM model is assumed. Also shown are the independent
results obtained from the CMB data (blue contours) and SNIa data
(green contours). A combined analysis using all the three data sets
yields a result of $\Omega_{\Lambda}=0.735\pm0.023$ and
$\Omega_k=-0.010\pm0.011$ at the 68.3 \% CL (the inner, orange
contours). From \XRayAllenZeroSeven.} {\epsfysize=2.8in
\epsfbox{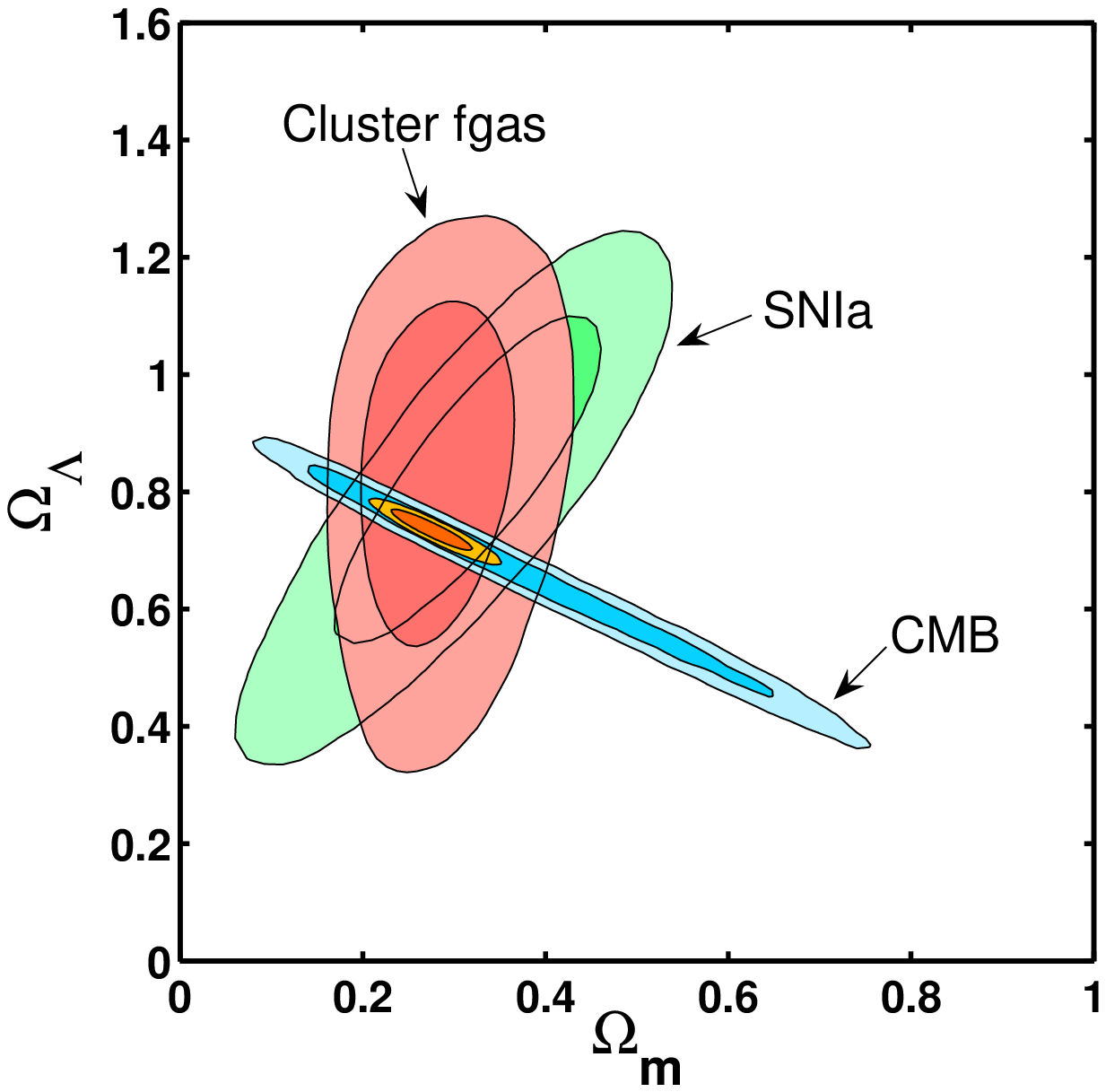}}

In galaxy clusters, the X-ray emitting intracluster gas is the
dominate component of baryonic mass content (it exceeds the mass of
optically luminous material by a factor $\sim$6 \XRayFukugita). In
addition, galaxy clusters also have the dark matter content
\refs{\XRayWhite,\XRayEke}. So the measurements of universal baryonic mass
function in the clusters, $f_b=\Omega_b/\Omega_m$, have been widely
used to determine the cosmic matter fraction $\Omega_m$ \XRayOmegaM.
In late 1990s, Sasaki \XRaySasaki\ and Pen \XRayPen\ proposed
that the dependence of $f_b$ on the angular diameter distances to
the clusters can also be used to constrain the geometry of the
universe. In 2002, Allen {\it et al.}
\refs{\XRayAllenZeroTwo,\XRayAllenZeroThree,\XRayEttori} carried out such
a test and obtained $\Omega_m=0.30\pm0.04$ using a small sample of
X-ray luminous galaxy clusters. In 2008, based on an improved
analysis using the data of 42 clusters from Chandra X-ray
observations, Allen {\it et al.} \XRayAllenZeroSeven\ found a detection
of dark energy at 4$\sigma$ CL, with $\Omega_{\Lambda}=0.86\pm0.21$
for a non-flat $\Lambda$CDM model (see \FigXRay).

The results of Refs. \refs{\XRayAllenZeroTwo,\XRayAllenZeroThree,\XRayEttori,\XRayAllenZeroSeven}
are obtained from the measurements of the apparent evolution of the cluster X-ray gas mass fraction, $f_{\rm gas}=M_{\rm gas}/M_{\rm tot}$.
As showed in \XRayAllenZeroSeven, the $\chi^2$ function for $f_{\rm gas}$ data can be calculated as
\eqn\XRayChiSquareT{\chi^2_{f_{\rm gas}}=\sum^{42}_{i=1}{[f^{\Lambda{\rm CDM}}_{\rm gas}(z_i)-f_{{\rm gas},i}]^2\over\sigma^2_{{\rm gas},i}},}
where the fitting formula of $f^{\Lambda{\rm CDM}}_{\rm gas}(z)$ is given by
\eqn\XRayfGas{f^{\Lambda{\rm CDM}}_{\rm gas}(z) = {KA\gamma b(z)\over 1+s(z)} \left({\Omega_b\over \Omega_m}\right)
\left[ {d^{\Lambda{\rm CDM}}_A\over D_A(z)}  \right]^{1.5}.}
Here $D_A$ and $D^{\Lambda{\rm CDM}}_A$ are the angular diameter
distances to the clusters in the current test model and reference
cosmologies, respectively. The factor $b(z) = b_0(1+\alpha_bz)$ is a
``biased factor'' with $0.65<b_0<1.0,\ -0.1<\alpha_b<0.1$; parameter
$s(z)=s_0(1+\alpha_sz)$ models the baryonic mass fraction in stars,
and $s_0=(0.16\pm0.05)h^{0.5}_{70}$, $-0.2<\alpha_s<0.2$. $\gamma$
models non-thermal pressure support in the clusters
($1.0<\gamma<1.1$). $K$ is a ``calibration'' constant arises from
the residual uncertainty in the accuracy of the instrument
calibration and X-ray modelling ($K=1\pm0.1$). The factor $A$ is
\eqn\XRayFactorA{A=\left({\theta^{\Lambda{\rm CDM}}_{2500}\over
\theta_{2500}}\right)^{\eta} \approx\left({H(z)D_A(z)\over
[H(z)D_A(z)]^{\Lambda{\rm CDM}}}\right)^{\eta}.} where
$\eta=0.214\pm0.022$. One can refer to \XRayAllenZeroSeven\ for
details about the origins of the parameters $b(z),\ s(z),\ K,$ and
$A$.

Like the SNIa and BAO data, the $f_{\rm gas}$ measurements can also probe the redshift-distence relation,
with the dependence $f_{\rm gas}\propto d_L(z)D_A(z)^{0.5}$ \XRayAllenZeroSeven.
It has been shown that \refs{\XRayAllenZeroSeven,\XRayRapetti,\XRayRapettiTwo} the $f_{\rm gas}$ data
is useful in breaking the degeneracy between some cosmological parameters such as $\Omega_m$, $w$ and $H_0$ when combined with other cosmological observations.
In addition to $f_{\rm gas}$,
measurements of the amplitude and evolution of matter fluctuations using X-ray observations
have also been applied to probe dark energy. One can refer to
\refs{\XRayAllenZeroThree,\XRayDetails} for more details.

\subsec{Hubble parameter measurements}

In 1929 Hubble \HzHubble\ discovered a linear correlation
between the apparent distances to galaxies $D$ and their recessional
velocities $v$
\eqn\HzHubbleLaw{v=H_0D,}
where $H_0$ is the
so-called Hubble constant. This discovery provided strong evidence
that our universe is in a state of expansion and opened the
era of the modern cosmology. Soon after, people find that this Hubble's
law is just an approximate formula, and $H_0$ should be replaced by
$H(z)$, a function of the redshift $z$. As mentioned above, $H(z)$
describes the expansion history of the universe, and plays a central
role in connecting dark energy theories and observations.

At the beginning, Hubble and Humason \HzHubbleHumason\ measured a value for $H_0$ of 500 km/s/Mpc,
which is much higher than the currently accepted value due to errors in their distance calibrations.
Since the launch of the Hubble Space Telescope (HST) \HzHST, the value of $H_0$ was estimated to be between 50 and 100 km/s/Mpc.
For example, using the HST key project, Freedman {\it et al.} \HzFreedman\ obtained $H_0=72\pm8$ km/s/Mpc,
and Sandage {\it et al.} \HzSandage\ advocate a lower $H_0=62.3\pm6.3$ km/s/Mpc.
In 2009, Riess {\it et al.} \HzRiess\ gave $H_0=74.2\pm3.6$ km/s/Mpc with a 4.8 \% uncertainty.
In a latest work \HzRiessnew, Riess {\it et al.} obtained $H_0=73.8\pm2.4$ km/s/Mpc, corresponding to a 3.3 \% uncertainty.
In the future with more precise observations from HST, Spitzer \HzSpitzer,
Global Astrometric Interferometer for Astrophysics satellite (GAIA) \HzGAIA\ and James Webb Space Telescope (JWST) \HzJWST,
an uncertainty of 1\% in the Hubble constant will be a realistic goal for the next decade \HzReview.

The precise measurements of $H_0$ will be helpful to break the
degeneracy between some cosmological parameters \HzReview.
For example, in \HzHu, Hu pointed out that a measurement of $H_0$ to the percent level,
when combined with CMB measurements with the statistical precision of the Planck satellite,
offers one of the most precise measurements of dark energy EOS at $z\sim0.5$.

\bigskip

\centerline{{\bf Table III : Values of $H_0$ Measured in \refs{\HzSimon,\HzStern}}}

$$\vbox{\halign{\bf#\hfil&
\quad\hfil {$\displaystyle{#}$}\hfil & \quad\hfil {$\displaystyle{#}$}\hfil & \quad\hfil {$\displaystyle{#}$}\hfil &
\quad\hfil {$\displaystyle{#}$}\hfil & \quad\hfil {$\displaystyle{#}$}\hfil & \quad\hfil {$\displaystyle{#}$}\hfil  &
\quad\hfil {$\displaystyle{#}$}\hfil  & \quad\hfil {$\displaystyle{#}$}\hfil  & \quad\hfil {$\displaystyle{#}$}\hfil  &
\quad\hfil {$\displaystyle{#}$}\hfil  & \quad\hfil {$\displaystyle{#}$}\hfil  & \quad\hfil {$\displaystyle{#}$}\hfil  \cr
\noalign{\hrule\medskip}
& z & 0.1 & 0.17 & 0.27 & 0.4 & 0.48 & 0.88 & 0.9 & 1.3 & 1.43 & 1.53 &1.75\ \ \ \cr
\noalign{\medskip\hrule\medskip}
& H(z)  & 69 & 83 & 77 & 95 & 97 & 90 & 117 & 168 & 177 & 140 & 202\ \ \ \cr
\noalign{\medskip\hrule\medskip}
& \sigma_H & 12 & 8 & 14 & 17 & 60 & 40 & 23 & 17 & 18 & 14 & 40\ \ \ \cr
\noalign{\medskip\hrule}}}$$

\bigskip

\centerline{{\bf Table IV: Values of $H_0$ Measured in \HzGazta}}

$$\vbox{\halign{\bf#\hfil&
\quad\hfil {$\displaystyle{#}$}\hfil & \quad\hfil {$\displaystyle{#}$}\hfil & \quad\hfil {$\displaystyle{#}$}\hfil & \quad\hfil {$\displaystyle{#}$}\hfil\cr
\noalign{\hrule\medskip}
&z & 0.24 & 0.34 & 0.43 \ \ \ \cr
\noalign{\medskip\hrule\medskip}
&H(z) & 79.69 & 83.80 & 86.45\ \ \ \cr
\noalign{\medskip\hrule\medskip}
&\sigma_{H,st}  & 2.32 & 2.96 & 3.27\ \ \ \cr
\noalign{\medskip\hrule\medskip}
&\sigma_{H,sys} & 1.29 & 1.59 & 1.69\ \ \ \cr
\noalign{\medskip\hrule}}}$$

\centerline{\vbox{\baselineskip12pt
\advance\hsize by -1truein\noindent\footnotefont{The inferred $H(z)$ with its statistical
and systematical errors for each redshift slice.}}}

\centerline{}

In addition to the direct measurements for the $H_0$,
the precise measurements of $H(z)$ are also useful in studying the cosmic acceleration.
In 2005, Simon {\it et al.} \HzSimon\ measured the Hubble parameter $H(z)$ at nine different redshifts from the differential ages of passively evolving galaxies.
In 2009, Stern {\it et al.} \HzStern\ extended this dataset to eleven data points (see Table III).
Soon after, by studying the clustering of LRG galaxies in the latest spectroscopic SDSS data releases,
Gazta$\tilde{n}$aga et al. \HzGazta\ obtained three more $H(z)$ data points (see Table IV).
Utilizing these $H(z)$ data, it is straightforwad to put constraint on dark energy parameters by calculating the corresponding $\chi^2$ as
\eqn\Hzchisquare{\chi^2_H=\sum^{14}_{i=1}{[H_{\rm obs}(z_i)-H(z_i)]^2\over\sigma^2_{hi}}.}
As shown in \TJZhangH, the current $H(z)$ data from direct measurements can provide valuable constraints on dark energy.
In the future, with the developments in the observational technique of LRGs,
the $H(z)$ measurements can provide useful complements to other cosmic observations \refs{\TJZhangHz,\TJZhangreview}.

\subsec{Cosmic age tests}

The cosmic age problem is a longstanding issue in cosmology and
provides an important tool for constraining the expanding history of
the Universe  \Chaboyer. Before the great discovery that our
universe is undergoing an accelerated expansion, the most popular
SCDM model (i.e., a flat universe with $\Omega_{m}=1$) is always
plagued by a longstanding puzzle: in this model the present age of
the universe is $t_0={2\over3 H_0} \simeq 9$ Gyr, while astronomers
have already discovered that many objects are older than 10 Gyr
\refs{\CosmicAgeA,\CosmicAgeB}. For example, based on the study of white
dwarf cooling, a lower limit of cosmic age $t_0=12.7\pm0.7$ Gyr have
been obtained \CosmicAgeC. The cosmic age problem becomes more acute
if one considers the age of the universe at a high redshift. For
instance, a 3.5 Gyr-old galaxy 53W091 at redshift $z=1.55$ and a 4
Gyr-old galaxy 53W069 at $z=1.43$ are more difficult to accommodate
in the SCDM model \CosmicAgeD. Along with the discovery of
accelerated expansion of the universe and the return of the
cosmological constant $\Lambda$, the cosmic age problem has been
greatly alleviated. The 7-year WMAP observations show that in the
$\Lambda$CDM model the present cosmic age is
$t_{0}^{\rm obs}=13.75\pm0.11$ Gyr. Besides, it is shown that the
$\Lambda$CDM model can also easily accommodate galaxies 53W091 and
53W069 \AlcanizLima. So the cosmic age problem is a ``smoking-gun''
of evidence for the existence of dark energy.

However, the cosmic age problem has not been completely removed by the introduction of dark energy.
By comparing photometric data acquired from the Beijing-Arizona-Taiwan-Connecticut system with up-to-date theoretical synthesis models,
Ma {\it et al.} \Ma\ obtained the ages of 139 globular clusters (GCs) in the M31 galaxy,
in which 9 extremely old GCs are older than the present cosmic age predicted by the 7-year WMAP observations \AgeSWone.
In addition, the existence of high-z quasar APM 08279+5255 at $z=3.91$ \Hasinger\ is also a mystery \refs{\HzMys,\HzBinWang,\HzBWone}.
Using the maximum likelihood values of the 7-year WMAP observations $\Omega_m=0.272$ and $h=0.704$,
the $\Lambda$CDM model can only give a cosmic age $t=1.63$ Gyr at redshift $z=3.91$, while the lower limit of this quasar's age is 1.8 Gyr.
To accommodate these anomalous objects,
some authors suggested that a lower $H_0$ should be advocated \refs{\HzSandage,\HzTammann},
while some authors suggested that more complicated cosmological model should be taken into account \refs{\AgeSWtwo,\CuiZhang}.
Therefore, the cosmic age puzzle still remains in the standard cosmology.

In addition, one can also use the ages of old galaxies to perform
the best-fit analysis on dark energy. In \HzSimon, Simon {\it et al.} established
these so-called ``lookback time-redshift'' (LT) data by estimating
the age of 32 old passive galaxies distributed over the redshift
interval $0.11 \leq z \leq 1.84$ and the total age of the universe
$t_o^{\rm obs}$. The galaxy samples of passively evolving galaxies are
selected with high-quality spectroscopy, and the method used to
determine ages of galaxy samples indicates that systematics are not
a serious source of error for these highredshift galaxies \HzSimon.
Utilizing these LT data, one can calculate the corresponding
$\chi^2$ as
\eqn\LookBackTimeChi{\chi^2_{age}(\theta)=\sum_{i=1}^{32}
{[t^{\rm obs}(z_i;\tau)-t(z_i;\theta)]^2\over\sigma^2_T}+{[t^{\rm obs}_0-t_0(\theta)]^2\over\sigma^2_{t^{\rm obs}_o}}.}
where $t(z;\theta)$ is the age of the universe at redshift z, given
by
\eqn\CosmicAgetz{t(z;\theta)=\int_z^\infty{d\tilde{z}\over(1+\tilde{z})H(\tilde{z};\theta)}.}
Here $t_0(\theta)=t(0;\theta)$ is the present age of the universe.
$\sigma_T^2\equiv\sigma_i^2+\sigma^2_{t^{\rm obs}_o}$, where $\sigma_i$
is the uncertainty in the individual lookback time to the $i^{\rm th}$
galaxy of the sample, $\sigma_{t^{\rm obs}_o}=0.7 Gyr$ stands for the
uncertainty in the total expansion age of the universe
($t^{\rm obs}_o$), and $\tau$ means the time from Big Bang to the
formation of the object.

\subsec{Growth factor}

In addition to the expansion history of the universe, the growth of
large-scale structure can also provide important constraints on dark
energy and modified gravity. The growth rate of large scale
structure is derived from matter density perturbation $\delta=\delta
\rho_m/ \rho_m$ in the linear regime that satisfies the simple
differential equation \GrowFacPeebles
\eqn\PertEq{\ddot{\delta}+2H\dot{\delta}-4\pi G_{eff}\rho_m\delta=0,}
where the effect of dark energy is introduced through the ``Hubble damping" term $2H\dot{\delta}$,
and the effect of modified gravity is introduced via the effective gravitational ``constant'' $G_{eff}$.
Eq. \PertEq\ can be written in terms of the logarithmic growth factor $f=d\ln \delta / d\ln a$ \DEReviewCarroll
\eqn\GrowFacEq{{df\over d\ln a}+f^2+\left({\dot H\over H^2}+2\right)f={3\over2}{G_{eff}\over G}\Omega_m(z).}

There is no analytical solution to the Eq. \GrowFacEq, and much
efforts have been paid to solve this equation
\refs{\GrowFacLuScSt,\DGPStructureFormation,\GrowFacSol}.
The growth function $f$ was shown to be well approximated by the
ansatz \refs{\CLWang,\GrowFacFry,\GrowFacLightman}
\eqn\GrowFacf{f=\Omega_m^\gamma(z),}
where $\gamma$ is the so called  ``growth index''. Based on this
ansatz, the approximate expressions of $f$ for various gravitational
theory have been obtained \refs{\GrowFacLahav,\GrowFacFormular}. Besides,
there has been much interest in exploring parameterization of the
growth index as a function of the redshift
\GrowFacredshift.

\centerline{{\bf Table V : The Growth Factor Data}}

$$\vbox{\halign{
\quad\hfil\rm#\hfil & \quad\hfil\rm#\hfil  & \quad\hfil\rm#\hfil  &
\quad\hfil\rm#\hfil \cr \noalign{\hrule\smallskip}
\noalign{\hrule\medskip} $z$ & $f_{\rm obs}$ & References \ \ \ \cr
\noalign{\smallskip\hrule\medskip}
$0.15$ & $0.49\pm 0.1$ & \GrowFacGuzzo \ \ \ \cr
$0.35$ & $0.7\pm0.18$ & \BAOTegmark \ \ \ \cr
$0.55$ & $0.75\pm0.18$ & \GrowFacRoss \ \ \ \cr
$0.77$ & $0.91\pm0.36$ & \GrowFacGuzzo \ \ \ \cr
$1.4$ & $0.9\pm0.24$ & \GrowFacJDa \ \ \ \cr
$3.0$ & $1.46\pm0.29$ & \GrowFacMcDonald \ \ \ \cr
$2.125 - 2.72$ & $0.74\pm0.24$ & \GrowFacVHS \ \ \ \cr
$2.2 - 3$ & $0.99\pm1.16$ & \GrowFacMViel \ \ \ \cr
$2.4 - 3.2$ & $1.13\pm1.16$ & \GrowFacMViel \ \ \ \cr
$2.6 - 3.4$ & $0.99\pm1.16$ & \GrowFacMViel \ \ \ \cr
$2.8 - 3.6$ & $0.99\pm1.16$ & \GrowFacMViel \ \ \ \cr
$3.0 - 3.8$ & $0.99\pm1.16$ & \GrowFacMViel \ \ \ \cr
\noalign{\medskip\hrule}}}$$

In Table V, we list the growth data that was converted in the work of \refs{\GrowFacPoAm,\GrowFacNesseris,\GrowFacGuzzo},
from either measurement of redshift distorsion parameter or from various power spectrum amplitudes from Lyman-$\alpha$ Forest data.
The list of the respective original references is also given in the Table V.
A caveat in using this data to constrain other cosmological models
is that in various steps in the process of analysing or converting the data, the $\Lambda$CDM model was assumed.
So if one wants to use the data to constrain other models, and in particular modified gravity models,
one than should redo all the steps assuming that model, starting from original observations.
For more studies concerning the growth factor and its application on the probe of dark energy,
see e.g. \refs{\GrowFacPoAm,\GrowFacNesseris,\GrowFacMasGalClu,\GrowFacWHu}.

\subsec{Other cosmological probes}

Sandage-Loeb test \refs{\SLSandage,\SLLoeb} is a method which
directly measures the evolution of the universe. The idea is to
measure the drifts of the comoving cosmological sources caused by
the cosmic acceleration/deceleration. For an object at redshift
$z_s$, after a period $\Delta t_0$ ($t_0$ stands for the cosmic age
at our position), there would be small variation of its redshift due
to the drift caused by the evolution of the universe
\eqn\SLDeltaz{\Delta z\approx \left[{\dot a(t_0)-\dot a(t_s)\over
a(t_s)}\right]\Delta t_0.} The variations of redshifts can be
obtained by direct measurements of the quasar Lyman-$\alpha$
absorption lines at sufficiently separated epochs (e.g., 10-30 yrs).
The Sandage-Loeb test is unique in its coverage of the ``redshift
desert" at $2\leq z\leq 5$, where other dark energy probes are unable to
provide useful information about the cosmic expansion history. Thus,
this method is expected to be a good complementary to other dark energy
probes \SLCorasaniti.

Gravitational waves (GW) observations also have potential to make interesting contributions to the studies of dark energy.
In 1986, Schutz \GWSchutz \ found that the luminosity distance of the binary neutron stars (or black holes) can be independently determined
by observing the gravitational waves generated by these systems.
If their redshifts can be determined, then they could be used to probe dark energy through the Hubble diagram \GWHHD.
This so-called GW ``standard sires'' has drawn a lot of attentions \GWSG.
Besides, dark energy can also leave characteristic features on the spectrum of primordial gravitational waves \GWSteinhardt,
which may be detected by future ground-based and space-based GW detectors \GWGrishchuk\GWZhang.

In addition, through astronomical observations,
the validity of general relativity can be tested from observations of solar systems,
BBN, CMB, LSS, gravitational waves, and so on.
In the following we will present a brief introduction to the tests coming from the solar system and BBN.
For more details about the tests of gravity theories, see \refs{\TESTGRAVConfGRExp,\TESTGRAVCosmoTestofGrav,\TESTGRAVUzan} and the references therein.

\ifig\FigTestG{Measurements of the coefficient $(1+\gamma)/2$ from light deflection and time delay measurements.
Its value in the Einstein gravity is unity.
The arrows at the top denote anomalously large values from early eclipse expeditions.
The Shapiro time-delay measurements using the Cassini spacecraft yielded on agreement with Einstein gravity to $10^{-3}$
percent, and VLBI (very-long-baseline radio interferometry) light
deflection measurements have reached 0.02 percent. Hipparcos denotes
the optical astrometry satellite, which reached 0.1 percent. From
\TESTGRAVConfGRExp.} {\epsfysize=3.5in \epsfbox{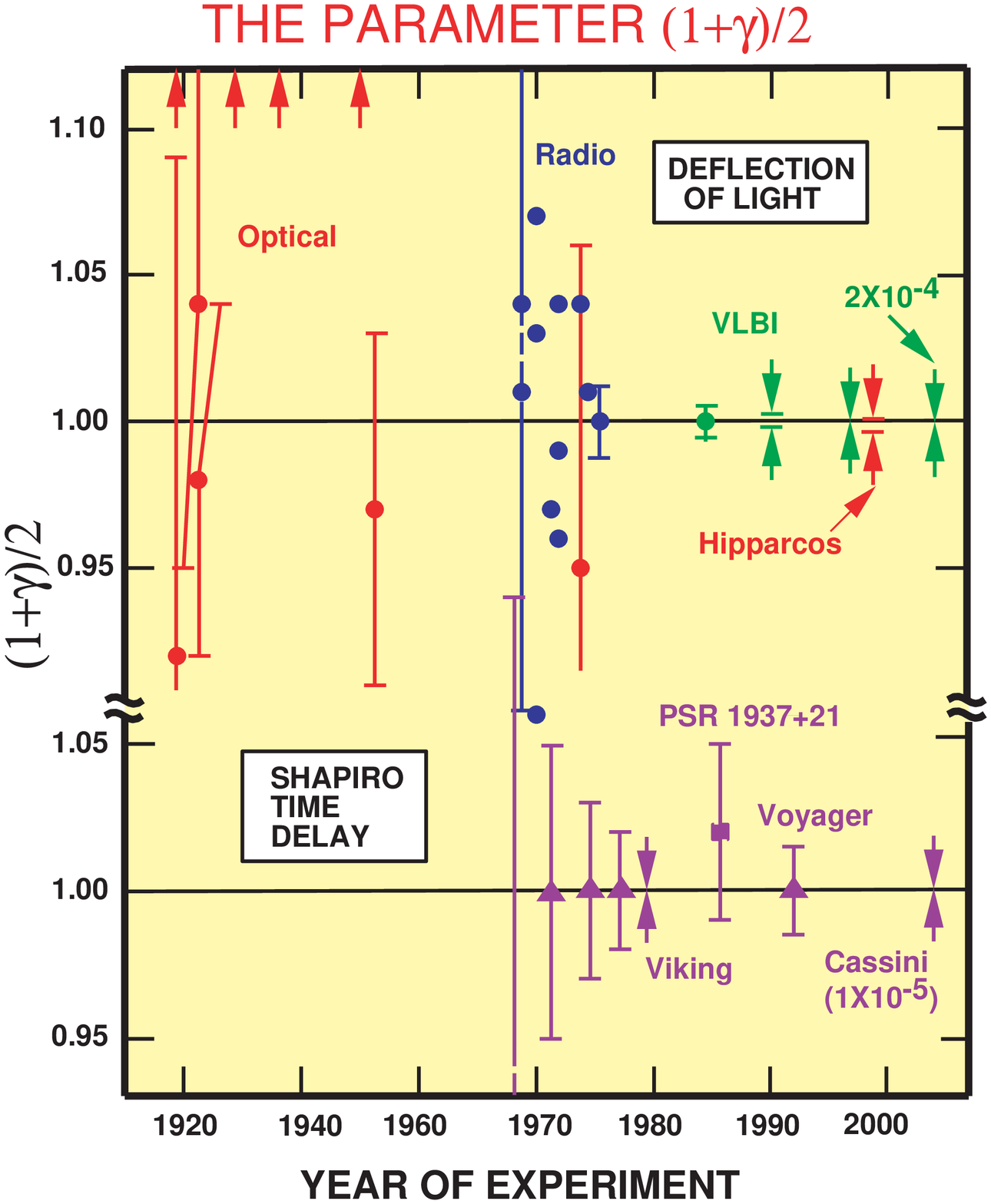}}

Solar system tests of alternate theories of gravity are commonly
described by the Parametrized Post-Newtonian (PPN) formalism
\TESTGRAVPPN. In this formalism, the metric is written as a
perturbation about the Minkowski metric
\eqn\PPNForm{ds^2=-(1+2\Psi-2\beta\Psi^2)dt^2+(1-2\gamma\Psi)d\vec{x}^2,}
here the potential $\Psi=-GM/r$ for the Schwarzschild metric. The
parameter $\beta$ stands for the nonlinearity in the superposition
law of gravity, and $\gamma$ describes the spacetime curvature
induced by a unit mass. The parameter $\gamma$ is the most relevant
PPN parameter for the modified gravity theories of interest. So far,
$\gamma$ has been tightly constrained from various observational
methods. The tightest constraint comes from time-delay measurements
in the solar system, specifically the Doppler tracking of the
Cassini spacecraft, which gives $\gamma-1=(2.1\pm2.3)\times10^{-5}$
\TESTGRAVCassini. Other tests include the observed perihelion
shift of Mercury's obit \TESTGRAVPeriShiftMercury\ and light
deflection measurement \TESTGRAVLightDeflection, which can
constrain $\gamma$ at the $10^{-3}$ and $10^{-4}$ level,
respectively (see \FigTestG).

Moreover, solar system observations can also test the modified gravity
models through their violations of the Strong Equivalence Principle
(SEP). That will result in a difference in the free-fall
acceleration of the earth and the moon towards the sun, named as the
Nordtvedt effect \TESTGRAVPeriShiftMercury. This effect is
detectable in the Lunar Laser Ranging (LLR). Current LLR data have
constrain PPN deviations from the Einstein gravity at the $10^{-4}$ level
\TESTGRAVConfGRExp. In the near future, the bound is expected to be
improved by an order of magnitude by the Apache Point
Observatory for Lunar Laser-ranging Operation (APOLLO) project
\TESTGRAVImproveLLR.

The BBN happened when the universe was $\sim$ 10-100 seconds old.
The abundance of light elements produced during BBN is very
sensitive to the Hubble parameter $H$ and the temperature $T$. On
the other hand, from the Friedmann Equation $H$ and $T$ are related
to the gravitational constant $G$ by (assume radiation domination
and zero curvature) \eqn\BBNTestG{H^2\sim Gg_\ast T^4,}
where $g_\ast$ is the number of relativistic species at BBN.
Based on this relation, a constraint on $G$ at the level of better than
10\% have been inferred from BBN \TESTGRAVBBNReview.

In addition, the influence of modified gravity on $H$ can also affect the epoch of
recombination, resulting a detectable effect on the damping of the
CMB power spectrum at high $l$. From current CMB observations, a
constraint on $G$ at $\sim$ 10\% level have been obtained
\TESTGRAVCMBConstrainG. The constraint from CMB can help to break the
high degeneracy of Hubble parameter $H$ and temperature $T$ in the BBN observations and improve
the constrain to $\sim$ 3\% \TESTGRAVCMBConstrainG. In the future,
it is expected that the Plank satellite can constrain $G$ at the
$\sim$ 1.5\% level \TESTGRAVCMBConstrainG.

\newsec{Dark energy projects}
\seclab\secDEProjects

Here we provide a brief overview of present and future projects to
probe dark energy. Although some high energy physics experiments (such as LHC
\LHC) might shed light on dark energy through discoveries about supersymmetry
or dark matter, here we only introduce the projects involving cosmological
observations. According to the DETF report \DETF, the dark energy projects
can be classified into four stages: completed projects are Stage I;
on-going projects, either taking data or soon to be taking data, are
Stage II; intermediate-scale, near-future projects belong to Stage
III; larger-scale, longer-term future projects belong to Stage IV.

Moreover, to compare various dark energy projects,  DETF \DETF\ also proposed
a quantity called figure of merit (FoM). Utilizing the famous Chevallier-Polarski-Linder (CPL)
parameterization \eqn\DETFFOM{w(a)=w_0+(1-a)w_a} a FOM is defined as
the reciprocal of the area of an error ellipse in the $w_0 - w_a$
plane. A conventional normalization takes for the FoM the square
root of the determinant of the $2 \times 2$ Fisher matrix for $w_0$
and $w_a$. Soon after, an extended version of FOM was proposed by
Wang \FOMWang. More advanced stages are expected to deliver tighter
dark energy constraints, and give larger FoMs. For examples, Stage III
experiments are expected to deliver a factor $\sim3$ improvement in
the FoM compared to the combined Stage II results, while Stage IV
experiments will improve the FoM by a factor of 10 compared to
Stage II.

In this section, we will introduce some most representative projects
of Stage II, Stage III and Stage IV. The corresponding dark energy projects
are shown in Fig 7. The Stage I experiments that have already
reported results, such as SNLS, ESSENCE, and WMAP, will not be
discussed here. Notice that some dark energy projects have already launched
since 2006, our classifications of dark energy projects are slightly
different from that in the DETF report. For a comprehensive list of
the dark energy experiments, see Ref. \ReviewofTurner.

\ifig\FigDEP{Dark energy projects introduced in this work. The black ones are
stage II projects, the blue ones are stage III projects, and the red
ones are stage IV projects.} {\epsfysize=3.5in \epsfbox{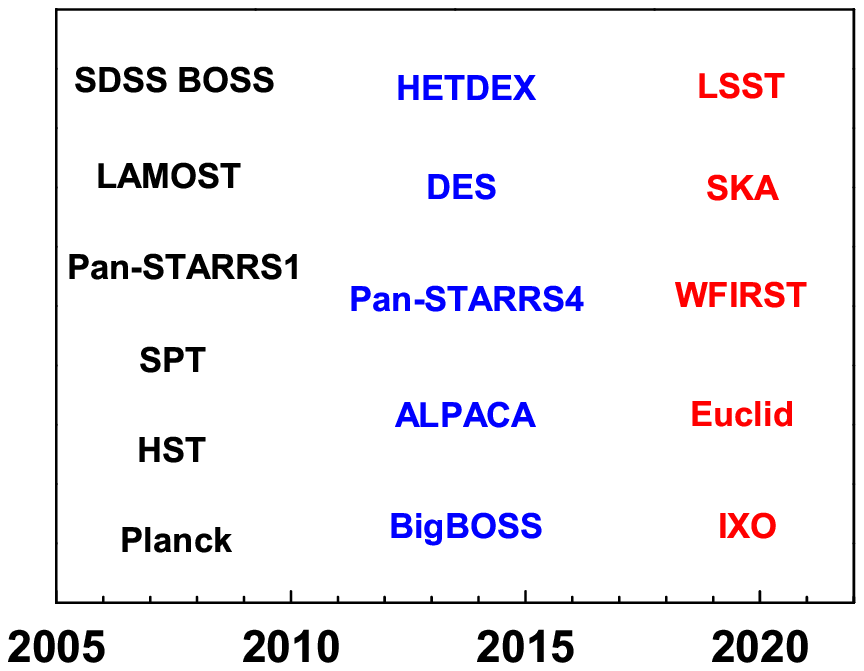}}

\subsec{On-going projects}

In this subsection we introduce the on-going projects. A list of
some most representative projects of Stage II is given in Table VI.

\

\centerline{{\bf Table VI : On-Going Projects}}

$$\vbox{\halign{
\quad\hfil\rm#\hfil & \quad\hfil\rm#\hfil  & \quad\hfil\rm#\hfil  &
\quad\hfil\rm#\hfil \cr \noalign{\hrule\smallskip}
\noalign{\hrule\medskip} Survey & Location & Description & Probes  \
\ \ \cr \noalign{\smallskip\hrule\medskip} SDSS BOSS & Sacramento
(USA) & Optical, 2.5-m & BAO  \ \ \ \cr LAMOST & Xinglong (China) &
Optical, 4-m & BAO  \ \ \ \cr Pan-STARRS1 & Hawaii (USA) & Optical,
1.8-m & SN, WL, CL \ \ \ \cr SPT & South Pole & Submillimeter, 10-m
& CL \ \ \ \cr HST & Low Earth orbit & Optical/near-infrared & SN
\ \ \ \cr Planck & Sun-Earth L2 orbit & SZE & CL \ \ \ \cr
\noalign{\medskip\hrule}}}$$

The SDSS Baryon Oscillation Spectroscopic Survey (BOSS)
\DEObjSDSSBOSS\ is one of the four surveys of SDSS-III, using a
dedicated wide-field, 2.5-meter optical telescope at Apache Point
Observatory in Sacramento Mountains. With a 5-year survey (2009 -
2014) of 1.5 million luminous red galaxies (LRGs), BOSS will achieve
the first measurement of the BAO absolute cosmic distance scale with
$1\% $ precision at redshifts z = 0.3 and z = 0.6. It will also
measure the distribution of quasar absorption lines at z = 2.5,
yielding a measurement of the angular diameter distance at that
redshift to an accuracy of $1.5\%$.

The Large Sky Area Multi-Object Fiber Spectroscopic Telescope
(LAMOST) \DEObjLAMOST\ is a National Major Scientific Project (NMSP) built
by the Chinese Academy of Sciences (CAS). It is a wide-field,
4-meter ground-based optical telescope located at the Xinglong
observing station of National Astronomical Observatories (NAO). The
main construction of this instrument was finished in 2008. After its
commissioning stage, LAMOST will study dark energy through the BAO technique.

The Panoramic Survey Telescope and Rapid Response System
(Pan-STARRS) \DEObjPanSTARRS\ is an international collaboration
program led by University of Hawaii. Its first stage, Pan-STARRS1,
is a 1.8-m wide-field telescope located at Haleakala in Hawaii. The
mirror has a 3 degree field of view and be equipped with a CCD
digital camera with 1.4 billion pixels. The regular observing of
Pan-STARRS1 has already started in 2009. As one of science goals,
Pan-STARRS1 study dark energy through the SN, the WL and the CL techniques.

The South Pole Telescope (SPT) \DEObjSPT\ is an international
collaboration program operating at the USA NSF South Pole research
station. As the largest telescope ever deployed at the South Pole,
SPT is a 10-m submillimeter-wave telescope with a 1000-element
bolometric focal plane array. Constructed between 2006 and 2007, it
is conducting a survey of galaxy clusters over 4000 deg$^2$ using
the Sunyaev - Zel'dovich effect (SZE). About 20,000 clusters are
expected to be discovered by the SPT in recent years.

The Hubble Space Telescope (HST) \DEObjHST\ is one of the most famous
telescopes in the world. It was built by the USA National
Aeronautics and Space Administration (NASA), with contributions from
the European Space Agency (ESA). HST has two mirrors, the primary
mirror diameter is 2.4 m, and the secondary mirror diameter is 0.3
m. Since launched in 1990, many HST observations have led to great
breakthroughs in astrophysics, such as the discovery of the
currently observed cosmic accelaration \refs{\riess,\perl}. So far,
utilizing the HST, several SNIa datasets have been obtained, such as
Gold \refs{\Goldfour,\Goldsix}, Union \Union, Constitution \Constitution\ and
Union2 \UnionTwo. The present SN survey project of HST, HST Cluster
Supernova Survey \DEObjHSTNew, is targeting supernovae in high
redshift galaxy clusters at $0.9 < z < 1.5$. It is expected that
more than 100 high redshift SNIa will be found in recent years.

The Planck \DEObjPLANCK\ is a CMB satellite, which is created as the
third Medium-Sized Mission of ESA's Horizon 2000 Scientific
Programme. The telescope is an off-axis, aplanatic design with two
elliptical reflectors and a 1.5 m diameter. It is designed to image
the CMB anisotropies and polarization maps over the whole sky, with
unprecedented sensitivity and angular resolution. While the WMAP
data reach 200 billion samples after its nine-year mission, just a
single year of observing by Planck will yield 300 billion samples.
Besides pinning down the cosmological parameters of dark energy,
Planck will also detect thousands of galaxy clusters using the SZE.
It has launched in 2009, and is expected to yield definitive data by
2012.

\subsec{Intermediate-scale, near-future projects}

In this subsection we introduce the intermediate-scale, near-future
projects. A list of some most representative projects of Stage III
is given in Table VII.

\bigskip

\centerline{{\bf Table VII : Intermediate-Scale, Near-Future Projects}}
$$\vbox{\halign{
\quad\hfil\rm#\hfil & \quad\hfil\rm#\hfil  & \quad\hfil\rm#\hfil  &
\hfil\rm#\hfil  \cr \noalign{\hrule\smallskip}
\noalign{\hrule\medskip} Survey & Location & Description & Probes  \
\ \ \cr \noalign{\smallskip\hrule\medskip}
HETDEX & Davis Mountains (USA) & Optical, 9.2-m & BAO
\ \ \ \cr DES & Cerro Pachon (Chile) & Optical, 4-m & BAO, SN, WL, CL
\ \ \ \cr Pan-STARRS4 & Hawaii (USA) & Optical, 1.8-m$\times4$ & SN, WL, CL
\ \ \ \cr ALPACA & Cerro Pachon (Chile) & Optical, 8-m & SN
\ \ \ \cr BigBOSS & Tucson and Cerro Pachon & Optical, 4-m & BAO \ \ \ \cr
\noalign{\medskip\hrule}}}$$

The Hobby-Eberly Telescope Dark Energy Experiment
(HETDEX)\DEObjHETDEX\ is an international collaboration program led
by University of Texas, Austin. It will use the 9.2-m Hobby-Eberly
Telescope at McDonald Observatory to measure BAO over two areas,
each 100 deg$^2$, using one million galaxies over the redshift range
$1.8 < z < 3.7$. It is expected that the survey will begin in 2012 and
last for 3 years.

The Dark Energy Survey (DES) \DEObjDES\ is an international
collaboration program led by Fermilab. It is a new 570 megapixel
digital camera (DECam) mounted on the 4-m Blanco Telescope of the
Cerro Tololo Inter-American Observatory (CTIO) in Chile. The DES
plans to obtain photometric redshifts in four bands, and the planned
survey area is 5000 deg$^2$. It will probe dark energy by using the BAO, the
CL, the SN, and the WL technologies. The proponents will begin at
2011 and take 5 years to complete.

The Pan-STARRS4 is the updated version of Pan-STARRS1 \DEObjPanSTARRS. It
is a large optical survey telescope to be sited on Mauna Kea in
Hawaii. It consists of an array of four 1.8-m telescopes, each
mirror will have a 3 degree field of view and be equipped with a CCD
digital camera with 1.4 billion pixels. The dark energy science goals of
Pan-STARRS4 include SN, WL and CL surveys.
The survey will continue for ten years.

The Advanced Liquid-mirror Probe for Astrophysics, Cosmology and Asteroids (ALPACA) \DEObjALPACA \
is a proposed wide-field telescope employing an 8-meter rotating liquid mirror.
It brings together the technologies of liquid-mirrors, lightweight conventional mirrors, and advanced detectors,
to make a novel telescope with uniquely-powerful capabilities.
Scanning a long strip of sky every night at CTIO,
ALPACA will discover $\sim 50,000$ SNIa and $\sim 12,000$ type II SN each year to a redshift $z \sim  0.8$.

The Big Baryon Oscillation Spectroscopic Survey
(BigBOSS) \DEObjBigBOSS\ is an USA NSF/DOE collaboration program.
As the updated version of SDSS BOSS, BigBOSS will probe dark energy through
the BAO and the redshift distortions techniques. It will build a new
4000-fiber spectrograph covering a 3-degree diameter field. This
instrument will be mounted on the 4-meter Mayall telescope at Kitt
Peak National Observatory (KPNO) for a 6-year run, then will be
moved to the 4-m CTIO Blanco Telescope for another 4-year run. After
10-year operation, BigBOSS will complete the survey of 50 million
galaxies and 1 million quasars from $0.2<z<3.5$ over 24,000 deg$^2$.
The construction of the instrument will begin in 2011, and the
operation of the survey will start in 2015.

\subsec{Larger-scale, longer-term future projects}

\

\centerline{{\bf Table VIII : Larger-Scale, Longer-Term Future Projects}}
\medskip
$$\vbox{\halign{
\quad\hfil\rm#\hfil & \quad\hfil\rm#\hfil  & \quad\hfil\rm#\hfil  &
\quad\hfil\rm#\hfil  & \quad\hfil\rm#\hfil \cr
\noalign{\hrule\smallskip} \noalign{\hrule\medskip} Survey & Location & Description & Probes  \ \ \ \cr
\noalign{\smallskip\hrule\medskip} LSST & Cerro Pachon (Chile) & Optical, 8.4-m & BAO, SN, WL
\ \ \ \cr SKA & Australia or South Africa & Radio, km$^2$ & BAO, WL
\ \ \ \cr WFIRST & Sun-Earth L2 orbit & Infrared, 1.5-m & BAO, SN, WL
\ \ \ \cr Euclid & Sun-Earth L2 orbit & Optical/NIR, 1.2-m & BAO, WL
\ \ \ \cr IXO & Sun-Earth L2 orbit & X-ray & CL  \ \ \ \cr
\noalign{\medskip\hrule}}}$$

In this subsection we introduce the larger-scale, longer-term future
projects. A list of some most representative projects of Stage IV
is given in Table VIII.

A most ambitious ground-based dark energy survey project is the Large
Synoptic Survey Telescope (LSST) \DEObjLSST, which is an USA NSF/DOE
collaboration program. LSST is an 8.4-meter ground-based optical
telescope (with a 9.6 deg$^2$ field of view and a 3.2 Gigapixel
camera) to be sited in Cerro Pachon of Chile. Over a 10-year
lifetime, it will obtain a database including 10 billion galaxies
and a similar number of stars. As one of the most important
scientific goals, LSST will study dark energy through a combination of the
BAO, the SN and the WL techniques. Since its compelling scientific
capacity and relatively low technical risk, LSST was selected as the
top priority large-scale ground-based project for the next decade of
astronomy in the Astro2010 report \AstroLatest. The appraised
construction cost of LSST is 465 million U.S. dollars. The project
is scheduled to have first light in 2016 and the beginning of survey
operations in 2018.

Another ambitious large-scale ground-based telescope is the
Square Kilometer Array (SKA) \DEObjSKA, which is an international
collaboration program. SKA has a collecting area of order one square
kilometer and a capable of operating at a wide frequency range (60
MHz - 35 GHz). It will be built in the southern hemisphere (either
in Australia or in South Africa), and the specific site will be
determined in 2012. SKA will probe dark energy by BAO and WL techniques via
the measurements of the Hydrogen line (HI) 21-cm emission in normal
galaxies at high redshift. The total Budget of SKA is 1.5 billion
U.S. dollars. Construction of the SKA is scheduled to begin in 2016
for initial observations by 2020 and full operation by 2024.

A most exciting space-based dark energy project is the Wide Field Infrared
Survey Telescope (WFIRST) \DEObjWFIRST, which is an USA NASA/DOE
collaboration program. WFIRST is a 1.5-meter wide-field
near-infrared space telescope, orders of magnitude more sensitive
than any previous project. The design of WFIRST is based on three
separate inputs (JDEM-Omega, the Microlensing Planet Finder, and the
The Near-Infrared Sky Surveyor) to Astro2010 \DEObjWFIRSTscience. It
will enable a major step forward in dark energy understanding,
provide a statistical census of exoplanets and obtain an ancillary
data set of great value to the astronomical community. The objective
of the WFIRST dark energy survey is to determine the nature of dark energy by
measuring the expansion history and the growth rate of large scale
structure. A combination of the BAO, the SN and the WL techniques
will be used to probe dark energy. Since its compelling scientific capacity
and relatively low technical risk, WFIRST was selected as the top
priority large-scale ground-based project for the next decade of
astronomy in the Astro2010 report \AstroLatest. The appraised
construction cost of WFIRST is 1.6 billion U.S. dollars. The project
is scheduled to launch in 2020 and has a 10-years lifetime.

Another exciting space-based dark energy project is the Euclid \DEObjEuclid, which is an ESA project.
Euclid is 1.2 m Korsch telescope, with optical and near-infrared (NIR) observational branch.
The primary goal of Euclid is to map the geometry of the dark universe,
and it will search galaxies and clusters of galaxies out to $z\sim2$,
in a wide extragalactic survey covering 20,000 deg$^2$, plus a deep survey covering an area of 40  deg$^2$.
The mission is optimised for two primary cosmological probes: BAO and WL.
It will also make use of several secondary cosmological probes
such as the ISW, CL and redshift space distortions to provide additional measurements of the cosmic geometry and structure growth.
After 5 years' survey,
Euclid will measure the DE EoS parameters $w_0$ and $w_a$ to a precision of $2\%$ and $10\%$, respectively,
and will measure the growth factor exponent $\gamma$ with a precision of $2\%$.

The last Stage IV project is the International X-ray Observatory
(IXO) \DEObjIXO, which is a partnership among the NASA, ESA, and the
Japan Aerospace Exploration Agency (JAXA). IXO is a powerful X-ray
space telescope that features a single large X-ray mirror assembly
and an extendible optical bench with a focal length of $\sim20$ m.
With more than an order of magnitude improvement in capabilities,
IXO will study dark energy through the WL technique. Because of IXO's high
scientific importance, it was selected as the fourth-priority
large-scale ground-based project in the Astro2010 report
\AstroLatest. The total Budget of IXO is 5.0 billion U.S. dollars.
The project is scheduled to launch in 2021.

\newsec{Observational constraints on specific theoretical models}
\seclab\secConsDEModels

In this section we will briefly review some research works concerning the cosmological constraints on the specific theoretical models.
These models can be classified into three classes:
(i) models that modify the energy-momentum tensor on the r.h.s. of the Einstein equation, i.e. dark energy models.
We will briefly review some numerical results of scalar field models, Chaplygin gas models, and holographic models.
(ii) models that modify the l.h.s. of the Einstein equation, i.e. modified gravity models.
We will briefly review some numerical results of the DGP braneworld model,
the $f(R)$ gravity, the Gauss-Bonnet gravity, the Brans-Dicke theory and the $f(T)$ gravity.
(iii) models that attempt to explain the apparent cosmic acceleration by assuring the inhomogeneities in the distribution of matter.
We will briefly review some numerical works on the inhomogeneous Lema\^\i tre-Tolman-Bondi model and backreaction model,
Lastly, we will summarize and compare these theoretical models.

\subsec{Scalar field models}

As is well known, the most popular phenomenological
models are models with rolling scalar fields $\phi$
\refs{\ZlatevTR,\CMBCaldwell}. These models are the direct generalization of
the cosmological constant and have been well-studied both
theoretically and numerically. In Sec. 11, we have introduced
the theoretical studies on the scalar field models. In this
subsection, we will discuss the scalar field models from the aspect
of observations.

\noindent $\bullet$ Reconstructing the scalar filed models from observations

\ifig\FigQuin{The reconstructed potential $V(z)$ and the kinetic
energy term $\dot \phi^2$ in units of the critical density
$\rho_{cr}=3M_p^2H_0^2$, based on 54 SNIa data of the SCP
\perl\ including the low-$z$ Calan Tololo sample \SNIaHamuy.
Also plotted are the two forms of $V(\phi)$ for this $V(z)$. The
value of $\phi$, known up to an additive constant, is plotted in
units of the Planck mass. The solid line corresponds to the best-fit
values of the parameters, the shaded area/dotted lines covers the
range of 68\%/90\% errors, respectively. The hatched area represents
the unphysical region $\dot\phi^2<0$. From \QuinSaini.}
{\epsfysize=2.7in \epsfbox{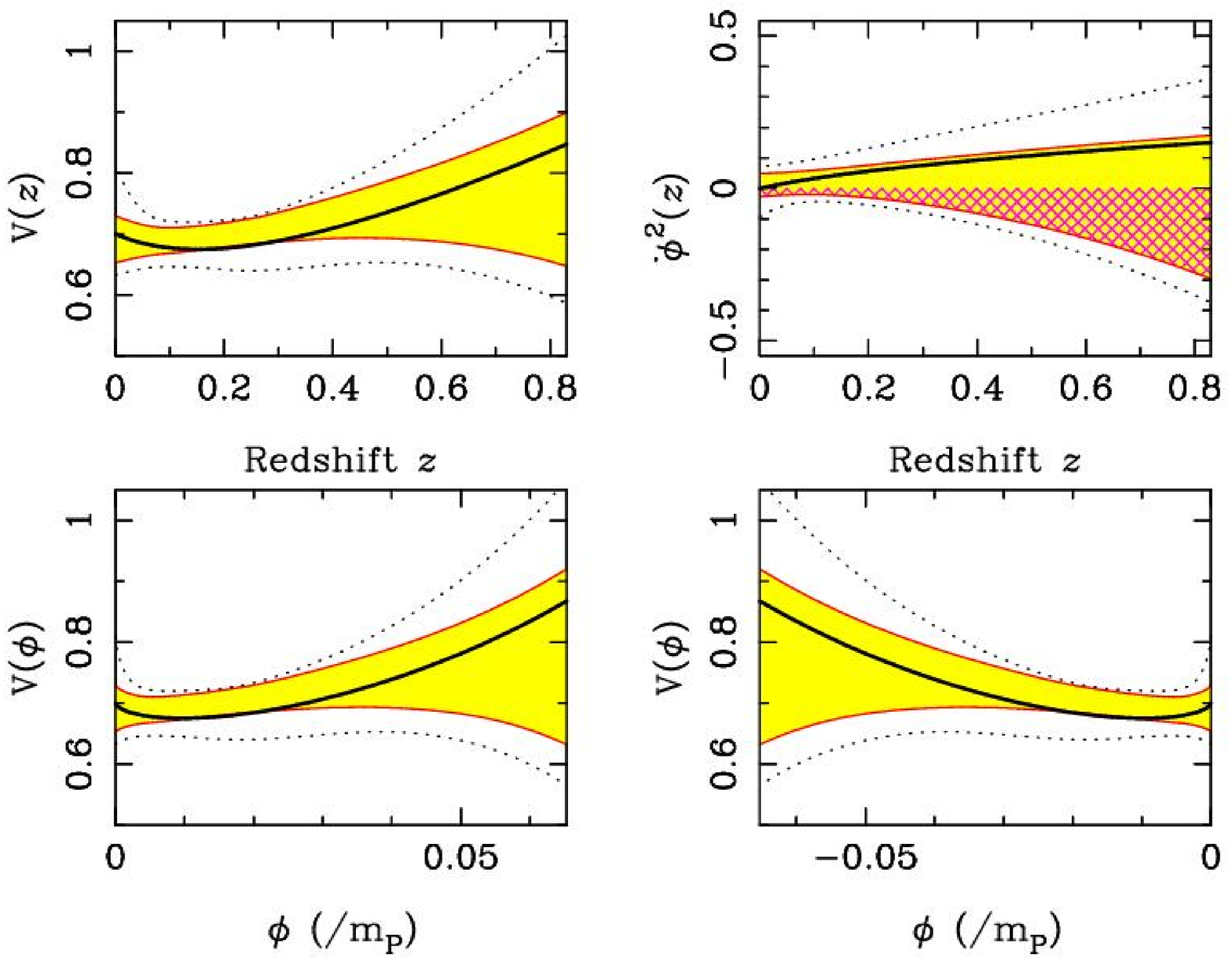}}

After the proposal of quintessence field as a candidate of dynamical dark energy \refs{\ZlatevTR,\CMBCaldwell},
a lot of numerical studies have been carried out to distinguish this model from the cosmological constant
(see \refs{\CLWang,\QuinStarobinsky,\QuinNakamura,\QuinSaini,\QuinChiba,\QuinEaryWorks,\DEReviewSahniB} for some early studies).
A widely used approach is directly reconstructing dark energy from the observational quantities like $d_L(z)$, $H(z)$, and so on.
In the quintessence model, the Einstein equations can be rewritten as \refs{\QuinStarobinsky,\QuinSaini,\QuinSahni}
\eqn\QuintA{{8\pi G \over 3H_0^2}V(x)={H^2\over H_0^2}-{x\over
6H_0^2}{dH^2\over dx}-{1\over2}\Omega_m x^3,}
\eqn\QuintB{{8\pi G\over 3H^2_0}\left(d\phi\over dx\right)^2={2\over
3H_0^2x}{d\ln H\over dx}-{\Omega_m x\over H^2},\ \ x\equiv1+z.}
One can determine $\phi (z)$, and thus its inversion $z(\phi)$ by
integrating Eq. \QuintB. The Hubble parameter $H(z)$ together with
its first derivative can be determined from observations, such as
the measurements of the luminosity distance $d_L(z)$
\refs{\QuinStarobinsky,\QuinNakamura,\QuinSaini,\QuinChiba,\QuinSahni}
\eqn\QuintC{H(z)=\left[{d\over dz}\left(d_L(z)\over
1+z\right)\right]^{-1}.}
Substituting $H(z)$ and $z(\phi)$ into Eq. \QuintA, one can
reconstruct the potential $V(\phi)$. In 2000 Saini {\it et al.}
\QuinSaini\  reconstructed $V(\phi)$ and $w_Q(z)$ based on 54 SNIa
given by Perlmutter {\it et al.} \perl. Their results are shown in \FigQuin.

In the same way one can also reconstruct other physical quantities from
observations. For instance, in \QuinSaini, Saini {\it et al.}
reconstructed the EOS $w(z)$ using the relation
\eqn\Quintw{w(z)={(2/3)(1+z)d\ln H/dz-1 \over 1-(H_0/H)^2\Omega_m (1+z)^3}.}
They obtained the 1$\sigma$ constraint
\eqn\QuintwResul{-1\leq w \leq -0.86,\ \ \ \ -1\leq w \leq -0.66,}
corresponding to the value of $w$ at $z=0$ and $z=0.83$,
respectively. So the result allows the possible evolution of dark energy,
while the cosmological constant with $w=-1$ is still consistent with the data.

In addition to the above method, a more widely used approach is to
reconstruct the quantities by a fitting ansatz which relies on a
small number of free parameters \HzSimon\QuinPara. We will discuss this
issue in the section 16.

Broadly speaking, by making use of the scalar fields one can reconstruct a dark energy component with any property.
So the issue of the observational tests of scalar field models is somewhat similar as the observational probe of the dynamical behavior of dark energy.
One can see
\refs{\CopelandWR,\DEBeanCaroll,\DEReviewPadmanabhanB,\Perivolaropoulos,\AASen,\QuinRefs}
and references therein
for more studies on the scalar field models from the aspect of observations.
In addition, there are also some theories using vector fields to describe dark energy \refs{\ArmendarizPiconPM,\NoScalar,\NonScalar,\NonnScalar}.

\noindent $\bullet$ Quintessence, phantom or quintom

Another interesting issue concerning the scalar filed models is the future evolution of dark energy and the fate of our universe.
Besides the quintessence field satisfying $w>-1$, another well-studied scalar field model is the phantom model satisfying $w<-1$.
In this model, the dark energy density will reach infinity and
lead to the ``big-rip'' singularity. The WMAP7 measurements
give $w=-1.10\pm0.14$ \WMAPSeven, the analysis of the Union2 SNIa dataset give
$w=-1.035^{+0.093}_{-0.097}$ \UnionTwo, and the analysis of the SDSS
DR7 give $w=-0.97\pm0.10$ \BAONewestData. So the current
observational data are still consistent with the cosmological constant, although
the possibilities of quintessence and phantom all exist. Besides,
the possibility of the quintom, where the EOS can cross
$w=-1$, can also provide a consistent fit to the observations. We
refer to \Quintom\ for related
numerical studies.

\noindent $\bullet$ Interaction between dark sectors

It is also worthwhile to consider the possible interaction between
the scalar field and the matter component. In
\AmendolaER, Amendola proposed the coupled
quintessence (CQ) model, in which the scalar field $\phi$ and the dark matter fluid
with each other through a source term in their respective covariant
conservation equation
\eqn\CQEqA{\nabla_\mu T^{\mu}_{\nu(\phi)}=-Q_{\nu};\ \ \ \nabla_\mu
T^{\mu}_{\nu(m)}=Q_{\nu},}
where $T^{\mu}_{\nu(\phi)}$ and $T^{\mu}_{\nu(m)}$ represent the
energy momentum tensors of $\phi$ and matter. In
\refs{\AmendolaER,\DEIntWetterich}, it was proposed that the source
term $Q_\nu$ assumes the form
\eqn\CQQ{Q_\nu=-\kappa\beta(\phi)T_{(m)}\nabla_\nu\phi,}
where $T_{(m)}$ is the trace of $T^{\mu}_{\nu(m)}$, and $\beta(\phi)$
(hereafter $\beta$) is the coupling function that sets the strength
of the interaction.

\ifig\FigCoupledQuin{Observational constraints on the coupling
between dark energy and dark matter using 71 SNIa from the fist year
SNLS \SNLS, the CMB shift parameter from the three-year WMAP
observations \WMAP, and the BAO peak found in the SDSS
\BAOEisenstein. The constant coupling $\delta=\Gamma/H$ is
considered, which leads to the Friedmann equation
$\dot\rho_m+3H\rho_m=\delta H\rho_m$. The left and right panels
shows observational contours in the ($w_X,\delta$) and
($\Omega_{X0},\delta$) planes, respectively (here ``$X$'' stands for
the dark energy component). The best-fit model parameters correspond to
$\delta=-0.03,w_X=-1.02$ and $\Omega_{X0}$=0.73. A stringent
constraint $-0.08<\delta<0.03$ at 95\% CL is obtained. From \CQGuo.}
{\epsfysize=2.6in \epsfbox{conall.eps} \epsfysize=2.6in
\epsfbox{conall2.eps}}

In \AmendolaER, Amendola also investigated the
evolution of the perturbations as well as the observational
signature of this model. Hereafter, a lot of studies were performed
to constrain the CQ model utilizing the various observational
techniques \refs{\CQNumWorks,\CQAmendolaD,\CQMaccio,\CQGuo,\CQBean}.
For instance, in \CQAmendolaD, from first-year WMAP observations,
Amendola {\it et al.} obtained a upper constraint $\sim 0.1$ for the
coupling parameter $\beta$. In \CQMaccio, Maccio {\it et al.}
performed the $N$-body simulations in CQ model and found that this
model is consistent with the growth of structure in the
non-linear regime. In \CQGuo, Guo {\it et al.} considered the following
kinds of interaction
\eqn\CQInt{\dot\rho_m+3H\rho_m=\delta H\rho_m.}
From a combination of SNIa, WMAP and BAO data, they put a stringent
constraints on the constant coupling mode
\eqn\CQdelta{-0.08<\delta<0.03,}
at the 2$\sigma$ CL (see \FigCoupledQuin). In \CQBean, Bean {\it et al.}
investigated a variety of CQ models and found that a combination of
SNIa, LSS and CMB data can constrain the strength of coupling between dark sectors to be less than 7\% of the coupling to gravity.
In all, the current observational data have already given tight constraints
on the interaction between dark sectors. One can see
\refs{\MBaldi,\CQOther} and references therein for more studies on this topic.
In addition, for the studies of dynamical system descriptions about the scalar dark energy, see \Urena.

\subsec{Chaplygin gas models}

In \KamenshchikCP, Kamenschchik {\it et al.} explained dark energy as a kind of fluid called Chaplygin gas (CG),
characterized by the EOS
\eqn\CGasp{p=-{A\over\rho},}
where $A$ is a constant.
Later, Bilic {\it et al.} \GCGasOrdinaryPaperA\ and Bento {\it et al.} \BentoPS\ proposed an extension of the original Chaplygin gas model,
called generalized Chaplygin gas (GCG) model, with the EOS
\eqn\CGasGCGp{p= - {A\over\rho^\alpha}.}

\noindent $\bullet$ Observational inspection of the Chaplygin gas models

\ifig\FigCG{{\it Left Panel}: A constraint on the standard Chaplygin
gas using SNIa, BAO and CMB. Clearly this model is a very poor fit
to the data. From \DEManyModelsTDavis. {\it Right Panel}: GCG as the
unification of dark matter and dark energy will produce inconsistent oscillations in
the mass power spectrum. The data points are the power spectrum of
the 2df galaxy redshift survey, and the curves from top to bottom
are GCG models with $\alpha$ = $-10^{-4}$, $-10^{-5}$, 0
($\Lambda$CDM), $10^{-5}$ and $10^{-4}$, respectively. From
\NCGasSandvik.} {\epsfysize=2.5in \epsfbox{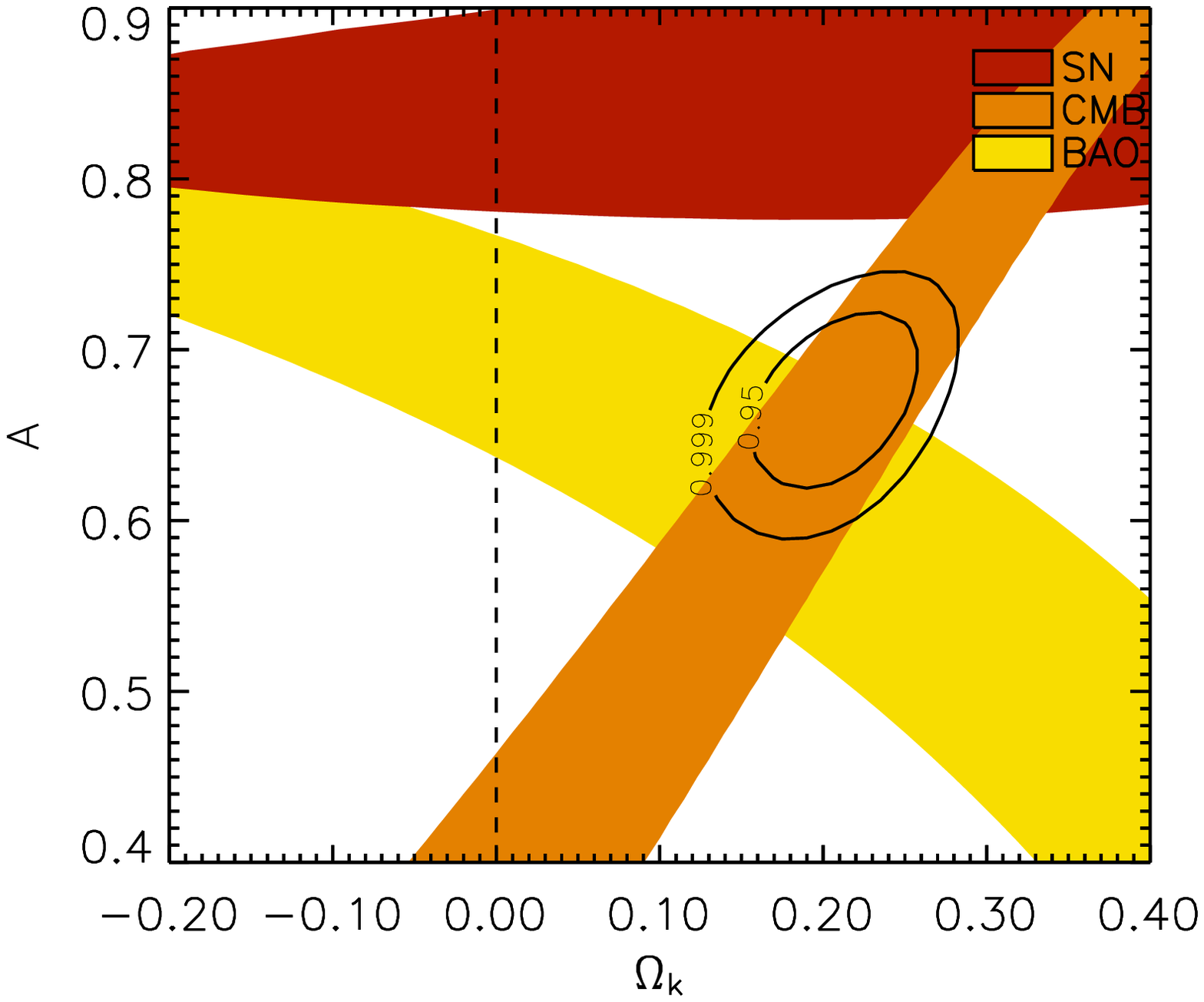} \epsfysize=2.5in
\epsfbox{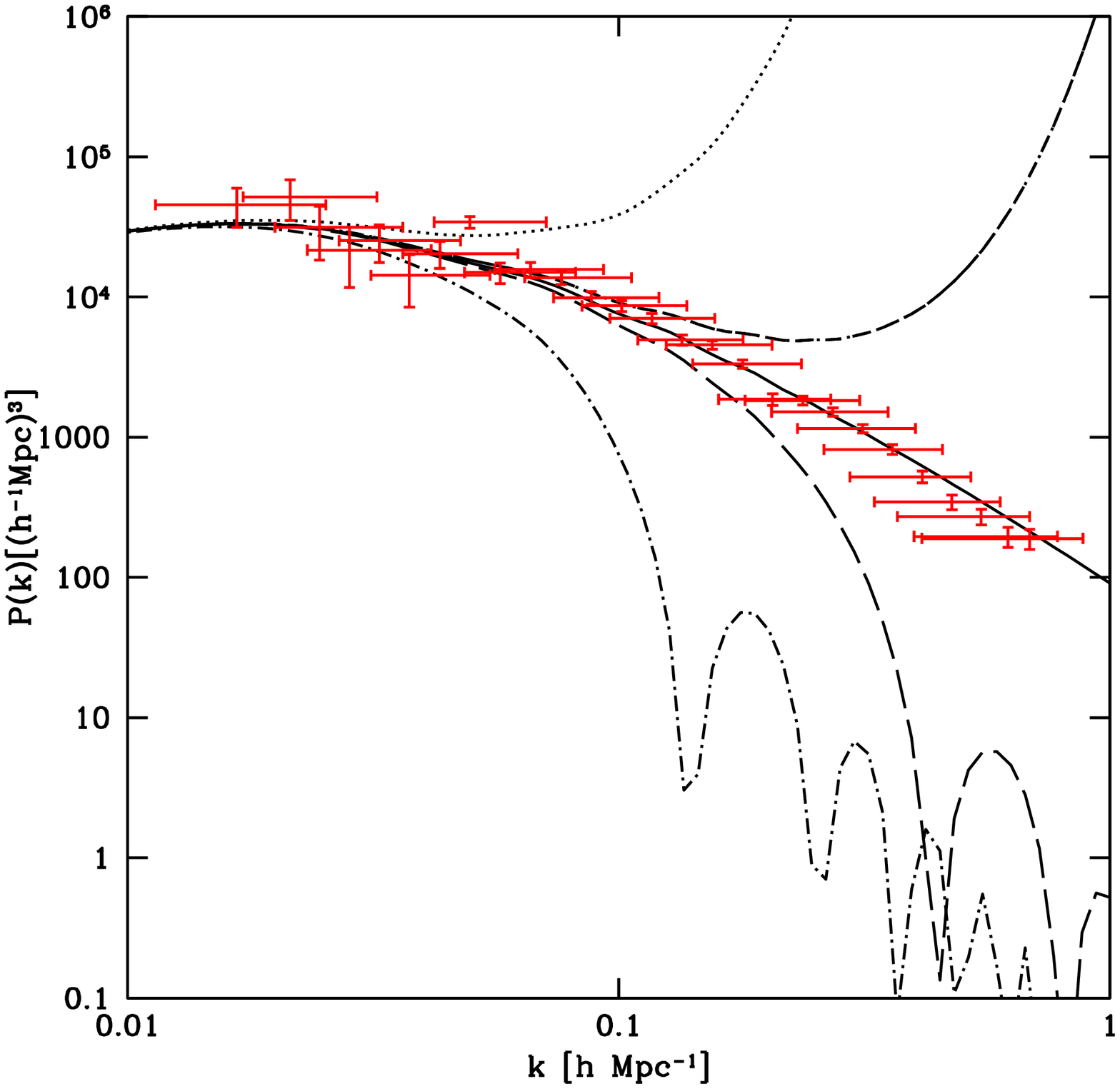} }

The observational inspection of the CG and GCG models have been
extensively investigated using various observational methods \refs{\DEManyModelsTDavis,\NCGasSandvik,\NCGasNumWorks,\NCGasZhu},
including the observations of SNIa, BAO, CMB, WL, X-ray, GRB, X-ray,
Fanaroff-Riley type IIb radio galaxies \NCGasDaly, and so on.
A common result of these studies is that the CG model has been ruled out by the observations.
For example, in \DEManyModelsTDavis, Davis {\it et al.} showed that the CG model is strongly disfavored by a combination of SNIa+BAO+CMB data,
since this model gives a $\chi^2_{\rm min}$ $\sim 100$ larger than that given by the $\Lambda$CDM model (see \FigCG\ for more details).
Similar result was obtained in \NCGasZhu,
where Zhu investigated the GCG model from the measurements of X-ray, the SNIa and Fanaroff-Riley typeIIb radio galaxies,
and got a constraint
\eqn\CGAmen{\alpha=-0.09^{+0.54}_{-0.33},}
at the 95\% CL.
Therefore, the CG model, which corresponds to $\alpha$=1, is ruled out at a 99\% CL.

\noindent $\bullet$ The GCG model as a unification of dark energy and dark matter

An attractive feature of the GCG model is that it can explain both
dark energy and dark matter in terms of a single component and has been refered to as ``unified dark matter'' (UDM) \refs{\MengJY,\RenNW,\NCGasMakerB}.
From Eq. \CGasGCGp\ one can obtain the energy density of GCG \NCGasMakerB
\eqn\CGasGCGRho{\rho(t)=\left[A+{B\over B^{3(1+\alpha)}}\right]^{1\over1+\alpha},}
where $B$ is an integration constant. Defining
\eqn\CGasOmegam{\Omega^\ast_m={B\over A+B},\ \ \rho_\ast=(A+B)^{1\over1+\alpha},}
then the Eq. \CGasGCGp\ becomes
\eqn\CGasrho{\rho(z)=\rho_0\left[(1-\Omega^\ast_m)+\Omega^\ast_m a^{-3(1+\alpha)}\right]^{1\over1+\alpha}.}
Obviously, the flat $\Lambda$CDM scenario with matter ratio $\Omega_m^\ast$ is recovered with
$\alpha=0$.

However, a problem with the UDM scenario is that it will produce oscillations
or exponential blowup of the matter power spectrum, which is
inconsistent with observation. In \NCGasSandvik, Sandvik {\it et al.}
found that the density fluctuation $\delta_k$ with wave vector $k$
evolves as
\eqn\NCGasdeltak{\delta^{\prime\prime}_k+[2+\zeta-3(2w-c^2_s)]\delta^\prime_k=\left[{3\over2}(1-6c^2_s+8w-3w^2)-\left({kc_s\over a H}\right)^2\right]\delta_k}
where the quantity $\zeta$, the EOS $w$ and the squared sound speed
$c^2_s$ take the form \NCGasSandvik
\eqn\NCGaszeta{\zeta\equiv{(H^2)^\prime\over2H^2}=-{3\over2}\left(1+(1/\Omega^\ast_m-1)a^{3(1+\alpha)}\right)^{-1},}
\eqn\NCGaswcs{w=-\left[1+{\Omega^\ast_m\over1-\Omega^\ast_m}a^{-3(1+\alpha)}\right]^{-1},\ \ \ c^2_s=-\alpha w, }
and prime denotes partial differentiation with respect to $\ln a$.
For Chaplygin gas, $\alpha>0$ will result in $c_s^2<0$, which leads
to a non-zero Jeans length $\lambda_J=\sqrt{\pi|c_s^2|/G\rho}$. In
this regime the fluctuations will oscillate instead of grow
polynomially. On the other hand, when $\alpha<0$, perturbations will
grow exponentially, which is also ruled out.

In \NCGasSandvik, Sandvik {\it et al.} further confirmed the above
arguments by numerical solving the equations (see \FigCG). By performing
a $\chi^2$ fit of the theoretically predicted power spectrum against
that observed by the 2dFGRS \BAOColless, they found that this
inconsistency excludes most of the previously allowed parameter
space of $\alpha$, leaving essentially only the standard
$\Lambda$CDM limit. This topic was also studied in
\refs{\NCGasUDM,\NCGasPark}. In \NCGasPark, based on an analysis including
the SNIa, the baryonic matter power spectrum, the CMB and the
perturbation growth factor, Park {\it et al.} found that the allowed
region for $\alpha$ is
\eqn\NCGasalpha{-5\times10^{-5}\leq\alpha\leq10^{-4}.} The
allowed parameter space is extremely close to the $\Lambda$CDM
model, so the possibility of the unification of dark matter and dark
energy in the GCG scenario has been ruled out by the
current cosmological observations.

\subsec{Holographic dark energy models}

In the following, we will introduce some numerical works about the holographic dark energy (HDE) model, which arises from the holographic principle.

\noindent $\bullet$ The HDE model with the future event horizon as the cutoff

The HDE has the form of energy density
\eqn\HDERho{\rho_{de}=3c^2M^2_pL^{-2},}
where $c$ is a number introduced in \LiRB.
In 2004, Li \LiRB\ proposed to take $L=1/R_h$,
where $R_h$ is the future event horizon defined as
\eqn\HDERh{R_h=a\int^\infty_a{da\over Ha^2}.}
That yields an EOS
\eqn\HDEEOS{w=-{1\over3}-{2\over3}{\sqrt{\Omega_{de}}\over c},}
which satisfies $w<-{1/3}$ and can accelerate the cosmic expansion.

In \HDEHuangSNIa, Huang and Gong first performed a numerical study on the HDE model.
Making use of the 157 gold SNIa data,
they obtained $\Omega_m=0.25^{+0.04}_{-0.03},\ \ w=-0.91\pm0.01$ at 1$\sigma$ CL for the $c=1$ case.
In \HDEZhangX, Zhang and Wu constrained the HDE model by performing a joint analysis of SNIa+CMB+LSS.
They found that the best fit results are $c=0.81$, $\Omega_m^0=0.28$, and $h=0.65$,
which implies that the HDE behaves as a quintom-type dark energy.
In \HDEZhangTJ, Yi and Zhang tested the HDE model by using the $H(z)$ data and the LT data.
They also verified that the HDE behaves like a quintom-type at 1$\sigma$ CL.
In addition, a lot of numerical studies were performed to test and constrain the HDE model \HDEWorks.
These works showed that the HDE model can provide a good fit to the
data. For example, in \HDECompareHDEModels, by using the combined Constitution+BAO+CMB data,
Li {\it et al.} obtained the following $\chi_{\rm min}^2$s for the $\Lambda$CDM and HDE models
\eqn\HDEChiSqu{\chi^2_{\rm \Lambda CDM}=467.775,\ \ \ \ \chi^2_{\rm HDE}=465.912.}
So the HDE model is consistent with the current observations.
Similar results have been obtained in e.g.
\refs{\HDEMaGongChen,\HDEYZMa,\HDEGong,\DEManyModelsMLi}. Therefore,
from the perspective of current observations, HDE is a competitive model.

\ifig\FigHDE{{\it Left Panel}: Constraints on the HDE model in
$\Omega_m$-$c$ plane from observational data including Constitution
SNIa sample \Constitution, the BAO data from SDSS \BAOEisenstein,
and the CMB data from WMAP5 measurements \WMAP. The data favor
$c<1$. For the constraint on $c$, the analysis gives
$0.818^{+0.196}_{-0.154}$ at the 68.3\% CL, and
$0.818^{+0.196}_{-0.097}$ at the 95.4\% CL. From
\HDECompareHDEModels. {\it Right Panel}: The contour maps of
$\alpha$ vs. $c$ for interacting HDE (IHDE) with 68\%, 95.5\% and
99.7\% CL, obtained from a joint analysis including the
Golden06+BAO+X-ray+GRB+CMB data. The black dot-dashed curve denotes
$w_{eff}=-1$ when $z\rightarrow -1$ with $\Omega_{de}=0.73$, and the
region below (over) it means $w_{eff}$ will (not) cross -1 during
infinite time. From \HDEYZMa} {\epsfysize=2.4in \epsfbox{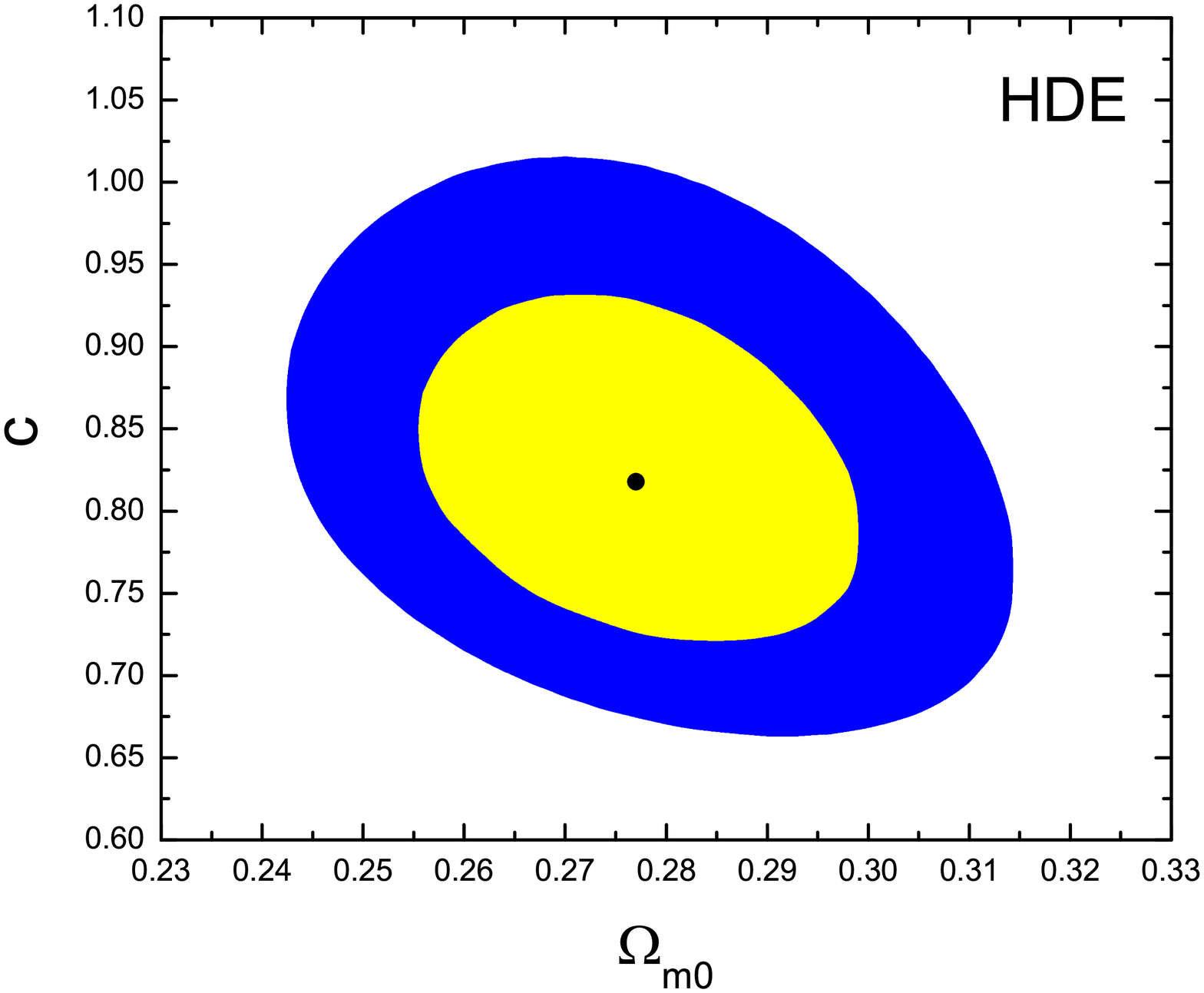} \epsfysize=2.6in
\epsfbox{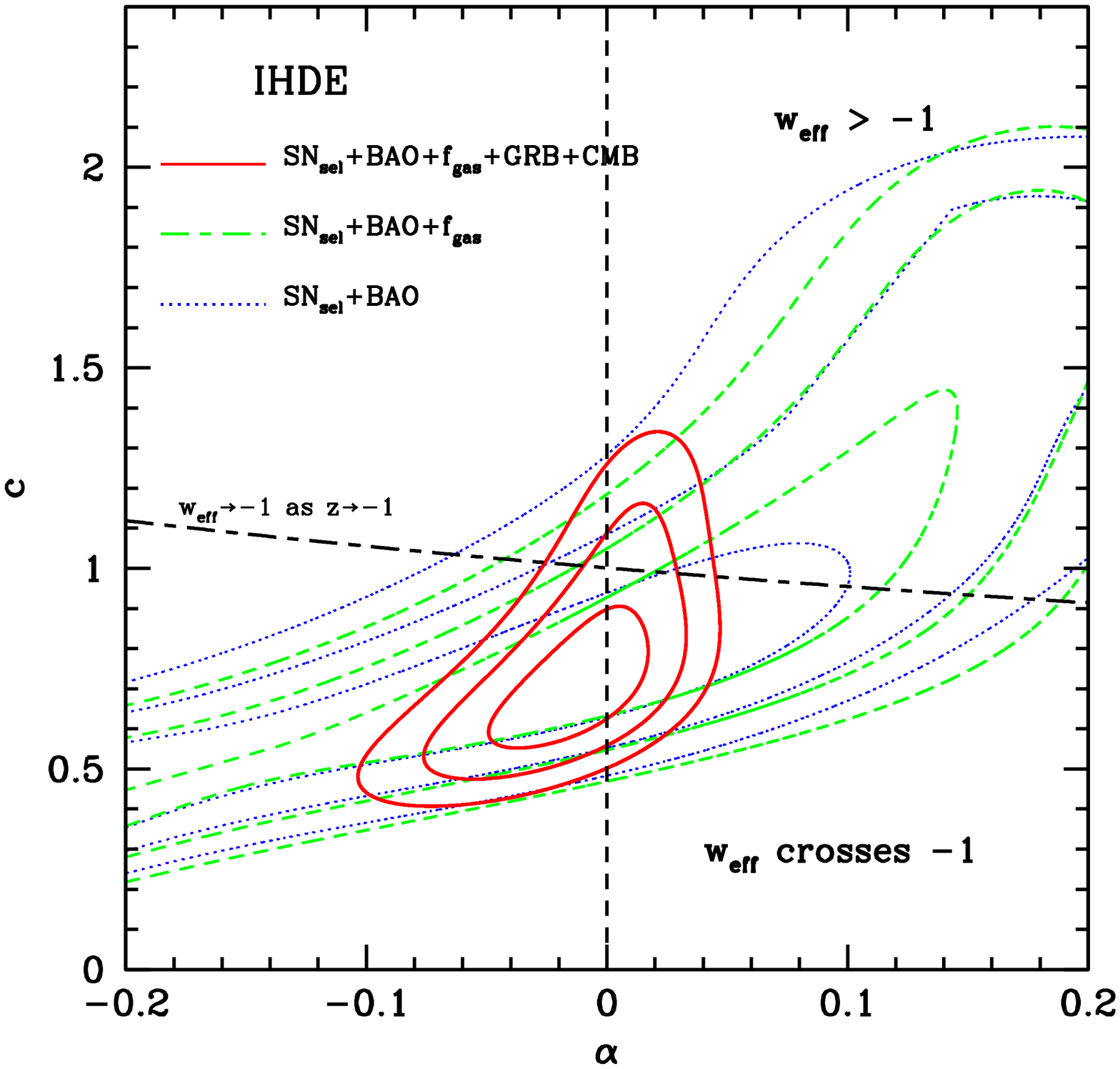}}

The parameter $c$ plays an essential role in determining the
evolution of the HDE. If $c=1$, the dark energy EOS will be asymptotic to that
of a cosmological constant and the universe will enter the de
Sitter phase in the future; if $c>1$, the EOS is always
greater than -1, and HDE  behaves as quintessence dark energy; if $c<1$,
initially the EOS of HDE is greater than -1, then it will
decrease and eventually cross the $w=-1$ line, leading to a phantom
universe with big rip as its ultimate fate. The numerical studies on
HDE showed that the cosmological observations favors $c<1$.
For example, Ref. \HDECompareHDEModels\ gave $c=0.818^{+0.113}_{-0.097}$  at the 68\% CL (see the left panel of \FigHDE),
while some later works \refs{\DEManyModelsHWei,\DEManyModelsMLi} with the improved data tighten the constraint to $c<0.9$ at the 95\% CL.

\noindent $\bullet$ Interactions between HDE and dark matter

In \refs{\WangJX, \HDEIntWangA}, Wang {\it et al.} first studied the interaction between dark matter and the HDE.
The introduction of interactions not only alleviates the cosmic coincidence problem
\foot{coincidence problem may also be solved by the anthropic constraints \Barrave},
but also avoids the future big-rip singularity \refs{\HzBinWang,\CuiZhang,\HDEYZMa,\HDEInt,\HDEIntXLC,\HDEIntFeng,\HDECurvInt}.
Phenomenologically, the interaction can be introduced by
\eqn\HDEInt{\dot\rho_m+3H\rho_m=Q,\ \ \
\dot\rho_{de}+3H(\rho_{de}+p_{de})=-Q.} where $Q$ is the interaction
term. Current observations have put tight constraints on the
interactions. In \HDEYZMa, with the assumption of a flat universe,
Ma considered the following $Q$ and obtained the corresponding constraint
\eqn\HDEInt{Q=3\alpha H\rho_{de},\ \ \
\alpha=-0.006^{+0.021}_{-0.024},} from a joint analysis (see the
right panel of \FigHDE). In \HDEIntFeng, Feng {\it et al.} considered
another class of $Q$ and obtained the corresponding constraint
\eqn\HDEInt{Q=3bH(\rho_m+\rho_{de}),\ \ \
b=-0.003^{+0.012}_{-0.013},} from a combination of
SNIa+BAO+CMB+Lookbacktime data. Later, in \HDECurvInt, Li {\it et al.}
revisted these two models and obtained
\eqn\HDEIntalphab{\alpha=(-6.1\times10^{-5})^{+0.025}_{-0.036},\ \ \
\ b=(-1.6\times10^{-4})^{+0.008}_{-0.009}.} from a joint analysis of
Constitution+BAO+CMB. They also found that there exists significant degeneracy
between the phenomenological interaction and the spatial curvature in the HDE model.

\noindent $\bullet$ The ADE and RDE models

In addition to the HDE model with future event horizon as the cutoff,
the Agegraphic dark energy (ADE) model \refs{\CaiUS,\WeiTY,\ADEPaperThree} and the Holographic Ricci dark energy (RDE) model \RDEPaper\
are also motivated by the holographic principle (the ADE model can
also be obtained from the K\'{a}rolyh\'{a}zy relation; see
\refs{\CaiUS\WeiTY} for details).
In these two models, the IR cutoff length scale is given by
the conformal time $\eta$ and the average radius of the Ricci scalar curvature $|{\cal R}|^{-1/2}$, respectively.
There have been some numerical studies on these two models
\refs{\HDECompareHDEModels,\RDEXu,\RDEZhang,\RDEFeng,\ADEWeiCai,\ADEKyoung}.
In general, these studies showed that the ADE and RDE models are not
favored by current observations. For example, in \HDECompareHDEModels, Li {\it et al.} obtained
\eqn\ADERDEChiSqu{\chi^2_{\rm ADE}=481.694,\ \ \chi^2_{\rm RDE}=483.130.}
The $\chi^2_{\rm min}$s of the ADE and RDE model are much larger than
that of the $\Lambda$CDM and HDE model listed in Eq. \HDEChiSqu,
showing that these two models are not favored by observations. The
results have been further confirmed in some later works \refs{\DEManyModelsHWei,\DEManyModelsMLi}.

\subsec{Dvali-Gabadadze-Porrati model}

The Dvali-Gabadadze-Porrati (DGP) braneworld model is a
theory where gravity is altered at immense distances by the slow
leakage of gravity off our three-dimensional universe \DGPDGP. In this
model, the Firedmann equation is modified as
\eqn\DGP{H^2-{H\over r_c}={8\pi G\over 3}\rho_m}
where $r_c=(H_0(1-\Omega_m))^{-1}$ is the length scale beyond which
gravity leaks out into the bulk. At early times, $Hr_c\gg1$, the
Firemann equation of general relativity is recovered. In the future,
$H\rightarrow1/r_c$, the expansion is asympototically de Sitter.

There is also a generalized phenomenological DGP model characterized by the Friedmann
equation \DGPGenDGP
\eqn\GenDGP{H^2-{H^\alpha \over r_c^{2-\alpha}}={8\pi G\over
3}\rho_m}
where $r_c=H_0^{-1}/(1-\Omega_m)^{\alpha-2}$. This model
interpolates between the pure $\Lambda$CDM model and the DGP model
with an additional parameter $\alpha$. $\alpha=1$ corresponds to the
DGP model and $\alpha=0$ corresponds to the $\Lambda$CDM model.

\ifig\FigDGP{{\it Left Panel}: Combined SNIa+BAO+CMB observational
constraint shows that DGP does not achieve an acceptable fit. The
areas of intersection of any pair are distinct from other pairs.
This is a strong signal that the DGP model is incompatible with the
observations. From \DEManyModelsDRubin. {\it Right Panel}:
Predictions for the power spectra of the CMB temperature
anisotropies $C_l^{\rm TT}$ of the best-fit DGP model (solid), a
quintessence model with the same expansion history as DGP
(short-dashed), and the $\Lambda$CDM model (dashed, conincident with
the quintessence model at low $l$). Obtained by fitting to SNLS+
WMAP5+HST assuming a flat universe. Bands represent the 68\% and
95\% cosmic variance regions for the DGP model. Points represent
WMAP5 measurements. It is clear that the best-fit DGP model over
predicts the low-$l$ modes anisotropy. From \DGPHorScalGrowGeo.}
{\epsfysize=2.4in \epsfbox{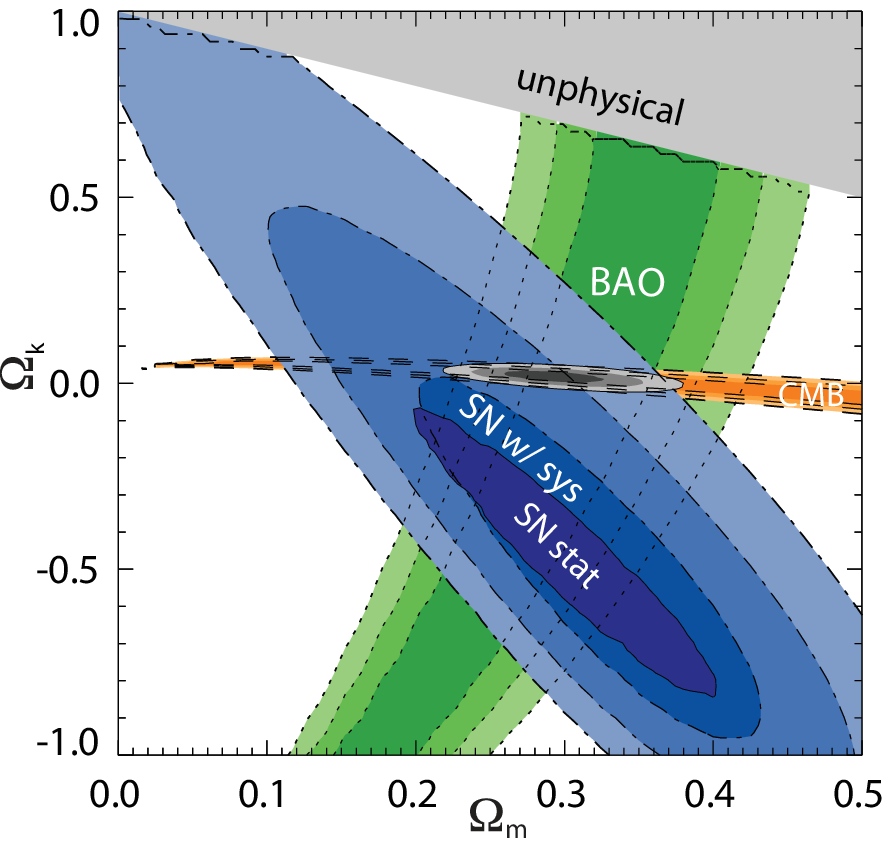} \epsfysize=2.5in
\epsfbox{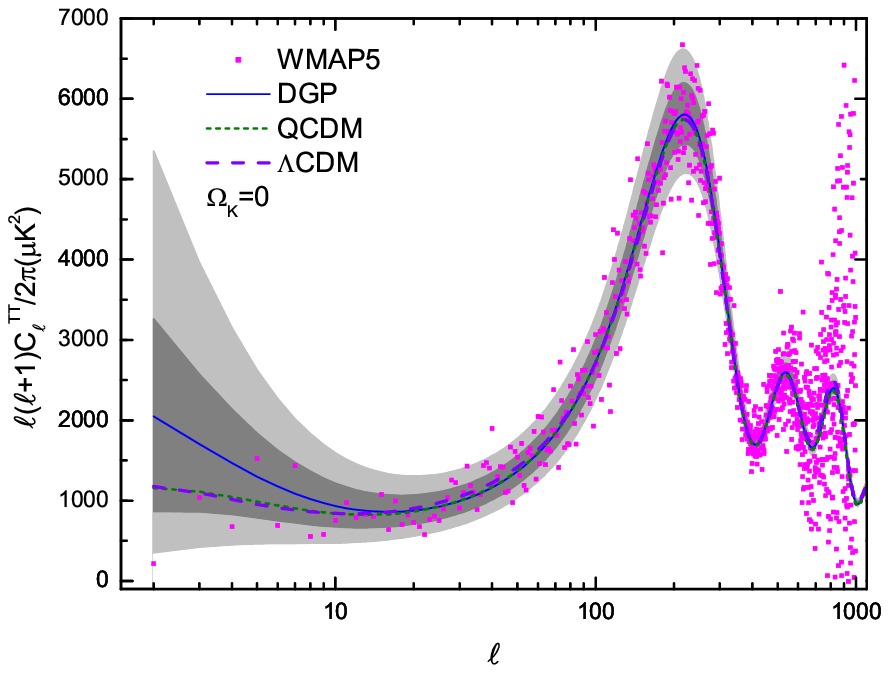}}

\noindent $\bullet$ The DGP model is disfavored by the observations

Although DGP is an attractive model allowing a self acceleration,
many research works show that it is disfavored by observations
\refs{\DEManyModelsDRubin,\DGPHorScalGrowGeo,\DGPGuo,\DGPFairbairn,\DGPmaartens,\DGPAlam}.
For examples, in \DEManyModelsDRubin,
from a joint analysis of SNIa+BAO+CMB,
Rubin {\it et al.} found that the DGP was disfavored by the data,
with a $\Delta \chi^2$=15 compared with the $\Lambda$CDM model (see the left panel of \FigDGP).
In \DGPGuo, from a combination of SNIa and BAO measurments, Guo {\it et al.} provide the constraints to the model parameters
\eqn\DGPEqGuo{\Omega_m=0.27^{+0.018}_{-0.017},\ \Omega_{r_c}=0.216^{+0.012}_{-0.013},\ \Omega_k=-0.350^{+0.080}_{-0.083},}
at 99.73\% CL.
This result is in contradiction to the WMAP results indicating a flat universe.
Moreover, the constraints to the generalized DGP model gives a small
$\alpha$, indicating that the DGP is incompatible with the
observations. In a recent work \DGPJQXia, Xia performed a joint
analysis including SNIa, BAO, CMB, GRB and the linear growth factor
of matter perturbations, and found a constraint
\eqn\DGPXia{\alpha=0.254\pm0.153,} at the 68\% CL, manifesting
that this model tends to collapse to the cosmological constant when
confronted with current observations.

\noindent $\bullet$ Testing the DGP model from the growth of structure

As a modified gravity scenario, the growth of structure in the DGP gravity differs from that in the $\Lambda$CDM scenario.
This can be used to test the DGP model \refs{\GrowFacLuScSt,\DGPReview}.
The perturbation theory in the DGP model has been studied
\refs{\DGPStru,\DGPLSTests,\DGPCosPert,\DGPPPF}.
These studies showed that the DGP gravity is disfavored by the observational data.
For examples, in \DGPLSTests, Song {\it et al.} showed that the constraints
from SNIa+CMB+$H_0$ exclude the simplest flat DGP model at about
3$\sigma$. Even including spatial curvature, best-fit open DGP model
is a marginally poorer fit to the data than the flat $\Lambda$CDM
model. In \DGPHorScalGrowGeo, Fang {\it et al.} showed that the
DGP model is excluded at 4.9$\sigma$ and 5.8$\sigma$
levels with and without curvature respectively (see the right panel
of \FigDGP). The corresponding $\chi^2_{\rm min}$s for the DGP and
$\Lambda$CDM model are
\eqn\DGPChiSqr{\chi^2_{\rm \Lambda CDM}=2777.8,\ \ \ \ \chi^2_{\rm DGP}=2805.6.}
The result is mainly due to the earlier beginning of the
acceleration and the additional suppression of growth in the DGP
scenario. In \DGPSelfConsis, by performing cosmological $N$-body
simulations of the DGP model, Schmidt found that, independently of
CMB constraints, the self-accelerating DGP model is strongly
constrained by WL and cluster abundance measurements. Compared with
the $\Lambda$CDM model, the abundance of halos above
$10^{14}M_{\odot}$ is suppressed by more than a half in the DGP
model. In all, when confronted with experiments, a lot of problems
will emerge in the DGP model, indicating that this theory is strongly disfavored by the observations

\subsec{$f(R)$ models}

$f(R)$ gravity is a simplest modification to the general relativity with the replacement $R\rightarrow f(R)$.
In this section, we will focus on the observational tests of this theory.
One can see \refs{\DEREviewTsujikawa,\SotiriouRP,\fRReviewA,\fRReviewB,\fRReviewC,\fRReviewD} and references therein for more details.

\noindent $\bullet$ $f(R)$ gravity and its viable conditions

At the beginning, it was proposed to explain the cosmic acceleration using the model with $f(R)=R-\alpha/R^n (\alpha>0,n>0)$
\refs{\fRTheoryCapozzielloA,\fRTheoryCapozzielloB,\fRTheoryNojiriA}.
However, later on, a lot of problems emerged in this model.
In \fRMatInst, it was shown that this model will lead to the matter instability.
In \fRSolarTest, it was also found that this model is unable to satisfy local gravity constraints and pass the solar system tests of gravity.
Much attentions have been paid to the analysis of the viable conditions of $f(R)$ models,
and a lot of valuable results have been obtained \refs{\fRCognola,\fRDynamical,\fRDynamicalHu,\fRMullerA,\fRAmendolaC,\fRStarobinB}.
For examples, to have a stable perturbation, the condition $f_{,RR}\equiv\partial^2f/\partial R^2<0$ is required;
to have a stable late-time de Sitter point, the condition $0<Rf_{,RR}/f_{,R}<1$ is also necessary.
In summary, the conditions for the viability of $f(R)$ dark energy
models include \refs{ \DEREviewTsujikawa,\fRAmendolaC,\fRStarobinB}:

\eqn\fRViableConOne{\eqalign{(1)\ f_{,R}>0\ {\rm and}\ f_{,RR}>0\
{\rm for}\ R\leq R_0;\ \ \ \ \ \ \ \ \ \ \ \ \ \ \ \ \ \ \ \ \ \ \ \
\ \ \ \ \ \ \ \ \ \ \ \ \ \ \ \cr (2)\ f(R)\rightarrow R-2\Lambda\
{\rm for}\ R{\ll}R_0;\ \ \ \ \ \ \ \ \ \ \ \ \ \ \ \ \ \ \ \ \ \ \ \
\ \ \ \ \ \ \ \ \ \ \ \ \ \ \  \ \ \ \ \ \ \ \  \cr (3)\
0<Rf_{,RR}/f_{,R}<1\ {\rm at\ the\ de\ Sitter\ point\ satisfying}\
Rf_{,R}=2f.}}

To be acceptable, an $f(R)$ model must satisfy the following
conditions. Some viable models satisfying all these requirements
have been proposed, and one can refer to e.g.
\refs{\DEREviewTsujikawa,\fRReviewB,\fRAmendolaC,\fRStarobinB,\fRAmendolaA,\frNojiriOdintsov,\fRNojiriOdintsov}
for more details about the viable conditions of $f(R)$ gravity.

\noindent $\bullet$ Cosmological tests of the $f(R)$ gravity

\ifig\FigfR{{\it Left panel}: Linear matter power spectrum for
several values of $B_0$ in the $\Lambda$CDM expansion history in the
model \refs{\fRSongA,\fRSongB}. $\sqrt{B}$ is of order the Compton
wavelength of the new gravitational degree of freedom, measured in
the unit of Hubble length. For length scale larger than
$\sqrt{B}/H$, the modified gravity effect is suppressed because of
the mass of the scalar gravitaton. For length scales smaller than
$\sqrt{B}/H$, gravity is modified. The modification of gravity leads
to an enhancement in the growth of perturbations and a corresponding
suppression for the decay of the gravitational potential. From
\fRSongB. {\it Right panel}: Contours of 2D marginalized 68\%, 95\%
and 99\% confidence boundaries using observational data from the
measurements of CMB, BAO, Hubble constant, and so on. For the
constraints from galaxy-ISW (gISW) cross correlation, the Compton
wavelength is rescaled as $B_0\rightarrow10^{-2}B_0$. From
\fRLombriser. } {\epsfysize=2in \epsfbox{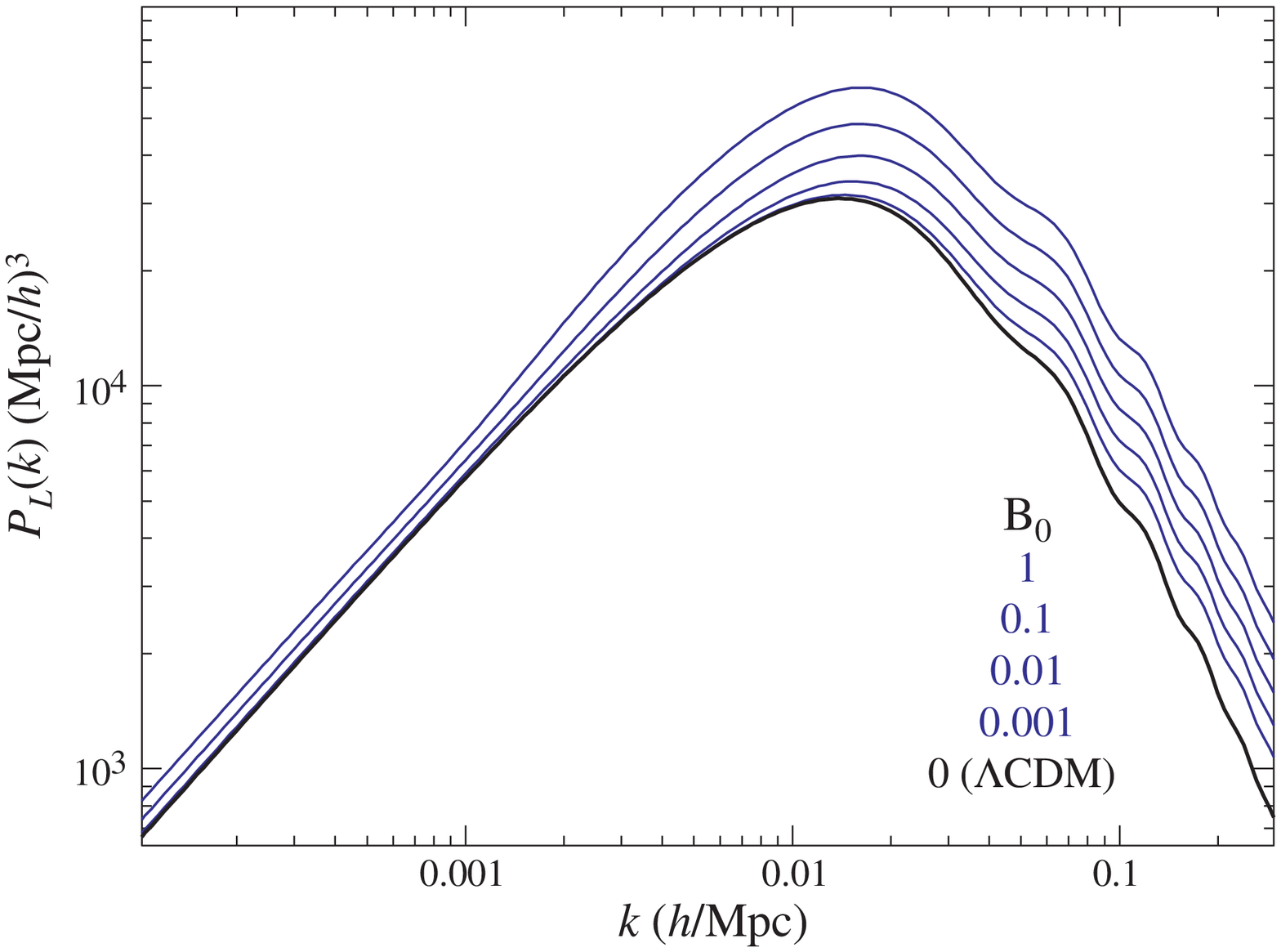}
\epsfysize=2.1in \epsfbox{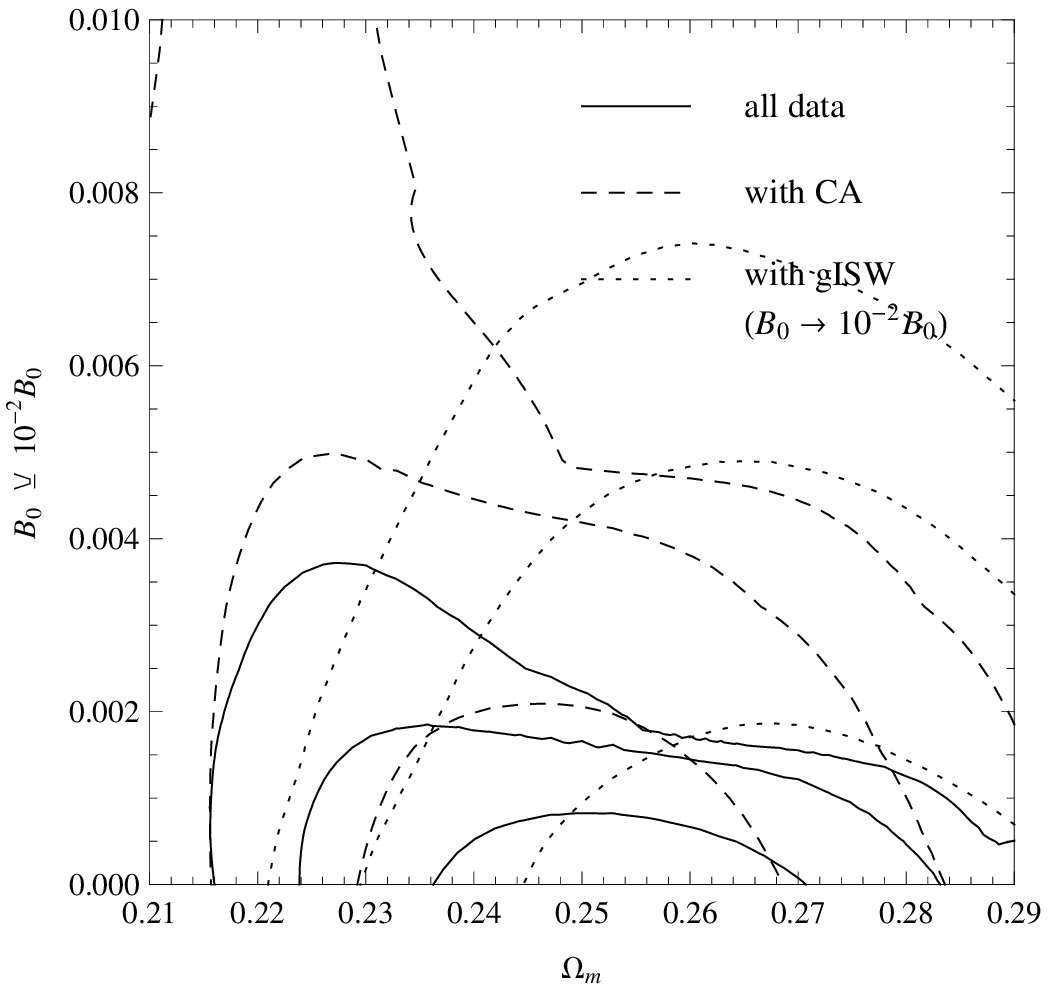} }

We have listed some general conditions required for an
$f(R)$ model to be valid. An $f(R)$ model satisfying these basic
requirements, furthermore, should be confronted with the
cosmological observations. A natural method is to test the $f(R)$
theories from the observations about the growth of structure, which
depends on the theory of gravity. This issue attracted a lot of
interests \refs{\fRDynamicalHu,\fRSongA,\fRSongB,\fRCosTest}.
Some observational signatures of the $f(R)$ models have been
presented. For example, in \refs{\fRSongA,\fRSongB}, Song {\it et al.} studied an
$f(R)$ model parameterized  by ``Compton wavelength parameter'' $B$,
which is proportional to the second order derivative $f_{,RR}$
\eqn\fRB{B={f_{,RR}\over f_{,R}}{H dR/d\ln a\over dH/d\ln a}.}
The $B<0$ branch violates the third condition of Eq.
\fRViableConOne. For the stable $B>0$ branch, it was found that this model will
predict a lower large-angle CMB anisotropy by reducing the ISW effect,
qualitatively change the correlations between the CMB and galaxy
surveys, and alter the shape of the linear matter power spectrum
(see the left panel of \FigfR\ for details).

To test of the modified gravity theory, it is worthwhile to include
various observational techniques probing gravity in different
scales \TESTGRAVCosmoTestofGrav. For example, in \fRSchmidt,
Schmidt {\it et al.} conduct the first, simulation calibrated, cluster
abundance constraints on a two-parameter modified action model
\eqn\fRSchmidt{f(R)=R-2a{R\over R+\mu^2}.}
They found that the local cluster abundance, when combined with the
data from CMB, SNIa, $H_0$ and BAO, can lead to a very tight
constraint to the model, improving the previous constraints by 3-4
orders of magnitude. The reason is that, the inclusion of cluster
abundance data improves the bounds on the range of force
modification from the several Gpc scale to tens of Mpc scale. In
\fRLombriser, Lombriser {\it et al.} revisited the model studied in \refs{\fRSongA,\fRSongB} and reported a strong constraint to
the current value of the Compton wavelength
\eqn\fRB{B_0<1.1\times10^{-3},}
at the 2$\sigma$ CL (see the right panel of \FigfR), mainly due to
the inclusion of data from cluster abundance.

Some other observational techniques have also been used to study the $f(R)$ models,
auch as the cosmic shear experiments \refs{\fRWL,\fRCamera},
the ``21 cm intensity mapping'' (detection of LSS in three dimensions without the detection of individual galaxies) \refs{\BAOTwentyOne,\fRMasui},
the variation of the fine structure ``constant'' \refs{\fRWebb,\fRfsc},
and so on.
Nowadays, the observational tests of $f(R)$ gravity has drawn increasingly attention.
For more details on the test of modify gravity theories from observations,
see \refs{\DEReviewUzan,\TESTGRAVCosmoTestofGrav,\TESTGRAV} and the references therein.

\noindent $\bullet$ The curvature singularity problem in $f(R)$ gravity

In \refs{\fRCurvSinProAba,\fRCurvSinProBris}, It was proposed that there is curvature singularity problem in $f(R)$ theories,
due to the dynamics of the effective scalar degree of freedom in the strong gravity regime.
This problem leads to the contradiction with the existence of the
relativistic stars (like neutron stars) \fRCurvSinProB.  Furthermore, in \fRCurSinProCure, it was shown that
this problem can be cured via the addition of $R^2$ term. One can
see Ref. \fRCurvPro\ for some studies on this topic.

\noindent $\bullet$ $f(R)$ theories with Palatini approach

The observational constraints on $f(R)$ theories within Palatini
approach was firstly performed by Amarzguioui \fRAmarzguioui {\it et al.}, where
they investigated the parameterization with the form of $f(R)=R+\alpha
R^\beta$ using the cosmological measurements SNIa+BAO+CMB. The best
fit values is found to be $\beta=0.09$ with the allowed values rage
\eqn\fRBeta{|\beta|<0.2,}
at 1$\sigma$ CL, and the previously commonly considered $1/R$
model corresponding to $\beta=-1$ is ruled out by the constraint. In
\fRTKoivisto, using the same form of parameterization, Koivisto
calculate the power spectrum in the Palatini formulation of $f(R)$
gravity. By comparing the results to the SDSS data \BAOSDSSTegmark,
it was found that the observational constraints reduce the allowed
parameter space to
\eqn\fRbetaB{|\beta|<\sim 10^{-4},}
which is a tiny region around the $\Lambda$CDM. Besides, the
Palatini $f(R)$ gravity is also faced with some problems associated
with non-dynamical nature of the scalar-field degree of freedom. One
can refer to \refs{\fRPalatini,\fRPalatinitwo,\fRPalatinithree} for
more studies about the Palatini $f(R)$ gravity.

\subsec{Other modified gravity models}

We have introduced the observational aspects of the DGP scenario and the $f(R)$ models.
In this section, we will discuss some other modified gravity theories,
including the Gauss-Bonnet gravity, the Brans-Dicke gravity and the $f(T)$ gravity.

\noindent $\bullet$ Gauss-Bonnet gravity

When compared with observations, many problems arise in the Gauss-Bonnet model
\GBTerm. In \refs{\GBKovistoA,\GBKovistoB}, from the data analysis including
the constraints from BBN, LSS, BAO and solar system data, Koivisto
and Mota found that this model is strongly disfavored by the observations. In \GBLi\ and \GBFelice, Li {\it et al.} and De Felice {\it et al.} investigated
the Gauss-Bonnet gravity, and found that the growth of perturbations gets
stronger on smaller scales. This is incompatible with the observed
galaxy spectrum, unless the deviation from the Einstein gravity is very small. Thus,
the Gauss-Bonnet models have been effectively ruled out.

\noindent $\bullet$ Brans-Dicke theory

Another well-known modified gravity model is the Brans-Dicke theory
\BransSX. For the Brans-Dicke theory, the PPN parameter $\gamma$
(see Eq. \PPNForm) takes the form
\eqn\BDGamma{\gamma=(1+\omega)/(2+\omega),}
where $\omega$ is the constant in Eq. \MGBransDicke.
From solar system and binary pulsar observations, the parameter $\omega$ has been tightly constrained to \TESTGRAVConfGRExp
\eqn\BD{\omega \geq 4\times10^4,}
and the cosmological effects of the scalar field are rendered insignificant.
Later on, the Brans-Dicke theory is generalized to the scalar tensor theory \AmendolaQQ.
One can see \refs{\DEREviewTsujikawa,\TESTGRAVCosmoTestofGrav,\fRReviewB,\BDWuFQ,\BDOther} and the references therein for more details on the test of this model.

\noindent $\bullet$ $f(T)$ theory

There have been some numerical studies on the recently
proposed $f(T)$ gravity \refs{\BengocheaGZ,\LinderPY} as an explanation of the cosmic
acceleration. In \fTWu, Wu and Yu examined the following models from the Union2 SNIa dataset together with the BAO and CMB data
\eqn\fT{f(T)=\alpha(-T)^n,\ \ \ f(T)=-\alpha T(1-e^{p T_0/T}).}
and obtained the constraint $n=0.04^{+0.22}_{-0.33},\ \ p=-0.02^{+0.31}_{-0.20},$
at the 95\% CL. They also compared the two models with the
$\Lambda$CDM by using the $\chi^2_{\rm min}/dof$ (dof: degree of freedom)
criterion. The results showed that $\Lambda$CDM is mildly favored by
the data. Later, the power-law model was revisited by Bengochea
\fTBengochea, with the inclusion of GRB and $H_0$ data into consideration.
In \fTWei, Wei, Ma, and Qi tried to constrain $f(T)$ theory by using the varying fine structure ``constant'';
it is found that the observational $\Delta\alpha/\alpha$ data make $f(T)$ theory almost indistinguishable from the $\Lambda$CDM model.
In addition, in \fTWeitwo, they also constrained $f(T)$ theory by using the varying gravitational ``constant'';
it is found that the allowed model parameter $n$ has been significantly shrunk to a very narrow range around 0.

In \fTBamba, Bamba {\it et al.} studied the cosmological evolutions of the EoS for dark energy $w_{de}$
in the exponential and logarithmic as well as their combination $f(T)$ theories.
They found that the crossing of the phantom divide line of $w_{de} = -1$ can be realized in the combined $f(T)$ theory
even though it cannot be in the exponential or logarithmic $f(T)$ theory.
Moreover, the crossing is from $w_{de} > -1$ to $w_{de} < -1$, which is favored by the recent observational data.

The perturbations in $f(T)$ gravity has been studied in
\refs{\fTChen,\fTDent,\fTZheng}. In \fTZheng, Zheng and Huang
derived the evolution equation of growth factor for matter
over-dense perturbation in $f(T)$ gravity. In addition, a problem in
$f(T)$ gravity was pointed out by Sotiriou {\it et al.} in
\fTSotiriou, where they showed that the Lorentz symmetry can not be
restored in $f(T)$ theories due to sensible dynamics.

\subsec{Inhomogeneous LTB and backreaction models }

The inhomogeneous models have gathered significant interest in
recent years as a scenario to explain the cosmological observations
without invoking dark energy . In this section, we
will briefly discuss the observational signature of the LTB model
and the backreaction model.

\noindent $\bullet$ LTB models

The LTB models \ChuangYI\LTBModel\ are the most commonly
considered inhomogeneous scenario to explain the cosmic acceleration
without introducing the dark energy component. For LTB models the structure
of our universe is described by the inhomogeneous isotropic
Lema\^{\i}tre-Tolman-Bondi (LTB) metric \refs{\LTBL,\LTBT,\LTBB}
\eqn\LTBmetric{ds^2=-dt^2+{R^\prime(r,t)^2\over
1+\beta(r)}dr^2+R^2(r,t)(d\theta^2+\sin^2\theta d\phi^2),} where the
prime denotes partial differentiation with respect to $r$.
$\beta(r)$ is a function of $r$. Notice that the FRW metric is
recovered by requiring $R=a(t)r$ and $\beta=-kr^2$. The Hubble
parameters at the transverse and radial direction are expressed as
\eqn\LTBH{H_{\bot}={\dot R^\prime\over R^\prime},\ \ H_{\|}={\dot
R\over R},} and the apparent cosmic acceleration can be explained by
choosing suitable form of $R(r,t)$, without the introduction of dark energy
\LTBTheory. In the simplest class of such models we live close to
the center of a huge, spherically symmetric Gpc scale void. Due to
the spatial gradients in the metric, our local region has a larger
Hubble parameter than the outer region
\LTBTom. In some more complicated scheme,
it have been proposed to reconstruct the cosmological
constant in an inhomogeneous universe
\LTBReCon. The idea is
that, since cosmological observations are limited on the light cone,
it is possible to reconstruct an inhomogeneous cosmological model
(indistinguishable from the homogeneous $\Lambda$CDM model) to
explain the cosmic acceleration without a cosmological constant. So
far there have been some studies on the observational tests of the
LTB model. In the following, we will briefly review some related works.

It has been shown that the void model is able to provide a good fit
to the SNIa data. For example, in \DEManyModelsSollerman, Sollerman
{\it et al.} tested two kinds of LTB models and compared them with the
$\Lambda$CDM model. The models they considered has the following
matter distribution
\eqn\LTBMassDisA{\Omega_m(r)=\Omega_{\rm out}+(\Omega_{\rm
in}-\Omega_{\rm out})e^{-(r/r_0)^2},}
\eqn\LTBMassDisB{\Omega_m(r)=\Omega_{\rm out}+(\Omega_{\rm
in}-\Omega_{\rm out})\left(1+e^{-r_0/\Delta
r}\over1+e^{(r-r_0)/\Delta r}\right),}
where $\Omega_{\rm in}$ is the matter density at the center of the
void, $\Omega_{\rm out}$ is the asymptotic value of the matter
density outside the void, and $r_0$ is the size the underdensity.
The second model \LTBBellidoB\ has a much sharper transition of
matter density than the first one, with the extra parameter
$\Delta_r$ characterizes the transition width. Using the first-year
SDSS-II SNIa data \SDSSSN\ together with the BAO and CMB
measurements, they found that $\chi^2$ values for the LTB fits are
comparable to that of the $\Lambda$CDM model
\eqn\LTBChi{\chi^2_{\rm \Lambda CDM}=233.2,\ \ \ \chi^2_{\rm LTB\ model1}=235.5,\ \ \ \chi^2_{\rm LTB\ model2}=237.6.}
while the extra parameters in the LTB models make the models fare
poorly in the information criteria tests.

\ifig\FigLTBBlueShift{{\it Left panel}: An off-center galaxy cluster
in a void will observe a dipole in the CMB. Since the expansion rate
inside the void is higher, photons arriving through the void (from
the right in the figure) will have a larger redshift ($\Delta
z_{in}$) than photons did not pass through the void (left, with
redshift $\Delta z_{out}$). From \LTBkSZ. {\it Right panel}: The
changes in the $\chi^2$ values as a function of the observer¡¯s
position. The SDSS-II data \SDSSSN\ combined with the CMB dipole
requirement is used. The $\chi^2$ value quickly increases as the
observer is displaced away from the center. See \LTBBlomqvist\ for
details about the meaning of the diamonds, stars, lines and arrows.
From \LTBBlomqvist.} {\epsfysize=2.1in \epsfbox{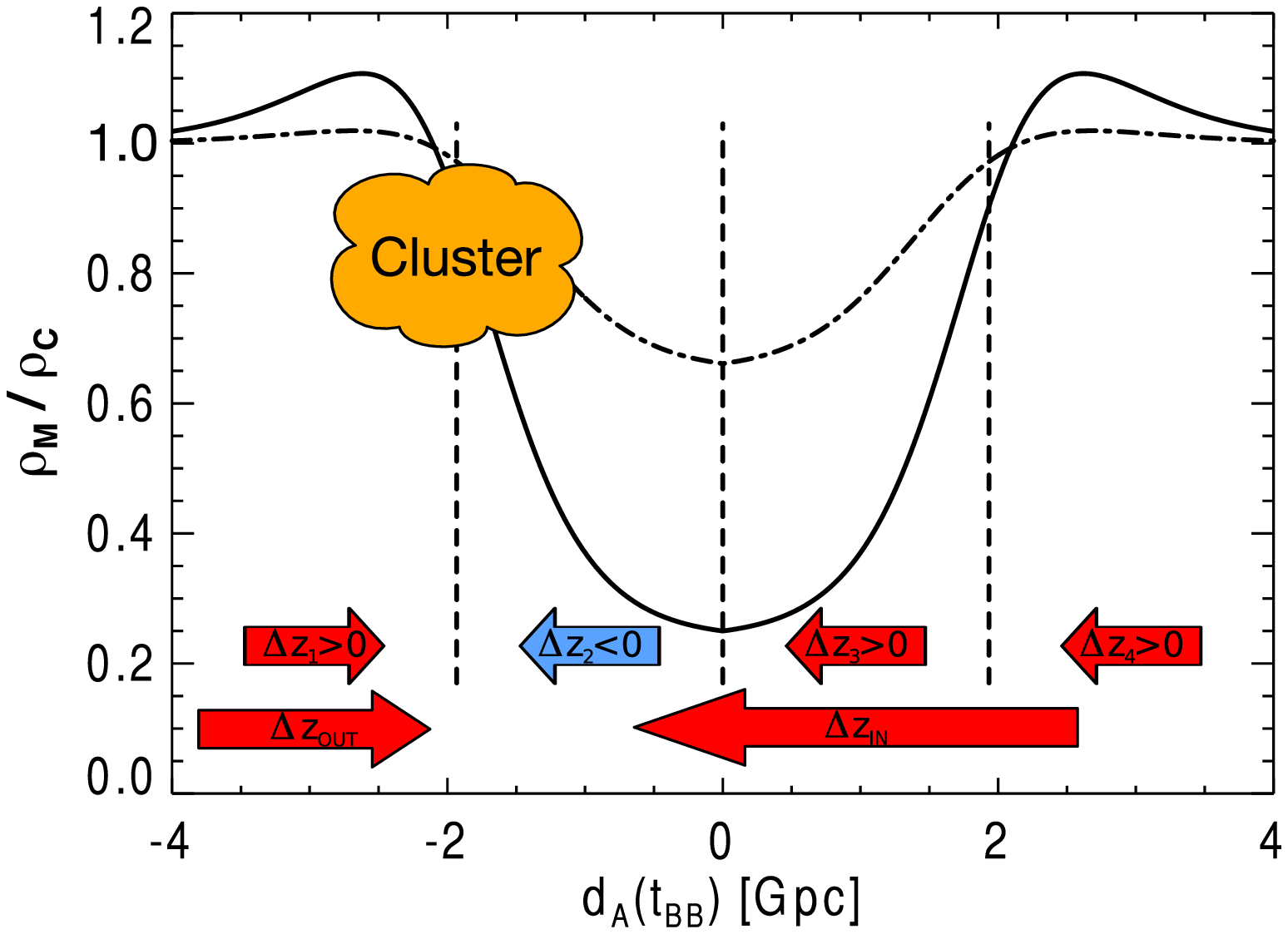}
\epsfysize=2.4in \epsfbox{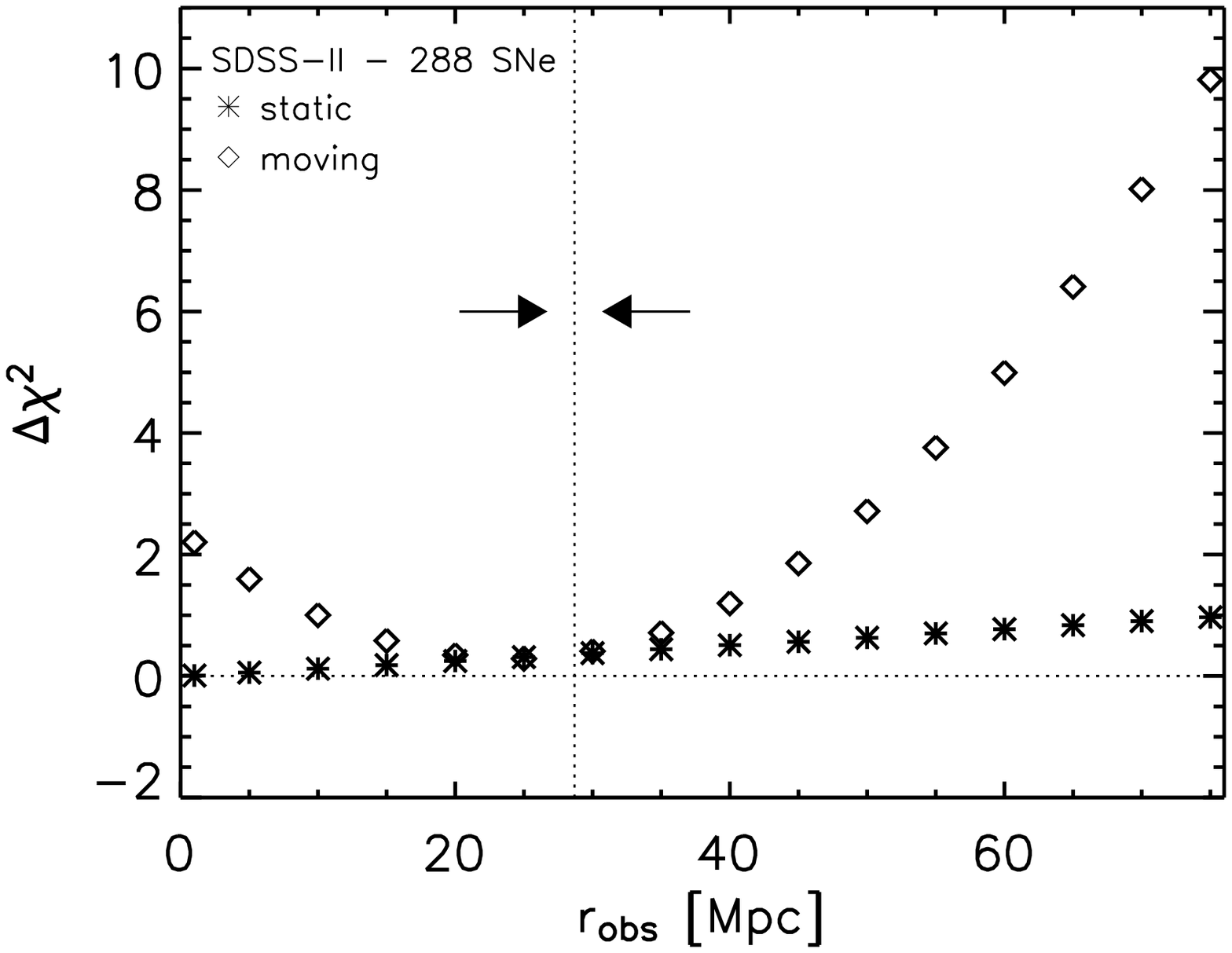}}

However, a lot of problems arise in the LTB model when confronted with some other cosmological observations.
In \LTBCenterA, Alnes {\it et al.} pointed out that a problem of the void model is that
it requires us to live {\it precisely} near the center of the void.
If there is a deviation between our position and the center of the void,
the observed CMB dipole would become much larger than that allowed by observations
(see the left panel of \FigLTBBlueShift\ for a brief description).
Currently, the maximum distance to the center have been constrained to be very small \refs{\LTBCenterB,\LTBBiswas,\LTBMoss}
\eqn\LTBrc{r_{\rm obs}<\sim 20 {\rm\ Mpc},}
which leads to a fine-tuning problem.
For example, in \LTBBlomqvist\  Blomqvist and M${\rm \ddot{o}}$rtsell tested the LTB described by Eq. \LTBMassDisA\
by using the SNIa and CMB dipole data.
They found that the position of the observer has been confined to within about one percent of the void scale radius (see the right panel of \FigLTBBlueShift).

Besides, even if we happen to live very close to the center of the
void, there should be some off-centered galaxy clusters where a
large CMB dipole can be observed in their reference frame. For us,
the relative motion between the CMB frame and the matter frame
manifests itself observationally as a kinematic Sunyaev-Zeldovich
effect. In \LTBkSZ, Bellido {\it et al.} demonstrated that the limited
observations of only 9 clusters with large error bars already rule
out LTB models with void sizes greater than $\sim$ 1.5 Gpc and a
significant underdensity.

Another problem with the void models is that when fitted to the
data, they slow local expansion values
\refs{\LTBBiswas,\LTBMoss,\LTBWiltshire,\LTBZibin,\LTBNadathur}
\eqn\LTBh{h_0\sim0.45-0.6,}
which seems to be in contradiction with the measurements of the
local Hubble constant. Recently, in \HzRiessnew, Riess {\it et al.}
obtained
\eqn\LTBH{H_0=73.8\pm2.4 {\rm\ km/s/Mpc},}
corresponding to a 3.3 \% uncertainty. They found that the void
models with $h_0\sim0.6$ is ruled out by the measurements in more
than 5 $\sigma$.

There have been some other useful methods proposed to
distinguish the void models between other inhomogeneous dark energy models.
In \LTBTimeDrift, Uzan {\it et al.} presented the redshift drift $\dot z$
in a general spherically symmetric spacetime, and demonstrated that
its observation would allow the test of Copernican principle.
In addition,
Yoo \LTBYoo\ and Quartin \LTBQuartin\ showed that the $dz/dt$ in void
modes is always negative, which is greatly different from other dark energy
models. Another interesting idea is, the ionized universe severs as
a mirror to reflect CMB photons in other regions of the universe to
us, and thus can tell us deviation from the Copernican principle \refs{\LTBGoodman,\LTBMirror}.
Utilizing this method, some models with largest voids have be
excluded \LTBMirror. There are also some other methods of testing
the LTB model, such as the scalar perturbations
\LTBZibin, the slope of low $z$ SNIa distance moduli \LTBDistantSN,
the constant curvature condition \LTBClarksonA, the small scale CMB
\LTBSmallScaleCMB, the cosmic neutrino background \LTBNeutrino, the cosmic age test \CosmicAgeLTB,
and so on.

In all, in recent years the research of the LTB model has drawn a
lot of interests. Although this model can provide a good explanation of
the SNIa data without introducing the mysterious dark energy component, when
confronted with the CMB, $H_0$ and some other cosmological
observations, a lot of problems emerged.
For more research works about the LTB model see \refs{\DEInhomUniverBuchert,\MarraPaakkonen,\LTBOther}.
For a recent review, see \MarraRev.

\noindent $\bullet$ Backreaction model

Like the LTB model, the backreaction model
\refs{\RasanenKI,\RasanenBE,\BRRasanen,\BRKolbA,\BRKolbB} is another model
in which the cosmic expansion is due to the effect of the
inhomogeneities of the universe. There have been some
works concerning the possible observational signature of this model
\refs{\BRNLi,\BRSeikel,\BRLarena}. For example, in \BRNLi, Li {\it et al.} showed a
non-trivial scale dependence of the Hubble rate in this model. In
\BRLarena, Larena {\it et al.} proposed to use a template metric to deal
with observations in backreaction context and found that averaged
inhomogeneous models can reproduce the observations of SNIa data and
the position of the CMB peaks. To provide evidence for the
backreaction mechanism, further studies are needed. One can see
Refs.
\refs{\DEInhomUniverBuchert,\BROther} for mode studies concerning the backreaction model.

\subsec{Comparison of dark energy models}

Facing so many dark energy candidates, it is very important to decide which one is more favored by the observational data.
So far there have been many works on the comparison of various dark energy models
\refs{\DEManyModelsHWei,\DEManyModelsTDavis,\DEManyModelsMLi,\DEManyModelsDRubin,
\DEManyModelsNesseris,\DEManyModelsSzydlowski,\DEManyModelsAKurek,\DEManyModels}.
We will briefly review the topic of model comparison in this subsection.

\noindent $\bullet$ Model selection and the information criteria

The $\chi^2$ statistics alone cannot provide an effective way to make a comparison between competing dark energy models.
To do this, we should
take into account the relative complexity of the models. To give a
blatant example, a 10th-order polynomial will always give an equal
or better fit than a straight line to any data set, but this does
not mean that any of the extra eight coefficients have any
significance. It juts means that a model with more parameters will
generally give an improved fit (always, if the simpler model is a
subclass of the more complex one) \DEManyModelsTDavis.

To enforce a model comparison, a general way is to employ the
information criteria (IC) to assess different models \refs{\ModelCompBIC,\ModelCompAIC,\ModelCompIC}. These
statistics favor models that give a good fit with fewer parameters.
The most frequently used IC including the Bayesian information
criterion (BIC)\ModelCompBIC\ and the Akaike information criterion
(AIC) \ModelCompAIC. They are defined as
\eqn\ModelCompBIC{{\rm BIC}=-2\ln{\cal L}_{\rm max}+k \ln N}
\eqn\ModelCompAIC{{\rm AIC}=-2\ln{\cal L}_{\rm max}+2k\ \ \ \ \ }
where ${\cal L}_{\rm max}$ is the maximum likelihood (under Gaussian
assumption $\chi^2_{\rm min}=-2\ln{\cal L}_{\rm max}$), $k$ is the number of
parameters, and $N$ is the number of data points used in the fit.
According to these criteria, models that give a good fit with fewer
parameters will be more favored. So these criteria embody the
principle of Occam's razor, ``entities must not be multiplied beyond
necessity". Generally, a $\Delta$BIC of more than 2 (or 6) is
considered positive (or strong) evidence against a model
\ModelCompBICLiddle. It should be noted that the IC alone can at
most say that a more complex model is not necessary to explain to
current data, since a poor information criterion result might arise
from the fact that the current data are too limited to constrain the
extra parameters in this complex model, and it might become
preferred with improved data.

In addition to the AIC and BIC, a more sophisticated method for model
selection is the so-called Bayesian evidence (BE), which considers
the increase of the allowed volume in the data space due to the
addition of extra parameters rather than simply counting parameters.
So it requires an integral of the likelihood over the whole model
parameter space
\eqn\ModelEqCompBE{BE=\int{\cal L}({\rm \bf d}|\theta,M){\rm \bf p}(\theta|M)d\theta}
In practice, the integral in Eq. \ModelEqCompBE\ can be calculated
using the nested sampling method \ModelCompNestA.
BE has already been applied in cosmology \refs{\ModelCompBE,\ModelCompBELZ}.
Since in most cases the simpler AIC and BIC methods are sufficient to employ a model comparison, they are more commonly used compared with BE.

\noindent $\bullet$ Comparison of different dark energy models

\ifig\FigDEComp{Graphical representation of the results of the IC
values of some popular dark energy models. $\Delta$AIC and $\Delta$BIC are
represented by the light and dark grey bars, respectively. From left
to right, the models are given in order of increasing $\Delta$AIC.
The crosses mark the number of free parameters in each model
({\it right-hand ordinate}). The ``unsupported" or ``strongly
unsupported" lines stand for a $\Delta$BIC of 2 and 6. Clearly the
flat $\Lambda$CDM is the most preferred model. From
\DEManyModelsTDavis.} {\epsfysize=3.0in \epsfbox{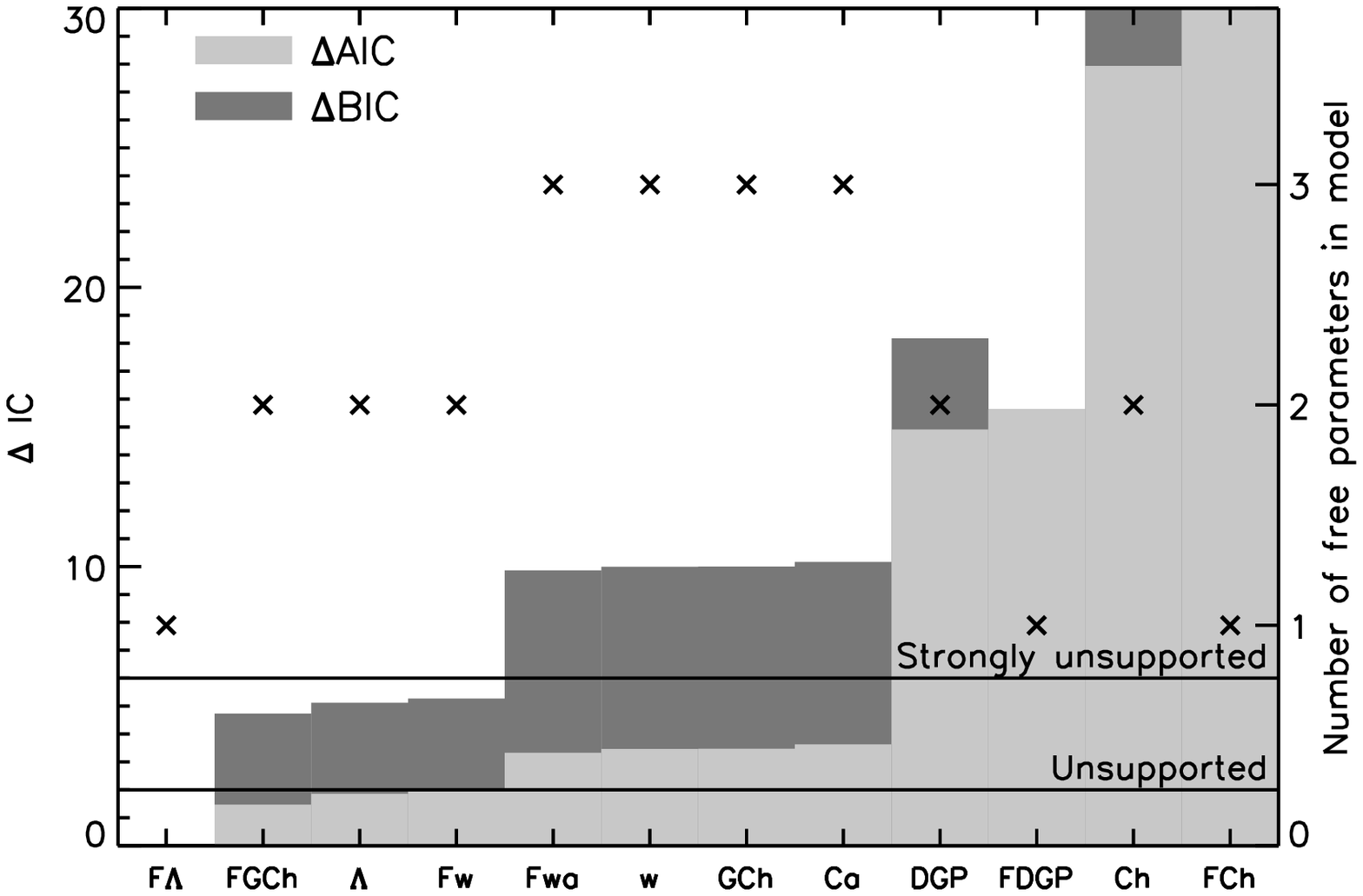}}

In \DEManyModelsTDavis, based on the observational data of SNIa, BAO
and CMB, Davis {\it et al.} scrutinized and compared a number of
dark energy models by using AIC and BIC. They found that the $\Lambda$CDM model
almost achieves the best fit of all the models despite its economy
of parameters. A series of models, including the XCDM model with
constant EOS of dark energy $w$, the Cardassian expansion model and
the CPL parametrization, can also provide comparably good fits but
have more free parameters. The DGP model and the standard CG model
with $\alpha=1$ are clearly disfavored. Based on their AIC and BIC
values, one can determine the ``rank'' of these dark energy
models (see \FigDEComp). From the figure, it is clear that the
$\Lambda$CDM model is best favored, while the DGP model and the
standard CG model is strongly disfavored.

By comparing the values of IC, similar results are obtained by
Szydlowski {\it et al.} \DEManyModelsSzydlowski, A. Kurek {\it et al.}
\DEManyModelsAKurek, Li {\it et al.} \DEManyModelsMLi\ and Wei {\it et al.}
\DEManyModelsHWei. In the above works, the authors all found that
the one-parameter flat $\Lambda$CDM performs best in the set of
models considered in the context. These results further solidified
the status of the $\Lambda$CDM scenario as the standard paradigm in
modern cosmology, although there are still some puzzling conflicts
between $\Lambda$CDM predictions and current observations \LCDMPerivolaropoulos.

To the contrary of the good performance of the $\Lambda$CDM model,
another common results of these works is that the DGP model and the
standard CG model usually performs badly and ranks worst in
the dark energy models considered thereof. So these two models faced with
high crisis when confronted with observations. Besides the DGP and the CG model, some
other models like the ADE and RDE models also performed badly in a
model comparison. For example, in \DEManyModelsMLi, Li {\it et al.}
obtained the following results of $\chi^2_{\rm min}$s from the
combination of Constitution+BAO+CMB+$H_0$ data
\eqn\DECompLi{\chi^2_{\rm \Lambda CDM}=468.461,\ \ \chi^2_{\rm RDE}=493.772,\ \ \chi^2_{\rm ADE}=503.039,\ \ \chi^2_{\rm DGP}=530.443.}
Compared with the $\Lambda$CDM model, the last three models have much larger BIC values
\eqn\DECompLiIC{\rm \Delta BIC_{RDE}=31.308,\ \
\Delta BIC_{ADE}=34.578,\ \ \Delta BIC_{DGP}=61.982.}

What should be mentioned is that a somewhat different result was obtained by Sollerman {\it et al.} in \DEManyModelsSollerman,
where they found that the flat DGP model performed even better than the flat $\Lambda$CDM model
from first-year SDSS-II SNIa dataset \SDSSSN\ analyzing using the MLSCS2k2 light-curve fitter.
Notice that in this work, the authors also took two kinds of LTB models into consideration,
which were not included in the model-comparison by the numerical studies listed above.
In addition, they showed that the extra parameters required by these two models are not supported by the IC tests
(their IC values are $\sim 5-25$ larger compared with $\Lambda$CDM model).

Some other models, such as the general DGP model, the HDE model and the
parameterizations like XCDM and CPL, can also provide rather
good fits to the observational data. From current observational data
it is hard to discriminate these models. Since they have more free
parameters, these models are all less favored by the $\Lambda$CDM
under the IC tests, indicating that given the current quality of the
data there is no reason to prefer more complex models. The resulted
$\Delta BIC$ values (with the $\Lambda$CDM model as a reference) of
these models given in \DEManyModelsMLi\ are (in a flat universe)
\eqn\DECompBICsA{\rm \Delta BIC_{XCDM}=5.862,\ \ \Delta
BIC_{GDGP}=5.897,} \eqn\DECompBICsB{\rm \Delta BIC_{CPL}=11.195,\ \
\Delta BIC_{HDE}=8.048.} Here the GDGP model refers to the
generalized DGP model described by Eq. \GenDGP. Notice that the CPL
model does not achieve a good performance due to its large number of
free parameters. Another interesting phenomenon is that most of the
above models (like the HDE model) can reduce to the
$\Lambda$CDM model; from their best-fit parameters, it is found that
they do tend to collapse to $\Lambda$CDM model
\refs{\DEManyModelsTDavis,\DEManyModelsMLi}. Therefore, the current
observational data are still too limited to distinguish which
theoretical model is better.

\newsec{Model-independent dark energy reconstructions}
\seclab\secReconDE

In the last section, we have introduced some representative numerical works on the specific dark energy models.
Due to the lack of a compelling fundamental theory to explain the dark energy, another route,
the model-independent dark energy reconstructions, have drawn more and more attentions
\refs{\NCGasDaly,\WellerAlbrecht,\GongWang,\HT,\Maor,\Wetterich,\JBP,\Starobinsky,\Wellergroup,\DEModInde}.

The dark energy reconstruction is a classic statistical inverse problem for the Hubble parameter
\eqn\HBasic{H(z) = H_{0} \sqrt{\Omega_{m0}(1+z)^{3}+(1-\Omega_{m0})f(z)},}
where $f(z)\equiv \rho_{de}(z) / \rho_{de}(0)$ is the dark energy density function.
Different reconstruction method will give different $f(z)$.
An ideal dark energy reconstruction should be sufficiently versatile to accommodate a large class of dark energy models.
The main target of a dark energy reconstruction is to detect the dynamical property of dark energy,
i.e. determine whether the accelerating expansion is consistent with a cosmological constant.

To begin with one should choose an appropriate quantity characterizing dark energy.
It is widely believed that the EOS of dark energy $w \equiv p_{de}/\rho_{de}$ holds essential clues for the nature of DE \TurnerWhite.
It should be mentioned that $w(z)$ is related with $f(z)$ through an integration \refs{\WangGarnavich,\WangTegmark,\WangFreese}
\eqn\fw{f(z) = \exp\left(3\int_{0}^{z}dz'{1+w(z')\over 1+z'}\right).}
Since the $\Lambda$CDM model always satisfy $w=-1$, the deviation from this constant EOS will reveal the variation of dark energy density.
Therefore, most researchers have chosen to study dark energy by constraining $w$ from observations.

\ifig\FigYW{A comparison between the $w$ parametrization and the $\rho_{de}$ parametrization using the same observational data.
The regions inside the solid and dashed lines correspond to 1$\sigma$ and 2$\sigma$ confidence regions, respectively.
Clearly the $\rho_{de}$ reconstruction can give much tighter constraint on dark energy compared with the $w$ parametrization.
From \WangFreese.} {\epsfysize=3.0in \epsfbox{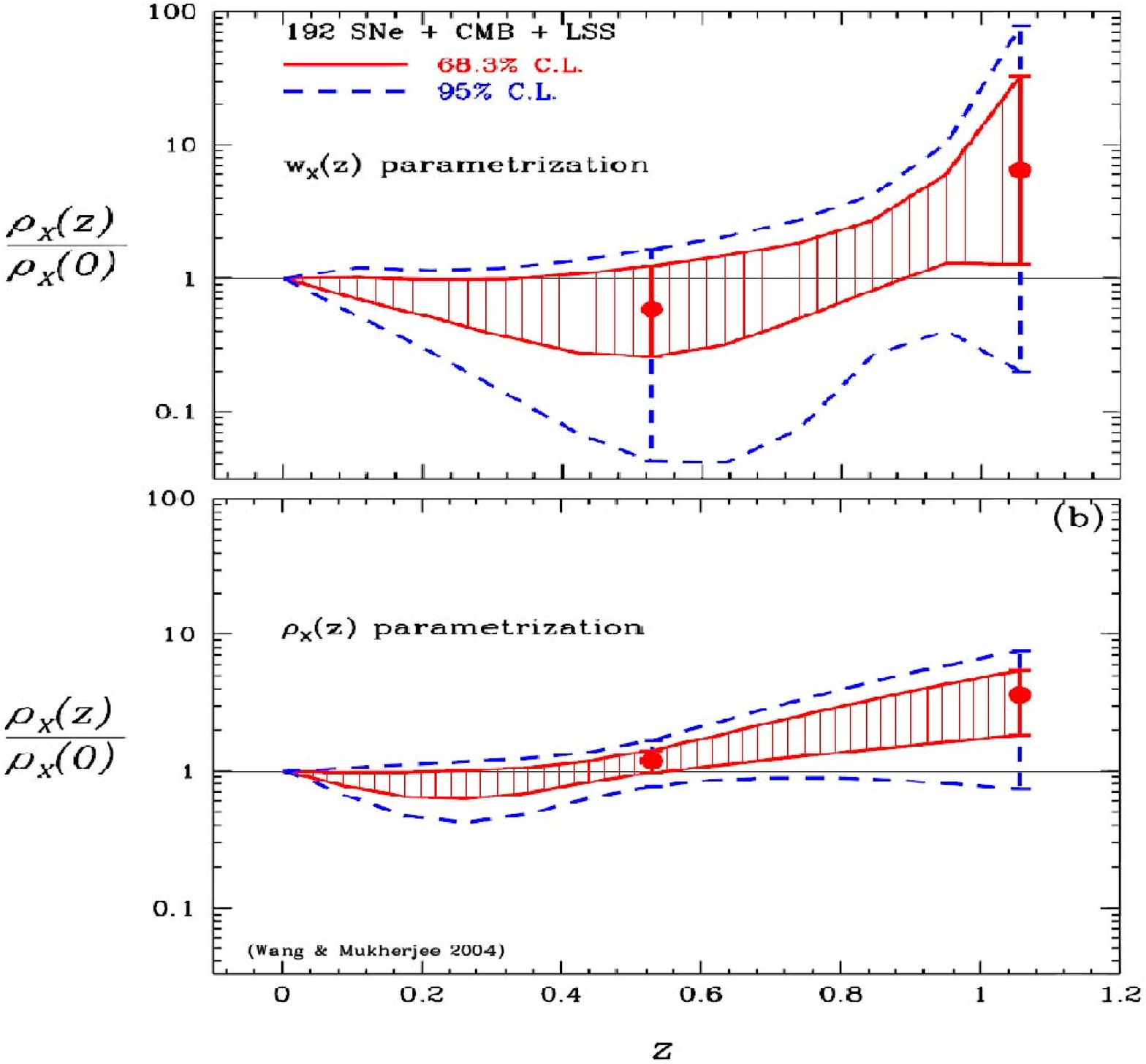}}

However, in a series of works \refs{\FOMWang,\WangGarnavich,\WangTegmark,\WangFreese,\WangMukherjeetwo,\YWangtwo,\YWanglatest},
Wang and collaborators argued that, due to the smearing effect \Maor\
arising from the multiple integrals relating $w(z)$ to the luminosity distance of SN $d_L(z)$,
it is difficult to constrain $w$ using the SN data \HT.
On the contrary, since using the dark energy density $\rho_{de}$ can minimize the smearing effect by removing one integral,
$\rho_{de}$ can be constrained more tightly than $w$ given the same observational data (see \FigYW\  for details).
It should be mentioned that there is still a debate on which quantity, $w$ or $\rho_{de}$, is better in describing dark energy \Linderargue.

In addition to $w$ and $\rho_{de}$,
some other quantities, such as the deceleration parameter $q$ \refs{\GongWang,\WDQ,\LWY,\CA}, the redshift of acceleration-deceleration transition $z_c$ \SCA,
the statefinder diagnostic $(r,s)$ \statefinder, the jerk parameter $j$ \refs{\Visser,\Rapetti} and the diagnostic $Om$ \refs{\DEModIndeZ,\Omone},
can also provide very useful information for the study of dark energy.
Although these quantities can accurately reconstruct some dark energy models, they have difficulty to discriminate between different models of dark energy \PanAlam.
So in this section, we will introduce the model-independent methods based on the reconstructions of $w$ and $\rho_{de}$.

The model-independent dark energy reconstructions can be divided into four classes:
(i) Specific Ansatz: assuming a specific parameterized form for $w(z)$ and estimating the associated parameters.
(ii) Binned Parametrization: dividing the redshift range into different bins and using a simple local basis representation for $w(z)$ or $\rho_{de}(z)$.
(iii) Polynomial Fitting: treating the dark energy density function $f(z) \equiv \rho_{de}(z)/\rho_{de}(0)$ as a free function of redshift
and representing it by using the polynomial.
(iv) Gaussian Process modeling: using a distribution over functions that can represent w(z) and estimating the statistical properties thereof.
These four classes of reconstruction methods and the related research works will be introduced in the following.

\subsec{Specific ansatz}

\

\

\ifig\FigWCDM{$68.3\%$, $95.4\%$, and $99.7\%$ confidence regions of
the ($\Omega_{m0}$,$w$) plane from SN combined with the constraints
from BAO and CMB both without (left panel) and with (right panel)
systematic errors. From \UnionTwo.} {\epsfysize=2.0in
\epsfbox{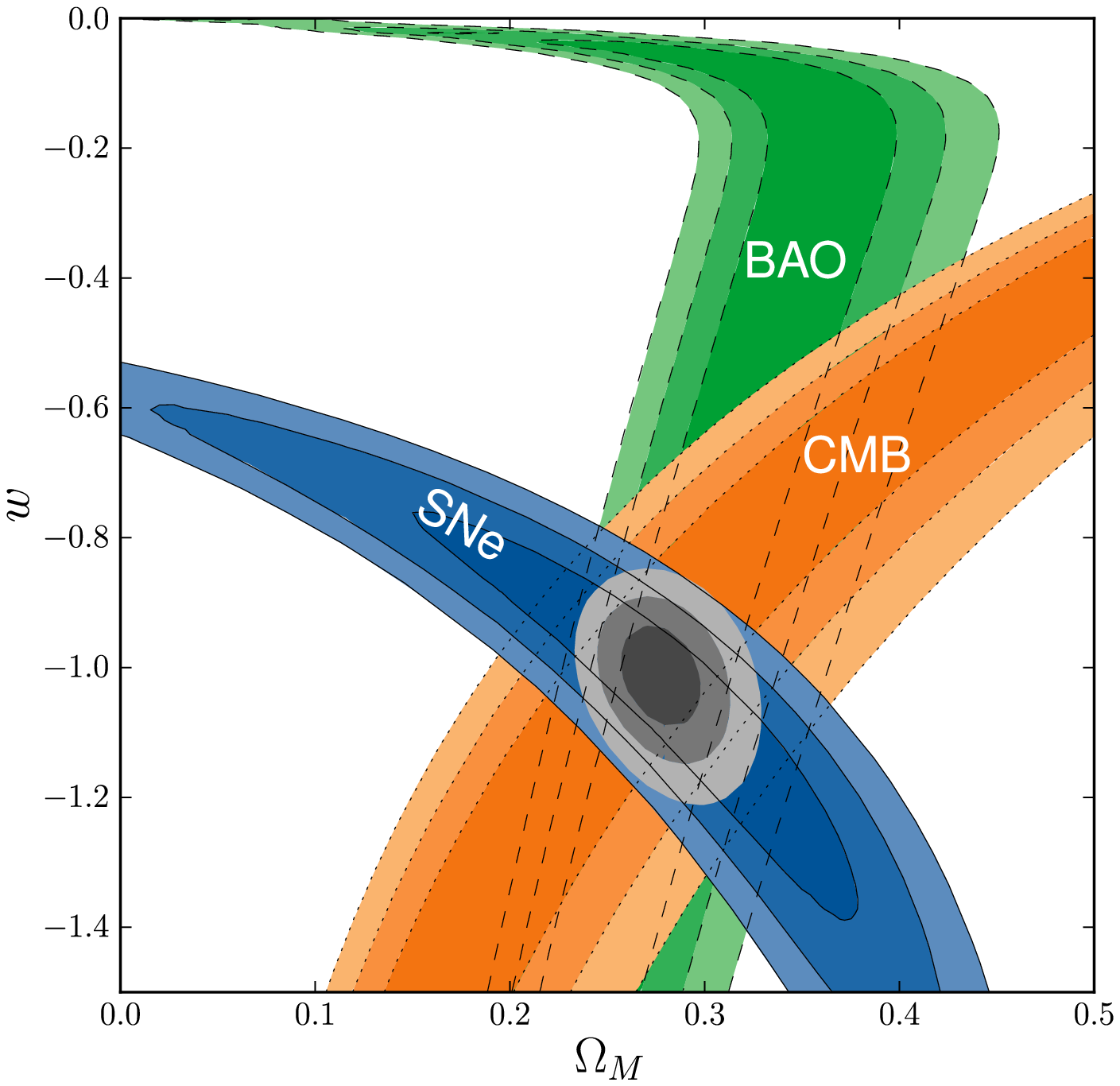} \epsfysize=2.0in
\epsfbox{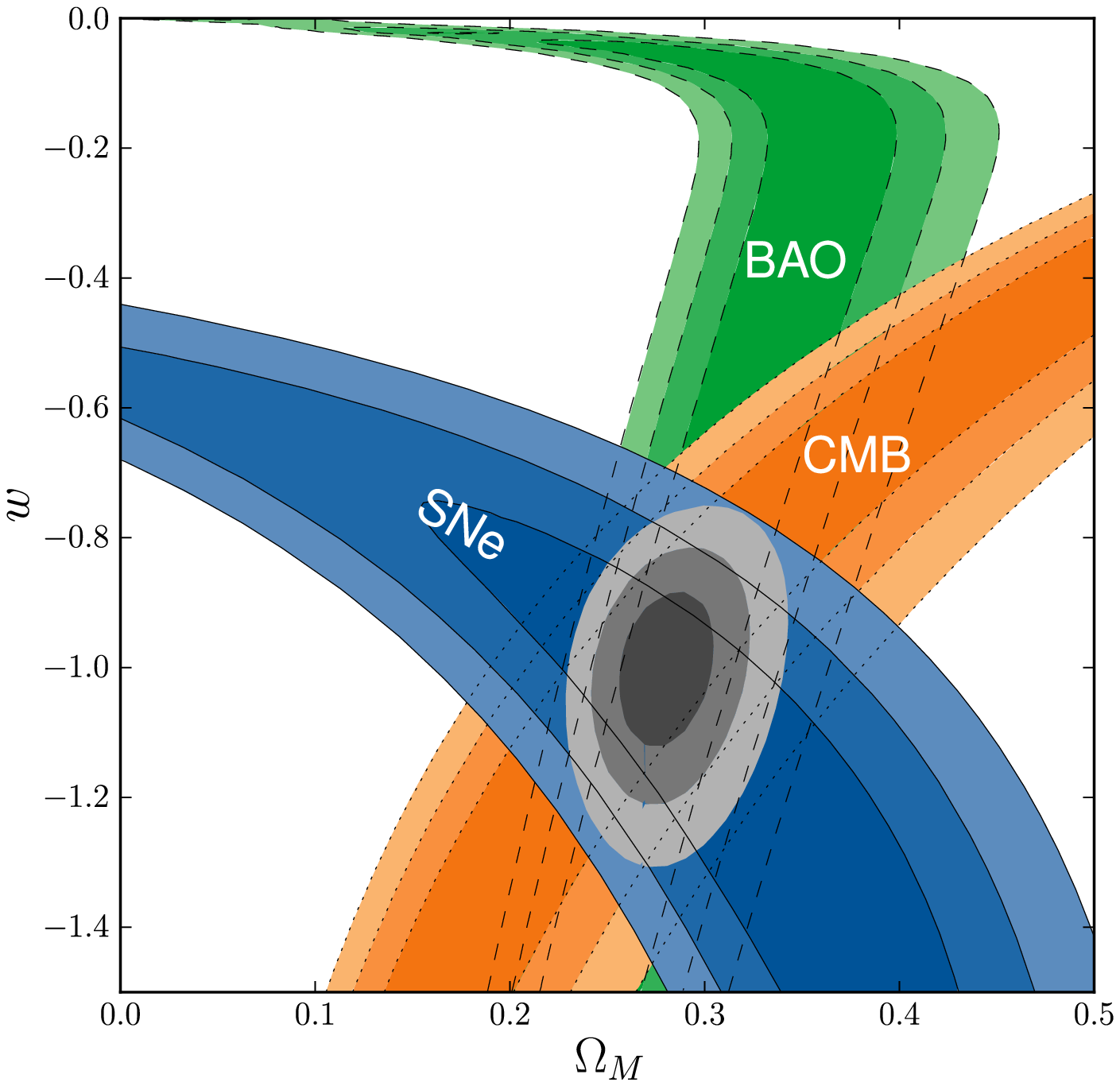}}

The ``specific ansatz'' is the most popular approach currently.
The key idea is assuming a specific parameterized form for $w(z)$ and estimating the associated parameters.
A simple and widely used ansatz is the XCDM ansatz, in which the EOS of dark energy is a constant, i.e. $w = const$.
This yields a simple form of $f(z)$
\eqn\fXCDM{f(z) = (1+z)^{3(1+w)}.}
In \WMAPFive, by combining the WMAP5 observations with BAO and SN data, Komatsu {\it et al.} obtained $w=-0.992_{-0.062}^{+0.061}$ at the 1$\sigma$ CL,
while in \WMAPSeven, a combination analysis of WMAP7+BAO+SN gave $w=-0.980_{-0.053}^{+0.053}$.
A more recent constraint on $w$ by the SCP team \UnionTwo\ also presented the consistent result (see \FigWCDM).
So the current observations still favor $w=-1$ (i.e. $\Lambda$CDM model).

\ifig\FigCPL{{\it Left panel}: $68.3\%$, $95.4\%$, and $99.7\%$
confidence regions of the ($w_0$,$w_a$) plane from Union2 SNIa
sample combined with the constraints from BAO and CMB both with
(solid contours) and without (shaded contours) systematic errors,
for a flat universe. Points above the dotted line ($w_0 + w_a = 0$)
violate early matter domination and are implicitly disfavored in
this analysis by the CMB and BAO data. From \UnionTwo. {\it Right
panel}: Joint constraints on the CPL model from the WMAP7
observations. The contours show $68.3\%$ and $95.4\%$ CL from
WMAP+$H_0$+SNIa (red), WMAP+BAO+$H_0$+SNIa (blue) and
WMAP+BAO+$H_0$+$D_{\Delta t}+$SNIa (black), for a flat universe.
``$D_{\Delta t}$'' denotes the time-delay distance to the lens
system B1608+656 at $z=0.63$ measured by \CPLLensSysSuyu. From
\WMAPSeven.} {\epsfysize=2.0in \epsfbox{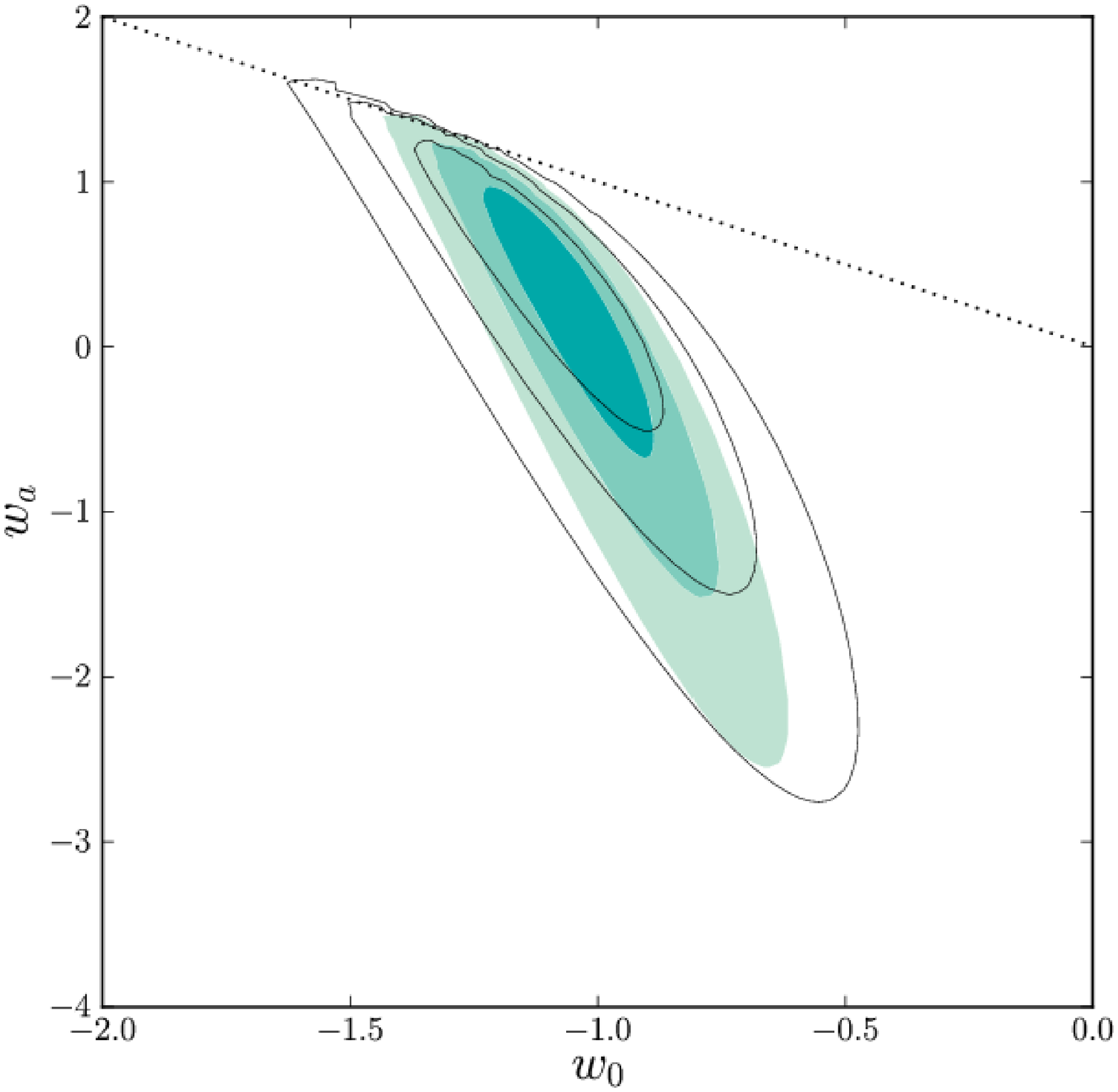}
\epsfysize=2.0in \epsfbox{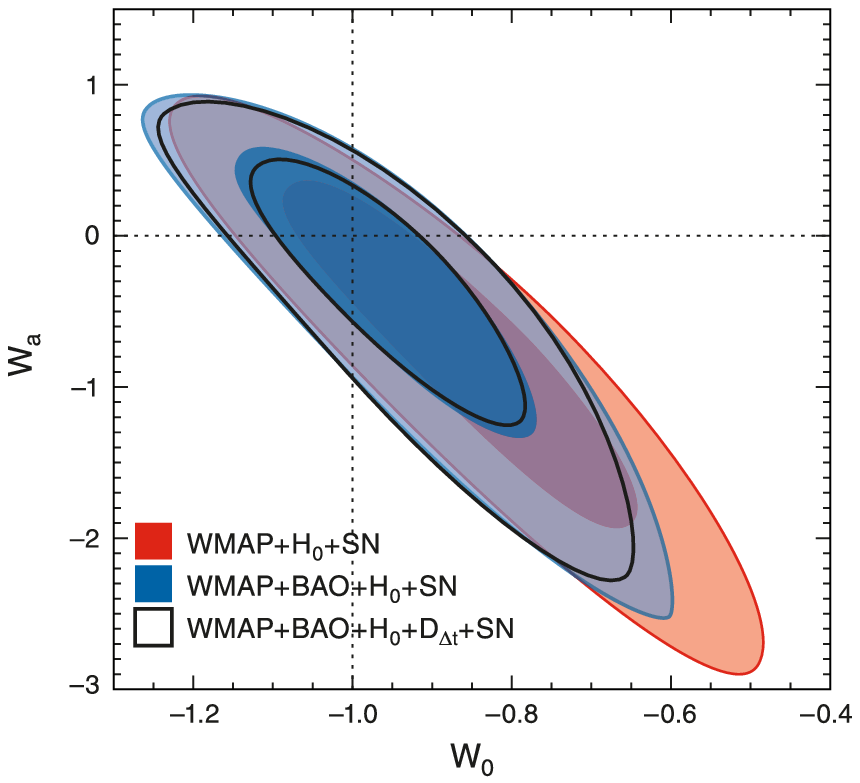}}

Besides, one can also assume that the EOS of dark energy is not a constant.
The most popular parametrization with dynamical $w$, which assume $w(z) = w_{0} + w_{a}z/(1 + z)$,
was firstly proposed by Chevallier and Polarski \CPLBE, then was used to explore the expansion history of the universe by Linder \CPL.
So this ansatz is often called CPL parametrization.
The corresponding $f(z)$ is given by
\refs{\CPLBE,\CPLtwo} \eqn\fCPL{f(z) = (1+z)^{3(1+w_0+w_a)}\exp\left(-{3w_{a}z\over1+z}\right).}
Here $w_0$ denotes the value of the present EOS, while $w_a$ denotes the variation of the EOS.
Because of its bounded behavior at high redshift and high accuracy in reconstructing many scalar field EOS \CPL,
the CPL parametrization has become one of the most popular methods to study dark energy \refs{\JBP,\CPLPerivolaropoulos}.
In \FigCPL\ we show the constraint on $(w_0, w_a)$ in the CPL ansatz given by the Union2 SNIa dataset \UnionTwo\ and by the WMAP7 observations \WMAPSeven.
Obviously, the current data favor the result of $w_0=-1$ and $w_a=0$, which is consistent with the cosmological constant.

In addition, some other ansatzs have also been proposed.
Using the principal component analysis, Linder and Huterer \CPLpar\ argued that 2 parameters,
involving a measure of the EOS value at some epoch (e.g. $w_0$) and a measure of the change in EOS (e.g. $w'$),
are most realistic in projecting dark energy parameter constraints.
Therefore, most ansatzs contain 2 parameters.
In \HutererTurner, Huterer and Turner proposed a linear parametrization $w(z) = w_{0} + w_{1}z$ to study the evolution of dark energy.
This ansatz can fit the low redshifts data well, but its dark energy component grows increasingly unsuitable at redshifts $z > 1$.
In \JBP\ Jassal {\it et al.} proposed a more general form of the CPL parametrization $w(z) = w_{0} + w_{a}z/(1 + z)^p$ and investigated the case of $p=1,2$.
In \Efstathiou, Efstathiou introduced another parametrization $w(z) = w_{0} + w_{1}\ln(1+z)$.
It should be mentioned that the parametrizations listed above have difficult to fit the rapidly varying dark energy models,
and some other parametrizations \refs{\Wetterich,\DeParUSeljak,\Upadhye,\YGGongtwo} have been proposed to fit a fast transition of $w(z)$.
For more research works of various parametrization forms,
see \refs{\Perivolaropoulos,\DEManyModelsNesseris,\WellerAlbrecht,\DeParUSeljak,\BABassett,\SilvaAlcaniz,\ansatzZhang} and references therein.

\subsec{Binned parametrization}

In addition to the specific ansatz, another popular approach is the binned parametrization.
The binned parametrization was firstly proposed by Huterer and Starkman \HutererStarkman\
based on the principal component analysis (PCA) \refs{\HutererStarkman,\HutererCooray}.
It often used to measure the EOS $w$ and the density $\rho_{de}$ of dark energy.
The key idea is dividing the redshift range into different bins and picking a simple local basis representation for $w(z)$ or $\rho_{de}(z)$.
The simplest way is setting $w(z)$ or $\rho_{de}(z)$ as piecewise constant in redshift.
For the case where $w$ is piecewise constant in redshift, $f(z)$ can be written as
\Sullivan
\eqn\fzwbinned{f(z_{n-1}<z \le z_n)=(1+z)^{3(1+w_n)}\prod_{i=0}^{n-1}(1+z_i)^{3(w_i-w_{i+1})},}
where $w_i$ is the EOS parameter in the ith redshift bin defined by an upper boundary at $z_i$.
This parametrization has been extensively studied \refs{\DEModIndeZ,\Qi,\Ournewpaper}.
For the case where dark energy density $\rho_{de}$ is piecewise constant in redshift, $f(z)$ can be written as
\eqn\fzbinned{f(z)=\cases{1 & $0\leq z \leq z_1$;\cr f_i & $z_{i-1} \leq z \leq z_{i}$ $(2\leq i \leq n)$.\cr}}
Here $f_i$ is a piecewise constant, and from the relation $E(0)=1$ one can easily obtain $f_1=1$.
For same number of redshift bins, the number of free parameters of piecewise constant $\rho_{de}$ parametrization
is one fewer than that of piecewise constant $w$ parametrization.

\ifig\FigCC{Uncorrelated estimates of the expansion history.
Using $n\Delta z = 40$, 20, and 15, respectively, 3, 4, or 5 bins'
independent measurements of $H(z)$ from the gold sample are plotted
in the top panel. The solid black line in this plane denotes the
prediction of the $\Lambda$CDM model with $\Omega_{m0}=0.29$. The
bottom panel shows the independent measurements of the kinematic
quantity $\dot{a}$ versus redshift. In this plane a positive or
negative sign of the slope of the data indicates deceleration or
acceleration of the expansion, respectively. From \Goldsix.}
{\epsfysize=2.5in \epsfbox{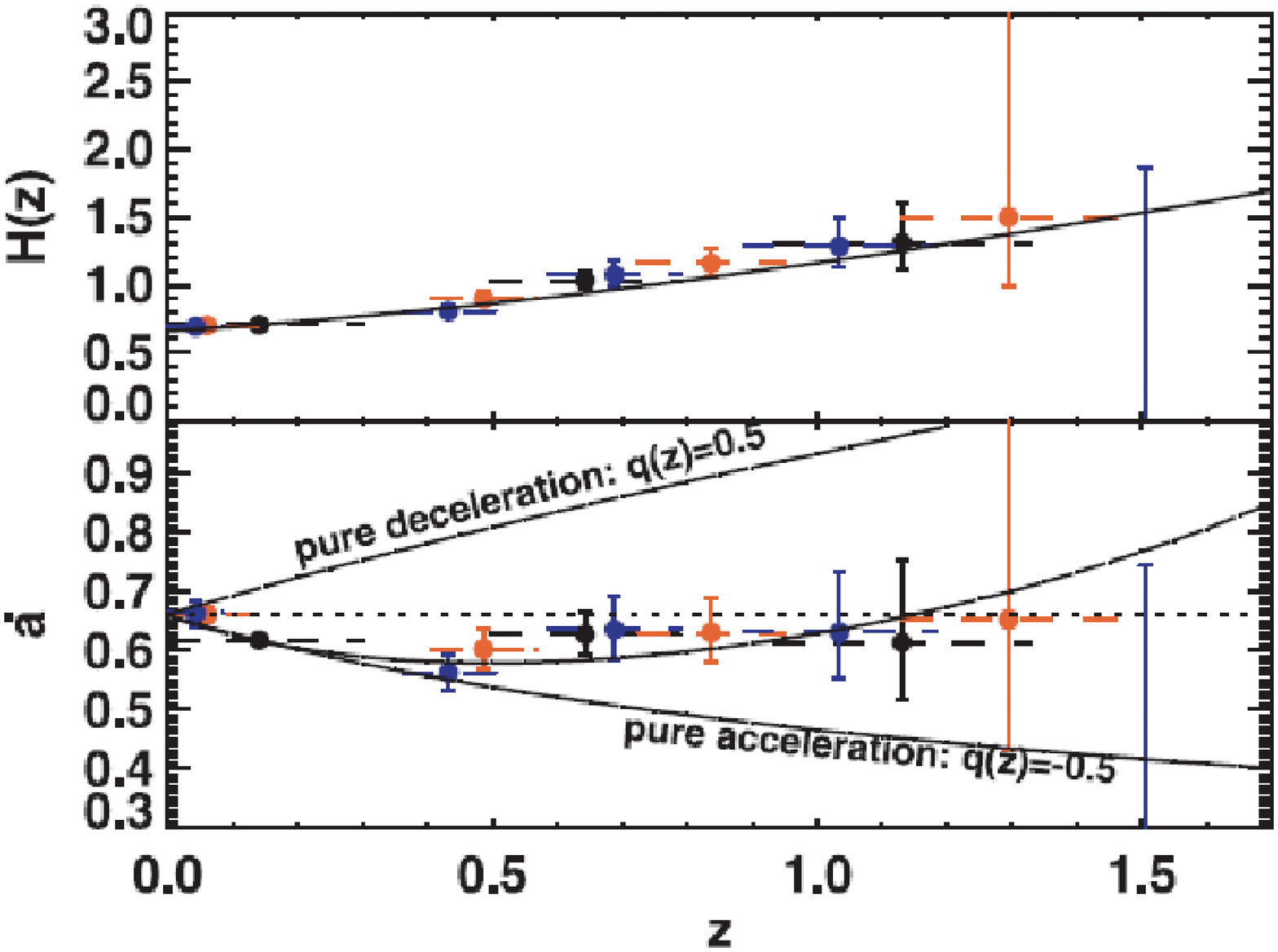}}

It should be mentioned that the optimal choice of redshift bins is
still in debate. In \Goldsix, Riess {\it et al.} proposed an uniform,
unbiased binning method, in which the number of SNIa in each bin
times the width of each bin is a constant (i.e. $n \Delta z=const$).
Using $n\Delta z = 40$, 20, and 15, respectively, they derived 3, 4,
or 5 bins' independent measurements of $H(z)$ and $\dot{a}$ from the
gold sample \Goldsix. (See \FigCC). This binning method has drawn a
lot of attention. For examples, by setting $n \Delta z \sim 30$ and
using the piecewise constant $w$ parametrization, Gong {\it et al.}
explored the Constitution dataset in \YGGongone, and analyzed the
Union2 dataset in \YGGongtwo.

\ifig\FigUNionBin{Constrains on the dark energy density from a joint
data set of SN, BAO, CMB, and $H_0$. The left panel is the case for
three bins, the right panel is the case for four bins. From
\UnionTwo.} {\epsfysize=2.0in \epsfbox{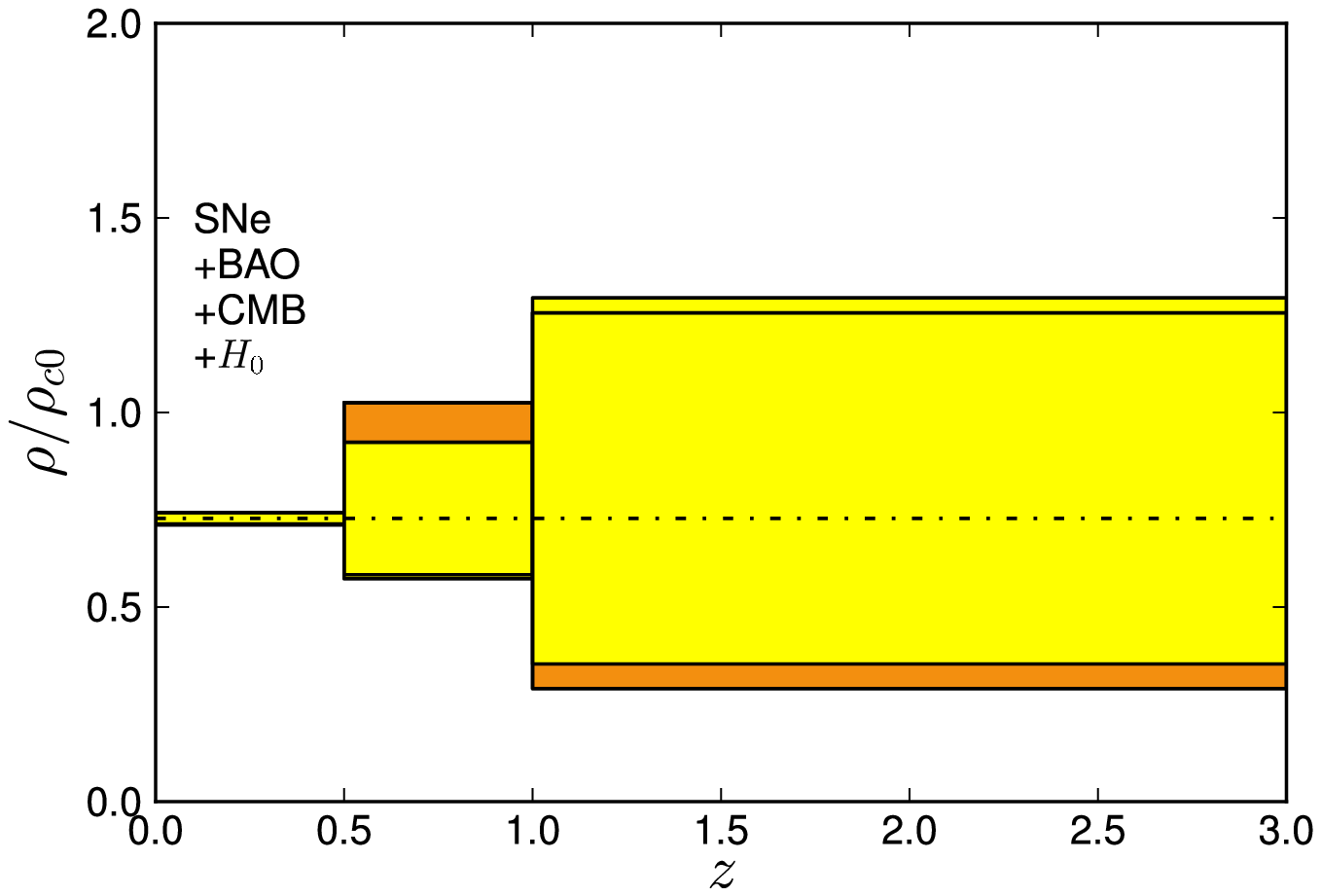}
\epsfysize=2.0in \epsfbox{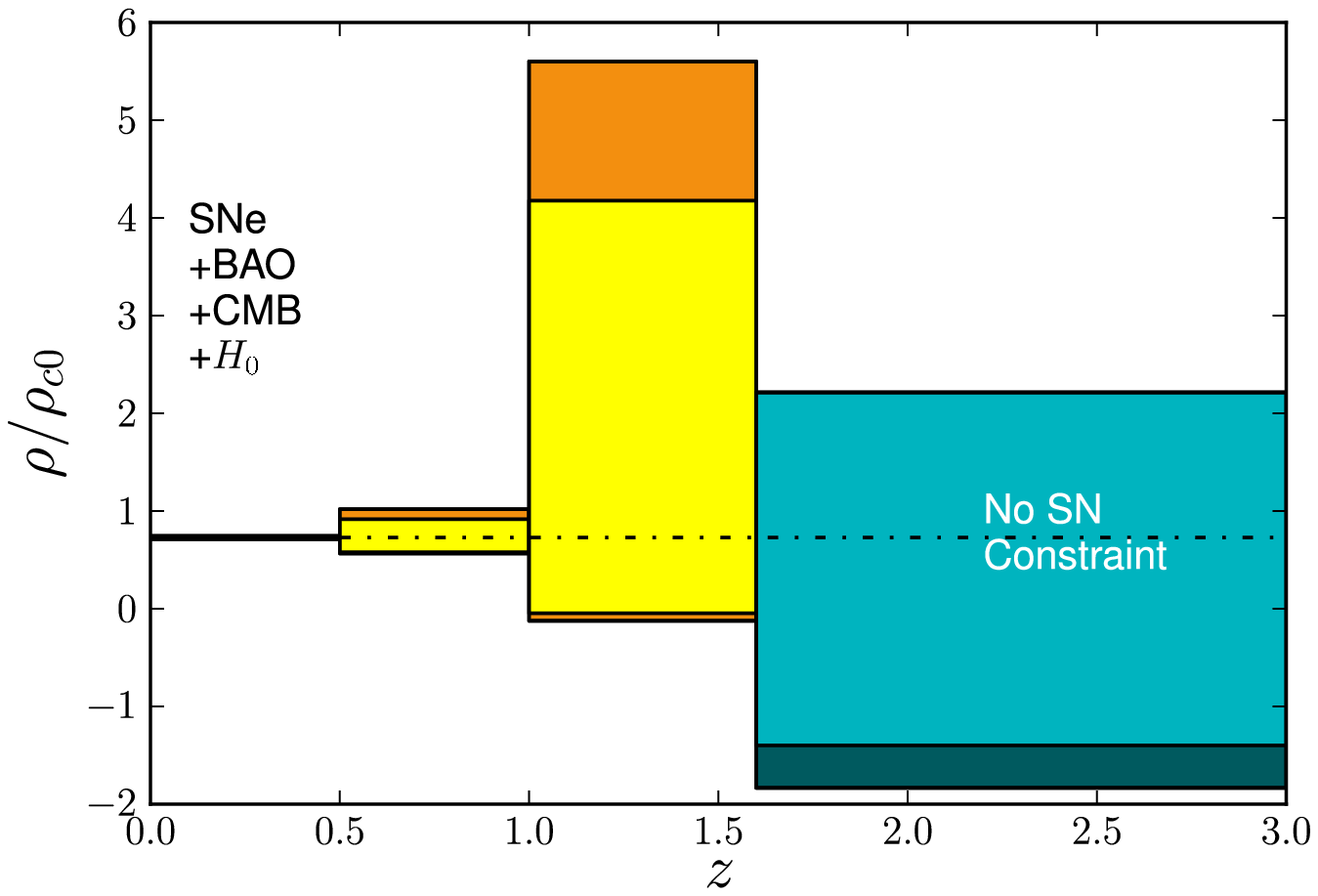}}

In \WangMPLA, Wang argued that one should choose a constant $\Delta
z$ for redshift slices. This is because for a galaxy redshift
survey, the observables are $H$ and $1/D_{A}$ (length scales
extracted from data analysis). Since these scales are assumed to be
constant in each redshift slice, the redshift slices should be
chosen such that the variation of $H$ and $1/D_{A}$ in each redshift
slice remain roughly constant with z. This binning method have also
been adopted by some experimental groups. For example, in \UnionTwo,
the SCP SNIa group explored the Union2 dataset by using this
constant binning method. They find that the current small sample of
SNIa cannot constrain the existence of dark energy above redshift 1 (see \FigUNionBin).

\ifig\FreeZi{The relationship between the $\chi_{\rm min}^{2}$ and the
discontinuity points of redshift ($z_1$ and $z_2$) for the 3 bins
piecewise constant $\rho_{de}$ parametrization. The left panel is
plotted by using the Union2 SNIa sample alone, and the right panel
is plotted by using the combined SNIa+CMB+BAO data. The x-axis
represents the redshift of the first discontinuity point $z_1$,
while the y-axis denotes the redshift of the second discontinuity
point $z_2$. Notice that the light-colored region corresponds to a
big $\chi^{2}$, and the dark-colored region corresponds to a small
$\chi^{2}$. Since $z_1\leq z_2$ must be satisfied, the bottom-right
region of the figure is always blank. From \FreeZitwo.}
{\epsfysize=2.0in \epsfbox{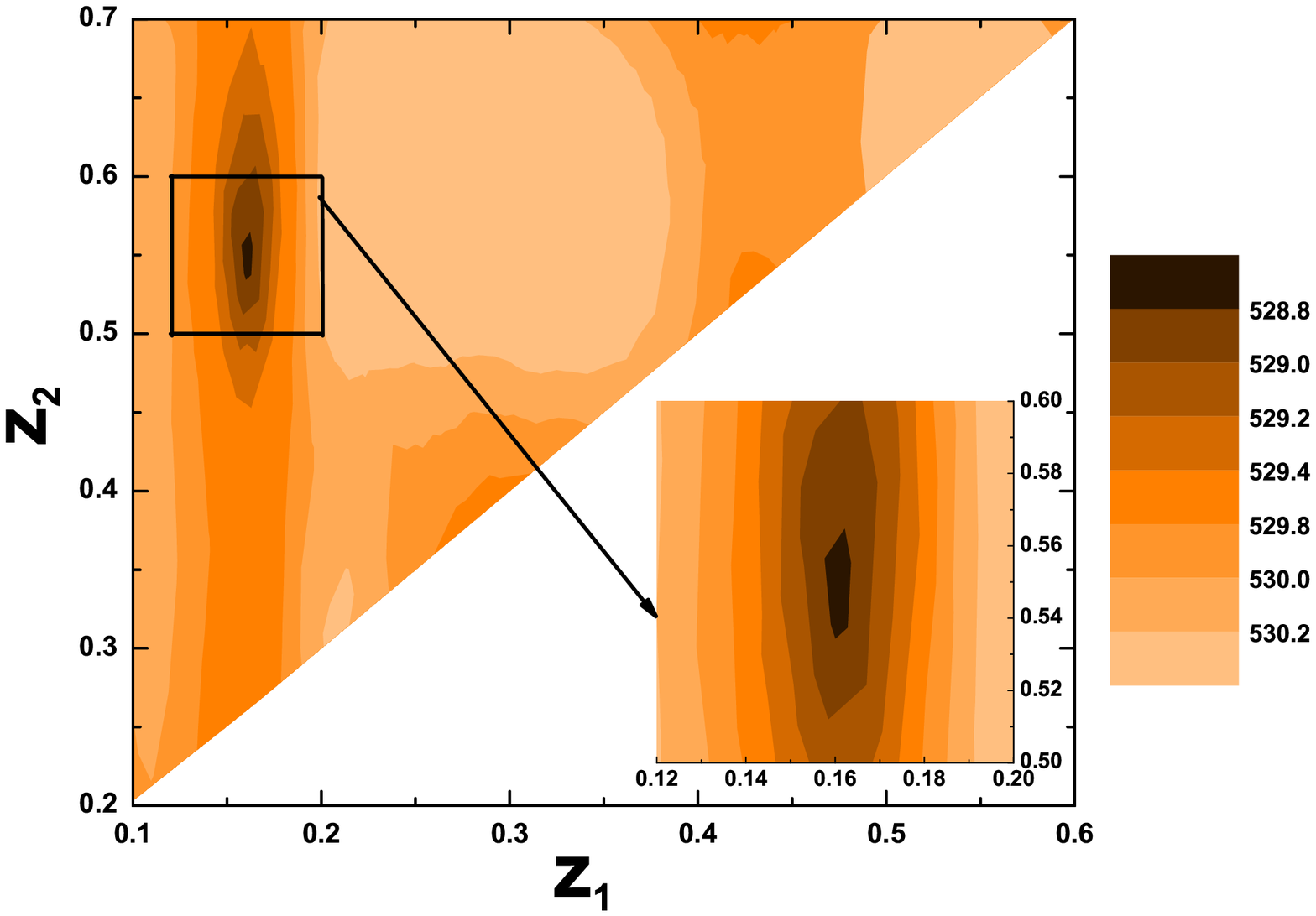} \epsfysize=2.0in
\epsfbox{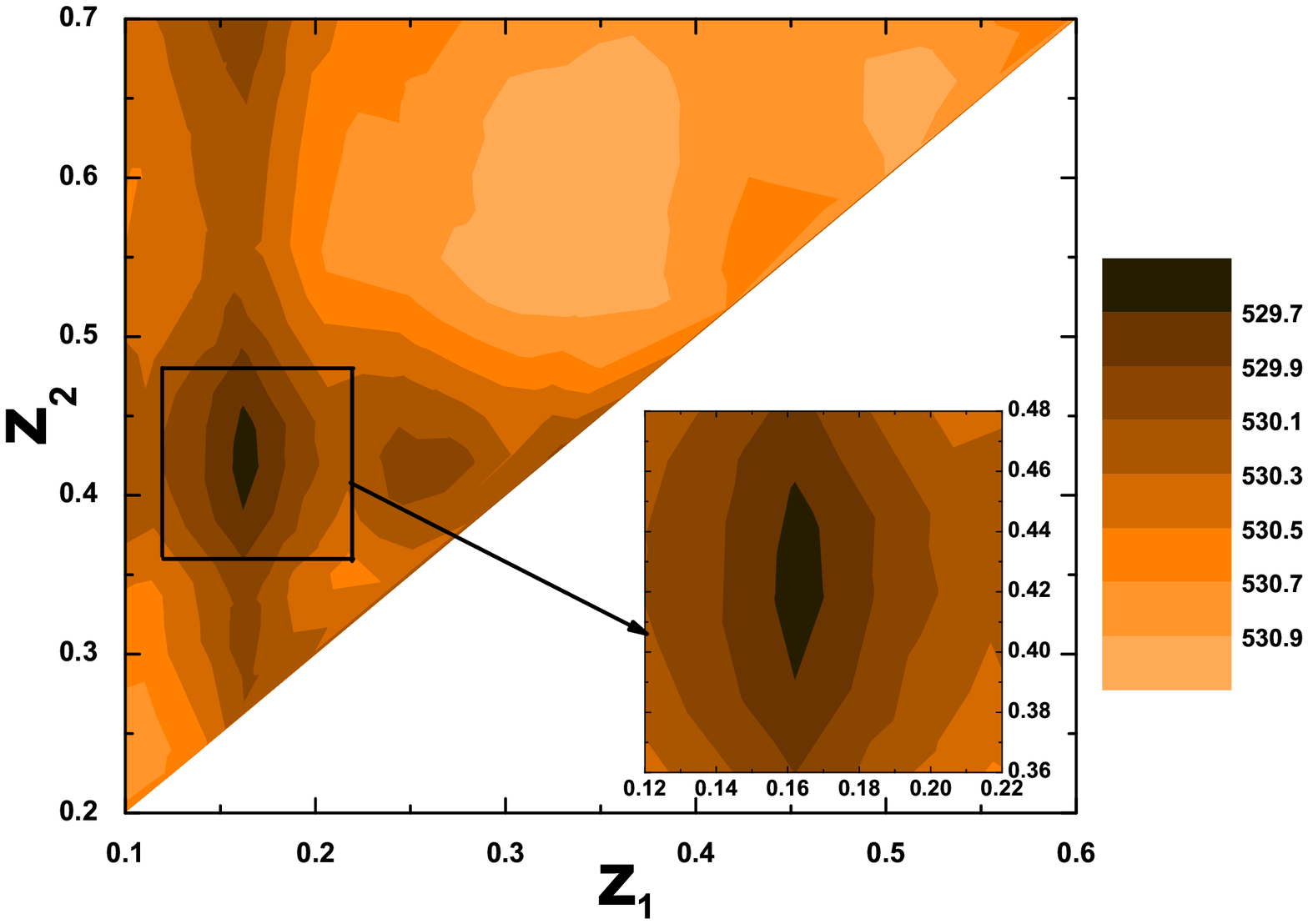}}

In \refs{\FreeZitwo,\FreeZione}, we presented a new binned parametrization
method. Instead of choosing the discontinuity points $z_i$ by hand,
one can treat $z_i$ as models parameters and let them run freely in the
redshift region of SNIa samples. Using the piecewise constant $w$
and the piecewise constant $\rho_{de}$ parametrization,
respectively, the Constitution SNIa dataset has been explored
\FreeZione. In addition, utilizing the piecewise constant $\rho_{de}$
parametrization, the Union2 SNIa dataset has also been analyzed
\FreeZitwo\ (the corresponding results are given in \FreeZi). These
works show that the Constitution dataset favors a dynamical dark
energy, while the Union2 dataset is still consistent with a
cosmological constant. Comparing with those two binning methods
listed above, the advantage of this binning method is that it can
achieve much smaller $\chi_{\rm min}^{2}$\foot{A simple comparison of
these three binning methods can be seen in \FreeZithree.}.

Besides the piecewise constant parametrization, some other
local basis representations for $w(z)$ or $\rho_{de}(z)$ are also
proposed, such as wavelet \wavelet\  and numerical derivatives
\refs{\NCGasDaly,\ShAlSaStr}.

\subsec{Polynomial fitting}

The third approach is the polynomial
fitting method. The key idea is treating the dark energy density function
$f(z)$ as a free function of redshift and representing it by using
the polynomial. Compared with the binned parametrization, the
advantage of the polynomial fitting parametrization is that the dark energy
density function $f(z)$ can be reconstructed as a continuous
function in the redshift range covered by the observational data.

\ifig\FigASSS{The evolution of w(z) with redshift for different values of $\Omega_m$.
The left panel corresponds to the case of $\Omega_m=0.2$,
the middle panel corresponds to the case of $\Omega_m=0.3$,
and the right panel corresponds to the case of $\Omega_m=0.4$,
In each panel, the thick solid line shows the best-fit, the light grey contour represents the 1$\sigma$ confidence level,
and the dark grey contour represents the 2$\sigma$ confidence level around the best-fit.
From \ASSS.}
{\epsfysize=1.8in \epsfbox{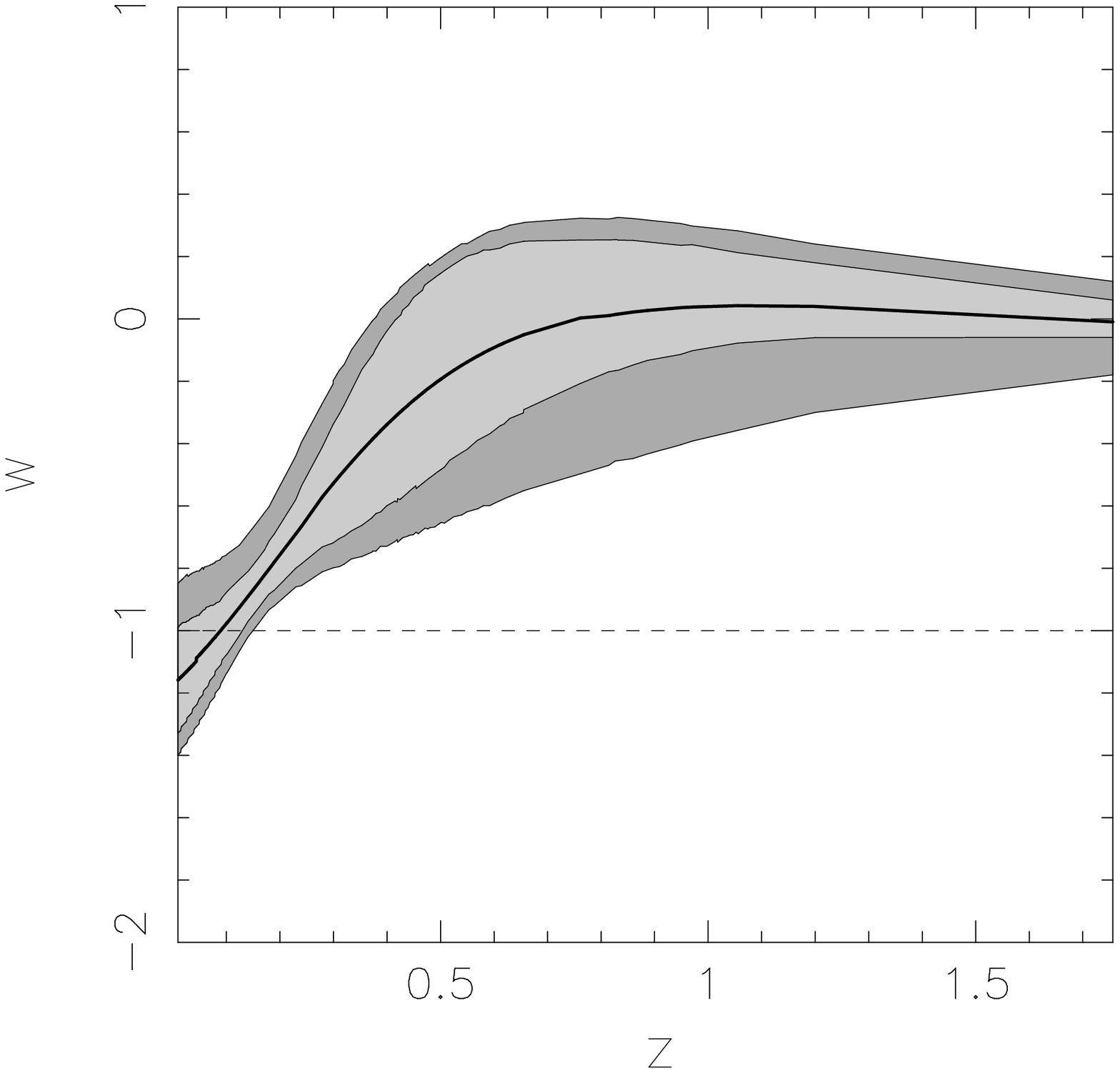} \epsfysize=1.8in \epsfbox{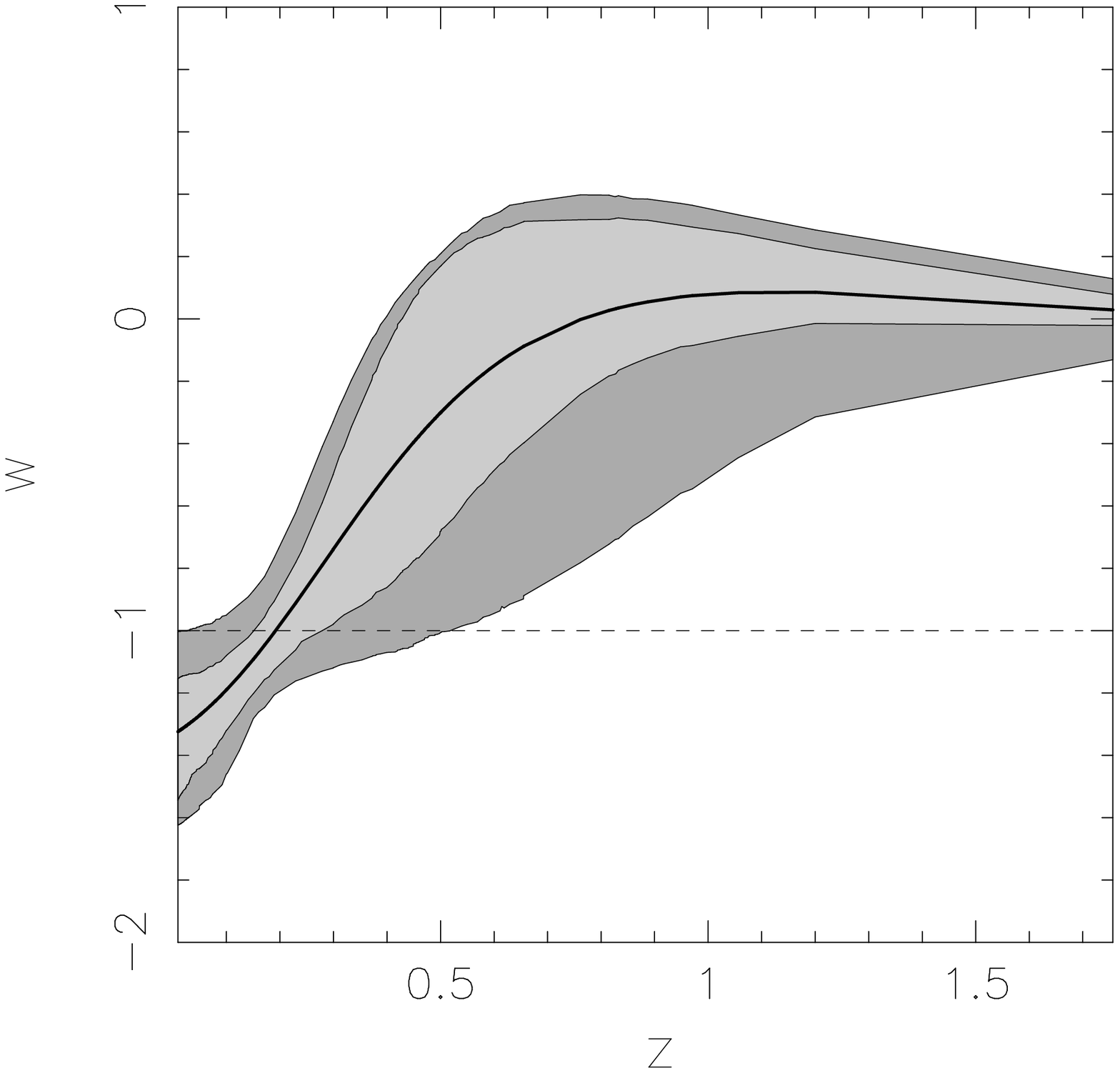} \epsfysize=1.8in \epsfbox{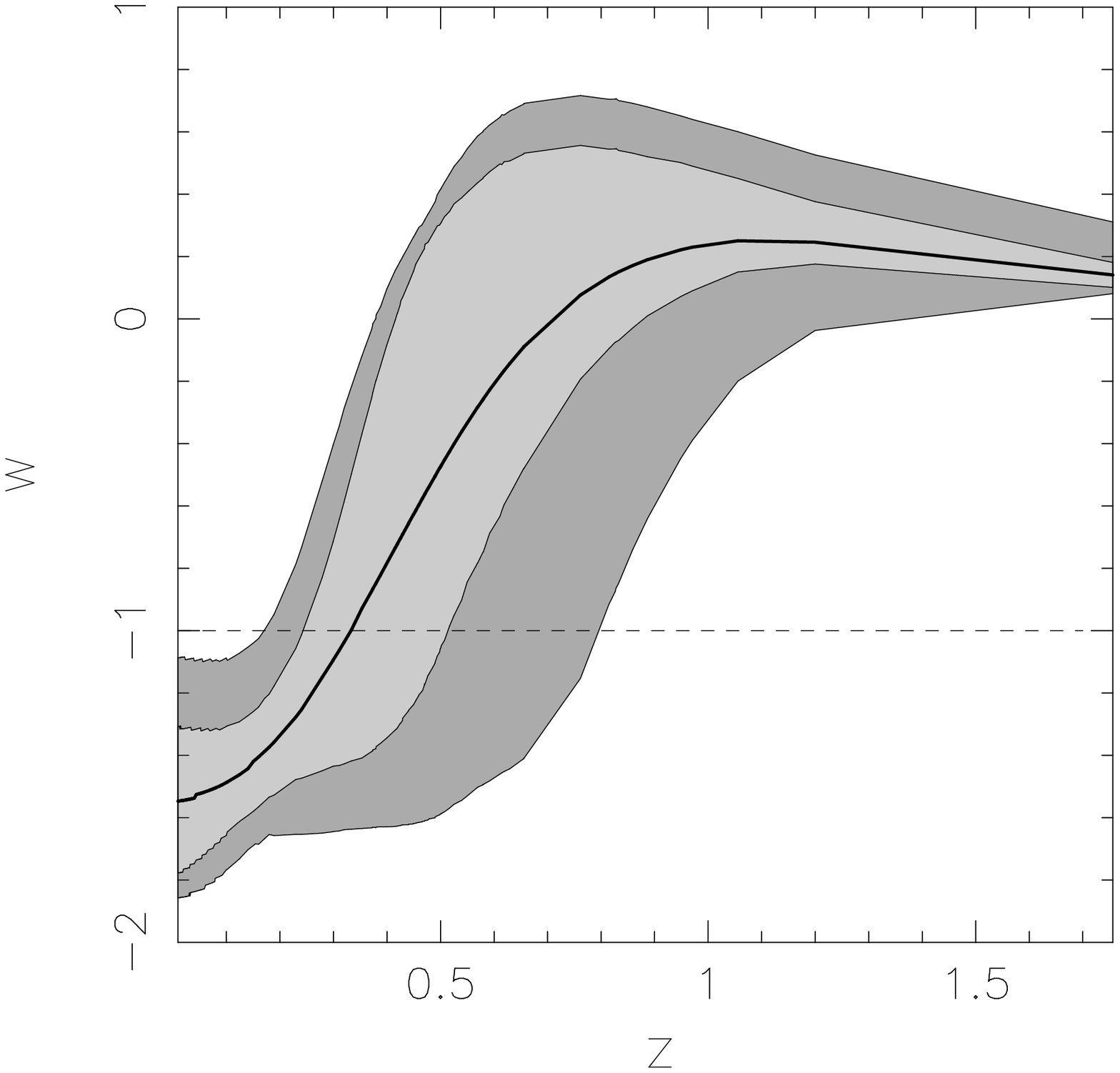}}

A simple polynomial fit to $f(z)$ was proposed by Alam {\it et al.}
\ASSSBefore, which is a truncated Taylor expansion
\eqn\Alam{f(z)=A_0 + A_1(1+z) + A_2(1+z)^2.}
This ansatz has only three free parameters ($\Omega_{m0},A_1,A_2$) since $A_0 + A_1 + A_2 = 1-\Omega_{m0}$ for a flat universe.
By using this ansatz, Alam {\it et al.} argued that the Tonry/Barris SNIa sample \refs{\Tonry,\Barris}
appear to favour dark energy which evolves in time \ASSS\ (see \FigASSS).
This conclusion will be modified if the effect of the CMB/LSS observations are taken into account \ASS.
It should be mentioned that there was a debate about the reliability of this ansatz \refs{\Jonsson,\ASSStwo}.

\ifig\FigYW{Dark energy density function $f(z)\equiv
\rho_X(z)/\rho_X(0)$ measured from combining SN Ia data with CMB,
BAO, GRB data, and imposing the SHOES prior on $H_0$. The 68\%
(shaded) and 95\% confidence level regions are shown. A flat
universe is assumed. The left panel is plotted by using the
Constitution set of 397 SNIa, while the right panel is plotted by
using the nearby+SDSS+ESSENCE+SNLS+HST data set of 288 SNIa. As seen
in this figure, using the nearby+SDSS+ESSENCE+SNLS +HST dataset gives
much more stringent constraints on dark energy than using the
Constitution dataset, and gives measurements that are closer to a
cosmological constant. Besides, flux-averaging has larger impact on
the results from using the Constitution set, and brings the
measurements closer to that predicted by a cosmological constant.
From \YWangtwo.} {\epsfysize=3.0in \epsfbox{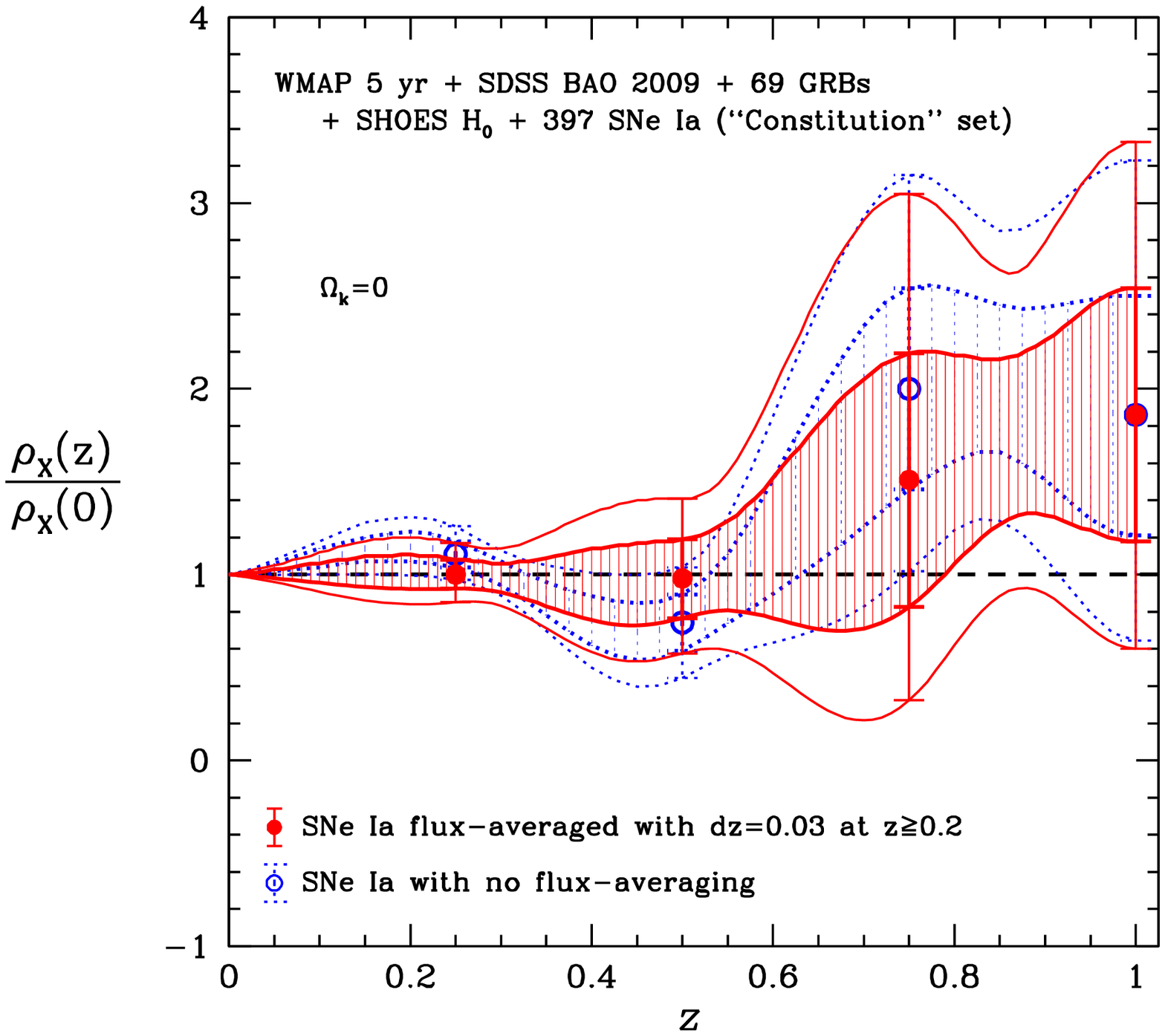}
\epsfysize=3.0in \epsfbox{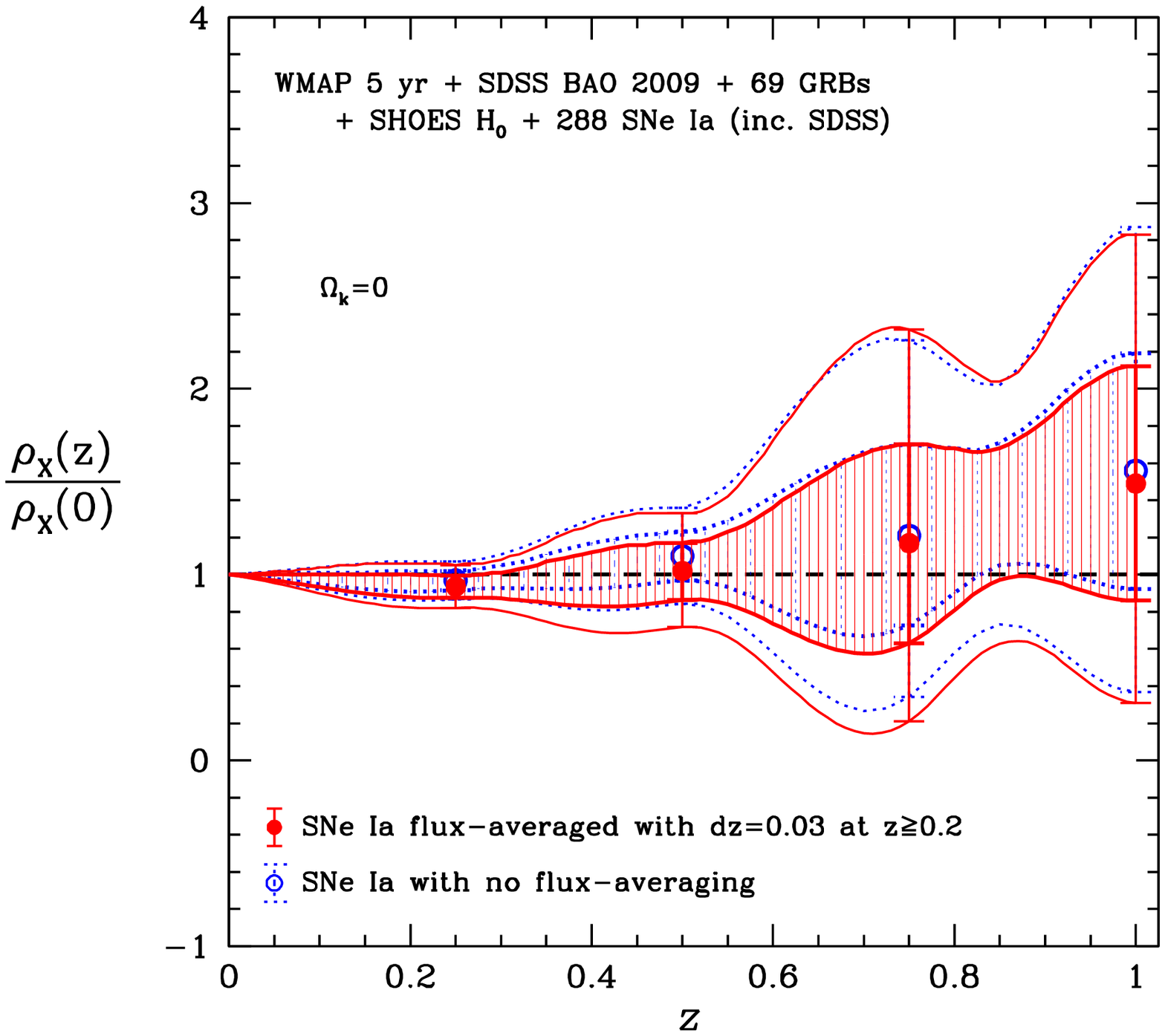}}

Another interesting polynomial fit is the polynomial interpolation,
which was proposed by Wang
\refs{\FOMWang,\WangMukherjee,\WangGarnavich,\WangTegmark,\WangFreese,\WangMukherjeetwo,\YWangtwo,\YWanglatest}.
It choose different redshift points $z_i = i\ast z_{\rm max}/n (i = 1,
2,\cdots, n)$, and interpolate f(z) by using its own values at these
redshift points. This yields \eqn\YUN{f(z)=\sum_{i=1}^{n}
f_i{(z-z_1)...(z-z_{i-1})(z-z_{i+1})...(z-z_n)\over(z_i-z_1)...(z_i-z_{i-1})(z_i-z_{i+1})...(z_i-z_n)}.}
Here $f_i=f(z_i)$ and $z_n=z_{\rm max}$. Based on the relation $f(0)=1$,
one parameter can be fixed directly, and only $n-1$ model parameters
need to be determined by the data. In \WangTegmark, Wang and Tegmark
made an accurate measurement of the dark energy density function $f(z)$ by
using the spectacular of the high redshift supernova observations
from the HST/GOODS program and previous supernova. In \WangFreese,
Wang and Freese demonstrated that $\rho_{de}(z)$ can be constrained
more tightly than $w(z)$ given the same observational data by using
the Tonry/Barris SNIa sample \refs{\Tonry,\Barris}. In \YWangtwo, by
utilizing the nearby+SDSS+ESSENCE+SNLS+HST set of 288 SNIa and the
Constitution set of 397 SNIa, Wang showed that flux-averaging of SNIa
can be used to test the presence of unknown systematic
uncertainties, and yield more robust distance measurements from SNIa
(see \FigYW\ for details). The latest Union2 set of 557 SNIa have also
been explored by using this polynomial interpolation method
\FreeZitwo.

\subsec{Gaussian process modeling}

\ifig\FigGP{Reconstruction of $w(z)$ based on GP modeling combined with MCMC.
The left panel uses a Gaussian covariance function $(\alpha \simeq 2)$, while the right panel uses an exponential covariance function $(\alpha = 1)$.
These two results are very similar, both are very close and in agreement with a cosmological constant (black dashed line).
The dark blue shaded region indicates the 68\% CL, while the light blue region extends it to 95 \% CL.
From \GPPRL.} {\epsfysize=2.0in \epsfbox{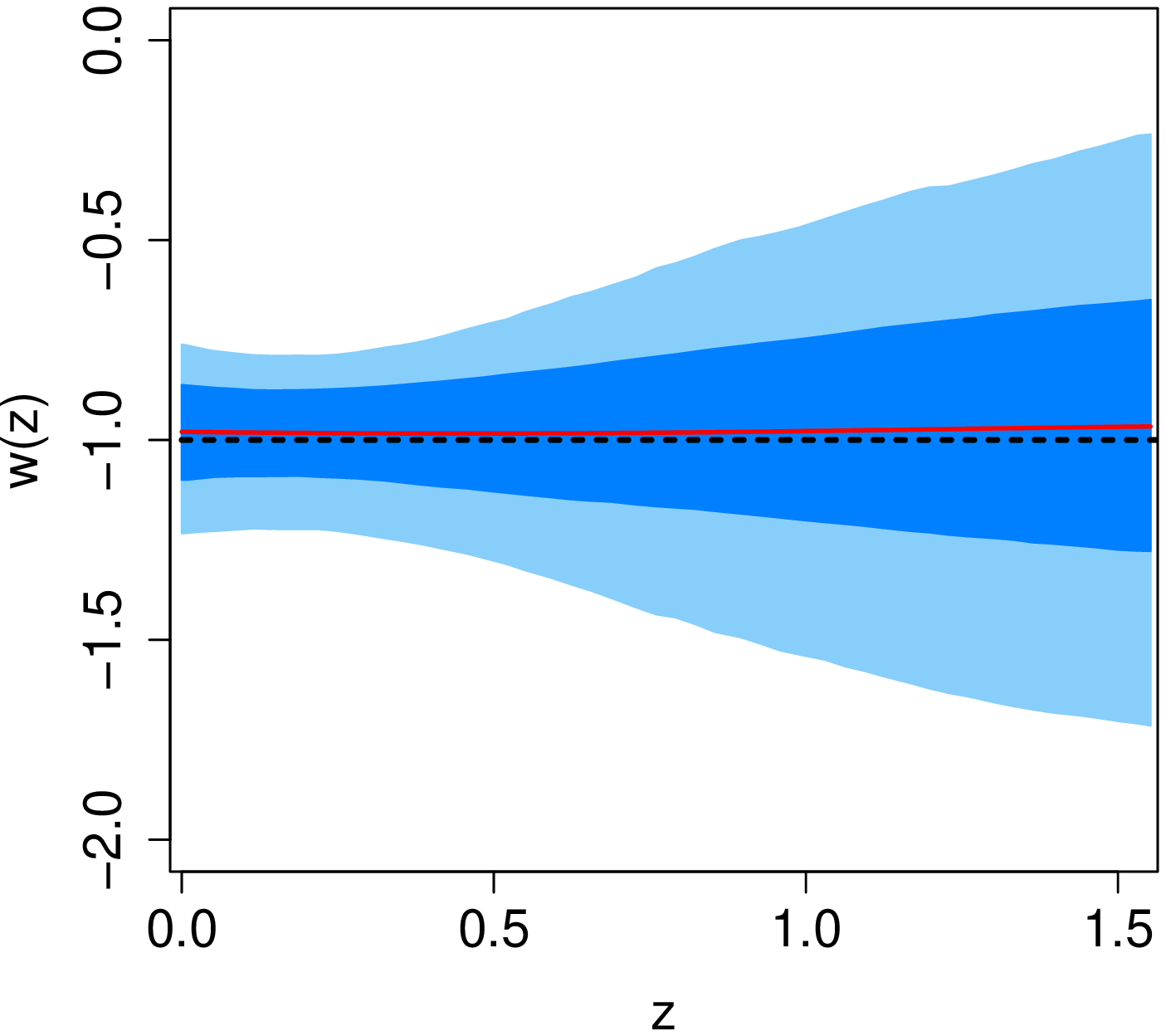} \epsfysize=2.0in \epsfbox{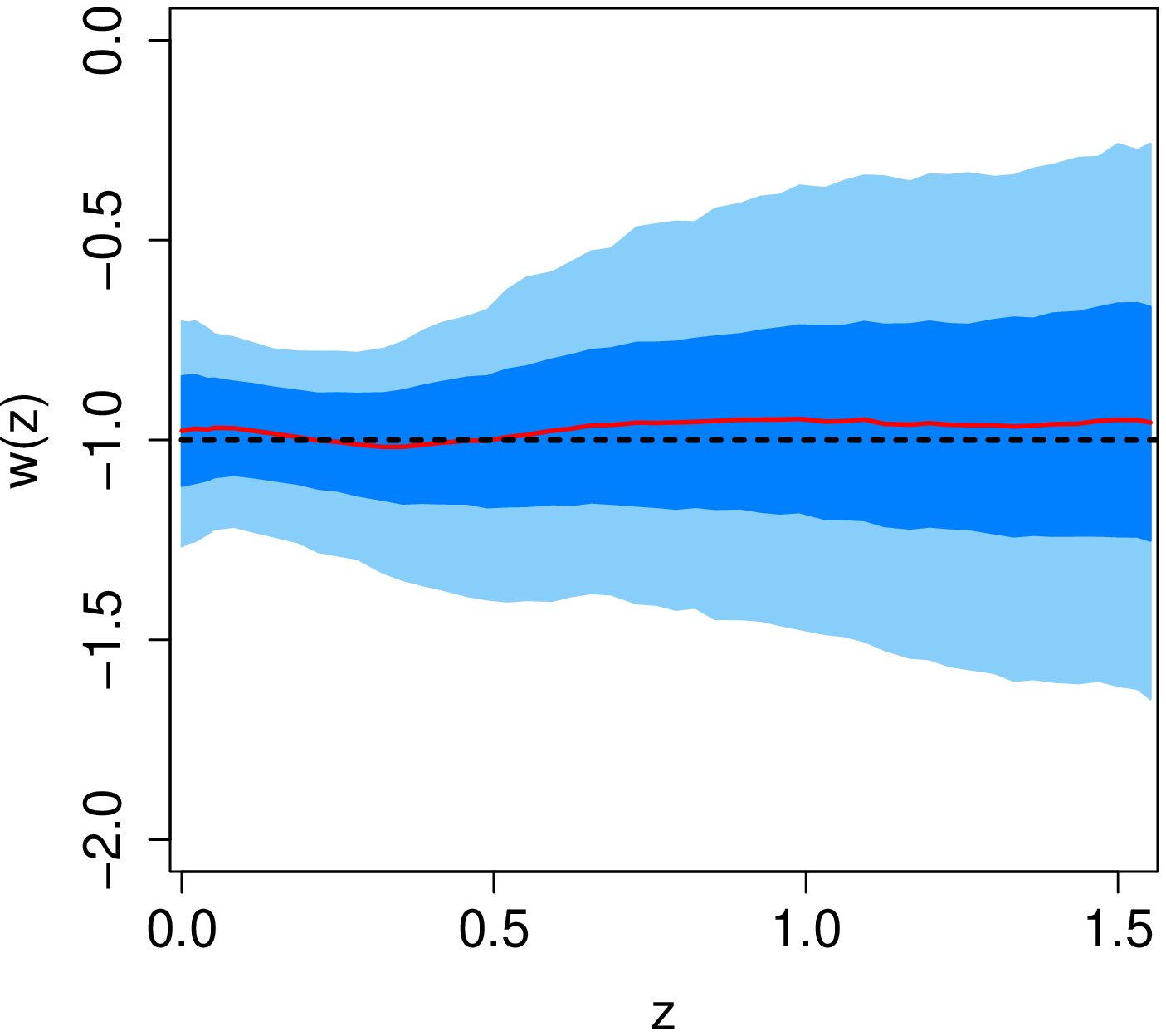}}

The fourth approach is the Gaussian Process (GP) modeling, which is proposed by Holsclaw {\it et al.} \refs{\GPPRL,\GPPRDOne}.
GP is a stochastic process, which is indexed by z.
The defining property of a GP
is that the vector that corresponds to the process at any finite collection of points follows a multivariate Gaussian distribution \GPBaRa.
GPs are elements of an infinite dimensional space, and can be used as the basis for a nonparametric reconstruction method.
They are characterized by a mean and a covariance function, defined by a small number of hyperparameters \refs{\GPBaRa,\GPPRDTwo}.

Based on the definition of a GP, one can assume that, for any collection $z_1$, ..., $z_n$,
$w(z_1)$, ..., $w(z_n)$ follow a multivariate Gaussian distribution with a constant negative mean and exponential covariance function written as
\eqn\GPone{K(z, z') = \kappa^2 \rho^{|z-z'|^{\alpha}}.}
The hyperparameters $\rho \in (0, 1)$ and $\kappa$, and the parameters defining the likelihood, are determined by the data.
The value of $\alpha \in (0, 2]$ influences the smoothness of the GP realizations:
for $\alpha=2$, the realizations are smooth with infinitely many derivatives,
while $\alpha=1$ leads to rougher realizations suited to modeling continuous non-(mean-squared)-differentiable functions.
Moreover, one can set up the following GP for $w$
\eqn\GPtwo{w(u) \sim GP(-1, K(u, u')).}
Making use of the Eq. \GPtwo,
one can take advantage of the particular integral structure of luminosity distance $d_L(z)$ expressed by $w(z)$ (see \refs{\GPPRL,\GPPRDOne} for details).

This new, nonparametric reconstruction method has the following
advantages: it avoids artificial biases due to restricted parametric
assumptions for $w(z)$, it does not lose information about the data
by smoothing it, and it does not introduce arbitrariness (and lack
of error control) in reconstruction by representing the data using a
certain number of bins, or cutting off information by using a
restricted set of basis functions to represent the data
\refs{\GPPRL,\GPPRDOne}. In \GPPRL, using this reconstruction method,
Holsclaw {\it et al.} reconstructed $w(z)$ utilizing the Constitution
dataset \Constitution. The obtained results are consistent with the
cosmological constant, with no evidence for a systematic mean
evolution in $w$ with redshift (see \FigGP).

\newsec{Concluding Remarks}

We have reviewed theoretical models as well as observational technologies and experimental projects and
numerical studies of dark energy. Numerous works and papers have been done and
written since the discovery of the accelerating expansion of the universe, it is
impossible to cover even a small part of the heroic endeavors of our community in any review
article.

However, the problem of understanding the nature of dark energy is as daunting as ever, or
perhaps some already hold the key to this understanding without being commonly accepted yet.
Clearly, there is a long long way to go for both theorists and experimentalists.

It is without any doubt that the process of detecting the nature of dark energy and understanding
its origin will prove to be one of the most exciting stories in modern science.

\bigskip

{\noindent \bf Acknowledgments}

We thank Robert Brandenberger, Yi-Fu Cai, Qing-Guo Huang, Richard Woodard, and Xin Zhang for useful discussions.
We also thank Robert Kirshner for helpful suggestions.
This research was supported by a NSFC grant
No.10535060/A050207, a NSFC grant No.10975172, a NSFC group grant
No.10821504 and Ministry of Science and Technology 973 program under
grant No.2007CB815401. YW acknowledges fellowships from FQI and IPP.


\listrefs
\end